\documentclass[reprint,aps,superscriptaddress,preprintnumbers,nofootinbib,prd]{revtex4-2}
\usepackage[english]{babel}
\usepackage{amsmath,amssymb,graphicx,bbm,xcolor}

\usepackage{mathrsfs}
\usepackage{booktabs}
\usepackage{mathtools}
\usepackage{dutchcal}
\usepackage[hidelinks]{hyperref}
\newcommand{\cdummy}{\cdot}
\newcommand{\mathd}{\mathrm{d}}
\newcommand{\bigintlim}{\int\limits}
\newcommand{\tmcolor}[2]{{\color{#1}{#2}}}

\newcommand{\tmmathbf}[1]{\ensuremath{\boldsymbol{#1}}}
\newcommand{\tmop}[1]{\ensuremath{\operatorname{#1}}}
\newcommand{\tmtextit}[1]{{\itshape #1}}

\makeatletter 
\renewcommand\onecolumngrid{
\do@columngrid{one}{\@ne}%
\def\set@footnotewidth{\onecolumngrid}
\def\footnoterule{\kern-6pt\hrule width 1.5in\kern6pt}%
}
\renewcommand\twocolumngrid{
        \def\footnoterule{
        \dimen@\skip\footins\divide\dimen@\thr@@
        \kern-\dimen@\hrule width.5in\kern\dimen@}
        \do@columngrid{mlt}{\tw@}
}%
\makeatother    

\makeatletter
\let\ORIbbl@fixname\bbl@fixname
\def\bbl@fixname#1{%
  \@ifundefined{languagealias@\expandafter\string#1}
    {\ORIbbl@fixname#1}
    {\edef\languagename{\@nameuse{languagealias@#1}}}%
}
\newcommand{\definelanguagealias}[2]{%
  \@namedef{languagealias@#1}{#2}%
}
\makeatother
\definelanguagealias{en}{english}

\begin{document}

\preprint{WUB/26-03}

\title{Entropy, area, and the choice of regulator during gravitational collapse}

\author{Jana N. Guenther}
\email[Email: ]{jguenther@uni-wuppertal.de}

\author{Christian Hoelbling}
\email[Email: ]{hch@uni-wuppertal.de}
\affiliation{Department of Physics, University of Wuppertal, Gau{\ss}strasse
20, D-42119 Wuppertal, Germany}

\author{Sophie Mutzel}
\email[Email: ]{sophie.mutzel@minesparis.psl.eu}
\affiliation{Laboratoire de Physique de l'{\'E}cole Normale Sup{\'e}rieure,
Mines Paris, Inria, CNRS, ENS-PSL, Centre Automatique et Syst{\`e}mes (CAS),
Sorbonne Universit{\'e}, PSL Research University, Paris, France}

\author{ Lukas Varnhorst}
\email[Email: ]{varnhorst@uni-wuppertal.de}
\affiliation{Department of Physics, University of Wuppertal, Gau{\ss}strasse
20, D-42119 Wuppertal, Germany}

\date{September 4, 2026}

\begin{abstract}
  We present a real time formalism to numerically describe the gravitational
  collapse of a scalar quantum field in the spherically symmetric case. We
  employ a Pauli-Villars regulator that is specifically designed to cancel the
  ultraviolet divergences in the energy-momentum tensor identified by covariant point splitting, including the logarithmic ones. Using this
  regulator, we find that the leading term of the entanglement entropy, which is proportional to the surface area, vanishes for a spherical region of flat spacetime. First numerical results for the dynamical case
  indicate that for a collapsing shell of a massless scalar field, the
  area normalized entropy is concentrated on the shell and has an
  approximately constant maximum value during time evolution. This maximum value appears to be finite in the
  continuum limit and only mildly dependent on the regulator mass. 
\end{abstract}

{\maketitle}

\section{Introduction}

As the seminal work of Bekenstein \cite{Bekenstein:1972tm,Bekenstein:1973ur}
and Hawking \cite{Hawking:1974sw,Hawking:1974rv} demonstrated, black holes
carry entropy and temperature. It was later found by Srednicki
\cite{Srednicki:1993im}, that the entanglement entropy of a spherical region is proportional to its surface area for a
free, massless scalar field even in flat spacetime, with a prefactor that diverges with the ultraviolet cutoff. In both cases, the entropy is generated when a pure
(vacuum) state is spatially truncated, either by a physical horizon or by
fiat. Since the Bekenstein entropy applies to the asymptotic final states of a
classical gravitational collapse, the question naturally arises, how it is
approached in the collapse itself, how it relates to the Srednicki entropy
of a scalar field and which part of the matter entanglement entropy survives once the energy-momentum tensor is properly renormalized.

In this paper, we present a formalism to numerically study the thermal
properties of a radially symmetric, free scalar field in a pure, gaussian
state undergoing gravitational collapse. As a regulator, we use either the
discretized radial coordinate itself or variants of the Pauli-Villars scheme
\cite{RevModPhys.21.434}, which was pioneered for semiclassical gravity in
\cite{Zeldovich1972,Vilenkin:1978wc} and applied to gravitational collapse
recently \cite{Berczi:2020nqy,Berczi:2021hdh,Berczi:2024yhb}. We derive the conditions on the regulator masses under which all ultraviolet divergences identified by covariant point-splitting, including the logarithmic ones, cancel in the coincidence limit, and argue that the resulting energy-momentum tensor verifies Wald's axioms. We perform an explicit construction of one such regulator, which we call the log-polynomial Pauli-Villars scheme.

We first apply our formalism to flat spacetime, where we reproduce Srednicki's
result and generically find an entropy that is proportional to the surface
area and the square of the regulator scale. In the specific case of the
log-polynomial Pauli-Villars regulator however, we find the constant of
proportionality to vanish so that the area term of the scalar field entropy upon removal
of a spherical region of space is exactly zero.

We then proceed to a first investigation of a dynamical system, where we
still neglect backreaction effects due to quantum fluctuations. We find
numerical evidence that the entropy deviates from its flat spacetime value in
a region around the inmoving shell of non-zero energy density only. Furthermore, for the log-polynomial
Pauli-Villars scheme the resulting entropy per unit area seems to have a finite continuum
limit and its dependence on the Pauli-Villars scale is not significant within
the limited scope of our numerical study.

\section{Dynamics of the system }\label{deriv}

We first derive the semiclassical Einstein equations in an effective radial form that we can then study numerically. Our starting point is an $N_c$ component, real scalar field $\overline{\phi}
(t, r, \theta, \varphi)$ and the metric
\cite{Christodoulou:1986zr,Christodoulou:1987vv,Goldwirth:1987nu,Choptuik:1992jv}
\[ \mathd \tau^2 = \alpha^2 (t, r) \mathd t^2 - a^2 (t, r) \mathd r^2 - r^2
   (\mathd \theta^2 + \sin^2 \theta \mathd \varphi^2) \]
that has a determinant $\sqrt{- g} = \alpha a r^2 \sin \theta$. The total
action of the system may thus be written as
\[ S = \bigintlim_{- \infty}^{\infty} \mathd t \, \bigintlim_0^{\infty} \mathd
   r \, \mathcal{L} \;,\]
with the effective radial Lagrangian density $\mathcal{L}=\mathcal{L}_g
+\mathcal{L}_k +\mathcal{L}_M$, where
\begin{align*}
  \mathcal{L}_k & = \frac{\alpha a r^2}{2} \int \mathd \Omega \,
  g^{\alpha \beta} \overline{\phi}^T_{, \alpha} \overline{\phi}_{, \beta}\\
  \mathcal{L}_M & = - \frac{M^2 \alpha a r^2}{2} \int \mathd \Omega \, 
  \overline{\phi}^T \overline{\phi}
\end{align*}
are the kinetic and mass terms of the scalar field and
\[ \mathcal{L}_g = - \frac{\alpha a r^2}{16 \pi} \int \mathd \Omega \, R = -
   \frac{\alpha a r^2}{4} R \]
is the gravitational part. Note that the spherical symmetry of the metric
implies a spherically symmetric curvature scalar $R$. However, since we will
promote $\phi$ to a quantum field, we will need to track its angular
excitations. We can therefore not assume that it is spherically symmetric at
this point.

\subsection{Angular momentum decomposition}

Let us write the scalar kinetic term more explicitly as
\[ \mathcal{L}_k = \frac{\alpha a r^2}{2} \int \mathd \Omega \, \left(
   \frac{1}{\alpha^2} \dot{\overline{\phi}}^T \dot{\overline{\phi}} -
   \frac{1}{a^2} {\overline{\phi}'}^T \overline{\phi}' - \frac{1}{r^2}
   \overline{\phi}^T L^2 \overline{\phi} \right)\;, \]
with the squared angular momentum operator
\[ L^2 = - \frac{1}{\sin \theta} \partial_{\theta} \sin \theta \partial_{\theta} -
   \frac{1}{\sin^{2} \theta} \partial_{\varphi}^2 \]
and overdots and primes indicating partial derivatives with respect to $t$ and
$r$, respectively. Decomposing our scalar field into angular momentum
components
\[ \overline{\phi} (t, r, \theta, \varphi) = \sum_{l = 0}^{\infty} \sum_{m = -
   l}^l \overline{\phi}_{l m} (t, r) Y_{l m} (\theta, \varphi) \]
with the spherical harmonics $Y_{l m} (\theta, \varphi)$ that fulfill the
orthogonality relations
\[ \bigintlim \mathd \Omega \, Y_{l m} (\theta, \varphi) Y^{\ast}_{l' m'}
   (\theta, \varphi) = \delta_{l l'} \delta_{m m'} \]
and
\[ \sum_{l = 0}^{\infty} \sum_{m = - l}^l Y_{l m} (\theta, \varphi)
   Y^{\ast}_{l m} (\theta', \varphi') = \frac{\delta (\varphi - \varphi')
   \delta (\theta - \theta')}{\sin \theta} \]
and are eigenfunctions of the squared angular momentum operator
\[ L^2 Y_{l m} (\theta, \varphi) = l (l + 1) Y_{l m} (\theta, \varphi)\;. \]
We can decompose the scalar field kinetic term into a sum over angular momentum
modes
\begin{equation*}
\begin{split}
\mathcal{L}_k = &\frac{\alpha a r^2}{2} \sum_{l = 0}^{\infty} \sum_{m = -
   l}^l \left( \frac{1}{\alpha^2} \dot{\overline{\phi}}_{l m}^{\dag}
   \dot{\overline{\phi}}_{l m}\right.\\
   & \left.- \frac{1}{a^2} \overline{\phi}^{\dag\prime}_{l m}
   \overline{\phi}'_{l m} - \frac{l (l + 1)}{r^2} \overline{\phi}_{l m}^{\dag}
   \overline{\phi}_{l m} \right)\;.
\end{split}
 \end{equation*}  
The mass term can also be trivially decomposed into angular momentum
contributions and reads
\[ \mathcal{L}_M = - \frac{M^2 \alpha a r^2}{2} \sum_{l = 0}^{\infty} \sum_{m
   = - l}^l \overline{\phi}_{l m}^{\dag} \overline{\phi}_{l m} \;. \]
The total effective radial Lagrangian density of our system therefore is
\begin{equation}
\begin{split}
  \mathcal{L}= &\frac{\alpha a r^2}{2} \sum_{l = 0}^{\infty} \sum_{m = - l}^l
  \left( \frac{1}{\alpha^2} \dot{\overline{\phi}}^{\dag}_{l m}
  \dot{\overline{\phi}}_{l m} - \frac{1}{a^2} \overline{\phi}^{\dag\prime}_{l m}
  \overline{\phi}'_{l m} \right.\\
  &\left.-\left( \frac{l (l + 1)}{r^2} + M^2 \right)
  \overline{\phi}_{l m}^{\dag} \overline{\phi}_{l m} \right) - \frac{\alpha a
  r^2}{4} R\;. \label{erld}
\end{split}
\end{equation}

\subsection{Classical field equations and Hamiltonian density}

Requiring the variation of the action $S$ with respect to the two parameters
$\alpha$ and $a$ to vanish results in the two classical field equations

\begin{displaymath}
\begin{split}
     G^t_t = \sum_{l, m} &\left( \frac{1}{\alpha^2}
     \dot{\overline{\phi}}_{l m}^{\dag} \dot{\overline{\phi}}_{l m} +
     \frac{1}{a^2} \overline{\phi}^{\dag\prime}_{l m} \overline{\phi}_{l m}'\right.\\
     &\left.+
     \left( \frac{l (l + 1)}{r^2} + M^2 \right) \overline{\phi}_{l m}^{\dag}
     \overline{\phi}_{l m} \right)
     \end{split}
   \end{displaymath}
and
\begin{displaymath}
\begin{split}
     G^r_r  = - \sum_{l, m} &\left( \frac{1}{\alpha^2}
     \dot{\overline{\phi}}_{l m}^{\dag} \dot{\overline{\phi}}_{l m} +
     \frac{1}{a^2} \overline{\phi}^{\dag\prime}_{l m} \overline{\phi}_{l m}' \right.\\
     &\left.-
     \left( \frac{l (l + 1)}{r^2} + M^2 \right) \overline{\phi}_{l m}^{\dag}
     \overline{\phi}_{l m} \right)
     \end{split}
   \end{displaymath}
with the relevant elements of the Einstein tensor given by
\begin{displaymath}
\begin{split}
     G^t_t &=  \frac{1}{r a^2} \left( \frac{a^2 - 1}{r} + 2 \frac{a'}{a}
     \right) \\
     G^r_r & =  \frac{1}{r a^2} \left( \frac{a^2 - 1}{r} - 2
     \frac{\alpha'}{\alpha} \right) \;.
     \end{split}
\end{displaymath}
Taking the difference and sum, respectively, of the two field equations and
replacing the $t$ derivative of the field with the canonically conjugate
momentum
\[ \overline{\Pi}_{l m} = \frac{\partial \mathcal{L}}{\partial
   \dot{\overline{\phi}}^{\dag}_{l m}} = \frac{a r^2}{\alpha}
   \dot{\overline{\phi}}_{l m} \]
we find
\begin{equation}
  \begin{split}
    \ln' (a \alpha) = & \sum_{l, m} \left( \frac{1}{r^3} \overline{\Pi} _{l
    m}^{\dag} \overline{\Pi}_{l m} + r \overline{\phi}^{\dag\prime}_{l m}
    \overline{\phi}_{l m}' \right)\\
    \left( \ln' \frac{a}{\alpha} + \frac{a^2 - 1}{r} \right) \frac{1}{r a^2}
    = & \sum_{l, m} \left( \frac{l (l + 1)}{r^2} + M^2 \right)
    \overline{\phi}_{l m}^{\dag} \overline{\phi}_{l m} \;.
  \end{split}  \label{feq1}
\end{equation}
From the effective radial Lagrangian density \eqref{erld} we also find the
Hamiltonian density
\[ \mathcal{H}_{\tmop{tot}} =\mathcal{H}+ \frac{\alpha a r^2}{4} R \]
where the scalar field Hamiltonian is given by
\begin{equation}
\begin{split}
  \mathcal{H}&= \frac{1}{2} \sum_{l, m}  \left( \frac{\alpha}{a r^2}
  \overline{\Pi}_{l m}^{\dag} \overline{\Pi}_{l m}
  \right.\\&\left.
  + \overline{\phi}_{l
  m}^{\dag} \left( \partial_r^T \frac{\alpha}{a} r^2 \partial_r + a \alpha r^2
  \left( \frac{l (l + 1)}{r^2} + M^2 \right) \right) \overline{\phi}_{l m}
  \right)\;.
  \end{split}
  \label{classhamdens}
\end{equation}
We may write this in the more compact form
\begin{equation}
  \mathcal{H}= \frac{1}{2} \overline{\psi}^{\dag} Q \left(\begin{array}{cc}
    K  & 0\\
    0 & \mathbbm{1}
  \end{array}\right) Q \overline{\psi} \label{hamori}
\end{equation}
by defining
\[ \overline{\psi} = \left(\begin{array}{c}
     \overline{\phi}\\
     \overline{\Pi}
   \end{array}\right) \;, \]
where the $\overline{\phi}$ and $\overline{\Pi}$ themselves are the direct sums of the component vectors
\[ \overline{\phi} = \bigoplus_{l = 0}^{\infty} \bigoplus_{m = - l}^l
   \overline{\phi}_{l m} \qquad \overline{\Pi} = \bigoplus_{l =
   0}^{\infty} \bigoplus_{m = - l}^l \overline{\Pi}_{l m} \]
and the metric factors are collected in the diagonal operator
\begin{equation}
  Q = Q^T = \mathbbm{1}\bigotimes\left(\begin{array}{cc}
    \sqrt{\frac{a}{\alpha}} r & 0\\
    0 & \sqrt{\frac{\alpha}{a}} \frac{1}{r}
  \end{array}\right) \label{rescop}
\end{equation}
that is identical in every block. The block diagonal operator
\[ K = \bigoplus_{l = 0}^{\infty} \bigoplus_{m = - l}^l K_l \qquad \]
is a direct sum of the kernel matrices
\begin{equation}
  K_l = q^T q + \alpha^2 \left( \frac{l (l + 1)}{r^2} + M^2 \right) \;,
  \label{kdef}
\end{equation}
where the operator $q$ is given by 
\begin{equation}
  q = \sqrt{\frac{\alpha}{a}} r \partial_r \sqrt{\frac{\alpha}{a}} \frac{1}{r} \;.
  \label{qop}
\end{equation}

\subsection{Williamson normal form}

According to the Williamson theorem
\cite{a6cda7d0-29df-361a-a64d-8dd6c762bce6}, a generic, even dimensional,
positive definite, symmetric matrix $M$ may be cast into a diagonal form
\[ \left(\begin{array}{cc}
     \mathrm{\omega} & 0\\
     0 & \mathrm{\omega}
   \end{array}\right) = B^T M B \]
by a symplectic transformation $B$, with $\mathrm{\omega}$ a diagonal,
positive definite matrix whose nonzero elements are known as the symplectic
eigenvalues of $M$. Generically, one may find the symplectic eigenvalues by a
standard diagonalization of the matrix $i \Omega M$, where
\[ \Omega = \left(\begin{array}{cc}
     0 & \mathbbm{1}\\
     -\mathbbm{1} & 0
   \end{array}\right) \]
as
\[ i \Omega M = U^{- 1} \left(\begin{array}{cc}
     \mathrm{\omega} & 0\\
     0 & -\mathrm{\omega}
   \end{array}\right) U \;.\]
For the special case of the Hamiltonian \eqref{hamori}, we may however find
the Williamson form
\begin{equation}
  \mathcal{H}= \frac{1}{2} \hat{\chi}^{\dag} \left(\begin{array}{cc}
    \mathrm{\omega} & 0\\
    0 & \mathrm{\omega}
  \end{array}\right) \hat{\chi} \label{hamwilcl}
\end{equation}
and the corresponding symplectic transformation $B$, where
\begin{equation*}
  \overline{\psi} = B \hat{\chi}
\end{equation*}
and
\begin{equation}
  \left(\begin{array}{cc}
    \mathrm{\omega} & 0\\
    0 & \mathrm{\omega}
  \end{array}\right) = B^T Q\left(\begin{array}{cc}
    K  & 0\\
    0 & \mathbbm{1}
  \end{array}\right)Q B \label{wiltrans}
\end{equation}
by diagonalizing the symmetric, positive semidefinite, block diagonal kernel operator
\[ K = V\mathrm{\omega}^2 V^T \qquad V^T = V^{- 1} \;. \]
This allows us to write the kernel in a diagonal form
\[ \left(\begin{array}{cc}
     \mathrm{\omega}^2 & 0\\
     0 & \mathbbm{1}
   \end{array}\right) = W^T \left(\begin{array}{cc}
     K & 0\\
     0 & \mathbbm{1}
   \end{array}\right) W \;,\]
with \[  W = \left(\begin{array}{cc}
     V & 0\\
     0 & V
   \end{array}\right) \]
which can be recast into the Williamson form
\[ \left(\begin{array}{cc}
     \omega & 0\\
     0 & \omega
   \end{array}\right) = S^{- 1} \left(\begin{array}{cc}
     \mathrm{\omega}^2 & 0\\
     0 & \mathbbm{1}
   \end{array}\right) S^{- 1} \]
by a simple diagonal rescaling operator
\[ S = S^T = \left(\begin{array}{cc}
     \sqrt{\mathrm{\omega}} & 0\\
     0 & \frac{1}{\sqrt{\mathrm{\omega}}}
   \end{array}\right) \;.\]
We thus find
\[ B = Q^{- 1} W S^{- 1} \]
which is symplectic, i.e.
\[ B \Omega B^T = \Omega \]
as are $Q$, $W$ and $S$ individually. Furthermore, $V$ inherits the block
diagonal structure of $K$, i.e.
\[ V = \bigoplus_{l = 0}^{\infty} \bigoplus_{m = - l}^l V_l \;. \]
\subsection{Quantization}

We work in the semiclassical approximation with scalar quantum fields and a
classical metric. Ignoring quantum fluctuations in the metric, we obtain the
semiclassical field equations from \eqref{feq1} by taking the expectation
values of the right hand sides in some yet to be specified scalar quantum
state. The Hamiltonian that governs the time evolution of the scalar field in
our system is quadratic, albeit with metric dependent coefficients. In a fixed
background metric, the time evolution of the scalar field is thus trivial and
we will perform it in the Schr{\"o}dinger picture, evolving the quantum state.
In particular, Gaussian states (see e.g.~\cite{RevModPhys.84.621}) will remain Gaussian during such a time
evolution. When changing the metric, the coefficients of the quadratic terms
in the Hamiltonian change. This will cause a change of the symplectic
transformation matrix $B$ in \eqref{wiltrans} that brings the Hamiltonian into
the Williamson normal form, but it does not affect the gaussianity of the
scalar field state. If we therefore start with the scalar field in a Gaussian
state, it will remain in a Gaussian state throughout the entire time
evolution. We will thus restrict our attention to scalar fields in a Gaussian
state, which are fully characterized by their averages and quadratic
fluctuations.

\subsubsection{Gaussian states}

Promoting fields and momenta to operators that fulfill the canonical
commutation relations
\[ [\hat{\tmmathbf{\chi}}, \hat{\tmmathbf{\chi}}^{\dag}] = i \Omega \]
the Hamiltonian in Williamson form \eqref{hamwilcl} becomes
\begin{equation}
  \underset{}{\tmmathbf{\mathcal{H}}} = \frac{1}{2}
  \hat{\tmmathbf{\chi}}^{\dag} \left(\begin{array}{cc}
    \mathrm{\omega} & 0\\
    0 & \mathrm{\omega}
  \end{array}\right) \hat{\tmmathbf{\chi}} \;.\label{hwifq}
\end{equation}
At the initial time $t_0$, where the Hamiltonian is given by
\[ \underset{}{\tmmathbf{\mathcal{H}}_0} = \frac{1}{2}
   {\hat{\tmmathbf{\chi}}^0}{}^{\dag} \left(\begin{array}{cc}
     \mathrm{\omega}^0 & 0\\
     0 & \mathrm{\omega}^0
   \end{array}\right) \hat{\tmmathbf{\chi}}^0 \]
we prepare a coherent state, which may be interpreted as the $\beta
\rightarrow \infty$ limit of a shifted thermal state with the density matrix
\[ \tmmathbf{\rho}= \frac{e^{- \beta
   \tmmathbf{\mathcal{H}}_0^{\hat{\chi}}}}{\tmop{Tr} (e^{- \beta
   \tmmathbf{\mathcal{H}}_0^{\hat{\chi}}})} \;,\]
where \[\underset{}{\tmmathbf{\mathcal{H}}_0^{\hat{\chi}}} = \frac{1}{2}
   (\hat{\tmmathbf{\chi}}^0 - \hat{\chi}^0)^{\dag} \left(\begin{array}{cc}
     \mathrm{\omega}^0 & 0\\
     0 & \mathrm{\omega}^0
   \end{array}\right) (\hat{\tmmathbf{\chi}}^0 - \hat{\chi}^0) \;. \]
Obviously, this is a Gaussian state. It is easy to see that the averages of
the field operators are given by
\[ \langle \hat{\tmmathbf{\chi}}^0 \rangle_{\tmmathbf{\rho}} = \tmop{Tr}
   (\hat{\tmmathbf{\chi}}^0 \tmmathbf{\rho}) = \hat{\chi}^0 \;, \]
while for the second moments we find
\[ \langle (\hat{\tmmathbf{\chi}}^0 - \hat{\chi}^0) (\hat{\tmmathbf{\chi}}^0 -
   \hat{\chi}^0)^{\dag} \rangle_{\tmmathbf{\rho}} = \frac{1}{2} \left(
   \left(\begin{array}{cc}
     \mathrm{\sigma}^0 & 0\\
     0 & \mathrm{\sigma}^0
   \end{array}\right) + i \Omega \right)\]
with
   \[\mathrm{\sigma}^0 =
   \frac{1 + e^{- \beta \omega^0}}{1 - e^{- \beta \omega^0}} = \coth \left(
   \frac{\beta \mathrm{\omega}^0}{2} \right) \]
or, in the $\beta \rightarrow \infty$ limit,
\[ \langle (\hat{\tmmathbf{\chi}}^0 - \hat{\chi}^0) (\hat{\tmmathbf{\chi}}^0 -
   \hat{\chi}^0)^{\dag} \rangle_{\tmmathbf{\rho}} = \frac{1}{2} (\mathbbm{1}+
   i \Omega) \;.\]
Performing an arbitrary symplectic basis transformation $T$
\[ \hat{\tmmathbf{\chi}} = T \hat{\tmmathbf{\chi}}^0 \]
we find that, in the new basis, the averages are given by
\[ 
\langle \hat{\tmmathbf{\chi}} \rangle_{\tmmathbf{\rho}} = T \hat{\chi}^0 \;, 
\]
while the second moments are
\[ \langle (\hat{\tmmathbf{\chi}} - \hat{\chi}) (\hat{\tmmathbf{\chi}} -
   \hat{\chi})^{\dag} \rangle_{\tmmathbf{\rho}} = \frac{1}{2} (\hat{C} + i
   \Omega) \]
with the symmetric covariance matrix
\[ \hat{C} = T \left(\begin{array}{cc}
     \mathrm{\sigma}^0 & 0\\
     0 & \mathrm{\sigma}^0
   \end{array}\right) T^T \;. \]
In the $\beta \rightarrow \infty$ limit, the covariance matrix simplifies to
\begin{equation}
  \hat{C} = T T^T \;. \label{covt}
\end{equation}
Note that the time evolution of a Gaussian state on a fixed background metric
is also described by a symplectic transformation. This can most easily be seen
by considering the Heissenberg equations for the field and momentum operators
in the diagonal basis of the Hamiltonian
\[ \dot{\hat{\tmmathbf{\chi}}} = - i [\hat{\tmmathbf{\chi}},
   \tmmathbf{\mathcal{H}}] = \left(\begin{array}{cc}
     0 & \omega\\
     - \omega & 0
   \end{array}\right) \hat{\tmmathbf{\chi}} \]
which can be integrated over a finite amount of time $\Delta t$ to give
\begin{equation*}
  \hat{\tmmathbf{\chi}} (t + \Delta t) = O \hat{\tmmathbf{\chi}} (t)
  \end{equation*}
with the time evolution operator
  \begin{equation}
  O = \left(\begin{array}{cc}
    \cos (\omega \Delta t) & \sin (\omega \Delta t)\\
    - \sin (\omega \Delta t) & \cos (\omega \Delta t)
  \end{array}\right) \label{evs}
\end{equation}
that is clearly symplectic.

\subsubsection{The semicalssical field equations}

Ignoring fluctuations in the metric, the semiclassical field equations are
obtained from \eqref{feq1} by promoting fields and momenta on the right hand
side to operators and then taking the expectation values over a Gaussian
quantum state. These expectation values may in turn be expressed by the
averages and covariances
\[ 
\begin{split}
\langle \overline{\tmmathbf{\psi}} \overline{\tmmathbf{\psi}}^T
   \rangle_{\tmmathbf{\rho}} &= B \langle \hat{\tmmathbf{\chi}}
   \hat{\tmmathbf{\chi}}^{\dag} \rangle_{\tmmathbf{\rho}} B^T \\
   &= \frac{1}{2} B
   (\hat{C} + i \Omega) B^T + B \hat{\chi} (B \hat{\chi})^{\dag} \\
   &= \frac{1}{2}
   (\overline{C} + i \Omega) + \overline{\psi} \overline{\psi}^{\dag} 
  \;, \end{split}
   \]
where we have introduced the short hand notation $\overline{C} = B \hat{C}
B^T$. As detailed in App.~\ref{apprs}, the radial symmetry of our problem
implies that all averages, except the $l = m = 0$ component, need to vanish.
In addition, the covariance matrix $\hat{C}$ has to be block diagonal in the
angular momentum $l$ and its components $m$, and all $2 l + 1$ blocks for the
same $l$ need to be identical. It has to be block diagonal in the component
index, too, with all blocks identical. The basis transformation
\eqref{wiltrans} and the time evolution \eqref{evs} preserve this structure,
since $Q$, $W$, $S$ and $O$ have precisely this block structure, too. This
allows us, in particular, to write the semiclassical form of the field
equations \eqref{feq1} for each $l$ mode and component separately as
\begin{equation*}
\begin{split}
    \ln' (a \alpha) = & \mathcal{h}_0\\
     \left( \ln' \frac{a}{\alpha} + \frac{a^2 - 1}{r} \right) \frac{r}{a^2} 
     = & \mathcal{m}
     \end{split}
\end{equation*}
with the densities
\begin{equation}
  \begin{split}
    \mathcal{h}_0 (r) = & \frac{1}{r^3} \left( | \overline{\Pi}_r |^2 + N_c
    \sum_{l = 0}^{\infty} \frac{2 l + 1}{2} \overline{C}_{l, r r}^{\Pi \Pi}
    \right) \\
    & + r \left( | \overline{\phi}'_r |^2 + N_c \sum_{l = 0}^{\infty}
    \frac{2 l + 1}{2} (\partial_r \overline{C}^{\phi \phi}_l \partial^T_r)_{r
    r} \right)\\
    \\
    \mathcal{m} (r) = & r^2 M^2 | \overline{\phi}_r |^2 \\
    &+ N_c \sum_{l =
    0}^{\infty} \frac{2 l + 1}{2} (l (l + 1) + r^2 M^2) \overline{C}^{\phi
    \phi}_{l, r r}
  \end{split} \label{h0m}
\end{equation}
that are linear combinations of the energy and radial pressure terms. Note
that both $\mathcal{h}_0$ and $\mathcal{m}$ do not explicitly depend on the
metric parameters. In addition, the number of field components has become explicit
in the covariance terms, while for the averages we have retained only a single
nonzero component. It is now convenient to introduce new metric parameters
\[ d = \frac{r}{a^2} \qquad \hat{\alpha} = \alpha a \]
that allow us to write the field equations in the decoupled form
\begin{equation}
  \begin{split}
    \ln' \hat{\alpha}  = & \mathcal{h}_0\\
    1 - d' = & d\mathcal{h}_0 +\mathcal{m}\;.
  \end{split} \label{feq}
\end{equation}
Note that in these new metric parameters the $l$ component of the kernel operator \eqref{kdef} reads
\[ K_l = q^T q + \frac{d \hat{\alpha}^2}{r} \left( \frac{l (l + 1)}{r^2} + M^2
   \right) \]
with
\[q = \sqrt{d \hat{\alpha} r} \partial_r \sqrt{\frac{d
   \hat{\alpha}}{r^3}} \]
and the diagonal rescaling operator \eqref{rescop} is given by
\begin{equation}
   Q = \mathbbm{1}\bigotimes\left(\begin{array}{cc}
    \sqrt{\frac{r^3}{d \hat{\alpha}}} &
    0\\
    0 & \sqrt\frac{d \hat{\alpha}}{r^3}
  \end{array}\right) \;.\label{scalh}
\end{equation}
One immediate problem of the densities \eqref{h0m} is that they have divergent
contributions and need to be regularized. This can be made explicit by e.g.~introducing a point splitting in the two point function. The resulting
correlator then has to be of Hadamard form \cite{DeWitt:1975ys}, which
contains divergent terms in the coincidence limit. As detailed in Sect.
\ref{reg} and App.~\ref{App:renormalizationSE}, we address these divergences
by utilising a specific Pauli-Villars regulator scheme. Additionally, all
dynamical data presented in this first, exploratory paper disregard the vacuum
contributions to the densities \eqref{h0m}, so that the time evolution of our
system is in fact classical. While this choice prohibits us from studying
backreaction effects, it does still allows us to track quantum observables
during the time evolution.

\subsection{Radial discretization and time evolution}

For a numerical treatment, we discretize the radial coordinate $r$. Here we
restrict ourselves to a discretization in $N_r$ equal steps of width $\Delta$,
so that $r_i = r_0 + i \Delta$. At the innermost coordinate, which we choose to be $r_0=0$, we impose
the boundary condition $d_0 = 0$.\footnote{Note that for any $r_0>0$ the boundary condition $d_0=0$ would correspont to a horizon at that coordinate.}
At the outermost coordinate $L=r_{N_r}$ we impose the boundary
condition $\hat{\alpha} (L) = 1$, which implies that outside the simulated volume we have  a Schwarzschild metric with Schwarzschild radius $L-d(r_N)$. We studied the implications of this finite volume cutoff for the $l = 0$ mode
in \cite{Guenther:2020kro}.

\subsubsection{Radial integration of the field equations}

Let us first consider the numerical integration of the field equations
\eqref{feq}. The first field equation is readily integrated to
\begin{equation}
  \hat{\alpha}_i = e^{- \sum_{j = i + 1}^{N_r} h^0_j} \label{intalhat}
\end{equation}
where we have defined
\begin{equation}
  h^0_i = \bigintlim_{r_{i - 1}}^{r_i} \mathcal{h}_0 \mathd r \;.\label{h0i}
\end{equation}
To integrate the second field equation, we have to make an assumption about
the distribution of the scalar field densities within the shells. Different assumptions will
differ in the discretization errors they introduce. We choose the scalar field densities to be
concentrated in thin shells just inside the discrete coordinates, i.e.
\begin{equation}
\begin{split}
  \mathcal{h}_0 (r) = &\sum_{i = 1}^{N_r} h^0_i \hat{\delta}_{\varepsilon} (r_i
  - r) \\
  m (r) = &\sum_{i = 1}^{N_r} m_i \hat{\delta}_{\varepsilon}
  (r_i - r) \label{din}
  \end{split}
\end{equation}
with
\[ \hat{\delta}_{\varepsilon} (x) = \left\{\begin{array}{lll}
     \frac{1}{\varepsilon} &  & 0 \leqslant x \leqslant \varepsilon\\
     0 &  & \tmop{else}
   \end{array}\right. \]
which excludes the appearance of coordinate singularities due to the inability of a horizon fully forming in finite time. With this choice,
we have $d' = 1$ in the range $r_{i - 1} \leqslant r \leqslant r_i -
\varepsilon$, so
\begin{equation}
  d (r_i - \varepsilon) = d_{i - 1} + \Delta  - \varepsilon \;.\label{deps}
\end{equation}
In the range $r_i - \varepsilon \leqslant r \leqslant r_i$, the general
solution for $h^0_i \neq 0$ is
\[ d (r) = \frac{\varepsilon - m_i}{h_i^0} + c \, e^{- \frac{h_i^0
   r}{\varepsilon}} \]
with $c$ a yet undetermined constant. Fixing the constant with \eqref{deps}, we
obtain
\begin{equation}
  d_i = d (r_i) = e^{- h_i^0} (d_{i - 1} + \Delta  - \varepsilon) +
  (\varepsilon - m_i) \frac{1 - e^{- h_i^0}}{h_i^0} \label{pree}
\end{equation}
and taking the $\varepsilon \rightarrow 0$ limit we arrive at the final result\footnote{The cardinal hyperbolic sine function is defined as
\[ \tmop{sinhc} (x) = \left\{\begin{array}{lll}
     \frac{\sinh (x)}{x} &  & x \neq 0\\
     1 &  & x = 0
   \end{array}\right. \]
and is well approximated, for small $x$, by the Horner polynomial
\[ 1 + \frac{x^2}{2 \cdummy 3} \left( 1 + \frac{x^2}{4 \cdummy 5} \left( 1 +
   \frac{x^2}{6 \cdummy 7} \left( 1 + \frac{x^2}{8 \cdummy 9} (1 + \cdots)
   \right) \right) \right) \]}
\begin{equation}
  d_i = e^{- h_i^0} (d_{i - 1} + \Delta ) - m_i e^{- \frac{h^0_i}{2}}
  \tmop{sinhc} \left( \frac{h^0_i}{2} \right) \label{intd}
\end{equation}
which is also valid for the $h^0_i = 0$ case. The relations \eqref{intalhat}
and \eqref{intd}, together with the boundary condition $d_0 = 0$, allow us to
radially integrate the metric for given $h^0$ and $m$.

In order to quantify discretization errors, let us also consider the case of
the densities being piecewise constant functions, which can be realized by
setting $\varepsilon = \Delta$ in \eqref{pree}. This choice would result in
\[ d_i = e^{- h_i^0} \left( d_{i - 1} + \Delta \frac{e^{h_i^0} - 1}{h_i^0}
   \right) - m_i e^{- \frac{h^0_i}{2}} \tmop{sinhc} \left( \frac{h^0_i}{2}
   \right) \]
so that the relative difference to \eqref{intd} is no worse than of order
$h_i^0$. We thus need to ensure  that the inetgrated density $h_i^0$ fulfills the necessary condition $h_i^0 \ll 1$for keeping discretization errors from the radial integration of the metric under
control.

\subsubsection{Time evolution of the scalar field}

\begin{figure}[htb]
  {\includegraphics[width=\columnwidth]{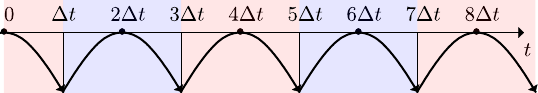}}
  \caption{\label{updit}Sketch of the time evolution of the system. The dots
  at even multiples of $\Delta t$ represent points where the metric and scalar
  field are compatible, while the vertical lines at odd multiples of $\Delta
  t$ represent the points where the metric parameters used for the scalar
  field time evolution are changed. Downward slopes indicate explicit scalar
  field integration, whereas upward slopes indicate implicit ones.}
\end{figure}

Ideally, we would like to time evolve the scalar field and the metric
parameters simultaneously. This would ensure that at every point in time the
metric parameters that govern the scalar field evolution are identical to
those that are obtained from integrating the scalar field densities
\eqref{h0m}. We call this property compatibility between metric and scalar
field. For a numerical treatment, however, we have to discretize coordinate\footnote{We are discretizing global simulation time which, by means of $\alpha_{N_r}=1$, is fixed to the coordinate time at the outer edge of the simulation volume. }
time in steps $\Delta t$. Our general scheme is to use the exact time
evolution \eqref{evs} of the scalar field on a fixed background metric and
alternate it, in a leapfrog-like scheme, with the integration \eqref{intalhat},
\eqref{intd} of the metric parameters. We initialize metric and scalar field in
a compatible way, but throughout the time evolution, this compatibility can
only be enforced at certain points in time. We choose to enforce this
compatibility every second time step $2 \Delta t$ by first performing an
explicit scalar field integration by a time step $\Delta t$ with the old
metric, followed by an implicit scalar field update by $\Delta t$, where the
metric at the end point is used and iterated. This scheme, which is
symbolically depicted in Fig.~\ref{updit}, is time reversal invariant.

Now let the system at time $t$ be represented by the metric parameters
$\alpha_t$ and $a_t$ (or, equivalently $\hat{\alpha}_t$ and $d_t$) as well as
the scalar field averages $\hat{\chi}_t^{\tau}$ and covariances
$\hat{C}^{\tau}_t$. Note that the scalar field averages and covariances here
carry an additional superscript index $\tau$. It indicates the Williamson
basis \eqref{wiltrans} of the Hamiltonian, corresponding to metric parameters
$\alpha_{\tau}$ and $a_{\tau}$, according to which the scalar field averages
and covariances are given. The relation to real space averages and covariances
is then given by
\[ \overline{\psi}_t = B^{\tau}  \hat{\chi}^{\tau}_t \qquad \overline{C}
   = B^{\tau} \hat{C}^{\tau}_t (B^{\tau})^T \;. \]
Suppose now that we start out with a compatible set of metric parameters
$\hat{\alpha}_t$, $d_t$ and scalar field averages $\hat{\chi}_t^t$ and
covariances $\hat{C}^t_t$. Carrying out our integration scheme, we first
evolve the scalar field by an explicit time step $\Delta t$, so that
\[ \hat{\chi}_{t + \Delta t}^t = O^t \hat{\chi}_t^t \qquad \hat{C}^t_{t
   + \Delta t} = O^t \hat{C}^t_t (O^t)^T \]
where $O^t$ is the time evolution operator \eqref{evs} in the Williamson basis
of the Hamiltonian at time $t$. Next comes the implicit integration step,
which requires us to first go into the Williamson basis of the (yet unknown)
Hamiltonian at time $t + 2 \Delta t$ by a basis transformation
\[ \begin{split}
    \hat{\chi}_{t + \Delta t}^{t + 2 \Delta t} = &(B^{t + 2 \Delta t})^{- 1} B^t
   \hat{\chi}_{t + \Delta t}^t \\
   \hat{C}_{t + \Delta t}^{t + 2
   \Delta t} = &(B^{t + 2 \Delta t})^{- 1} B^t \hat{C}^t_{t + \Delta t} ((B^{t
   + 2 \Delta t})^{- 1} B^t)^T 
   \end{split}\]
followed by the integration step itself
\[ \begin{split}\hat{\chi}_{t + 2 \Delta t}^{t + 2 \Delta t} = &O^{t + 2 \Delta t}
   \hat{\chi}_{t + \Delta t}^{t + 2 \Delta t} \\
   \hat{C}_{t + 2
   \Delta t}^{t + 2 \Delta t} = &O^{t + 2 \Delta t} \hat{C}_{t + \Delta t}^{t +
   2 \Delta t} (O^{t + 2 \Delta t})^T 
   \end{split}\]
which has to be iterated with the metric integration at time $t + 2 \Delta t$.
All together, the symplectic transfer operator for a scalar field update by a
time step $2 \Delta t$ is thus given by
\begin{equation}
  T_{t + 2 \Delta t, t} = O^{t + 2 \Delta t} (B^{t + 2 \Delta t})^{- 1} B^t
  O^t \label{tmatp}
\end{equation}
and the total time evolution operator is
\begin{equation}
  T_t = T_{t, t - 2 \Delta t} T_{t - 2 \Delta t, t - 4 \Delta t} \ldots T_{t_0
  + 2 \Delta t, t_0} \;.\label{tevp}
\end{equation}

\subsubsection{Initial conditions}

We choose to start with a coherent state at time $t_0$, i.e.~$C^{t_0}_{t_0}
=\mathbbm{1}$. Therefore, the covariances are simply
\begin{equation}
  \hat{C}^t_t = T_t T_t^T \qquad \overline{C}_t = B^t T_t T_t^T (B^t)^T\;,
  \label{cortm}
\end{equation}
while for the averages
\begin{equation}
  \hat{\chi}_t^t = T_t \hat{\chi}_{t_0}^{t_0} \qquad \overline{\psi}_t =
  B^t T_t \hat{\chi}_{t_0}^{t_0} \label{avtm}
\end{equation}
we must still choose an initial state $\hat{\chi}_{t_0}^{t_0}$. In principle,
we may construct it by choosing arbitrary real space averages
$\overline{\psi}$. From these, we can compute the densities \eqref{h0m}, which
may in turn be integrated via \eqref{intalhat} and \eqref{intd} to yield the
initial metric parameters $\hat{\alpha}$ and $d$ and thus also the initial
Williamson basis. We would, however, prefer to set up the initial
configuration for the classical part of the scalar field in such a way, that
it has nonzero energy densities in a finite interval only, which we call the
window. In addition, we would like the energy densities to be initially
inmoving as much as possible. In order to achieve this, we start with the
classical Hamiltonian \eqref{classhamdens}. If we identify the averages
$\overline{\phi}$ and $\overline{\Pi}$ with the classical fields, we may
express the classical Hamiltonian density as a linear combination
\[ \mathcal{H}= \frac{\hat{\alpha}}{2} (d h^{0, c} + m^c) \]
of the classical part of the energy densities \eqref{h0m}
\[ h^{0, c} = \frac{1}{r^3} | \overline{\Pi} |^2 + r | \overline{\phi}' |^2
   \qquad m^c = r^2 M^2 | \overline{\phi} |^2 \;. \]
On a fixed background metric, both the classical field and the averages obey
the canonical equations of motion
\[ \dot{\overline{\phi}} = \frac{\hat{\alpha} d}{r^3} \overline{\Pi} \qquad
   \dot{\overline{\Pi}} = \hat{\alpha} r (d \overline{\phi}'' (r) - r M^2 
   \overline{\phi}) \;.\]
One can easily show that the quantity
\[ \tilde{h} = \frac{1}{2} \left( \frac{1}{r^4} | \overline{\Pi} |^2 + |
   \overline{\phi}' |^2 + \frac{r}{d} M^2  | \overline{\phi} |^2
   \right) = \frac{1}{\hat{\alpha} d r} \mathcal{H} \;,\]
which is proportional to the Hamiltonian density, obeys the continuity
equation
\[ \tilde{h}_{, t} - \tilde{p}_{, r} = 0 \]
on a fixed background metric, where
\[ \tilde{p} = \frac{\hat{\alpha} d}{r^3} | \overline{\Pi} | |
   \overline{\phi}' | \;. \]
Due to the continuity equation, $\tilde{p}$ can be interpreted as an effective
radial momentum. We thus construct the initial state such that $\tilde{p}$ is
nowhere positive. We do this by choosing a smooth window function $f (r)$,
which vanishes outside the window and has a single maximum. In the massless
case, we may then set
\begin{equation}
    \overline{\phi}' (r) = f (r) \qquad \overline{\Pi} (r) = - r^2 f (r) \label{m0init}
\end{equation}
and integrate $\overline{\phi}'$ numerically from e.g.~an initial
$\overline{\phi} (0) = 0$ outwards. This particular choice has the advantage
that $\tilde{p} = - \tilde{h}$ in the small field limit, i.e.~for $d \rightarrow r$ and
$\hat{\alpha} \rightarrow  1$. In the massive case, such a procedure still produces a
vanishing $h^{0, c}$ outside the window, but there will be a nonvanishing
$m^c$ when the field average $\overline{\phi}$ itself does not vanish. We may however initialize the field average as
\[ \overline{\phi} (r) = f (r) \]
and then set
\[ \overline{\Pi} (r) = - r^2\overline{\phi}' (r) = - r^2 f' (r) \]
so that $\tilde{p}$ can only take negative nonzero values. In this case, since both $\overline{\Pi}$ and
$\overline{\phi}'$ are given by the derivative of $f (r)$, the density $h^{0,
c}$ will have a double peak structure. Since in this study we only investigate the massless case, we avoid a double peak by using \eqref{m0init} to initialize our field averages.

\subsection{Thermal observables}

Horizon formation in our system is characterized by $d \rightarrow 0$, which
implies $\alpha / a \rightarrow 0$, so that the derivative terms in the scalar
field Hamiltonian \eqref{classhamdens} become negligible. Since this term
provides the only coupling between different radial layers, it is thus evident
that the dynamics of the system decouples the outside region upon horizon
formation. Apart from this natural decoupling, we may however artificially
divide our system into an inner and an outer part at any time and at any given radius. In the spirit of
\cite{Srednicki:1993im} we may then investigate the thermodynamic properties
of the artificially decoupled subsystems. Let us, for this purpose, introduce
orthogonal projection operators $P^i_{\pm}$. Let $P^i_+$ project onto all
radial shells $r_j > r_i$ and $P^i_-$ to all radial shells $r_j \leqslant
r_i$. Per definition
\[ (P^i_{\pm})^2 = P^i_{\pm} \qquad P^i_+ P^i_- = 0 \qquad P^i_+ + P^i_-
   =\mathbbm{1} \]
and we may define the truncated covariance matrices
\[ \overline{C}^i_{\pm} = P^i_{\pm} \overline{C} P^i_{\pm} \]
as well as the truncation of the scalar field Hamiltonian kernel \eqref{hamori}
\[ h^i_{\pm} = P^i_{\pm} h P^i_{\pm} \;. \]
As detailed in App.~\ref{thermorev}, we may now compute various thermal
observables of the truncated states. The most straightforward is the state
purity, which according to \eqref{pos} is given by
\[ \gamma^i_{\pm} = \frac{1}{\sqrt{\det \overline{C}^i_{\pm}}} \]
and the related linear entropy \eqref{slin}. From a symplectic diagonalization
of $\overline{C}^i_{\pm}$, we may obtain the corresponding symplectic
eigenvalues $\sigma_{\pm j}^i$. From these, the von Neumann entropy can be
calculated according to \eqref{vne} as
\begin{equation}
\begin{split}
  S_{\pm}^i = \sum_j  &\left( \left( \frac{\sigma^i_{\pm j}}{2} + \frac{1}{2}
  \right) \ln \left( \frac{\sigma^i_{\pm j}}{2} + \frac{1}{2} \right)\right.\\
  &\left.- \left(
  \frac{\sigma^i_{\pm j}}{2} - \frac{1}{2} \right) \ln \left(
  \frac{\sigma^i_{\pm j}}{2} - \frac{1}{2} \right) \right) \;. \label{entdef}
\end{split}
\end{equation}
Although in this paper we will focus entirely on the entropy, there are
additional thermal observables that are potentially interesting. One may e.g.
compute the pseudo occupation number \eqref{aon} associated with every
eigenmode of the truncated states
\[ \overline{n}_{\pm j}^i = \frac{\sigma^i_{\pm j} - 1}{2} \;.\]
Note, that these do not in general correspond to eigenmodes of the truncated
Hamiltonian, but rather to a fictitious Hamiltonian with respect to which the
truncated states are shifted thermal states. In the special case where the
truncated system is thermalized, the eigenmodes of the truncated covariance
matrix and the truncated Hamiltonian will agree. In order to get an estimate
of how far a truncated state is from being thermal, we can bring the truncated
Hamiltonian into Williamson form via the symplectic transformation
\[ h^i_{\pm} = D_{\pm}^i \left(\begin{array}{cc}
     \omega^i_{\pm} & 0\\
     0 & \omega^i_{\pm}
   \end{array}\right) (D^i_{\pm})^T \;.\]
Applying this same symplectic transformation to the truncated covariance
matrix we obtain
\begin{equation}
  \tilde{C}^i_{\pm} = (D^i_{\pm})^{- 1} \overline{C}^i_{\pm} 
  ((D^i_{\pm})^T)^{- 1} \label{csytr}
\end{equation}
which is nondiagonal in general. Defining the projector
\[ P_j = \left(\begin{array}{cc}
     e_j & 0\\
     0 & e_j
   \end{array}\right) \qquad (e_j)_{a b} = \delta_{a j} \delta_{b j} \]
which projects onto the pair of canonically conjugate variables of the
diagonalized Hamiltonian with eigenfrequency $\omega^i_{\pm j}$, we obtain the
truncated $2 \times 2$ covariance matrix
\[ \tilde{C}^i_{\pm j} = P_j \tilde{C}^i_{\pm} P_j \]
whose symplectic eigenvalue we denote by $\tilde{\sigma}^i_{\pm j}$. In the
case of $\overline{C}^i_{\pm}$ representing a shifted thermal state, the
symplectic transformation \eqref{csytr} will diagonalize the covariance matrix
and hence $\tilde{\sigma}^i_{\pm j} = \sigma^i_{\pm j}$, provided that we have
consistently ordered the corresponding modes.

\subsection{Regularization and Renormalization}\label{reg}

Both the entanglement entropy \eqref{entdef} and the vacuum expectation value
of the energy-momentum tensor \eqref{h0m} contain UV divergences, even in flat
spacetime. In particular, the corresponding two-point functions become
singular in the coincidence limit as the radial lattice spacing is taken to zero. It is therefore imperative to
introduce a regulator. In the very simplest case, we may regard the finite
spacing in the radial coordinate $\Delta$ itself as the regulator: we call
this the \tmtextit{brick wall} scheme, borrowing the term from 't Hooft \cite{tHooft:1984kcu}. One has to keep in mind though, that a
constant radial coordinate difference $\Delta$ does not translate into a
constant physical separation of the consecutive shells. Indeed, with a
coordinate spacing of $\Delta$, the physical length scale of the radial cutoff
is given by
\[ a \Delta = \sqrt{\frac{r}{d}} \Delta \;.\]
In addition to the UV divergencies, the brick wall scheme therefore has the additional  undesirable feature that for constant $\Delta$ it provides a radially dependent cutoff scale. 

Several standard prescriptions have been developed for the renormalization of
the energy-momentum tensor in curved spacetime. For instance, Hadamard
point-splitting allows one to subtract the universal short-distance singularities
in a fully covariant way, and explicit expressions are available which can be
applied to the specific spacetime studied (see e.g.~\cite{Decanini:2005eg}).
However, these turn out to be impractical for numerical applications, since
the geometric terms appearing in the subtraction include higher derivatives of
the spacetime metric. This is not only numerically costly but will likely
introduce large numerical errors and instabilities due to discretization
effects. A similar difficulty arises in a Lagrangian formulation, where the UV
divergences are absorbed by introducing higher order curvature counterterms,
which again require higher derivatives of the metric. At present, there
appears to be no generally efficient and stable renormalization scheme
suitable for numerical implementations of semiclassical gravity (see also the
recent discussion in \cite{delrio_backreaction_2025}).

From a numerical point of view, a desirable renormalization prescription
would keep the covariant structure without requiring higher order derivatives of the metric. In principle, a
Pauli-Villars regulator could offer these advantages, provided that the masses
and multiplicities of the Pauli-Villars fields are tuned to cancel the divergent contributions to the
two-point function. The idea of Pauli-Villars regularization for the
energy-momentum tensor was originally introduced in
\cite{Zeldovich1972,Vilenkin:1978wc} and has recently been applied in the
context of numerical simulations of semiclassical gravitational collapse
\cite{Berczi:2020nqy,Berczi:2021hdh}. Here we propose a new scheme with
additional regulator fields, such that all divergences in the energy-momentum
tensor are canceled.

More precisely, consider a system consisting of $n$ real scalar and
Pauli-Villars regulator fields with different masses $M_i$ and integer
multiplicities $p_i \in \mathbb{Z}$. We take the physical field to be the
first field, so that $p_1 = 1$ and, for the massless case, $M_1 = 0.$ A
positive multiplicity $p_i > 0$ denotes $p_i$ real scalar fields of mass
$M_i$, while a negative multiplicity $p_i < 0$ implies $- p_i$ regulator
fields with the corresponding mass. In order to be able to cancel UV
divergences at least in principle, one needs to have the same number of scalar
and regulator fields, i.e., we need to impose the condition
\begin{equation}
  \sum_{i = 1}^n p_i = 0 \;.\label{p0}
\end{equation}
It has been suggested in
\cite{BERNARD1977201,Vilenkin:1978wc,Berczi:2021hdh} to also impose the
conditions
\begin{equation}
  \sum_{i = 1}^n p_i M_i^2 = \sum_{i = 1}^n p_i M_i^4 = 0 \label{pm20}
\end{equation}
which cancel the terms proportional to $M^2$ and $M^4$ contributions that
naturally appear in the Hadamard form of the propagator as prefactors of UV
divergent terms \cite{BERNARD1977201}. These conditions provide a
Pauli-Villars scheme in which the power-law UV divergences are canceled. One
particularly straightforward, nontrivial combination of multiplicities and
masses that fulfills these constraints is given in Tab.~\ref{poltab}. We call
this scheme the \emph{polynomial Pauli-Villars} regulator.

\begin{table}
  \begin{tabular}{crcc}
  \hline\hline
    $i$ & $p_i$ & $M_i^2 / \Lambda^2$ & $\kappa_i^2$\\
    \midrule
    1 & 1 & 0 & 0\\
    2 & -1 & 4 & 4\\
    3 & 2 & 3 & 3\\
    4 & -2 & 1 & 1\\
    \hline\hline
  \end{tabular}
  \caption{\label{poltab}Multiplicity and squared masses in terms of the
  cutoff scale $\Lambda$ for our polynomial Pauli-Villars regulator scheme.
  For later convenience we also provide the ratio $\kappa_i$ of the mass to
  the lightest regulator mass, which in this scheme is identical to $M_i /
  \Lambda$.}
\end{table}

The conditions \eqref{p0} and \eqref{pm20} are however not sufficient to
remove all UV contributions to the energy-momentum tensor. In particular, the
short-distance expansion of the two-point function in general contains
logarithmic terms. As we show explicitly in App.~\ref{App:renormalizationSE}, the polynomial Pauli-Villars
regulator succeeds in canceling the quadratic and quartic
divergences of $\langle T_{\mu \nu} \rangle$ through the conditions \eqref{p0}
and \eqref{pm20}, but fails to cancel logarithmic divergences of the form
\begin{equation}
  \langle T_{\mu \nu} \rangle_{\textrm{log}} \sim a_{2 }^{\mu \nu} (x) \left(
  \gamma + \frac{1}{2} \ln \left| \frac{1}{4} \mu^2 {\sigma }  \right| \right)
\end{equation}
where $a^{\mu \nu}_{2 } (x)$ is a local curvature tensor built from the
DeWitt--Hadamard coefficient $a_2 (x)$, $\sigma (x, x')$ is Synge's world
function, equivalent to half the squared geodesic separation between $x$ and
$x'$, and $\mu$ is an infrared (mass) regulator introduced to render the
proper--time integral finite at large $s$. Physically, $\mu$ defines an
arbitrary renormalization scale. Moreover, the introduction of massive regulator fields generates additional logarithmic divergences of the form $M^2 \log (M^2 \sigma
)$ and $M^4 \log (M^2 \sigma )$. Hence, canceling the power-law terms alone
does not guarantee that the regulated energy-momentum tensor has a well-defined
limit as the regulator masses are taken to infinity. The polynomial
Pauli-Villars scheme therefore provides a useful numerical regulator, but by
itself does not yet define a cutoff-independent renormalization prescription \footnote{In \cite{Berczi:2024yhb} the authors work within the polynomial scheme, i.e.~they impose conditions \eqref{p0} and \eqref{pm20} for finite Pauli-Villars masses. This leaves the logarithmic mass sums, so the regularized energy-momentum tensor contains finite remainders proportional to $g_{\mu\nu}$, $G_{\mu\nu}$ and curvature-squared tensors. The first two are removed by covariant counterterms fixing $\langle T_{\mu\nu}\rangle=0$ in Minkowski and redefining the bare Newton constant. No counterterms are added for the remainders proportional to the curvature-squared tensors, which would require higher-derivative terms in the action. Since these terms are not canceled, the regulator masses must be kept finite.}.

We therefore propose an alternative Pauli-Villars regulator that is designed to cancel these logarithmically divergent terms, too. In addition to Eq.
\eqref{pm20} we impose the conditions \cite{Asorey:2003uf,Fulling:2018qcn}
\begin{equation}
  \sum_{i = 1}^n p_i M_i^2 \ln \left(
  \frac{M_i^2}
  {\Lambda^2}
  \right)
  = \sum_{i = 1}^n p_i M_i^4 \ln \left(
  \frac{M_i^2}
  {\Lambda^2}
  \right) = 0
  \label{pm2log}
\end{equation}
and, for a nonzero physical mass $M_1$
\[ \sum_{i = 1}^n p_i \ln \left(
  \frac{M_i^2}
  {\Lambda^2}
  \right) = 0 \;. \]
In the case of a vanishing physical mass $M_1 = 0$, the last condition cannot
be fulfilled, and may instead be replaced by
\cite{PhysRevD.14.2490,PhysRevD.17.946,Asorey:2003uf}
\begin{equation}
  \sum_{i = 2}^n p_i \ln \left(
  \frac{M_i^2}
  {\Lambda^2}
  \right) = 0 \label{log0},
\end{equation}
where $\Lambda$ denotes the overall regulator scale (see Tab.~\ref{poltablog}).
The last condition fixes the finite part of the subtraction (the residual ambiguity allowed by Wald's axioms) and defines a
renormalization prescription in which all geometric divergences, including the
universal logarithmic ones, are canceled.

One minimal solution to the set of equations (\ref{p0}, \ref{pm20},
\ref{pm2log}, \ref{log0}) is given in Tab.~\ref{poltablog}. We call this
particular choice of multiplicities and masses the \emph{log-polynomial
Pauli-Villars} regulator.

\begin{table}
\begin{ruledtabular}
\begin{tabular}{lrcc}
    $i$ & $p_i$ & $M_i^2 / \Lambda^2$ & $\kappa_i^2$\\\midrule
    1 & 1 & 0 & 0\\
    2 & -1 & 106.8287106024297 & 33.00157557833226\\
    3 & 2 & 20.17193513394254 & 6.231523699293622\\
    4 & -2 & 3.2370791009314686 & 1\\
    5 & 3 & 84.93712615897233 & 26.23881700466686\\
    6 & -3 & 60.61745998016981 & 18.72597427808519
\end{tabular}
\end{ruledtabular}
\caption{\label{poltablog}Multiplicity and squared masses in terms of the
  cutoff scale $\Lambda$ for our logarithmic Pauli-Villars regulator scheme.
  For later convenience we also provide the ratio $\kappa_i$ of the mass to
  the lightest regulator mass.}
\end{table}

In App.~\ref{App:renormalizationSE}, we give a detailed derivation of this
scheme in the covariant point-splitting approach by Christensen
\cite{PhysRevD.14.2490,PhysRevD.17.946}. This refinement
should allow us to consistently take the limit $M_i \rightarrow \infty$ for
the fictitious regulator masses, thus obtaining a renormalized theory
independent of the ultraviolet cutoff. Crucially, the computational overhead
of this method grows at most by a constant prefactor determined by the number of Pauli-Villars fields introduced. It may in practice turn out to be even smaller due to a universality in the effective mass, that we will discuss below, making it especially suitable for numerical computations. We believe that the scheme thus provides a practical and covariant renormalization procedure for numerical studies of backreaction and semiclassical dynamics, and could close a gap that persists in existing formulations \cite{delrio_backreaction_2025}.

\section{Numerical results in flat spacetime}

\subsection{The Srednicki case: Flat spacetime with a brick wall cutoff}

\begin{figure}[htb]\centering
  \includegraphics[width=\columnwidth]{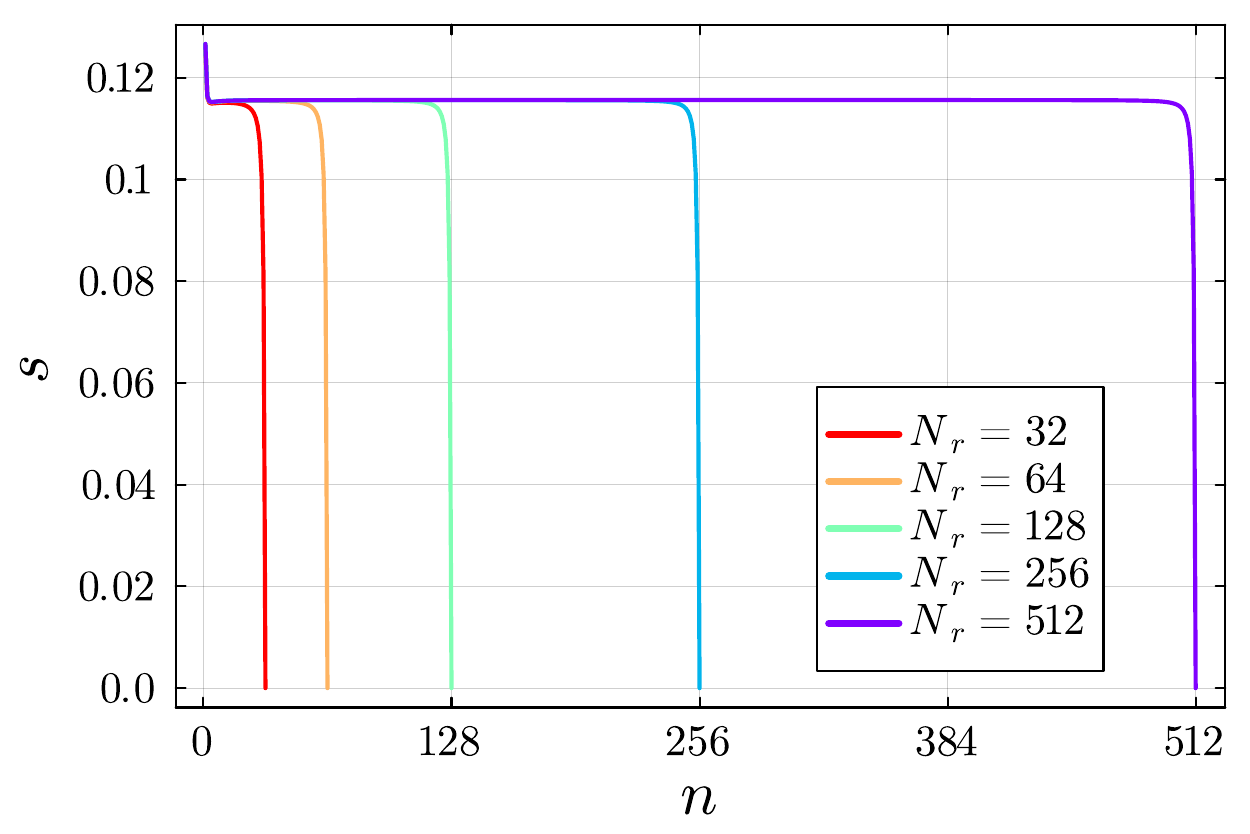}\par\medskip
  \includegraphics[width=\columnwidth]{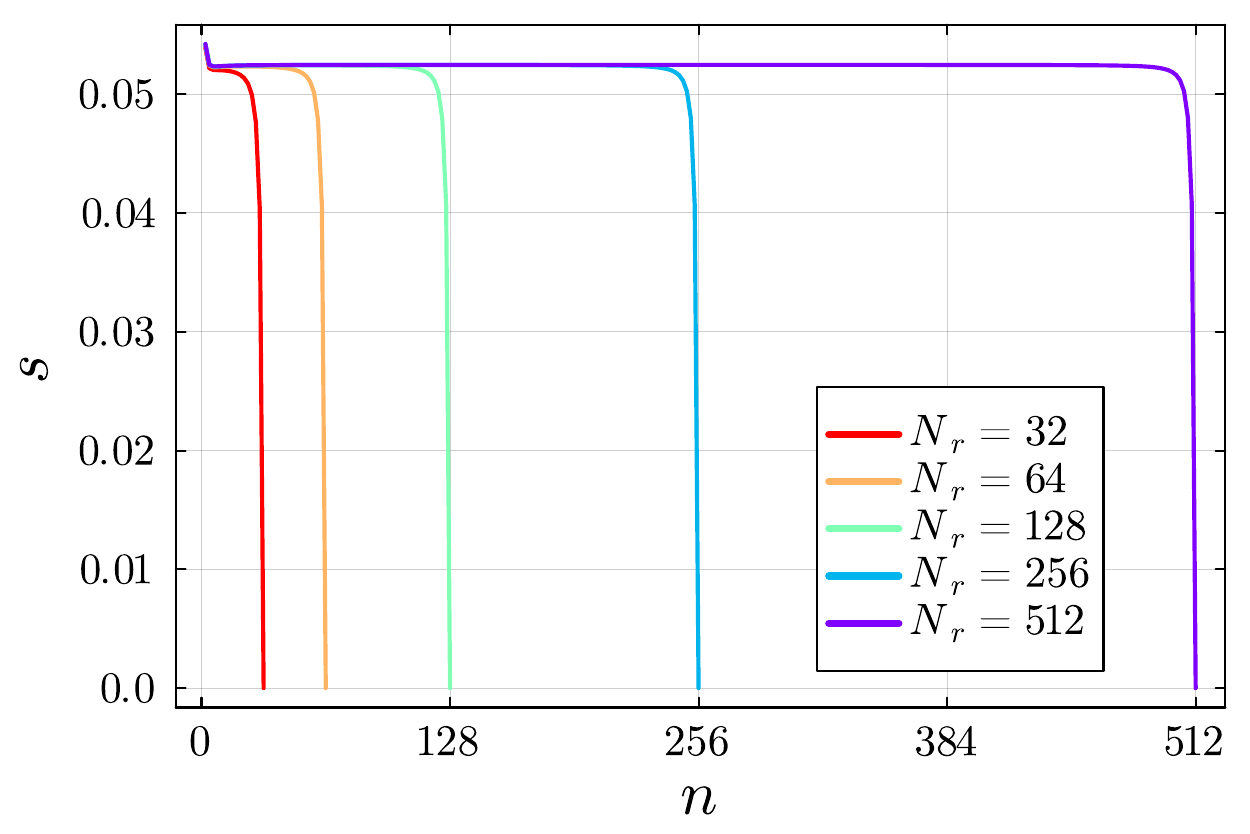}
  \caption{\label{sredent}The normalized entanglement entropy $s$ vs. the
  coordinate index $n$ at fixed anisotropy factor $c = 1$ (top) and $c = 0.5$
  (bottom) for five systems with different radial extent $N_r$. In the bottom
  panel only every second radial point is plotted to fulfill the relation
  \eqref{lnscal} exactly.}
\end{figure}

Before turning to gravitational collapse, as a first step, we check that our formalism correctly reproduces results for the entanglement entropy in flat space that were first derived by Srednicki \cite{Srednicki:1993im}. This also allows us to identify discretization and boundary effects. We will show that each angular momentum mode behaves like a one-dimensional field with an effective mass that depends on its angular momentum and on the distance from the origin, and that the total entropy is a sum over modes of one universal function of that effective mass. This particular structure will prove useful in the rest of the paper because massive regulator fields introduced for the Pauli-Villars regularization and a curved background will turn out to essentially act by changing the effective mass.
\begin{figure}[htb]\centering
  \includegraphics[width=\columnwidth]{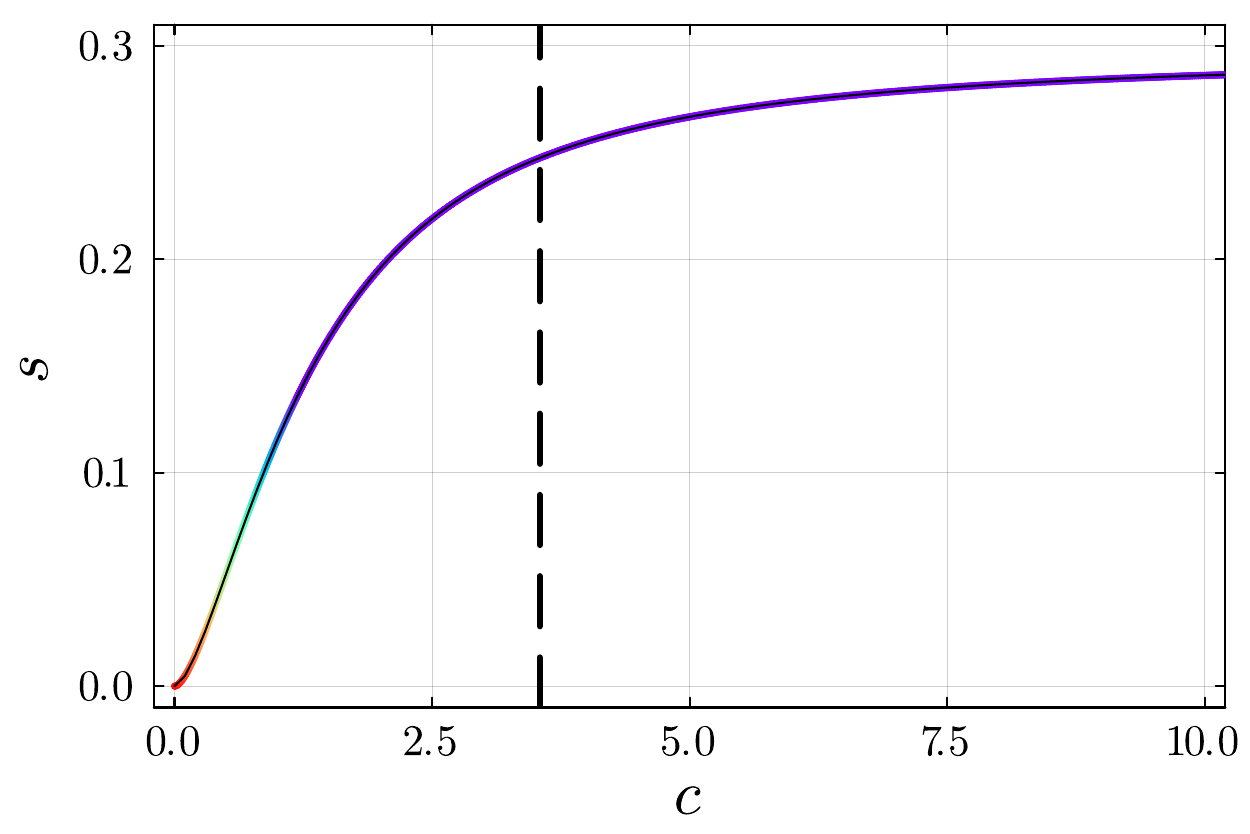}\par\medskip
  \includegraphics[width=\columnwidth]{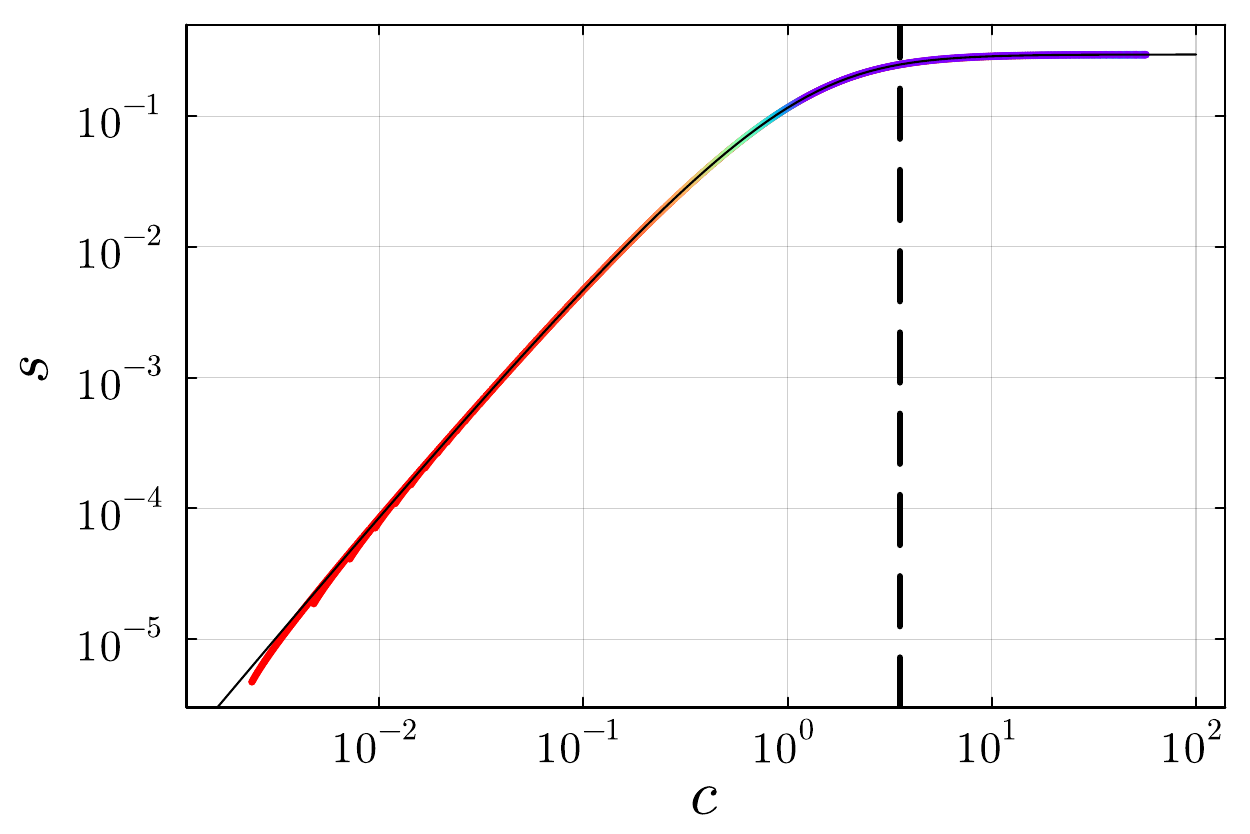}
  \caption{\label{scalsiz}The normalized entanglement entropy $s$ vs. the
  anisotropy factor $c$, in linear (top) and double logarithmic (bottom)
  representation for $N_r = 512$. The individual points correspond to all
  combinations of $n \in \{ 9, \ldots, 420 \}$ and $l_{\max} \in \{ 0, 8,\ldots,
  512 \}$. The color indicates the number of angular momenta, from red
  ($l_{\max} = 0$) to violet ($l_{\max} = 512$). The vertical dashed line
  represents the isotropic case $c = 2 \sqrt{\pi}$ and the black curves
  correspond to the prediction \eqref{sapprox}. Some hidden data points have
  been pruned to reduce the file size of the plot.}
\end{figure}

Srednicki employed a brick wall cutoff and found
the entropy to be proportional to $r^2$ with a proportionality constant that
depended on the discretization scale $\Delta$ as $\Delta^{- 2}$. Motivated by
this result, we first define a normalized entanglement entropy
\begin{equation}
  s = \frac{\Delta^2}{r^2} S \label{norment}
\end{equation}
that, very generically, is a function of the radial coordinate $r_n = \Delta
n$, the cutoff $\Delta$, the system size $r_{\max} = \Delta N_r$ and the
maximum number $l_{\max}$ of angular momenta taken into account. Due to the
scale invariance of the massless theory in the free case, the cutoff $\Delta$
does not explicitly enter any of the steps when calculating the entanglement
entropy according to \eqref{entdef}. Consequently, the entanglement entropy at
a given radial coordinate index $n$ is independent of $\Delta$, as is its
rescaled variant, so $s$ does in fact only depend on the three variables $n$,
$l_{\max}$ and $N_r$. Let us now consider a system with matched resolutions in
radial and angular directions. According to \eqref{constaniso}, we need to
scale
\begin{equation}
  (l_{\max} + 1) = c n \label{lnscal}
\end{equation}
to achieve this, where $c$ is an anisotropy factor. In Fig.~\ref{sredent} we
plot the normalized entropy vs. the radial index for two choices of the
anisotropy factor $c$ and various radial extents $N_r$. One can clearly
observe that, except for some boundary effects, the normalized entanglement
entropy is independent of the radial coordinate. For our largest $N_r=512$ system, the maximum relative difference in the central half of the coordinate range is $5\times10^{-5}$. There is thus no clear evidence of subleading corrections to the entropy that are not proportional to the area and generically appear in the continuum \cite{Casini:2009sr}. Our data are thus compatible with the existence of a finite, normalized entanglement entropy for the infinite system in the continuum limit, which only depends on the anisotropy factor $c$.

\begin{figure}[htb]\centering
  \includegraphics[width=\columnwidth]{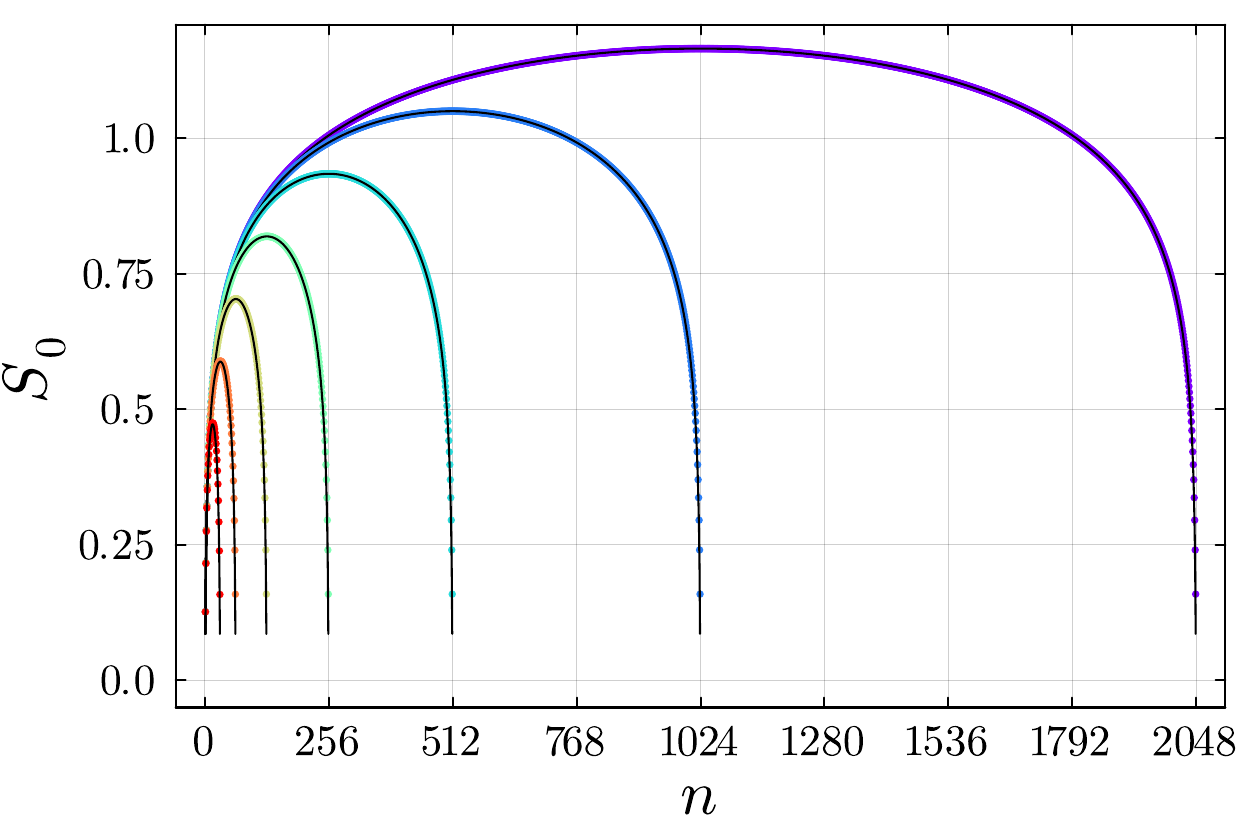}\par\medskip
  \includegraphics[width=\columnwidth]{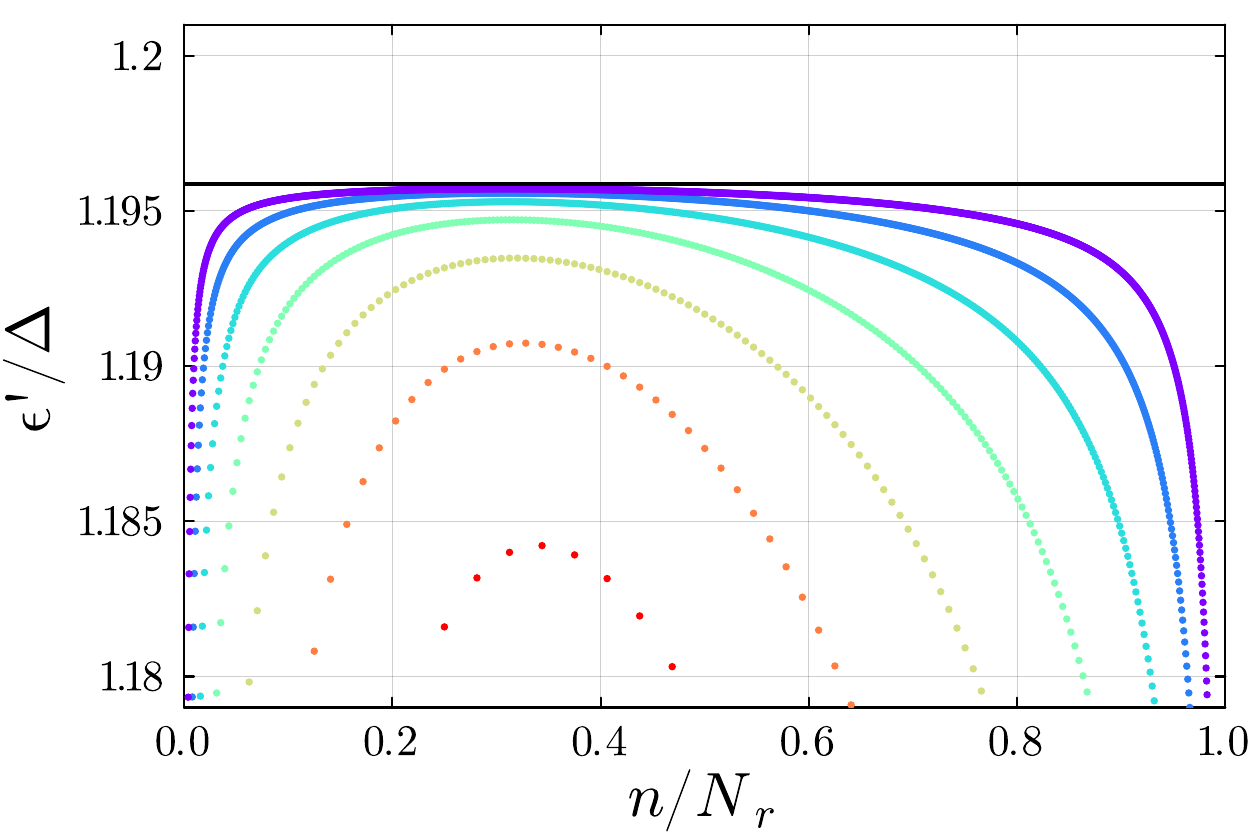}
  \caption{\label{hocl0}The $l = 0$ contribution to the entropy (dots)
  compared to the prediction \eqref{sourunits} with the UV cutoff set to
  $\varepsilon' \simeq 1.1959 \Delta$ (solid lines). In the top panel,
  data from radial discretizations of size $N_r = 32$ (red) to $N_r = 2048$
  (violet) are plotted vs. the radial index. In the bottom panel, the
  effective $\varepsilon' / \Delta$ for the same data are plotted vs. the
  radial index normalized by the total system size.}
\end{figure}
\begin{figure}[htb]\centering
  \includegraphics[width=\columnwidth]{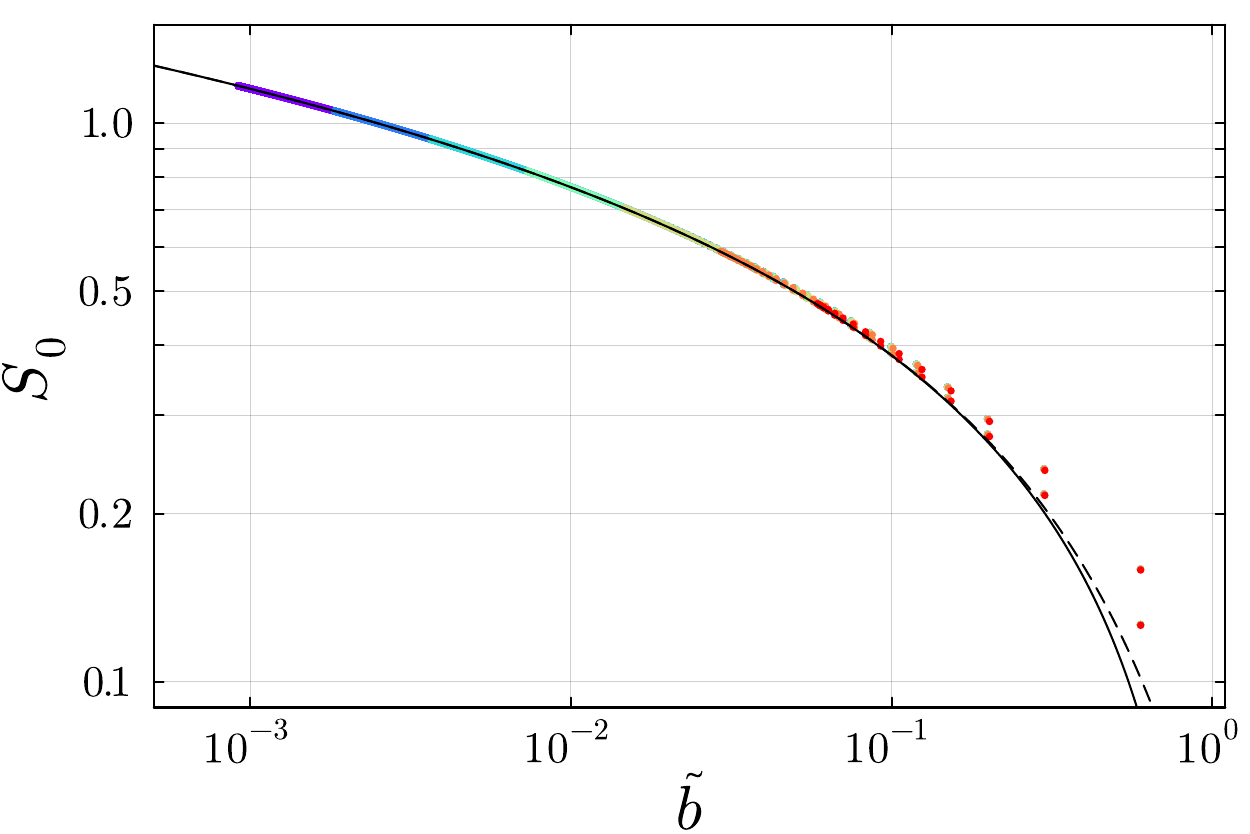}
  \caption{\label{S0btilde}The same data as in the top panel of Fig.
  \ref{hocl0}, plotted vs. the effective mass term \eqref{defctilde}. The
  dashed line represents the heuristic interpolation \eqref{masterformraw}
  with $\tilde{\mu} = 0$, while the solid line is the original prediction
  \eqref{sourunits}.}
\end{figure}
In order to find the universal dependence of $s$ on $c$, we look at individual
points on our largest $N_r = 512$ system. We discard the innermost 8 and the
outermost 192 points in order to eliminate boundary effects, so $n \in \{ 9,
\ldots, 420 \}$. For all other points we vary $l_{\max}$ between $0$ and $512$
in steps of $8$ and plot the resulting normalized entanglement entropy $s$ vs.
the anisotropy factor $c$ for that combination of $n$ and $l_{\max}$. The
result is displayed in Fig.~\ref{scalsiz}. One can clearly see a universal
behavior throughout the entire range of parameters, which span more than two
orders of magnitude in the anisotropy factor $c$. If we choose our brick wall cutoff $\Delta=\ell_P$ to be one Planck length, and impose an equal cutoff in all directions, which corresponds to an anisotropy factor $c = 2 \sqrt{\pi}$, we find the entanglement entropy to be
\begin{equation}
  S \simeq 0.248 \frac{r^2}{\ell_P^2} \;. \label{univdisc}
\end{equation}
As we will show in the remainder of this section, we also reproduce the $c\rightarrow\infty$ prefactor from \cite{Srednicki:1993im}.
For this purpose, it is instructive to first understand the approximate constancy of the normalized entanglement entropy over the radial coordinates at a given anisotropy factor on a more technical level. We note that we may rewrite the normalized entanglement entropy \eqref{norment} at the $n^{\tmop{th}}$ radial coordinate as
\[ s = \frac{S}{n^2} \;.\]
Considering that there are $(2 l + 1)$ modes per $l$ contributing an equal
amount $S_l$ to towards the entropy, we may write this relation as
\[ s = \sum_{l = 0}^{l_{\max}} \frac{2 l + 1}{n^2} S_l \;.\]
Substituting
\begin{equation}
  \tilde{c}^2 = \frac{l (l + 1)}{n^2} \qquad \mathd \tilde{c}^2 = \frac{2 l +
  1}{n^2} \mathd l \label{ctilde}
\end{equation}
and going to large radial coordinates $n$, we find that
\begin{equation}
  s \rightarrow \bigintlim_0^{\tilde{c}_{\max}^2} \mathd \tilde{c}^2 \,
  \tilde{S} (\tilde{c}^2) \;,\label{findiss}
\end{equation}
where we have defined the integrand
\[ \tilde{S} \left( \frac{l (l + 1)}{n^2} \right) = S_l \label{stildeint} \;.\]
In order to obtain the upper integration limit, we substitute either
$l_{\max}$ or $l_{\max} + 1$ into \eqref{ctilde} and express it in terms of
the anisotropy factor $c = (l_{\max} + 1) / n$, resulting in
\[ \tilde{c}_{\max}^2 = c^2 \left( 1 \mp \frac{1}{c n} \right) \;,\]
so that in the $n \rightarrow \infty$ limit we ultimately find
\begin{equation}
  s \rightarrow \bigintlim_0^{c^2} \mathd \tilde{c}^2 \, \tilde{S} (\tilde{c}^2)\;.
  \label{fusimo}
\end{equation}
In order for the normalized entropy to be a function of $c$ only, the
integrand $\tilde{S} (\tilde{c})$ may thus only depend on $\tilde{c}$ and $c$.
However, since $c$ requires knowledge of $l_{\max}$, which $\tilde{S}$ lacks,
we conclude that $\tilde{S}$ has to be a function of $\tilde{c}$ alone. This
somewhat heuristic statement can be made more precise by realizing that
$\tilde{c}$ is just the radially dependent effective mass of the Hamiltonian
\eqref{classhamdens} in terms of the cutoff $\Delta$
\begin{equation}
  \tilde{\mu}^2 = \Delta^2 M^2 + \tilde{c}^2 \label{meff}
\end{equation}
in the case of a massless ($M=0$) theory. In fact, for large radial coordinates, the relative
correction to a radially constant effective mass vanishes locally, and one may use
results of one dimensional field theory or the harmonic oscillator chain to
obtain predictions for the entropy contribution $\tilde{S}$. Specifically, one
may construct a hopping expansion (see App.
\ref{hop}) in curved space time, which in the free case and for large radial coordinates goes over to the 1+1 dimensional case flat case. It has an expansion parameter
\[ x = \frac{1}{4 (2 + \tilde{\mu}^2)} \]
and is convergent for $\tilde{\mu} > 0$. Consequently, in the infinite volume
limit and for large $n$, $\tilde{S}$ is indeed a
function of the effective mass only. This function $\tilde{S} (\tilde{\mu}^2)$
is known analytically for both the small and large $\tilde{\mu}$ limits. The
latter is covered by the hopping expansion, detailed in App.~\ref{hop}, which
to leading order is given by
\begin{equation}
    \tilde{S} (\tilde{\mu}^2)
    \xrightarrow{\tilde{\mu} \rightarrow \infty}{}
    \left(1+x^2 \right)
    \ln \left( 1+x^2 \right)
    - x^2
    \ln x^2
  \label{lohop}
\end{equation}
or, expanded to leading order in $\tilde{\mu}^{-1}$,
\begin{equation*}
    \tilde{S} (\tilde{\mu}^2)
    \xrightarrow{\tilde{\mu} \rightarrow \infty}
    \frac{1+4\ln(2\tilde{\mu})}{16\tilde{\mu}^4}\;.
\end{equation*}
For small $\tilde{\mu}$, there is an analytical result \eqref{sassyinf} from
$1 + 1$ dimensional field theory (see App.~\ref{hocapp} for details)
\begin{equation}
  \begin{aligned}
    \tilde{S} (\tilde{\mu}^2)
    \xrightarrow{\tilde{\mu} \rightarrow 0}{}&
    \frac{1}{12} \Gamma \left( 0,
    \frac{\varepsilon^2}{\Delta^2} \tilde{\mu}^2 \right)\\
    ={}& \frac{1}{12} \left(
    \ln \frac{1}{\tilde{\mu}^2} 
    - \ln \frac{e^{\gamma}\varepsilon^2}{\Delta^2}
    + O (\tilde{\mu}^2) \right)\;.
  \end{aligned}
  \label{snobo}
\end{equation}
For vanishing $\tilde{\mu}$, which corresponds to the $l = 0$ mode of the
massless theory, this expression is logarithmically divergent, which is
consistent with the failure of the infinite volume hopping expansion
to converge. In this case, an alternative infrared cutoff may be provided by a
finite volume. For large radial coordinates, we may then apply the well known result in the literature for
the entanglement entropy of $1 + 1$ dimensional field theory \eqref{l0e},
which should describe our system up to discretization and boundary effects.
\footnote{The harmonic oscillator chain also has a boundary, which we map to the central region of our space around  $r=0$. Since the large $n$ approximation is not valid in that region, we expect there to be some boundary effects that we cannot describe very well analytically.}In
terms of our discretized quantities, this relation reads
\begin{equation}
  S_0^H = \frac{1}{12} \ln \left( \frac{1}{\tilde{b}^2} \right)
  \label{sourunits}
\end{equation}
where
\begin{equation}
  \tilde{b} = \frac{\varepsilon'}{\Delta} \left( \frac{2 N_r}{\pi} \sin \left(
  \frac{\pi n}{N_r} \right) \right)^{- 1} \label{defctilde}
\end{equation}
may be interpreted as an effective mass term originating from boundary
effects. As argued in App.~\ref{hocapp}, the form of the two exact results
\eqref{snobo} and \eqref{sourunits} suggests a heuristic interpolation
\eqref{sugint}, which we may write as
\begin{equation}
  \tilde{S} \approx \frac{1}{12} \Gamma \left( 0,
  \frac{\varepsilon^2}{\Delta^2} \tilde{\mu}^2 + e^{- \gamma} \tilde{b}^2
  \right) \label{masterformraw}
\end{equation}
that has the correct limiting behavior for both $\tilde{\mu} \rightarrow 0$
and $\tilde{b} \rightarrow 0$. Taking the argument of this function, we may
more generically define a boundary corrected effective mass term
\[ \tilde{\mu}^2 + \frac{\Delta^2}{\varepsilon^2} e^{- \gamma} \tilde{b}^2. \]
While this does not provide a single scaling variable that $\tilde{S}$ depends
on exclusively in the finite volume case, it can still substantially improve the modelling of boundary efffects.

Comparing these predictions with our numerical data for the entanglement
entropy, we first look at the $l = 0$ contribution in Fig.~\ref{hocl0}. In
the top panel, we compare the numerical data for a range of system sizes
$N_r$ with the prediction \eqref{sourunits} for a fixed $\varepsilon' / \Delta
\simeq 1.1959$, revealing a very good qualitative agreement. In the bottom panel, we zoom in on the small discrepancies by equating the prediction
\eqref{sourunits} to the numerical data for all $N_r$ and $n$ and solving for
$\varepsilon' / \Delta$. In this way, we obtain an effective ratio
$\varepsilon' / \Delta$ which shows a very nice plateau with a continuum
extrapolated value of $\varepsilon' / \Delta \simeq 1.1959$.

\begin{figure}[htb]\centering  
  \includegraphics[width=\columnwidth]{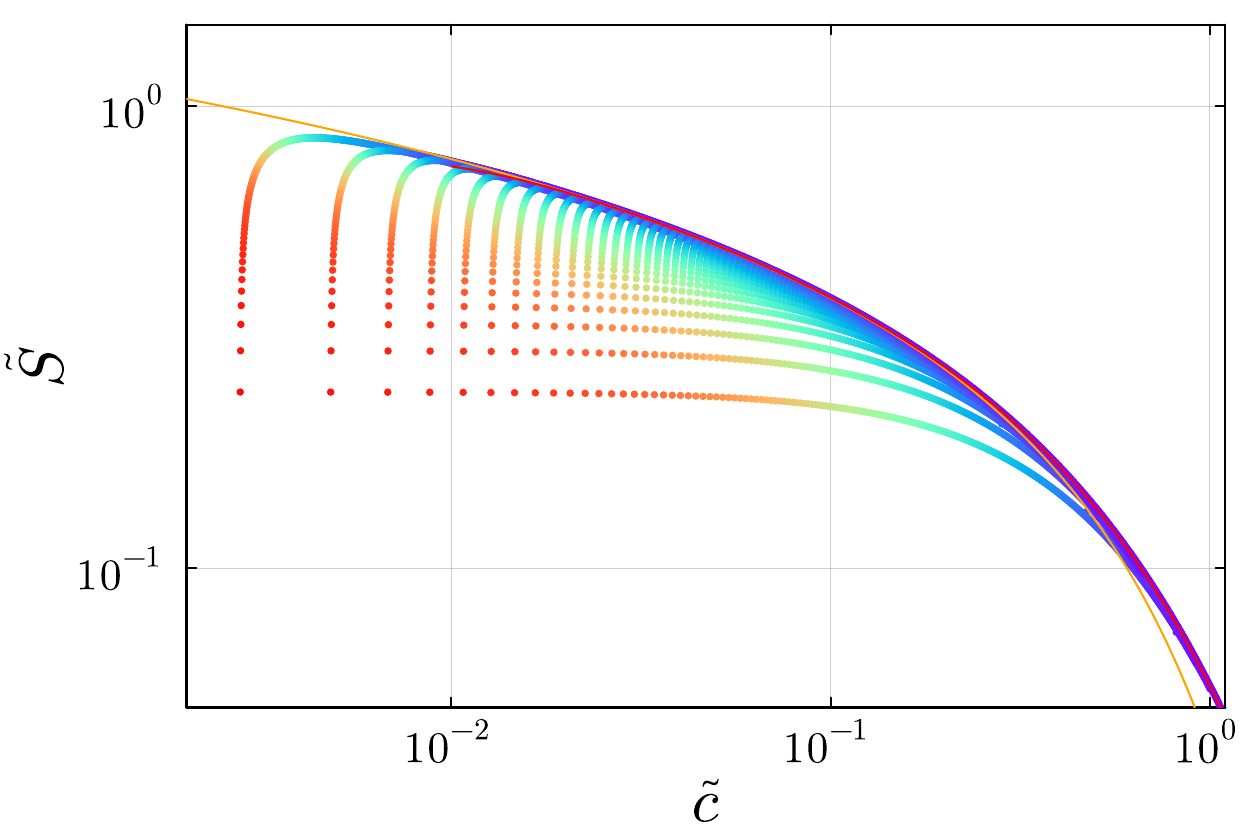}
  \includegraphics[width=\columnwidth]{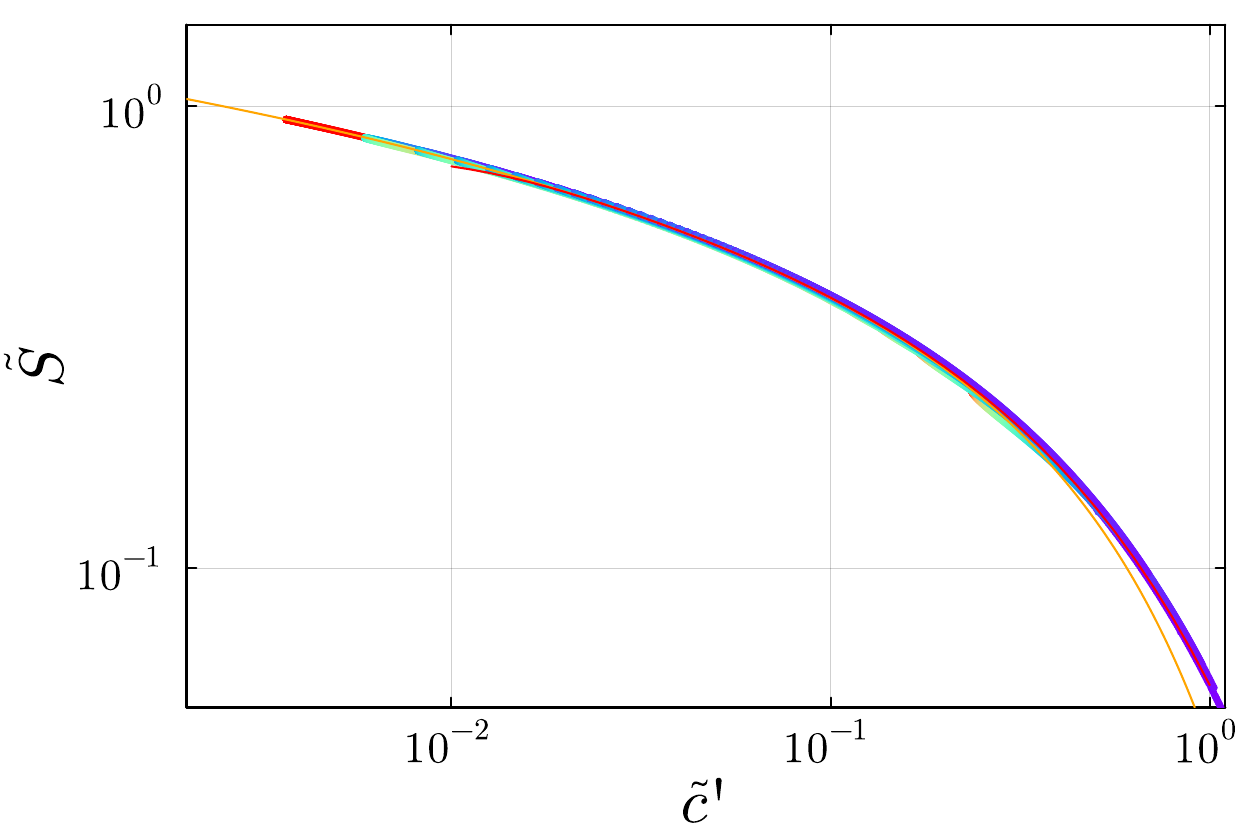}
  \caption{\label{bcstilde}The single mode entropy contribution $\tilde{S}$
  for $N_r = 512$ vs. $\tilde{c}$ (top) and the boundary corrected
  $\tilde{c}'$ (bottom) for $n \in \{ 3, \ldots, 510 \}$ and $l \in \{ 0,
  \ldots, 512 \}$ in the range displayed. The colors of the dots indicate the
  ratio $\tilde{c} / \tilde{b}'$ of the infinite volume scaling variable to
  the boundary correction term, from red (zero) to violet (large). The orange
  curves are the asymptotic form \eqref{snobo}, while red curves represent the
  partially resummed hopping expansion to $256^{\tmop{th}}$ order (see App.
  \ref{hop}).}
\end{figure}

Plotting the same data vs.~$\tilde{b}$ (Fig.~\ref{S0btilde}) reveals nice
consistency with both the exact continuum prediction \eqref{sourunits} and the
heuristic interpolation \eqref{masterformraw} except for points very close to
the boundary. In fact, we may go one step further and define an effective mass
term that accounts for boundary effects via inverting the relation
(eqref{sourunits} on our numerical data for the $l = 0$ entropy contribution
$S_0$ as
\begin{equation}
  \tilde{b}' = e^{- 6 S_0} \;.\label{boundcr}
\end{equation}
While this definition provides no further information in the $l = 0$ case,
it allows us to define a boundary corrected scaling variable
\[ \tilde{c}'{}^2 = \tilde{c}^2 + \frac{\Delta^2}{\varepsilon^2} e^{- \gamma}
   \tilde{b}'{}^2 \;,\]
which can be used to more precisely extract a scaling function $\tilde{S}
(\tilde{c}^2)$ from numerical data at finite lattices. To utilize this relation, we need to determine the ratio $\varepsilon / \Delta$. In
principle, this could be achieved by comparing data with negligible boundary
corrections to the asymptotic relation \eqref{snobo}. In practice, however,
this is complicated by the fact that in the $\tilde{c} \rightarrow 0$ limit
boundary effects play a larger role, so that \eqref{snobo} is no longer
applicable. It turns out, however, that within the numerical accuracy of our
data, the simple choice
\begin{equation}
  \varepsilon / \Delta = e^{- \gamma / 2} \label{epsdelrat}
\end{equation}
is adequate so that we may set
\[ \tilde{c}'{}^2 = \tilde{c}^2 + \tilde{b}'{}^2 \;.\]
Results from performing the hopping expansion to high order also seem to corroborate this simple choice (see Fig.~\ref{fighopres}).

In Fig.~\ref{bcstilde} we plot our numerical data for the single mode entropy
$\tilde{S}$ vs. $\tilde{c}$ and $\tilde{c}'$ for $N_r = 512$ in the relevant
range of small $\tilde{c}$ resp. $\tilde{c}'$. It is clearly visible that
boundary corrections are significantly reduced by using $\tilde{c}'$ and
numerical data lie close to a universal curve.

\begin{figure}[htb]\centering
  \includegraphics[width=\columnwidth]{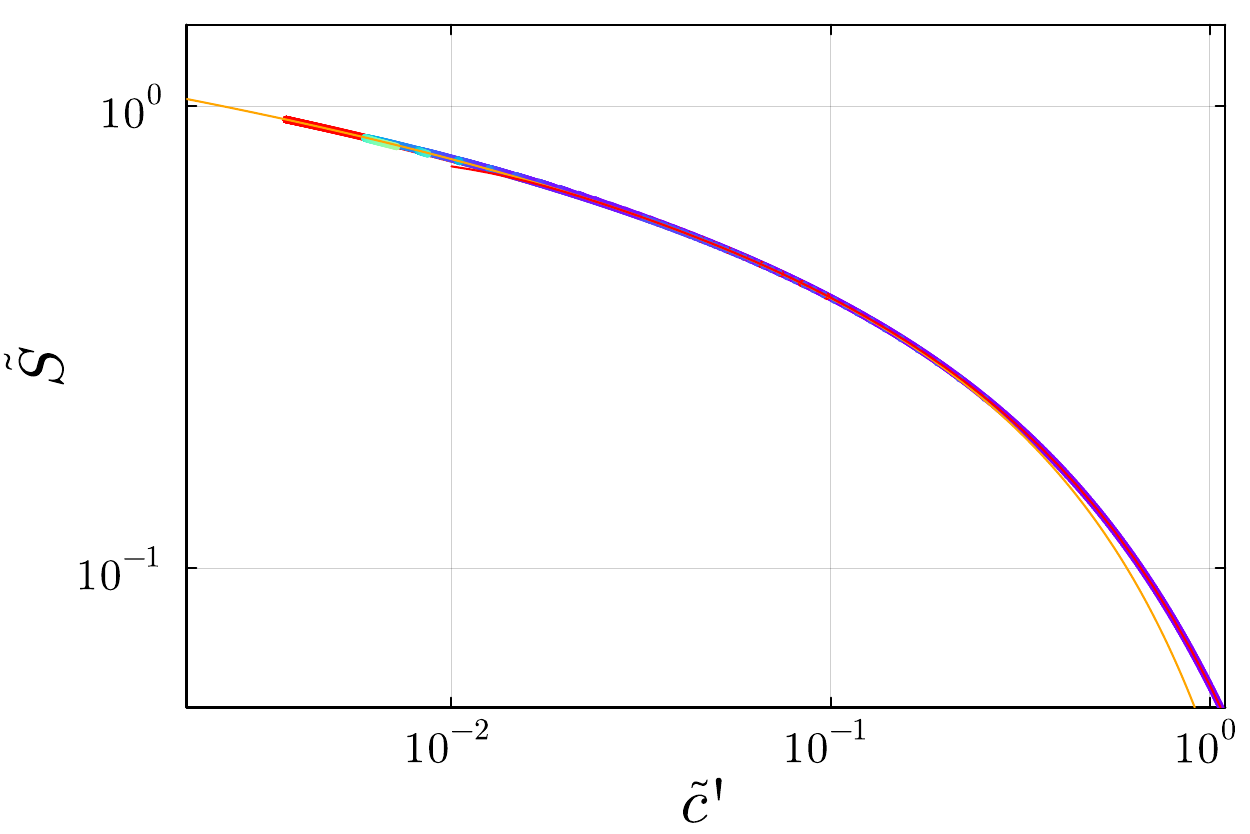}\par\medskip
  \includegraphics[width=\columnwidth]{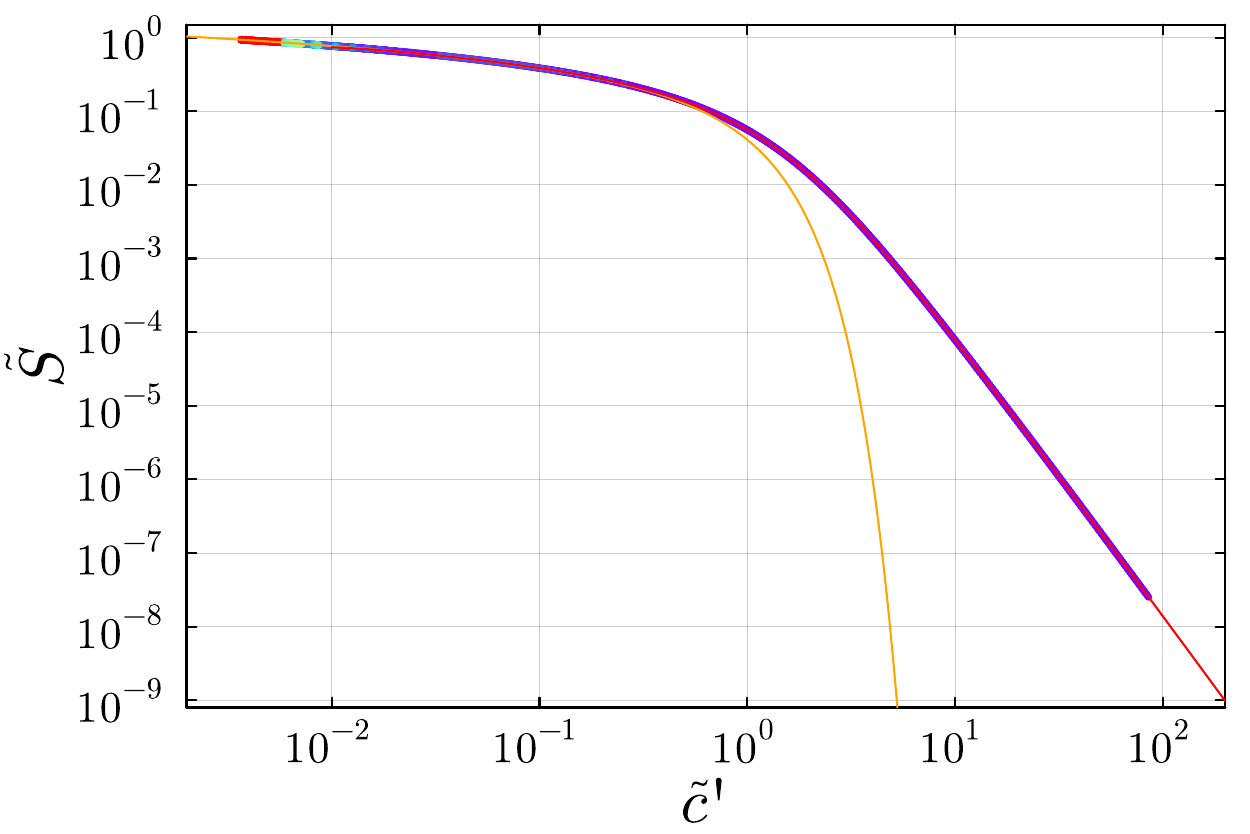}
  \caption{\label{brickscal}Same as the bottom panel of Fig.~\ref{bcstilde},
  but with radal index restricted to the range $n \in \{ 5, \ldots, 411 \}$
  (top) and additionally extended plot range to cover all data points
  (bottom).}
\end{figure}

Universal behavior is even more pronounced after pruning points close to the
boundary, as in Fig.~\ref{brickscal}. We also see clear numerical confirmation
for both the asymptotic form \eqref{snobo} as well as the hopping expansion
(App.~\ref{hop}) in their respective domains of applicability. As demonstrated in App.~\ref{hop}, the asymptotic forms actually make contact with sub per mil accuracy, which allows us to interpolate between them \eqref{stili}. In this manner, we obtain an accurate
approximation $\tilde{S}^i (\tilde{c}^2)$ of the universal function $\tilde{S}
(\tilde{c}^2)$ over the entire range $\tilde{c} > 0$. Using this
approximation, we can now integrate \eqref{fusimo} to obtain an approximation
to the brick wall normalized entropy as a function of the anisotropy factor
$c$ as
\begin{equation}
  s^i (c^2) = \bigintlim_0^{c^2} \mathd \tilde{c}^2 \, \tilde{S}^i (\tilde{c}^2)\;.
  \label{sapprox}
\end{equation}
As shown in Fig.~\ref{scalsiz}, we find excellent agreement with the direct
determination of $s$, except in the region of extremely small anisotropy
factors, where the direct method is problematic due to relatively large
discretization errors.

We accurately reproduce \eqref{univdisc} and obtain a
$c \rightarrow \infty$ limit of the normalized entropy of $s=0.29543144(20)$ with the error estimate dominated by the number of points on which we evaluated the hopping expansion. This result is in
good agreement with Srednicki's original value \cite{Srednicki:1993im} and the value from the $S_{\log}$ fit in \cite{Lohmayer:2009sq}, but less so with their value $S_{\mathrm{lin}}$. This is to be expected, as the universal function is valid in the $n\rightarrow\infty$ limit, where nonuniversal corrections vanish. In fact, subleading corrections to $s$, as those found in \cite{Casini:2009sr}, are absent in the strict $n\rightarrow\infty$ limit. However, there may be subtle order of limits effects influencing the subleading terms, which we will not investigate further in this paper.

\begin{figure}[htb]\centering
  \includegraphics[width=\columnwidth]{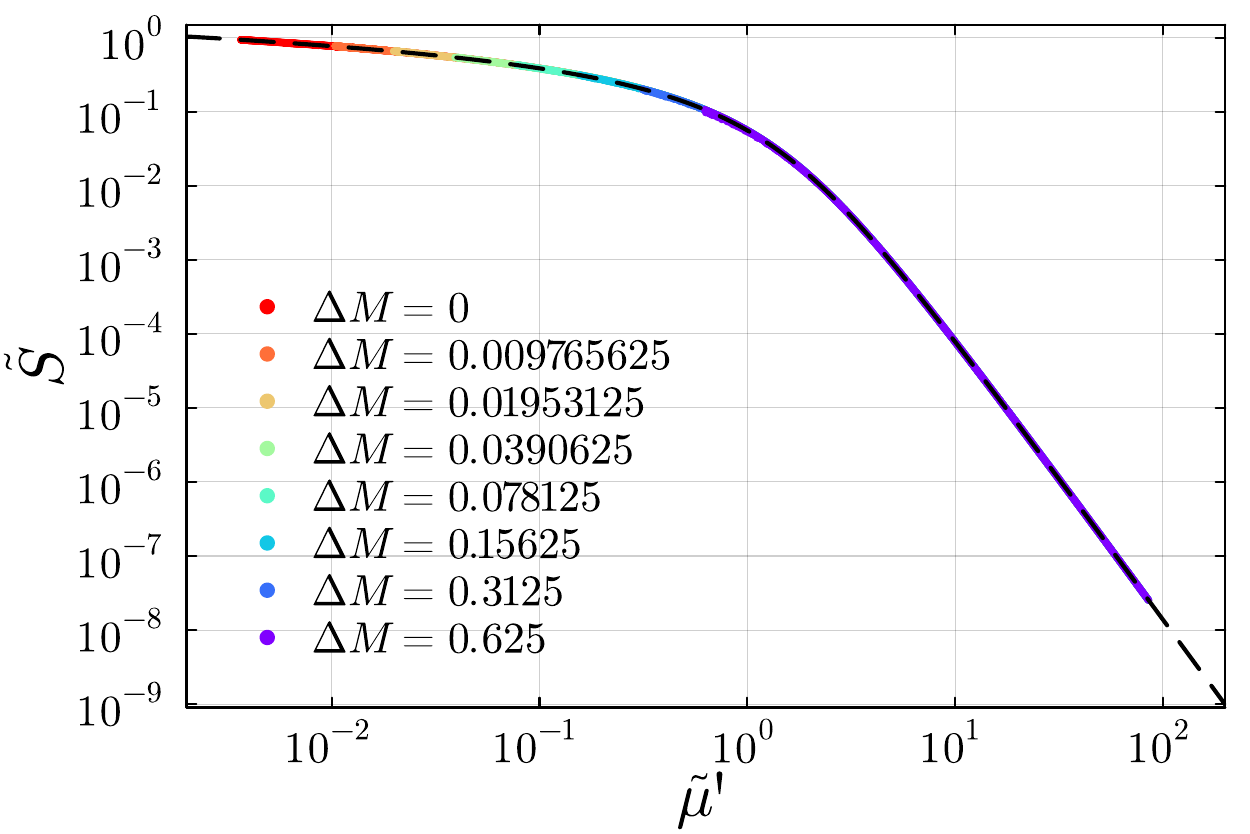}\par\medskip
  \includegraphics[width=\columnwidth]{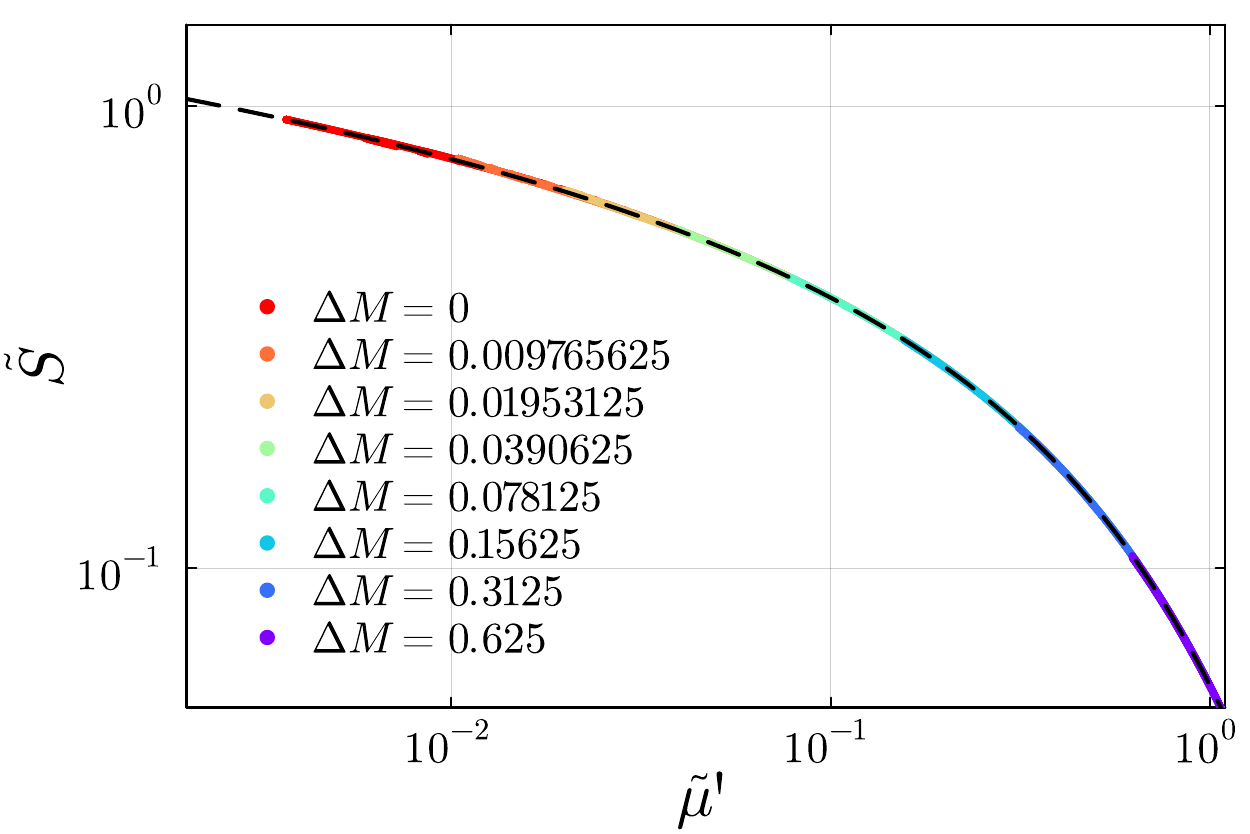}
  \caption{\label{massbw}The single mode brick wall entropy plotted vs. the
  boundary corrected effective mass \eqref{meffcor} for $N_r = 512$ for
  various values of the mass. The top panel shows all data points $n \in \{ 5,
  \ldots, 411 \}$, while the bottom panel is a zoom for the largest
  contributions with $n \in \{ 50, \ldots, 411 \}$. In both panels, the dashed
  line is the approximation $\tilde{S}^i (\tilde{\mu}^2)$ to the universal
  function obtained from $M = 0$ data.}
\end{figure}

\begin{figure}[htb]\centering
  \includegraphics[width=\columnwidth]{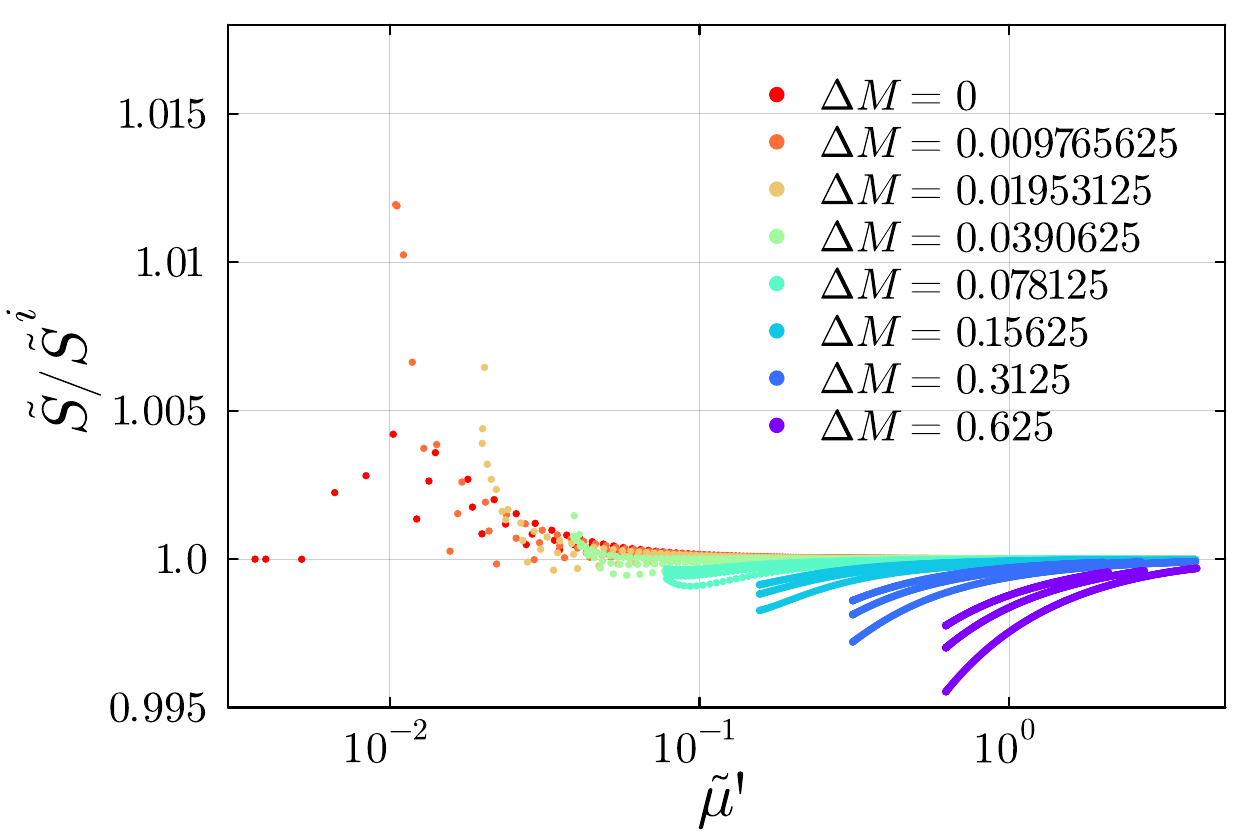}\par\medskip
  \includegraphics[width=\columnwidth]{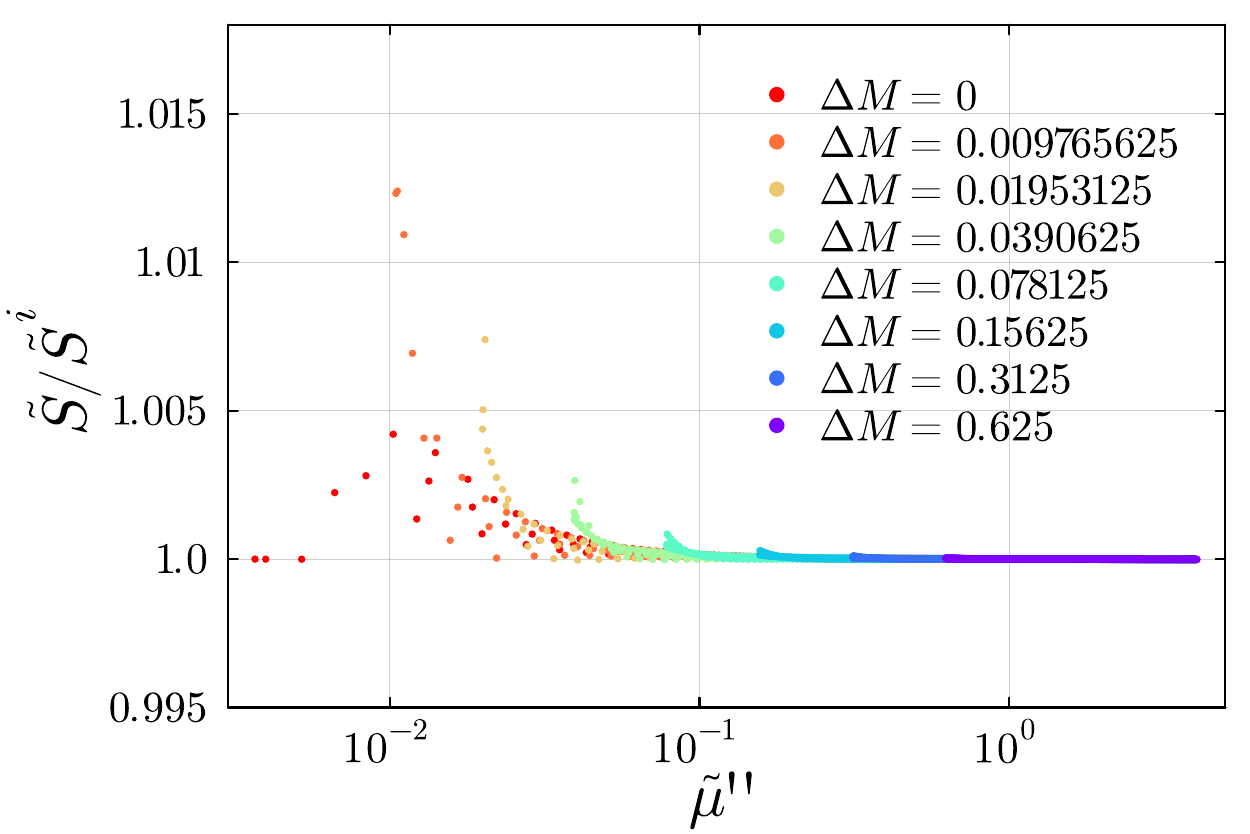}
  \caption{\label{massbwdetail}Ratio of the single mode brick wall entropy to the approximate universal function $\tilde{S}^i (\tilde{\mu}^2)$ vs. the boundary corrected effective masses $\tilde{\mu}'$ \eqref{meffcor} (top panel) and $\tilde{\mu}''$ \eqref{cmefffree} (bottom panel) for $N_r = 512$, radial indices $n\in\{128,192,256\}$
  and various values of the mass. Clearly there is less universality violation in $\tilde{\mu}''$ than in $\tilde{\mu}'$}.
\end{figure}
To compare our numerical data at finite mass  to the universal function, we add explicit boundary correction
terms in the same manner as for the massless case, defining
\begin{equation}
  \tilde{\mu}'{}^2 = \Delta^2 M^2 + \tilde{c}'{}^2 = \Delta^2 M^2 +
  \tilde{c}^2 + \tilde{b}'{}^2 \;.\label{meffcor}
\end{equation}
As one can see in Fig.~\ref{massbw}, the entropies for all masses fall nicely onto a
universal curve, which is consistent with $\tilde{S}^i (\tilde{\mu}^2)$.

To check how precisely the data follow the approximate universal curve $\tilde{S}^i (\tilde{\mu}^2)$, we plot the ratio of the data to the curve in the top panel of Fig.~\ref{massbwdetail} for three values of the radial coordinate and several masses. Apart from some percent level spread at low $\tilde{\mu}'$, there is a systematic, mass dependent deviation. Interestingly, modifying the mass term in \eqref{meffcor} by a factor
\[ M^2 \rightarrow \left( 1 + \frac{1}{n} \right) M^2 \]
seems to substantially improve the universality for the entire range of radial coordinates $n$. We thus propose a boundary corrected mass term
\begin{equation}
  \tilde{\mu}''{}^2 = \Delta^2 \left( \frac{l (l + 1)}{r^2} +
  \left( 1 + \frac{1}{n} \right) M^2 \right) + \tilde{b}'{}^2 \label{cmefffree}
\end{equation}
which shows an improved universality as shown in the bottom panel of Fig.~\ref{massbwdetail}. We note that the correction vanishes in the large $n$ limit and is, in this sense, a discretization artifact.  Its specific form seems to hint at a discretization artifact of a finite difference operator, but we are not able at this point to give any rigorous argument for its appearance.

\subsection{Flat spacetime with simple Pauli-Villars regularization}

As a next step, we study flat spacetime with regularization provided by a
single Pauli-Villars field. The mass of the Pauli-Villars field
$M_{\tmop{PV}}$ now provides an additional scale, and we define the
Pauli-Villars normalized entropy
\begin{equation}
  \hat{s} = \frac{S}{M_{\tmop{PV}}^2 r^2} \label{normentpv}
\end{equation}
by replacing the brick wall scale in the definition of the normalized
entanglement entropy \eqref{norment} with the Pauli-Villars mass. We also
introduce the dimensionless cutoff ratio
\begin{equation}
  k = \Delta M_{\tmop{PV}} \label{cur}
\end{equation}
in terms of which the effective mass for the Pauli-Villars field \eqref{meff}
reads
\[ \tilde{\mu}^2_{\tmop{PV}} = \tilde{c}^2 + k^2 \;. \]
In the infinite volume limit and for large $n$, the single mode entropy
$\tilde{S}_k (\tilde{c})$ of the Pauli-Villars regulated system at cutoff
ratio $k$ is thus given by
\begin{equation}
  \tilde{S}_k (\tilde{c}^2) = \tilde{S} (\tilde{c}^2) - \tilde{S} (\tilde{c}^2
  + k^2)\;. \label{stildek}
\end{equation}
It is now helpful to introduce the ratio
\[ \hat{c} = \frac{\tilde{c}}{k} \]
and rewrite the single mode entropy contribution in terms of this new variable
as
\begin{equation}
  \hat{S}_k (\hat{c}^2) : = \tilde{S}_k (k^2 \hat{c}^2) = \tilde{S} (k^2
  \hat{c}^2) - \tilde{S} (k^2 (\hat{c}^2 + 1)) \;.\label{shatk}
\end{equation}
Integrating this single mode entropy contribution analogous to \eqref{fusimo},
we find the Pauli-Villars normalized entropy
\begin{equation}
  \hat{s} = \frac{s}{k^2} = \bigintlim_0^{c^2 / k^2} \mathd \hat{c}^2 \,
  \hat{S}_k (\hat{c}^2)\;. \label{shat}
\end{equation}
For $\tilde{c} = k \hat{c} \gg 1$ we may approximate $\hat{S}_k (\hat{c}^2)$
by the hopping expansion and, using \eqref{lohop}, we find that asymptotically
\begin{equation}
  \tilde{S}_k (\tilde{c}^2) \xrightarrow{\tilde{c} \rightarrow \infty} k^2
  \frac{\ln (2 \tilde{c})}{2 \tilde{c}^6}\;. \label{simpvasy}
\end{equation}
The physically more interesting case however is the $\tilde{c} = k \hat{c} \ll
1$ region. Specifically, the asymptotic relation \eqref{snobo} implies that
\begin{equation}
  \hat{S}_k (\hat{c}^2) \xrightarrow{k \rightarrow 0} \frac{1}{12} \ln \left(
  \frac{1}{\hat{c}^2} + 1 \right) \label{asyskhat}
\end{equation}
which corresponds to the continuum limit at constant $M_{\tmop{PV}}$. The
single mode entropy of the Pauli-Villars regulated, infinite volume continuum
theory
\[ \hat{S} (\hat{c}^2) = \lim_{k \rightarrow 0} \hat{S}_k (\hat{c}^2) \]
is thus analytically known
\begin{equation}
  \hat{S} (\hat{c}^2) = \frac{1}{12} \ln \left( \frac{1}{\hat{c}^2} + 1
  \right) \;.\label{daform}
\end{equation}
This is an exact, parameterless prediction that
follows directly from \eqref{sassyinf}. Furthermore, plugging this result into
the Pauli-Villars normalized entropy \eqref{shat}, we find that the integral
is logarithmically divergent for large $\hat{c}$. In this sense, even the
regularized entropy diverges in the continuum limit, making the simple
Pauli-Villars regularization particularly unsuited for our purposes.

\begin{figure}[htb]\centering
  \includegraphics[width=\columnwidth]{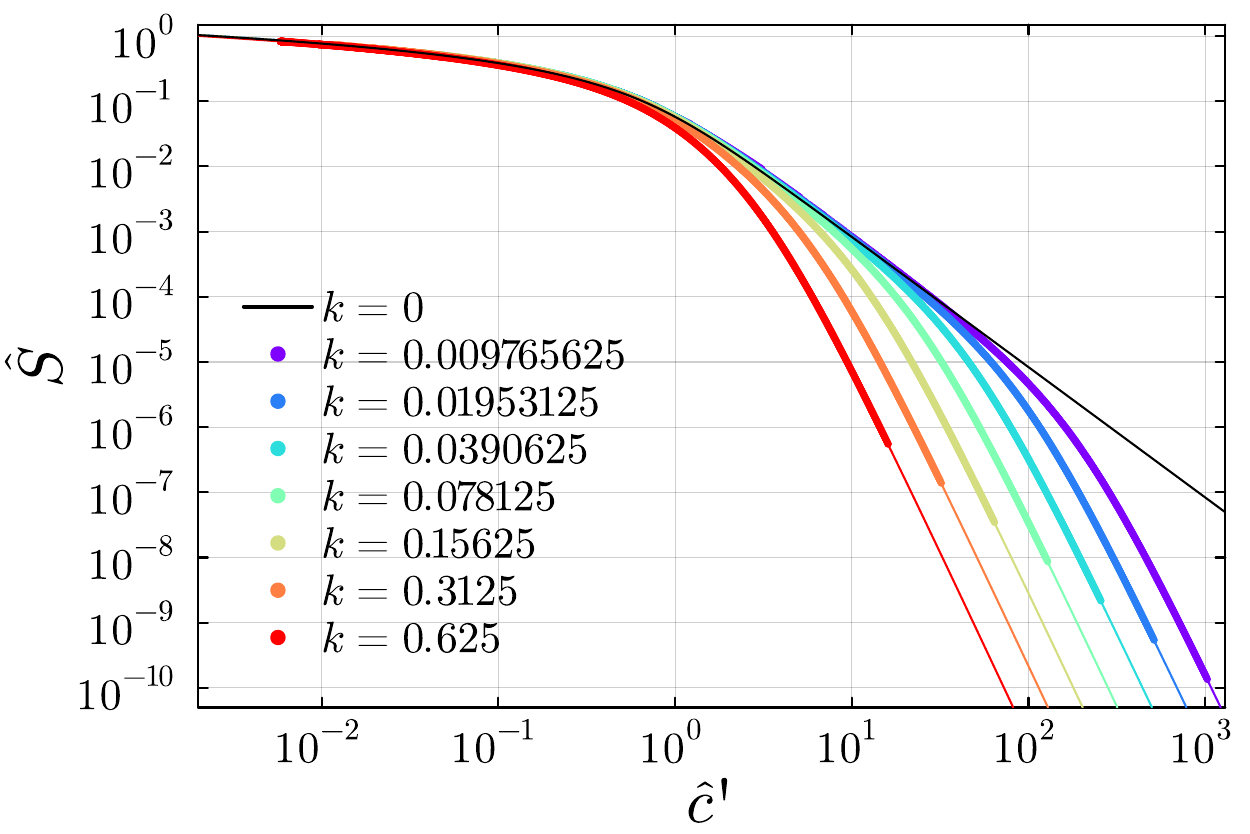}\par\medskip
  \includegraphics[width=\columnwidth]{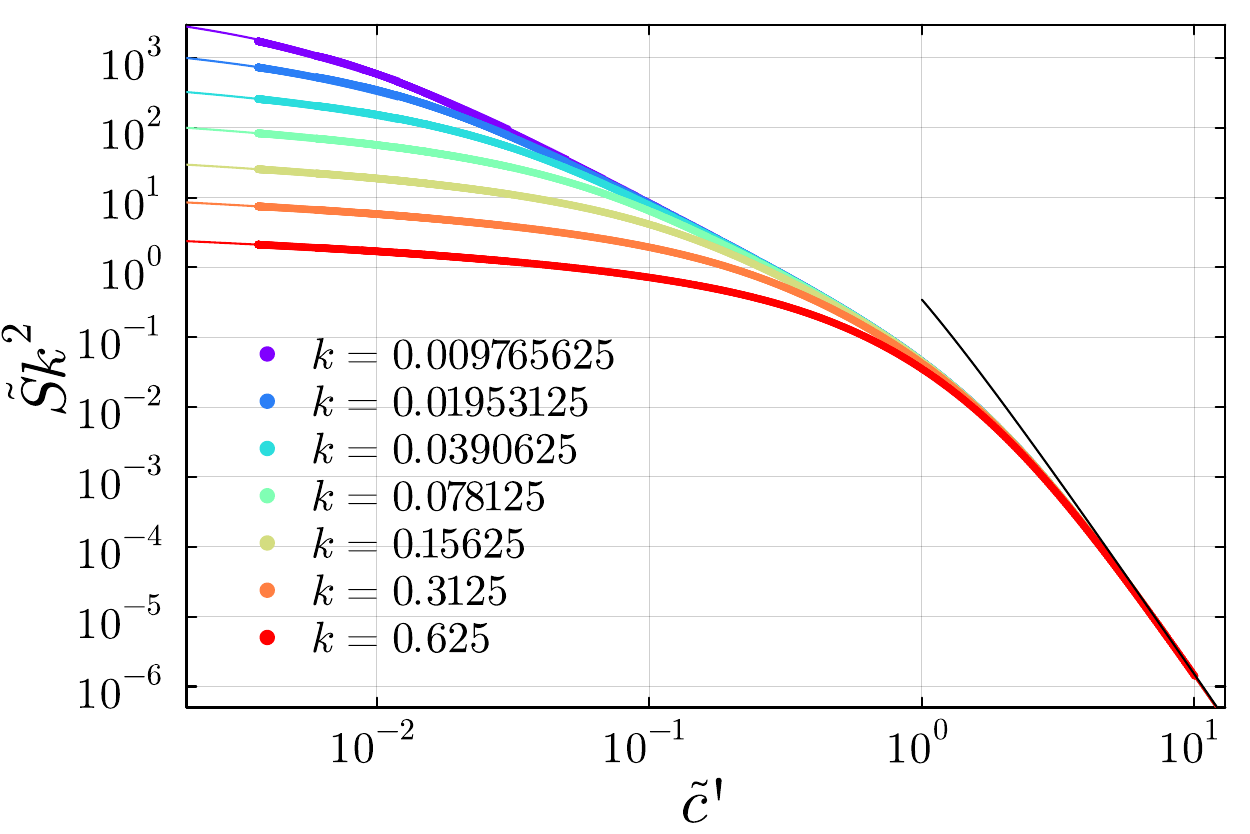}
  \caption{\label{simpvlim}The single mode contribution towards the
  entanglement entropy for the simple Pauli-Villars regularization and a range
  of cutoff ratios $k$. Colored lines are the predictions \eqref{skint}, while
  the dots represent numerical data for coordinates $n \in \{ 51, \ldots, 311
  \}$ of a discretization with radial size $N_r = 512$. In the top panel,
  data are plotted vs. $\tilde{c}$ to compare with the asymptotic form
  \eqref{daform} in the continuum limit (black line). In the bottom panel, the
  same data are plotted vs. $\tilde{c}$ and multiplied with $k^2$ for
  comparison with the asymptotic form \eqref{simpvasy} predicted by the
  hopping expansion (black line).}
\end{figure}

\begin{figure}[htb]\centering
  \includegraphics[width=\columnwidth]{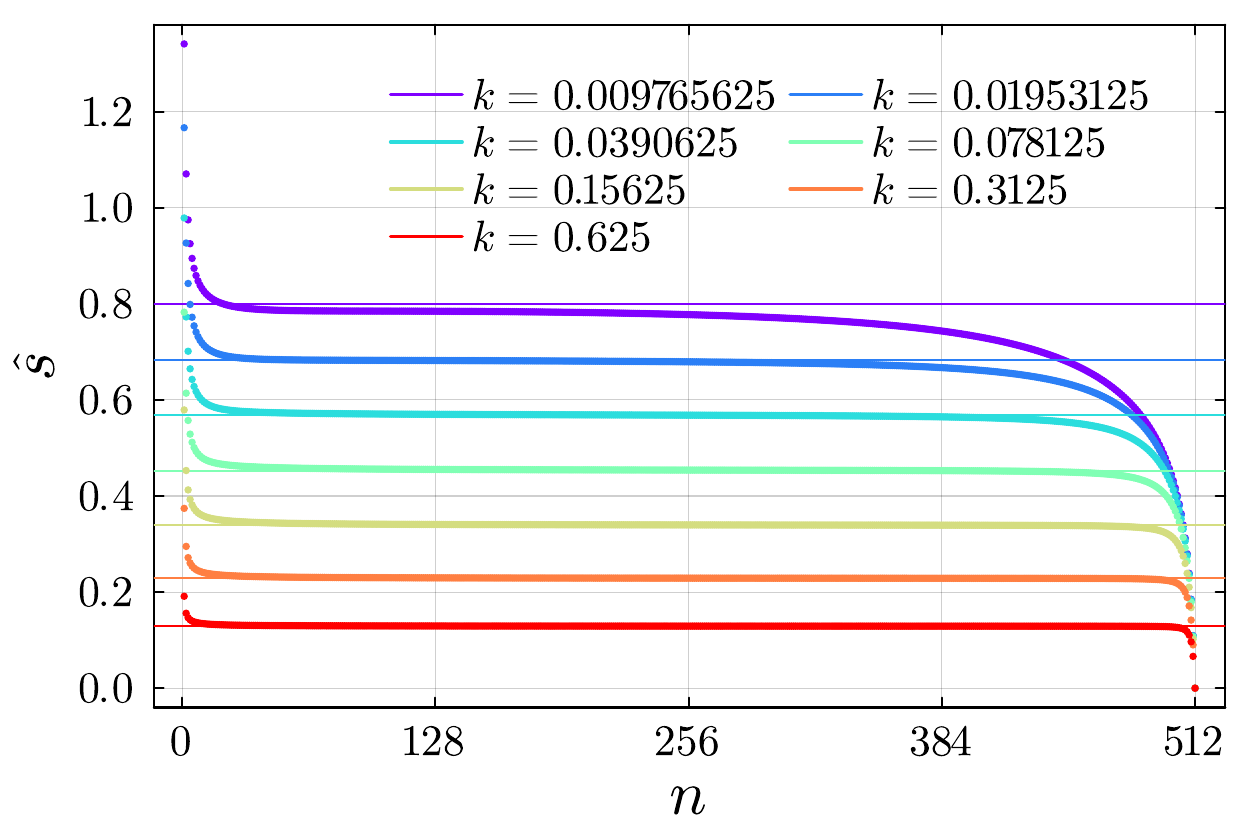}\par\medskip
  \includegraphics[width=\columnwidth]{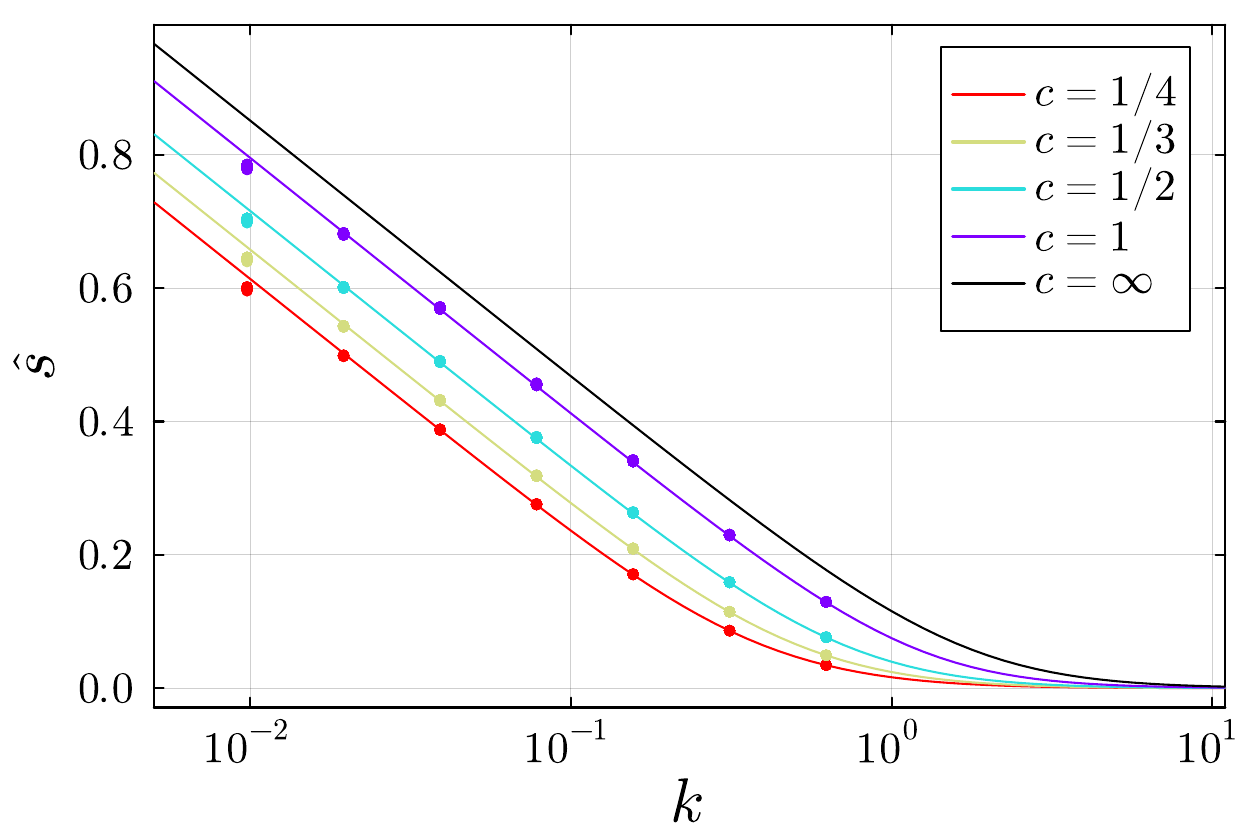}
  \caption{\label{simpshat}The Pauli-Villars normalized entropy $\hat{s}$ for
  systems of size $N_r = 512$ and with a variety of cutoff ratios $k$ and a
  single Pauli-Villars field. The lines result from evaluating the integral
  \eqref{shat} with the approximation \eqref{skint} of the universal function
  based single mode entropy, whereas the dots correspond to the direct entropy
  data. In the top panel, we plot $\hat{s}$ vs. the radial index $n$ for
  a number of different $k$ and constant $c = 1$. There is a good overall
  agreement between the two methods, but the lack of boundary corrections is
  clearly visible in the direct determination, especially towards the
  continuum limit. In the bottom panel, we plot the $k$-dependence for
  various $c$, with the direct data points restricted to the range $n \in \{
  65, \ldots, 255 \}$ to eliminate boundary effects. The logarithmic
  divergence for $k \rightarrow 0$ is clearly visible.}
\end{figure}

\begin{figure}[htb]\centering
  \includegraphics[width=\columnwidth]{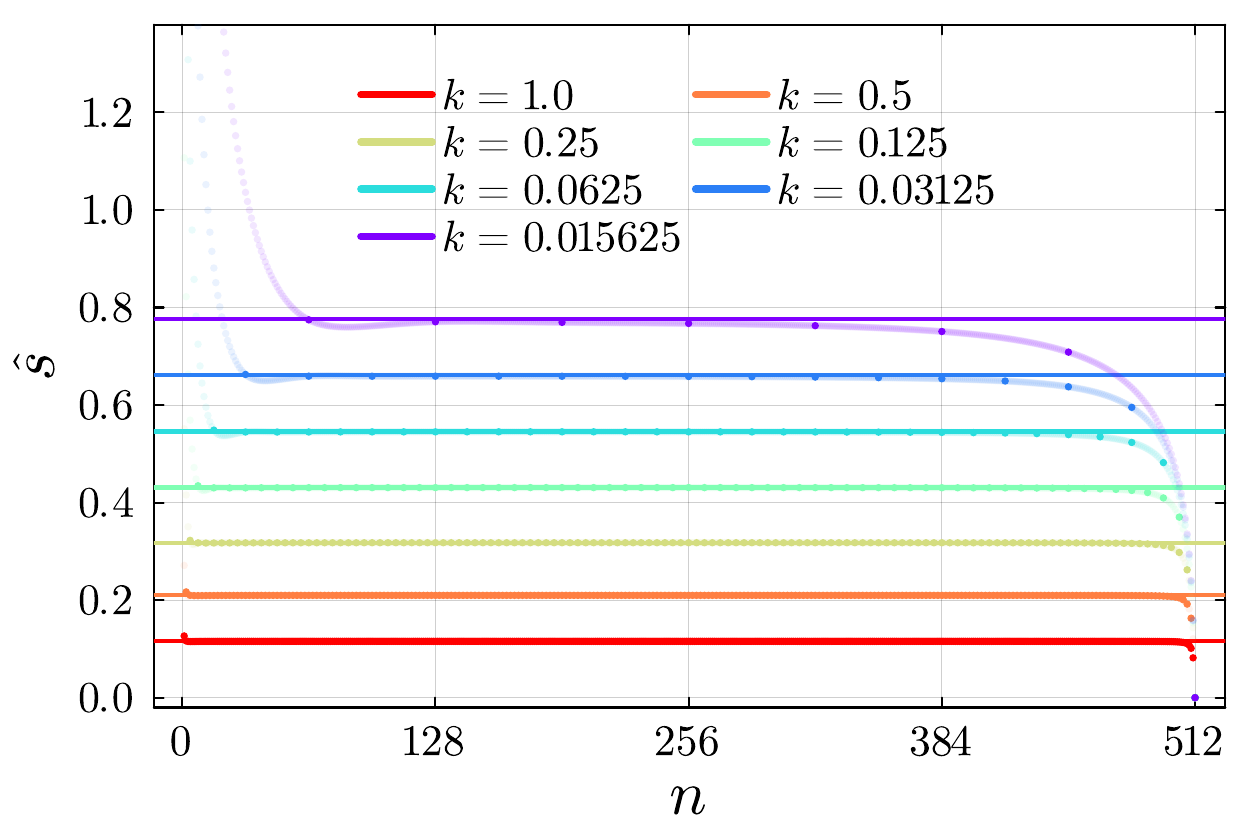}\par\medskip
  \includegraphics[width=\columnwidth]{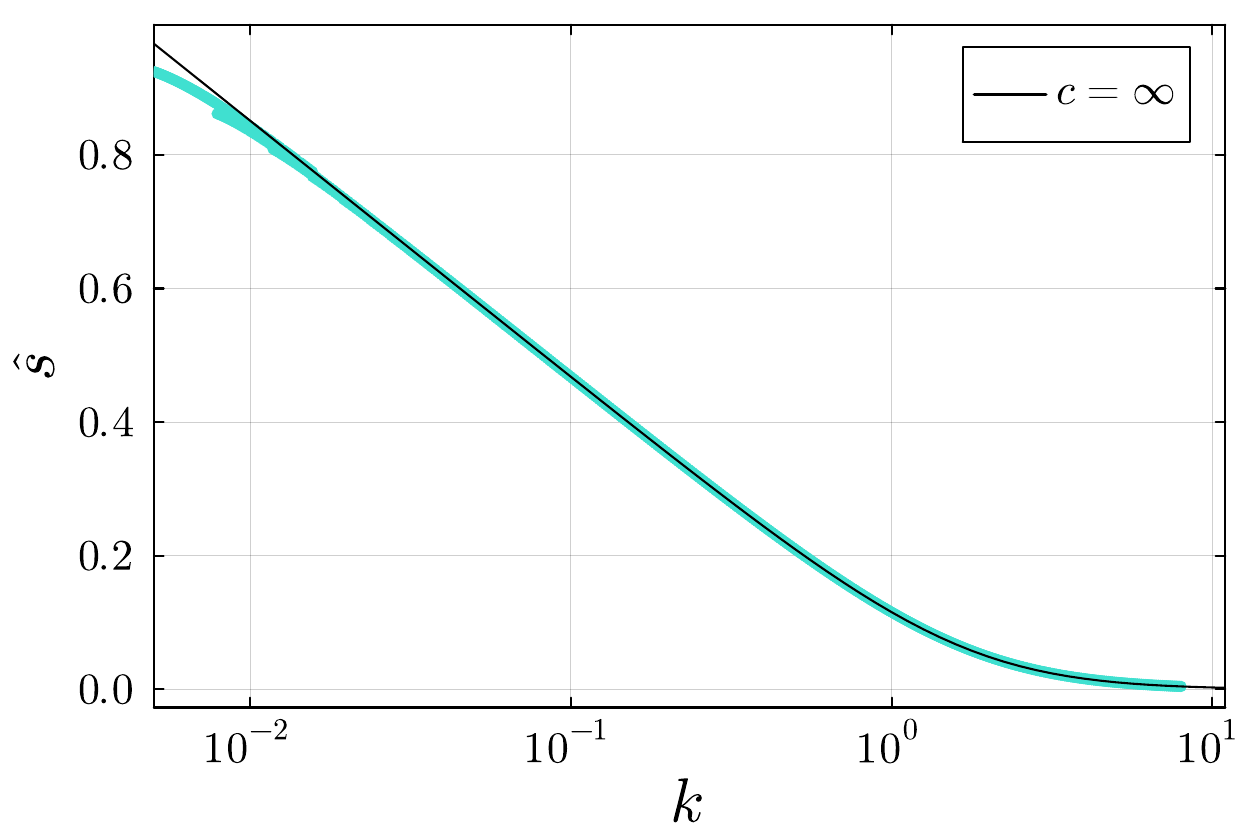}
  \caption{\label{pseudosimp}Comparison of the pseudo Pauli-Villars normalized
  entropy (\ref{pseudos}, \ref{pseudok}) (dots) with the large $n$ prediction
  \eqref{pseudoPVint} (lines). In the top panel, $\hat{s}$ is plotted vs.
  the radial index for various cutoff ratios $k$. The dark dots represent
  points where the relation \eqref{pseudok} has an exact integer solution in
  $l_{\max}$, while the lighter points are cubic spline interpolations for
  cases where it does not. In the bottom panel all available $\hat{s}$ for
  $n \in \{ 65, \ldots, 255 \}$ are plotted vs. $k$. One can observe good
  agreement except for boundary effects, which are comparable to those found
  for the regular Pauli-Villars case (compare Fig.~\ref{simpshat}).}
\end{figure}

To compare the analytical predictions to numerical data, we define the
boundary corrected
\[ \hat{c}'{}^2 = \frac{\tilde{c}'{}^2}{k^2} = \hat{c}^2 +
   \frac{\tilde{b}'{}^2}{k^2} \]
as in the brick wall case. Using the approximation $\tilde{S}^i
(\tilde{\mu})$ described in the previous section, we may also define
\begin{equation}
  \hat{S}_k^i (\hat{c}^2) = \tilde{S}^i (k^2 \hat{c}^2) - \tilde{S}^i (k^2
  (\hat{c}^2 + 1)) \label{skint}
\end{equation}
as an accurate approximation to \eqref{shatk} for finite $k$, which has the
correct asymptotics \eqref{simpvasy} and \eqref{daform}. As shown in Fig.~\ref{simpvlim}, the numerical data for a range of cutoff ratios $k$ are
excellent agreement with this approximation as well as with both asymptotic
forms.
Combining \eqref{skint} with \eqref{shat} we compute the Pauli-Villars
normalized entropy at an arbitrary cutoff ratio $k$ and anisotropy factor $c$
of the underlying brick wall discretization. The result is displayed in Fig.~\ref{simpshat}. We can see good agreement of this indirect determination with
the direct one, which clearly suffers from far larger boundary effects. The
logarithmic divergence in the $k \rightarrow 0$ limit is also clearly exposed.

There is an interesting alternative way of evaluating the Pauli-Villars
normalized entropy integral \eqref{shat}. Decomposing the entropy into the
contributions of the physical and the regulator field, we may write
\[ \hat{s} = \frac{s}{k^2} = \frac{1}{k^2} \left( \bigintlim_0^{c^2} \mathd
   \tilde{c}^2 \,  \tilde{S} (\tilde{c}^2) - \bigintlim_0^{c^2} \mathd
   \tilde{c}^2 \, \tilde{S} (\tilde{c}^2 + k^2) \right) \;. \]
With a simple change of integration variable and the substitution
$\tilde{\mu}^2 = \tilde{c}^2 + k^2$ in the second integral, this may be
written as
\[ \begin{aligned}
   \hat{s} ={}& \frac{1}{k^2} \Biggl(
   \bigintlim_0^{c^2} \mathd \tilde{\mu}^2 \, \tilde{S} (\tilde{\mu}^2)- \bigintlim_{k^2}^{k^2 + c^2} \mathd \tilde{\mu}^2 \,
   \tilde{S} (\tilde{\mu}^2) \Biggr)\\
   ={}& \frac{1}{k^2} \Biggl(
   \bigintlim_0^{k^2} \mathd \tilde{\mu}^2 \, \tilde{S} (\tilde{\mu}^2)- \bigintlim_{c^2}^{k^2 + c^2} \mathd \tilde{\mu}^2 \,
   \tilde{S} (\tilde{\mu}^2) \Biggr)\;.
   \end{aligned} \]
In the $c \rightarrow \infty$ limit, i.e.~when taking all angular momentum
modes into account, the second integral vanishes according to \eqref{lohop}.
In this case we thus obtain
\begin{equation}
  \hat{s} = \frac{1}{k^2} \bigintlim_0^{k^2} \mathd \tilde{\mu}^2 \,  \tilde{S}
  (\tilde{\mu}^2) \label{pseudoPVint}
\end{equation}
as the Pauli-Villars normalized entropy of the system. On a technical level,
this expression is just a rescaled version of the brick wall normalized
entropy \eqref{fusimo} with a different upper integration limit. We may thus
obtain the Pauli-Villars normalized entropy at infinite anisotropy factor
directly from the the massless field alone. Thus, instead of performing the
integral \eqref{shat} of the approximate Pauli-Villars scaling function
\eqref{skint} up to $c = \infty$, the identical result can be obtained by
integrating \eqref{pseudoPVint} with the approximate brick wall scaling
function $\tilde{S}^i (\tilde{c}^2)$ to the upper limit $k^2$. Furthermore,
since \eqref{fusimo} is the large $n$ approximation for the directly
measured brick wall normalized entropy, we may obtain the \emph{pseudo Pauli-Villars}
entropy from
\begin{equation}
  \hat{s} = \frac{s}{k^2} = \frac{1}{k^2} \sum_{l = 0}^{l_{\max}} \frac{2 l +
  1}{n^2} S_l \label{pseudos}
\end{equation}
with the relation
\begin{equation}
  (l_{\max} + 1) = k n \;,\label{pseudok}
\end{equation}
replacing the anisotropy relation \eqref{lnscal}. In Fig.~\ref{pseudosimp}, we
compare the behavior of this directly determined normalized entropy with the
prediction \eqref{pseudoPVint}. One can clearly see that the direct data
match the prediction well, except for very small $k$ where boundary effects
are stronger. Numerically, this approach is much cheaper, as it does not
require the Pauli-Villars field at all. In addition, results can be obtained
for a number of different cutoff ratios $k$ by varying the number of
$l$-modes, and it is cheapest to just consider the limit of an infinite
anisotropy factor.

\subsection{Flat spacetime with polynomial Pauli-Villars regularization}
\begin{figure}[htb]\centering
  \includegraphics[width=\columnwidth]{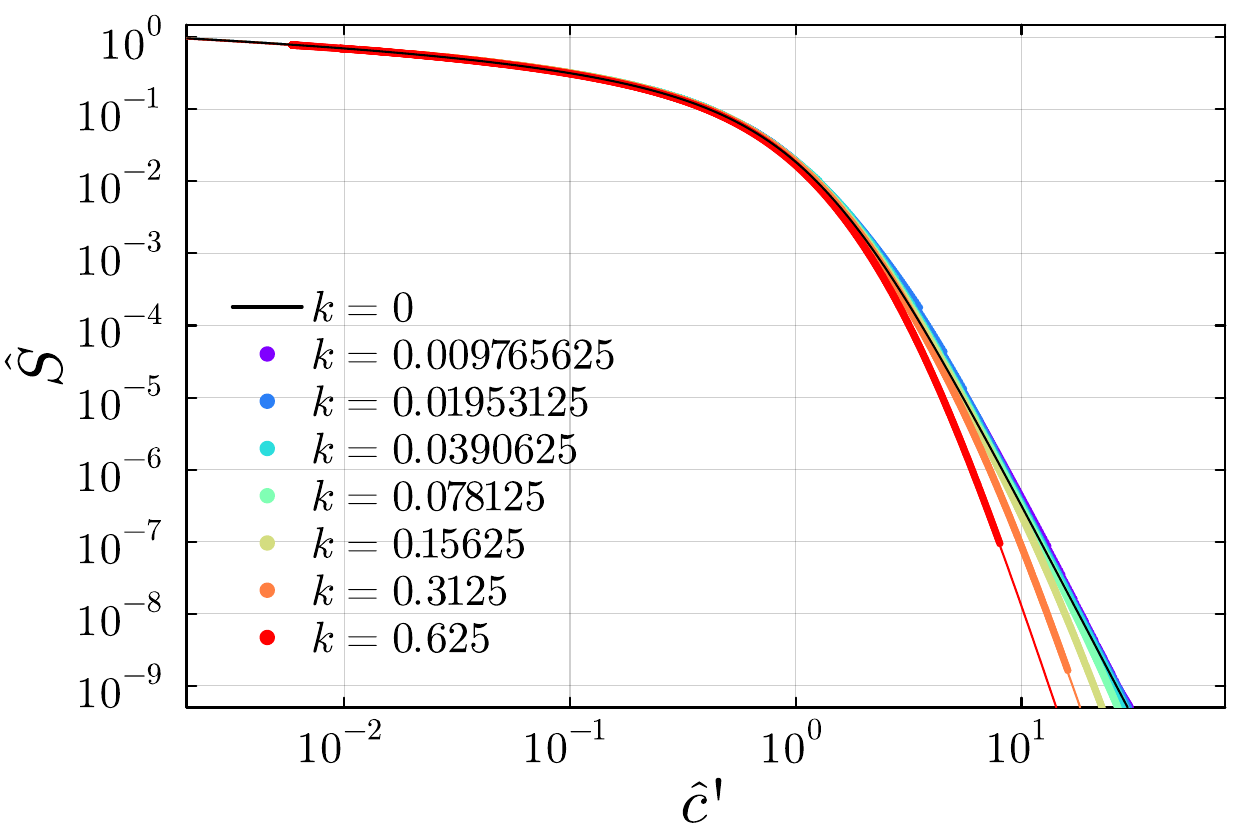}\par\medskip
  \includegraphics[width=\columnwidth]{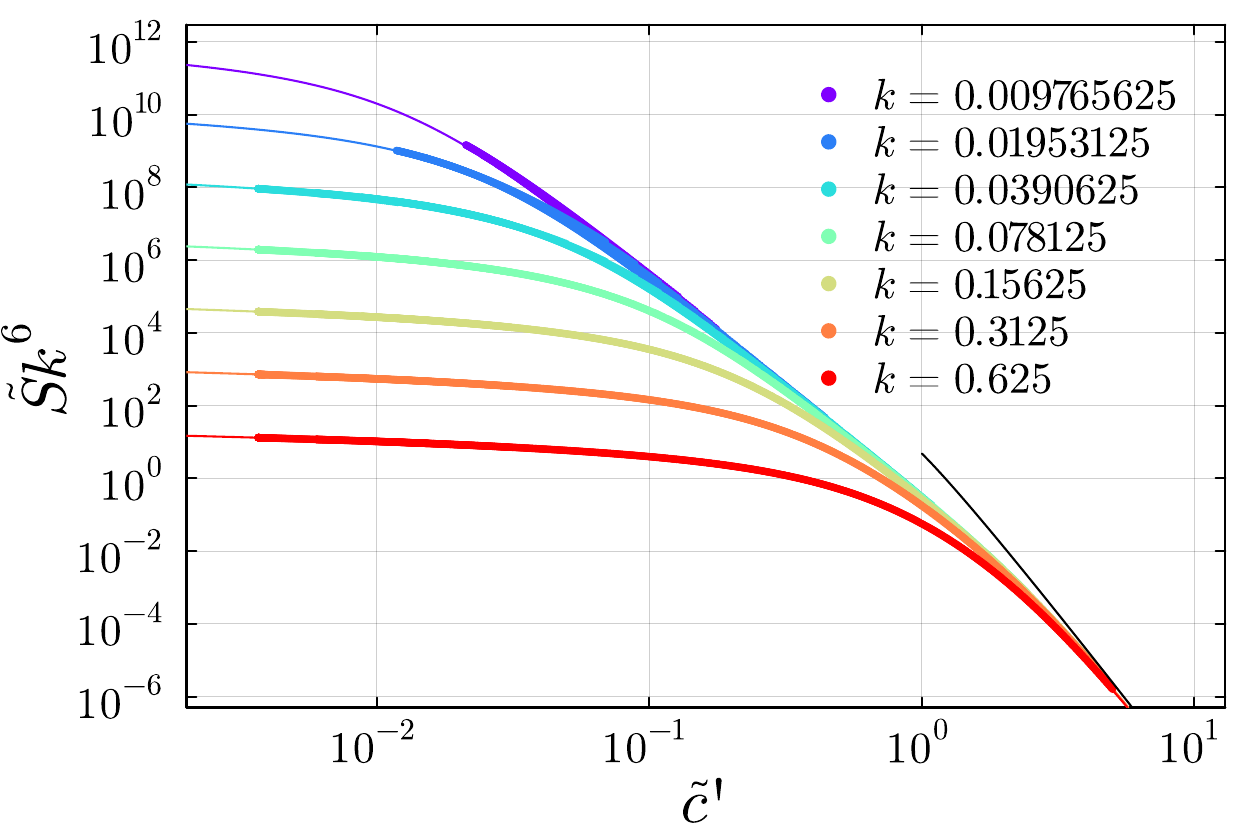}
  \caption{\label{polyvlim}The single mode contribution towards the
  entanglement entropy for the polynomial Pauli-Villars regularization and a
  range of cutoff ratios $k$. Colored lines are the predictions \eqref{pkint},
  while the dots represent numerical data for coordinates $n \in \{ 51,
  \ldots, 311 \}$ of a discretization with radial size $N_r = 512$. In the
  top panel, data are plotted vs. $\hat{c}$ to compare with the asymptotic
  form \eqref{polyform} in the continuum limit (black line). In the bottom
  panel, the same data are plotted vs. $\tilde{c}$ and multiplied with $k^6$
  for comparison with the asymptotic form \eqref{polyvasy} predicted by the
  hopping expansion (black line). For the two smallest $k$, a lower cut on $l$
  of $l_{\min} = 6$ resp. $l_{\min} = 3$ was introduced to eliminate points
  that are severely affected by boundary effects.}
\end{figure}

Due to the logarithmic divergence of the properly normalized entropy in the
continuum limit, a single Pauli-Villars field does not provide a satisfactory
regulator even for the free theory. As argued in Sect.~\ref{reg},
one may augment this by additional regulator fields. We first investigate the
polynomial Pauli-Villars regulator as described in Tab.~\ref{poltab}, which
was pioneered in the context of semiclassical collapse in
\cite{Berczi:2021hdh}. We again use \eqref{normentpv} and \eqref{cur} to
define normalized entropy $\hat{s}$ and a cutoff ratio $k$, where it is
implied that $M_{\tmop{PV}}$ refers to the lightest of the regulator fields,
i.e.~the field with the index $i = 4$ in Tab.~\ref{poltab}. This definition is
convenient in the sense that all Pauli-Villars masses lie at or above the
regulator scale. One has to keep in mind though that the heaviest
Pauli-Villars mass coincides with the brick wall scale $\Delta^{- 1}$ already
at $k = 0.5$ instead of $k = 1$. In analogy to \eqref{stildek} we now obtain the single mode entropy
contribution
\[ \tilde{S}_k^P (\tilde{c}^2) = \sum_i p_i \tilde{S} (\tilde{c}^2 +
   \kappa_i^2 k^2) \qquad \kappa_i = \frac{M_i}{M_{\tmop{PV}}} \]
with the $p_i$ and $M_i$ from Tab.~\ref{poltab}. This can be expressed in
terms of the single mode entropy contributions \eqref{stildek} of the simple Pauli-Villars
regularization as
\[ \tilde{S}_k^P (\tilde{c}^2) = - \sum_{i > 1} p_i \tilde{S}_{k_i}
   (\tilde{c}^2) \;.\]
\begin{figure}[htb]\centering
  \includegraphics[width=0.96\columnwidth]{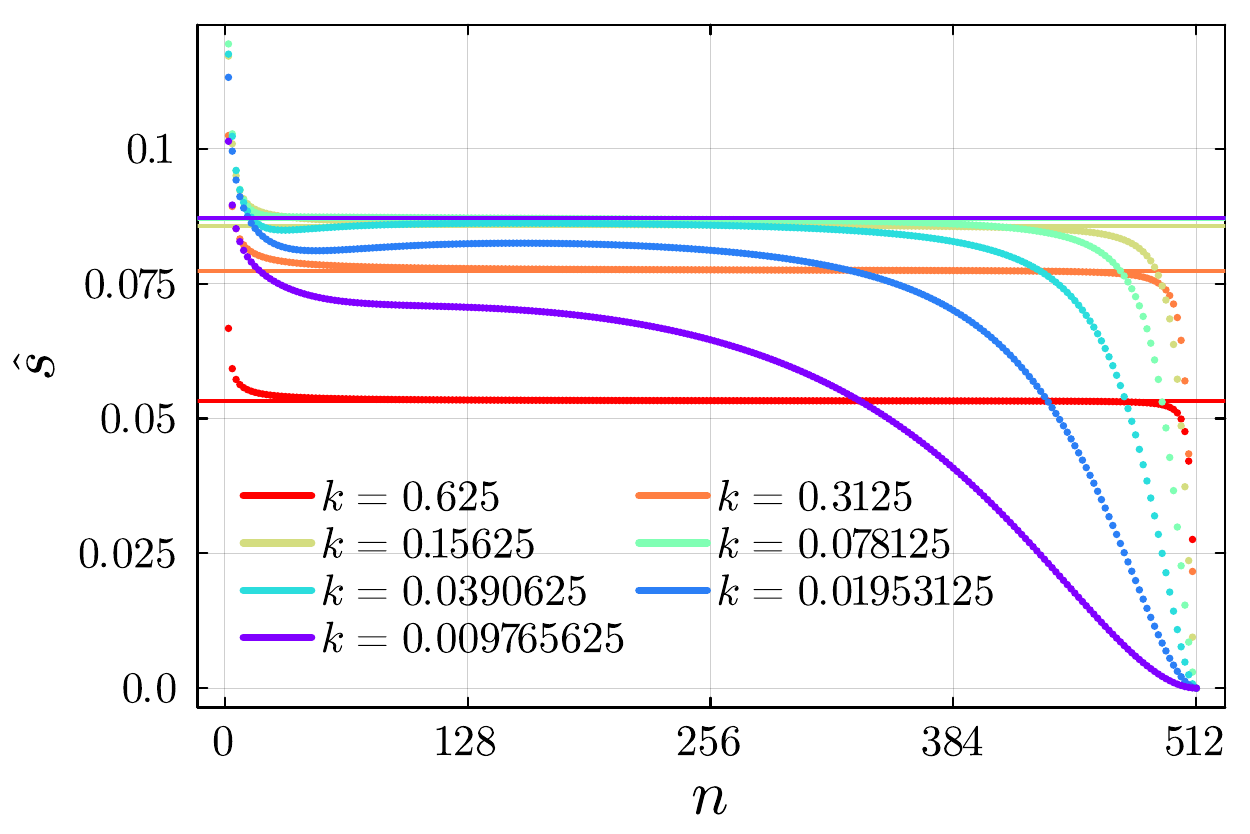}\par\medskip
  \includegraphics[width=0.96\columnwidth]{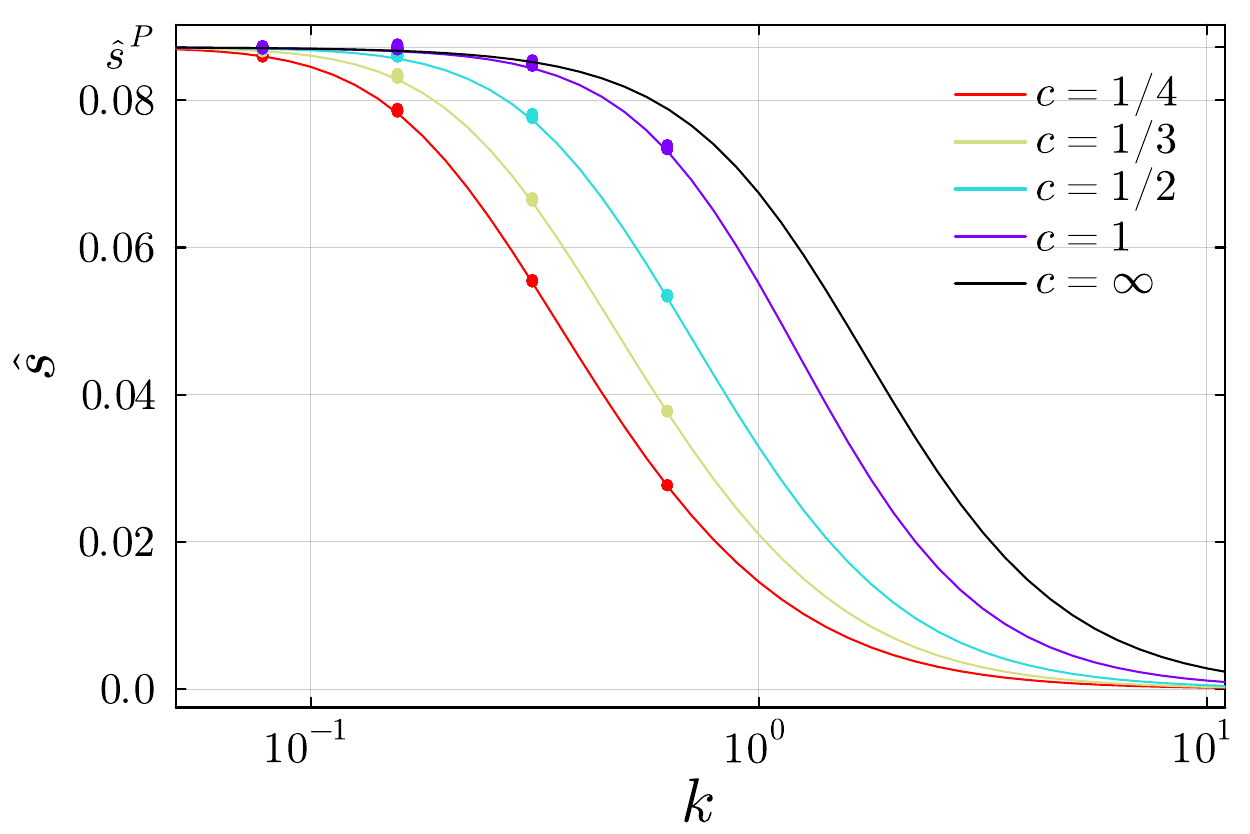}
  \caption{\label{polyshat}The Pauli-Villars normalized entropy $\hat{s}$ for
  systems of size $N_r = 512$ and with a variety of cutoff ratios $k$ and in
  the Pauli-Villars scheme. The lines result from evaluating the integral
  \eqref{shat} with the approximation \eqref{pkint} of the universal function
  based single mode entropy, whereas the dots correspond to the direct entropy
  data. In the top panel, we plot $\hat{s}$ vs. the radial index $n$ for
  a number of different $k$ and constant $c = 1 / 2$. There is a good overall
  agreement between the two methods, except for the smallest $k$, where
  boundary corrections are clearly visible in the direct determination. This
  effect is more pronounced than in the simple Pauli-Villars case due to
  overall larger cancellations. In the bottom panel, we plot the
  $k$-dependence for various $c$, with the direct data points restricted to
  the range $n \in \{ 65, \ldots, 255 \}$ to eliminate boundary effects.
  Contrary to the simple Pauli-Villars case, the continuum limit $k
  \rightarrow 0$ of $\hat{s}$ is finite. The the analytic value
  \eqref{polycontlimshat} indicated in the figure.}
\end{figure}

Plugging in the $\tilde{c} \gg 1$ asymptotic form \eqref{simpvasy} and noting
that \eqref{pm20} implies
\[ \sum_{i > 1} p_i \kappa_i^2 = \sum_{i > 1} p_i \kappa_i^4 = 0 \]
one finds that the leading and in fact also the next to leading order term
cancels. To obtain the $\tilde{c} \gg 1$ asymptotic form we expand \eqref{lohop} to higher orders, which results in
\begin{equation}
  \tilde{S}_k (\tilde{c}^2) \xrightarrow{\tilde{c} \rightarrow \infty} -
   k^6 \frac{24 \ln (2 \tilde{c}) -
  7}{\tilde{c}^{10}}\sum_{i > 1} \frac{p_i \kappa_i^6}{24} \label{polyvasygen}
\end{equation}
for any renormalization scheme fulfilling \eqref{pm20}. For the polynomial Pauli-Villars scheme we find
\[ - \sum_{i > 1} \frac{p_i \kappa_i^6}{24} = \frac{1}{2} \;,\]
so
\begin{equation}
  \tilde{S}_k^P (\tilde{c}^2) \xrightarrow{\tilde{c} \rightarrow \infty} k^6
  \frac{24 \ln (2 \tilde{c}) - 7}{2 \tilde{c}^{10}}\;. \label{polyvasy}
\end{equation}
To investigate the physically more interesting $\tilde{c} \ll 1$ limit, we
define, in analogy to the simple Pauli-Villars case
\[ \hat{S}_k^P (\hat{c}^2) = \tilde{S}_k^P (k^2 \hat{c}^2) = - \sum_{i > 1}
   p_i \hat{S}_{k_i} \left( \frac{\hat{c}}{\kappa_i} \right) \;.\]
Plugging in the asymptotic relation \eqref{asyskhat} and defining the single
mode entropy in the continuum limit
\[ \hat{S}^P (\hat{c}^2) = \lim_{k \rightarrow 0} \hat{S}^P_k (\hat{c}^2) \]
we then find
\begin{equation}
  \hat{S}^P (\hat{c}^2) = - \frac{1}{12} \sum_{i > 1} p_i \ln \left(
  \frac{\kappa_i^2}{\hat{c}^2} + 1 \right) \;.\label{genform}
\end{equation}
With the explicit values in Tab.~\ref{poltab}, this can be simplified to
\begin{equation}
  \hat{S}^P (\hat{c}^2) = \frac{1}{12} \ln \left( 1 + \frac{4}{\hat{c}^2 (3 +
  \hat{c}^2)^2} \right)\;. \label{polyform}
\end{equation}
The total Pauli-Villars normalized entropy $\hat{s}$ at an arbitrary cutoff
ratio $k$ can now be computed by plugging $\hat{S}_k^P (\hat{c})$ into
\eqref{shat}. Interestingly, due to the quadratic cancellation in
\eqref{pm20}, the continuum limit may now be taken and the corresponding
Pauli-Villars normalized entropy at vanishing brick wall cutoff $k = 0$ is
\begin{equation}
  \hat{s} = \frac{1}{12} \sum_{i > 1} p_i \kappa_i^2 \ln (\kappa_i^2)
  \label{shatpvpoly}
\end{equation}
in the general case and
\begin{equation}
  \hat{s}^P = \frac{1}{2} \ln (3) - \frac{2}{3} \ln (2) \simeq 0.087208
  \label{polycontlimshat}
\end{equation}
for the polynomial Pauli-Villars regulator of Tab.~\ref{poltab} \footnote{Restoring dimensions in Eq.~\eqref{shatpvpoly}, we have $S=\frac{A}{48\pi}\sum_ip_iM_i^2\ln M_i^2/\Lambda^2$, with $A=4\pi r^2$, the leading term proportional to the area of the Pauli-Villars regulated entropy. It was also calculated in \cite{Demers:1995dq} for a non-extremal Reissner-Nordström black hole, where they showed it to be equal to the one-loop contribution to the Bekenstein-Hawking entropy. Vanishing of the prefactor would therefore imply that the matter fields do not renormalize Newton's constant.}.

We may also introduce an approximate single mode entropy function at finite
cutoff ratio $k$ based on \eqref{skint} for the simple Pauli-Villars
subtraction
\begin{equation}
  \hat{S}_k^{P, i} (\hat{c}^2) = - \sum_{i > 1} p_i \hat{S}^i_{k_i} \left(
  \frac{\hat{c}^2}{\kappa_i^2} \right) \;. \label{pkint}
\end{equation}
In Fig.~\ref{polyvlim} we compare these predictions, as well as the asymptotic forms \eqref{polyvasy} and \eqref{polyform}, to boundary corrected numerical data and find good agreement.

\begin{figure}[htb]\centering
  \includegraphics[width=\columnwidth]{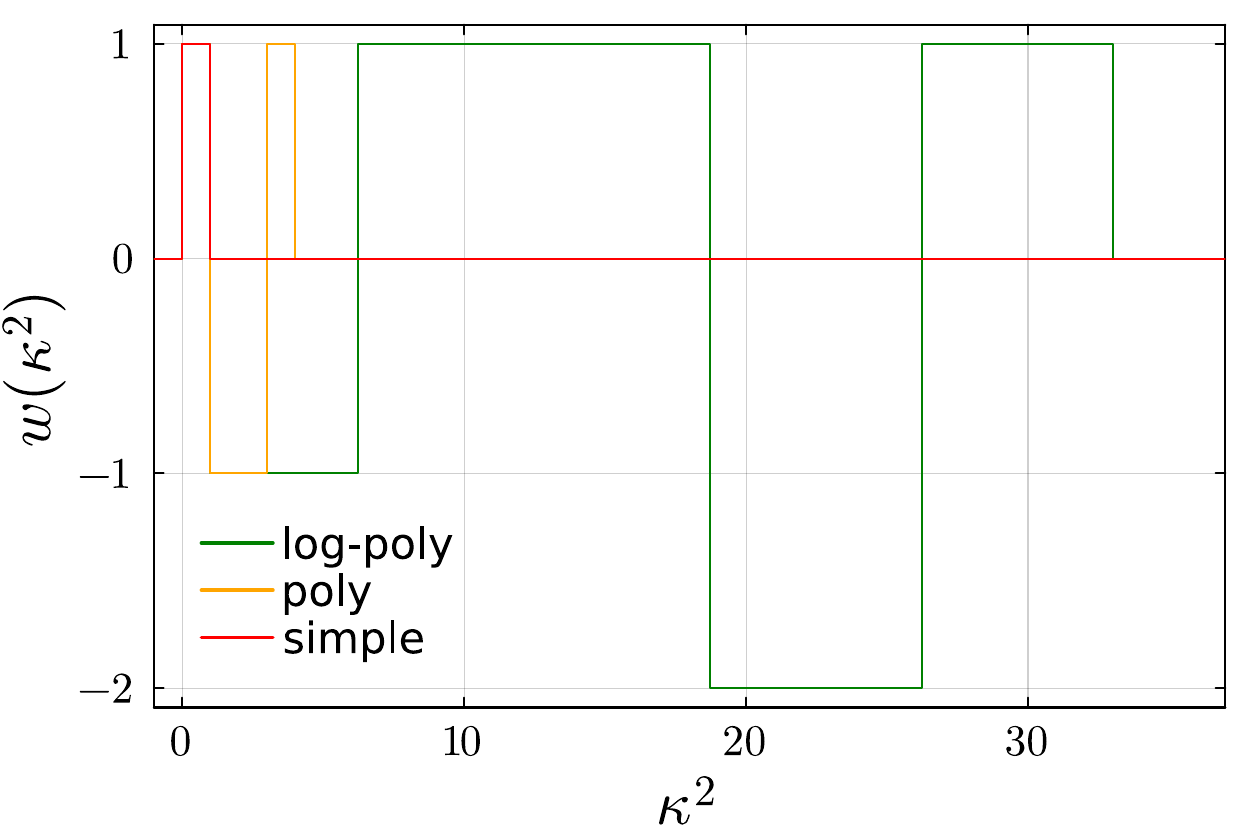}
  \caption{\label{figweights}A comparison of the weight function
  \eqref{pvweights} for the simple, polynomial (Tab.~\ref{poltab}) and
  log-polynomial (Tab.~\ref{poltablog}) Pauli-Villars regulators.}
\end{figure}

The Pauli-Villars normalized entropy for any cutoff ratio $k$
and anisotropy factor $c$ can now be obtained by integrating \eqref{pkint} according to \eqref{shat}. The
result is displayed in Fig.~\ref{polyshat}. Due to the substantially larger
cancellations than in the simple Pauli-Villars case, the direct data show
substantially more boundary effects, which renders them useless for very small
$k$. Still, the approach of the normalized entropy $\hat{s}$ to the finite
continuum limit \eqref{polycontlimshat} is clearly visible even in the direct
data. The polynomial Pauli-Villars scheme thus has the same qualitative
behavior as the brick wall scheme, with an entanglement entropy that diverges
quadratically in the cutoff scale, although the proportionality constant is
analytically known and smaller than in the brick wall case.

As in the case of the simple Pauli-Villars regulator, we may decompose the
normalized entropy into separate integrals over the physical and regulator
fields. We then obtain
\[ \hat{s} = - \frac{1}{k^2} \sum_{i > 1} p_i \bigintlim_0^{\kappa_i^2 k^2}
   \mathd \tilde{\mu}^2 \,  \tilde{S} (\tilde{\mu}^2) \]
as a straightforward generalization of \eqref{pseudoPVint} for any number of
regulator fields, provided only that the sum over the $p_i$ vanishes. We may
rewrite this integral as
\begin{equation}
  \hat{s} = \frac{1}{k^2} \bigintlim \mathd \tilde{\mu}^2 \, w \left(
  \frac{\tilde{\mu}^2}{k^2} \right) \tilde{S} (\tilde{\mu}^2)
  \label{shatpseudoPV}
\end{equation}
where $w (\kappa^2)$ is a weight function dependent on the particular
regulator scheme, which may be written as
\begin{equation}
  w (\kappa^2) = \sum_{i = 1}^n p_i \Theta (\kappa^2 - \kappa^2_i)\;.
  \label{pvweights}
\end{equation}
For better visualization, Fig.~\ref{figweights} compares the weight function
for the various regulators used in this work. Note in particular, that the
conditions \eqref{pm2log} imply a vanishing total integral over terms constant
and linear in $\tilde{\mu}^2$. Applied to the asymptotic form \eqref{snobo},
we thus find
\begin{equation}
  \hat{s} \xrightarrow{k \rightarrow 0} \frac{1}{12} \sum_{i > 1} p_i
  \kappa_i^2 \ln (\kappa_i^2) + O (\kappa_i^6 k^4) \label{shatpseudolim}
\end{equation}
in perfect agreement with \eqref{shatpvpoly}.

\begin{figure}[htb]\centering
  \includegraphics[width=\columnwidth]{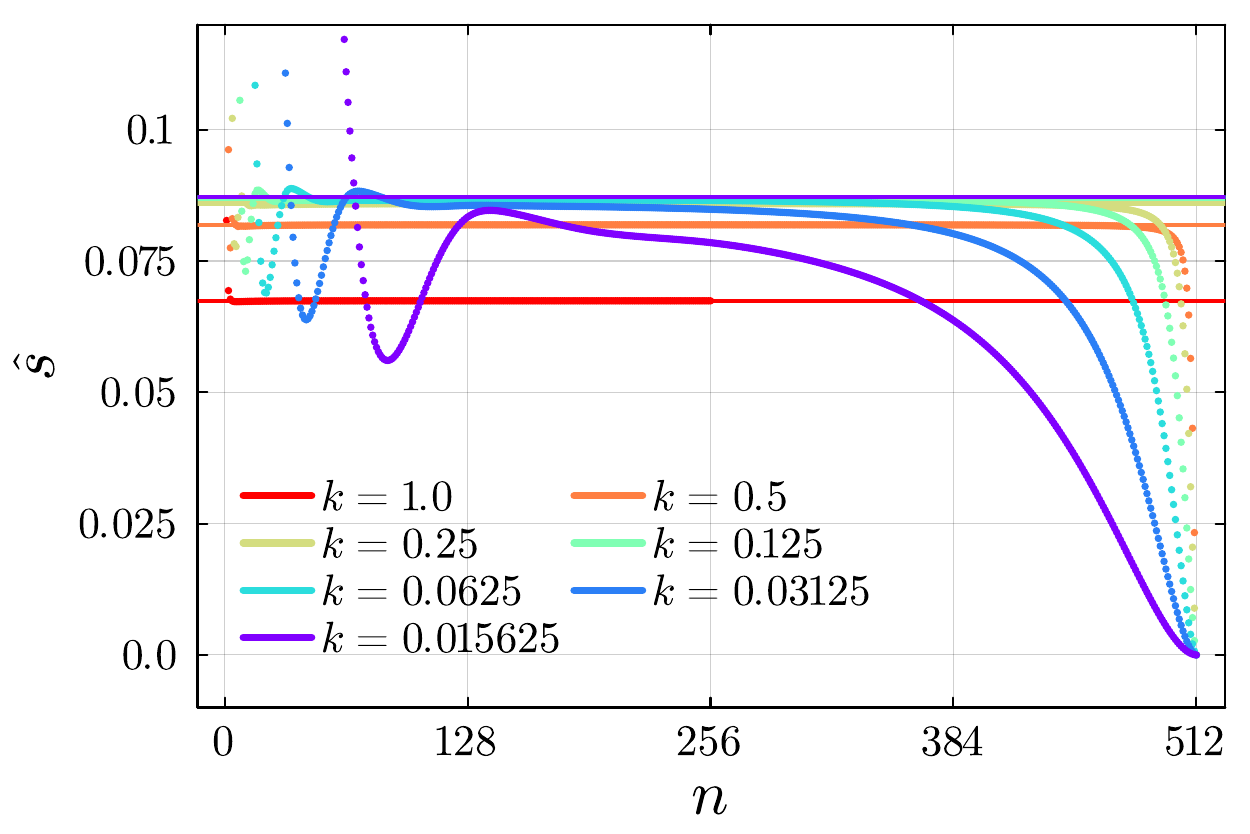}\par\medskip
  \includegraphics[width=\columnwidth]{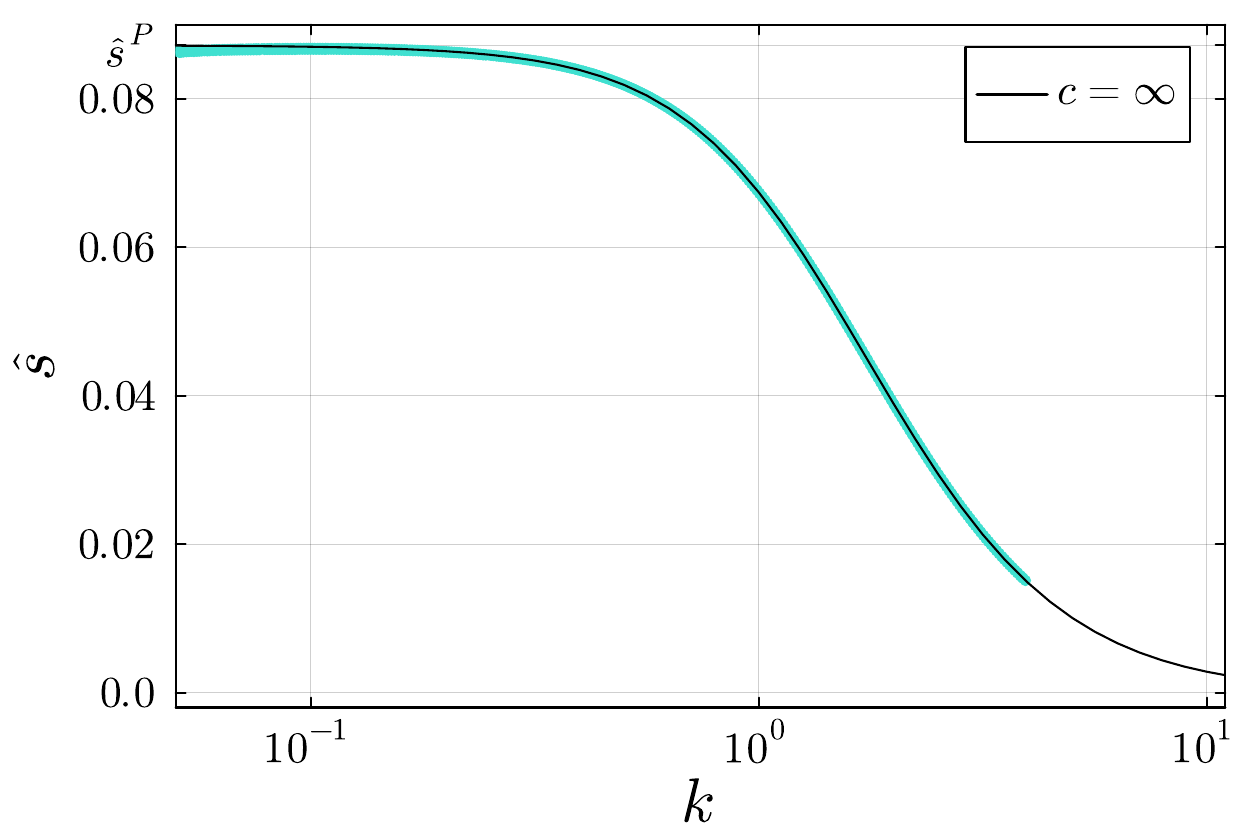}
  \caption{\label{pseudoPVpoly}Predictions \eqref{shatpseudoPV} (lines) and
  direct data \eqref{shatpseudodirect} (dots) for the normalized entropy in
  the polynomial pseudo Pauli-Villars scheme. In the top panel, the
  normalized entropy for various cutoff ratios $k$ is plotted vs. the radial
  index $n$, while in the bottom panel the normalized entropy is plotted
  vs. $k$ for the radial indices $n \in \{ 65, \ldots, 255 \}$. Note that
  since we have only tracked $512$ angular momentum modes in our simulation,
  we only have data for radial indices $n \leqslant 512 / (\kappa_{\max} k) =
  256 / k$.}
\end{figure}

In Fig.~\ref{pseudoPVpoly} we compare the prediction \eqref{shatpseudoPV} to
the direct measurement of the pseudo Pauli-Villars normalized entropy. This
direct measurement can in principle proceed as a straightforward extension of
the simple Pauli-Villars case \eqref{pseudos}. Denoting the brick wall
normalized entropy at a certain radial index $n$ and maximum number of angular
momenta $l_{\max} = \tilde{k} n - 1$ as $s_{n, \tilde{k}}$, we can in
principle compute the pseudo Pauli-Villars normalized entropy at a cutoff
ratio $k$ as
\begin{equation}
  \hat{s} = - \frac{1}{k^2} \sum_{i > 1} p_i s_{n, k \kappa_i}\;.
  \label{shatpseudodirect}
\end{equation}
However, generically $k \kappa_i n$ is not an integer for all $\kappa_i$
simultaneously, so at least some of the $s_{n, k \kappa_i}$ have to be
interpolated. The direct measurement data presented in Fig.~\ref{pseudoPVpoly}
correspond to such interpolations at a fixed $n$. The agreement between
prediction and direct data is comparable to that of the full Pauli-Villars
case, except for points at small $k$ near the inner boundary, where
discretization effects are more pronounced. This is no surprise, since for the
smallest $k$ displayed in the figure, $k n = 1$ is only reached at $n = 64$.
Thus for all $n$ that are not substantially larger than 64, we are dealing
with fractions of the entropy carried by the lowest angular momentum modes
where interpolation artifacts are supposed to be very prominent. We expect
that better, more physically motivated interpolators than the simple cubic
splines we used might provide substantially better results, but we will not
investigate this here.

\subsection{Flat spacetime with log-polynomial Pauli-Villars regularization}

\begin{figure}[htb]\centering
  \includegraphics[width=\columnwidth]{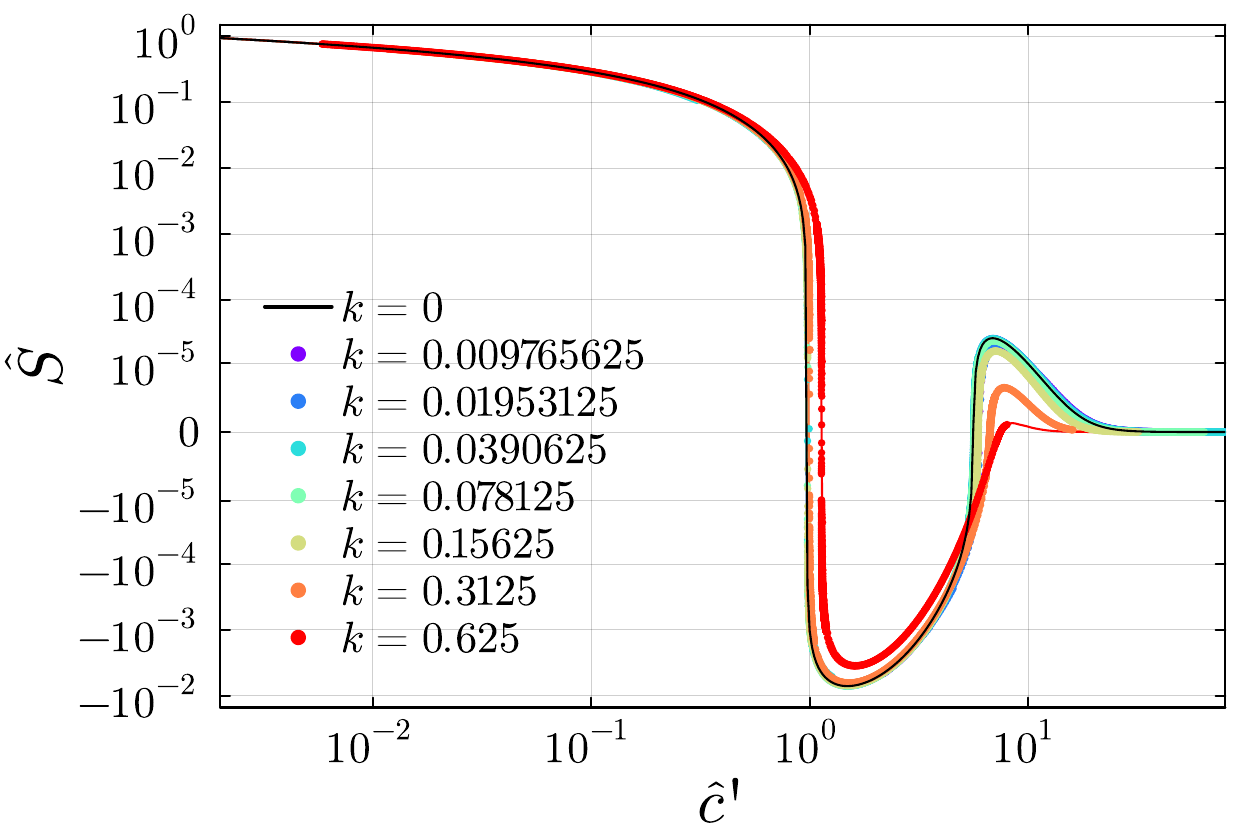}\par\medskip
  \includegraphics[width=\columnwidth]{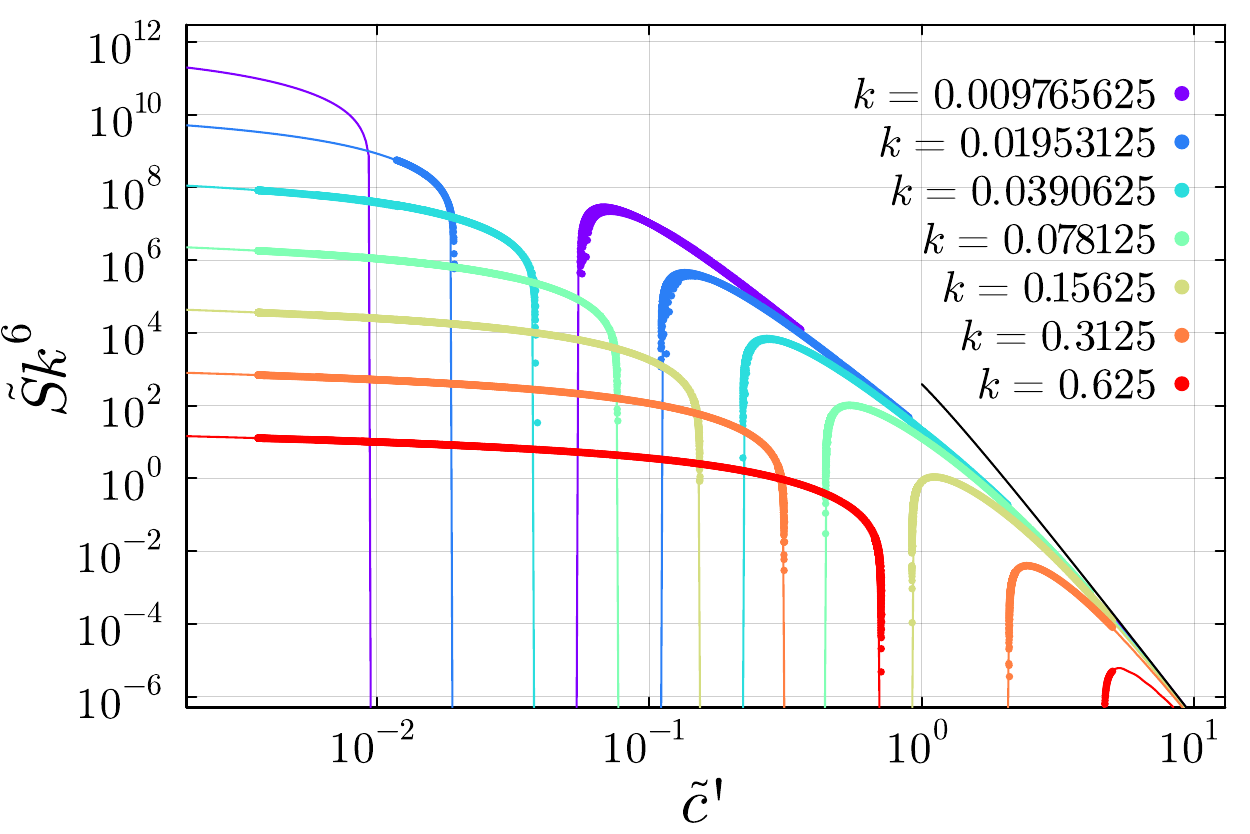}
  \caption{\label{logpolyvlim}The single mode contribution towards the
  entanglement entropy for the log-polynomial Pauli-Villars regularization and
  a range of cutoff ratios $k$. Colored lines are the predictions
  \eqref{lkint}, while the dots represent numerical data for coordinates $n
  \in \{ 51, \ldots, 311 \}$ of a discretization with radial size $N_r = 512$.
  In the top panel, data are plotted vs. $\hat{c}$ to compare them with the
  continuum asymptotic form \eqref{logpolyform} (black line). In
  the bottom panel, the same data are plotted vs. $\tilde{c}$ and multiplied
  with $k^6$ for comparison with the asymptotic form \eqref{logpolyvasy}
  predicted by the hopping expansion (black line). For the two smallest $k$, a
  lower cut on $l$ of $l_{\min} = 6$ resp.~$l_{\min} = 3$ was introduced to
  eliminate points that are severely affected by boundary effects.}
\end{figure}
The introduction of the polynomial Pauli-Villars scheme was motivated by a
partial cancellation of the divergent terms in the energy-momentum tensor, as
exposed by the Hadamard form. The obvious next step is a scheme that provides
a full cancellation of those terms, which is achieved by the log-polynomial
Pauli-Villars scheme, as defined by (\ref{p0}-\ref{log0}). We choose the
explicit realization given in Tab.~\ref{poltablog}, which consists of 6
scalar and 6 regulator fields, each arranged in multiplets of $1$, $2$ and $3$
masses respectively. The mass ratios are no more rational, as they are
solutions of transcendental equations, which we had to find numerically. We
again define the cutoff ratio $k = \Delta M_{\tmop{PV}}$ with $M_{\tmop{PV}}$
referring to the mass of the lightest regulator field, which happens to be the
one with index $i = 4$ in Tab.~\ref{poltablog}.
\begin{figure}[htb]\centering
  \includegraphics[width=\columnwidth]{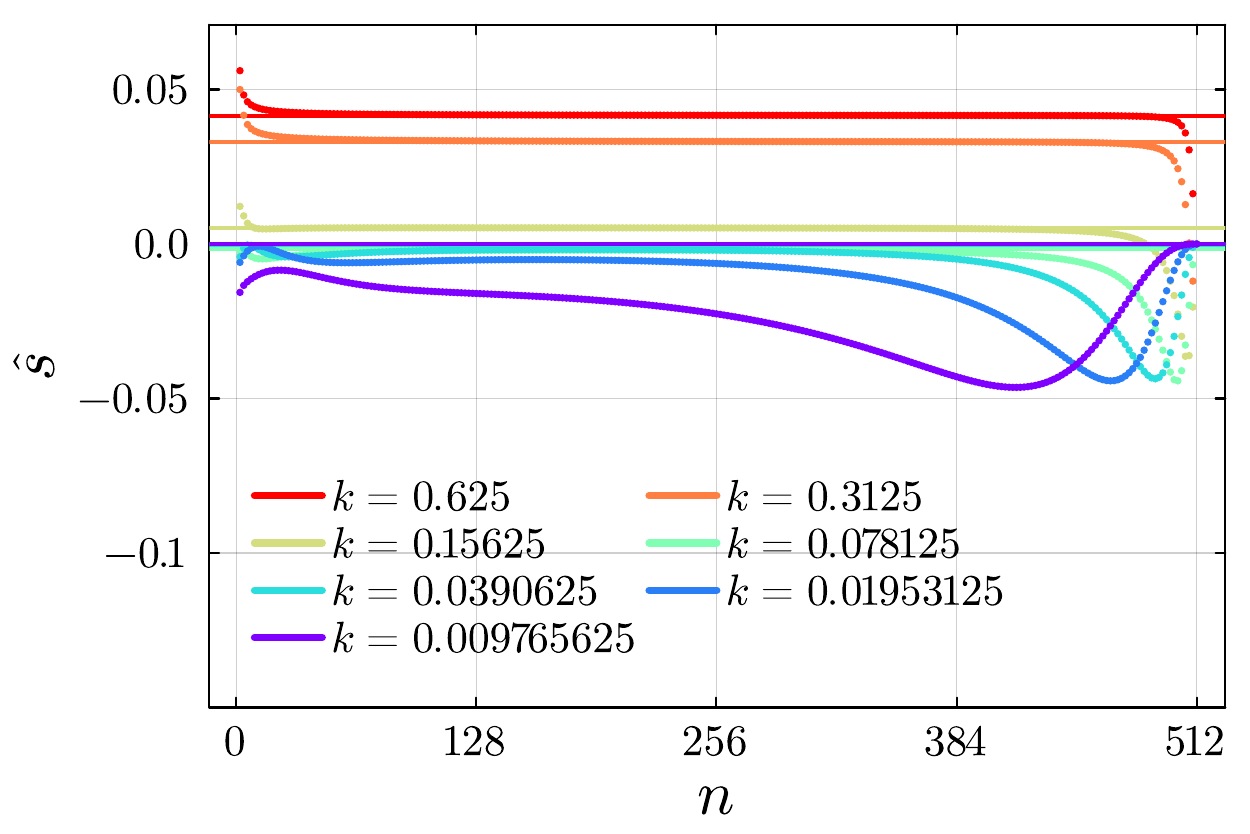}\par\medskip
  \includegraphics[width=\columnwidth]{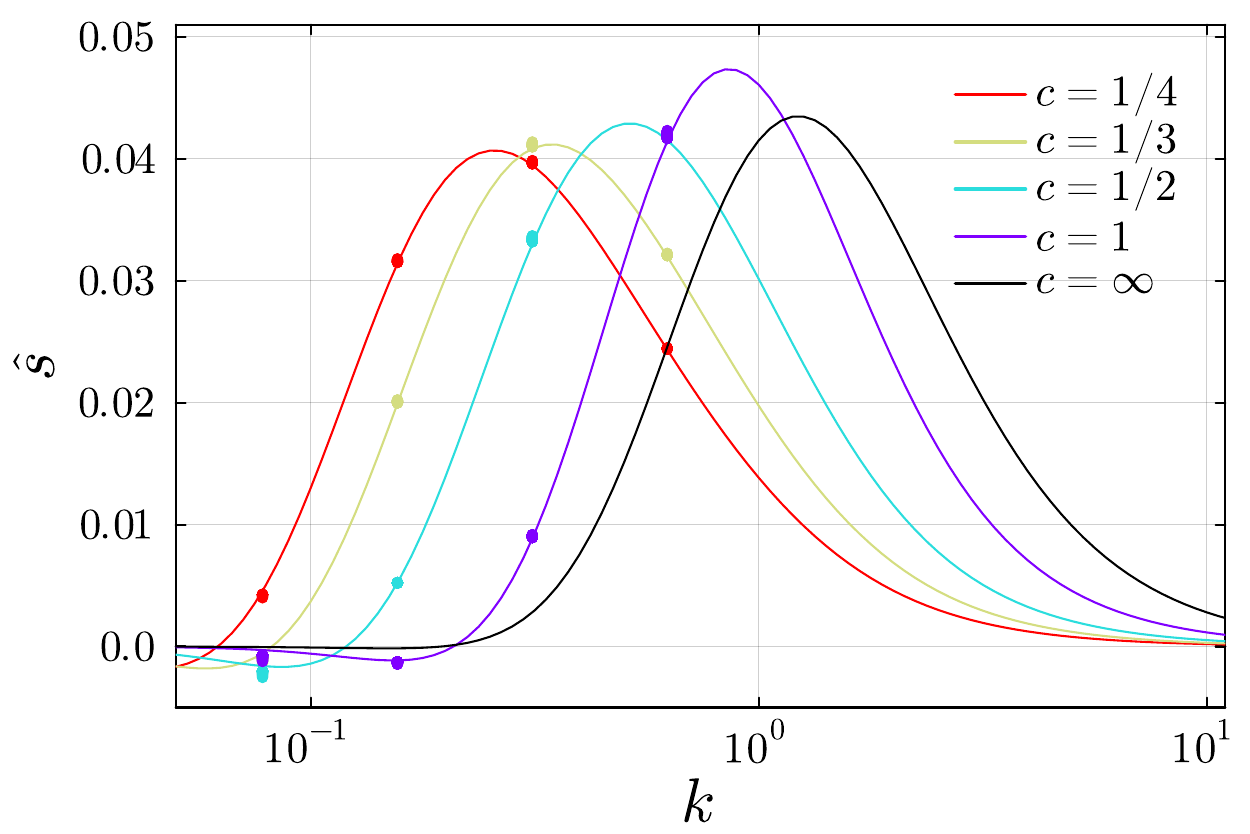}
  \caption{\label{logpolyshat}The Pauli-Villars normalized entropy $\hat{s}$ in the log-polynomial regularization
  for systems of size $N_r = 512$ at a variety of cutoff ratios $k$. The lines represent the integral
  \eqref{shat} over the approximation \eqref{lkint} of the universal function, whereas the dots correspond to direct entropy
  data. In the top panel, we plot $\hat{s}$ vs. the radial index $n$ for
  a number of different $k$ and constant $c = 1 / 2$. There is a good overall
  agreement between the two methods, except for the smallest $k$, where
  boundary corrections are clearly visible in the direct determination. This
  effect is more pronounced than in the simple Pauli-Villars case due to
  overall larger cancellations. In the bottom panel, we plot the
  $k$-dependence for various $c$, with the direct data points restricted to
  the range $n \in \{ 65, \ldots, 255 \}$ to eliminate boundary effects. The
  vanishing of $\hat{s}$ in the continuum limit $k \rightarrow 0$ is clearly
  visible.}
\end{figure}
\begin{figure}[htb]\centering
  \includegraphics[width=\columnwidth]{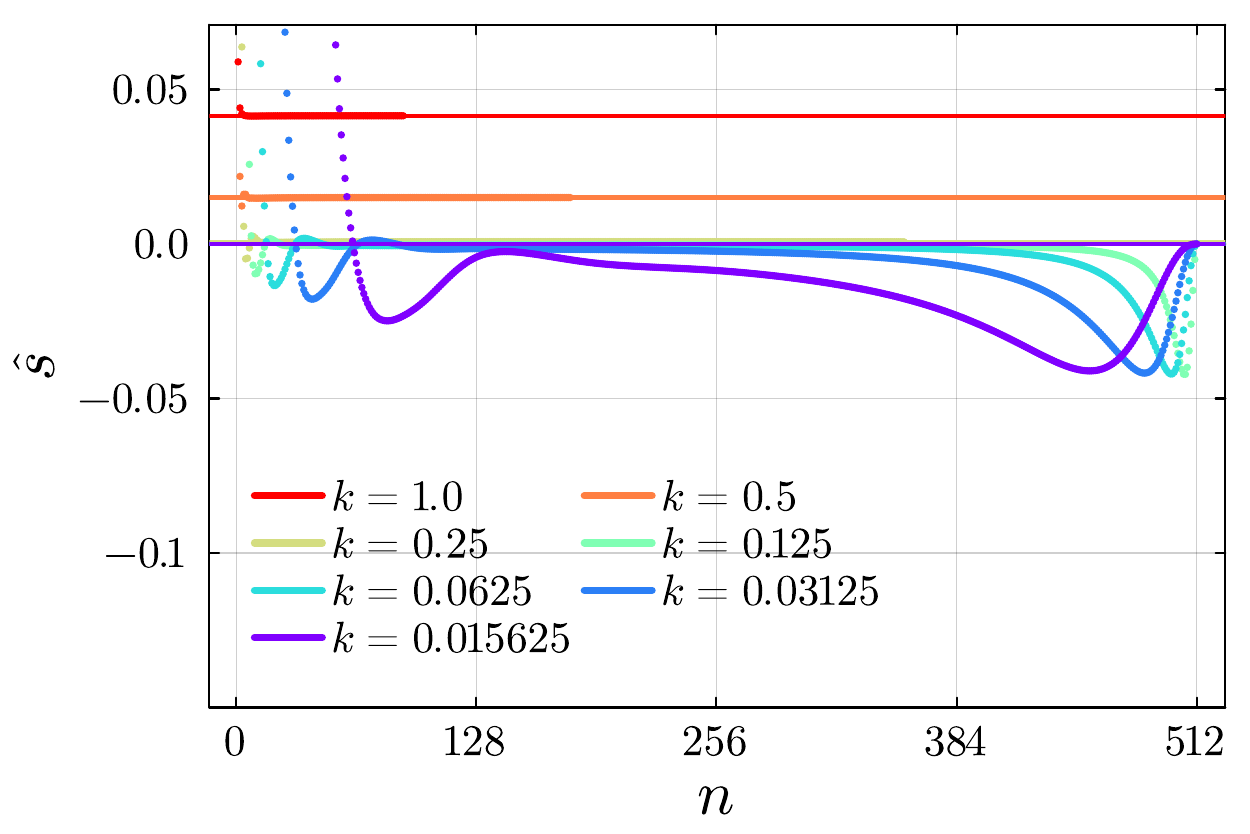}\par\medskip
  \includegraphics[width=\columnwidth]{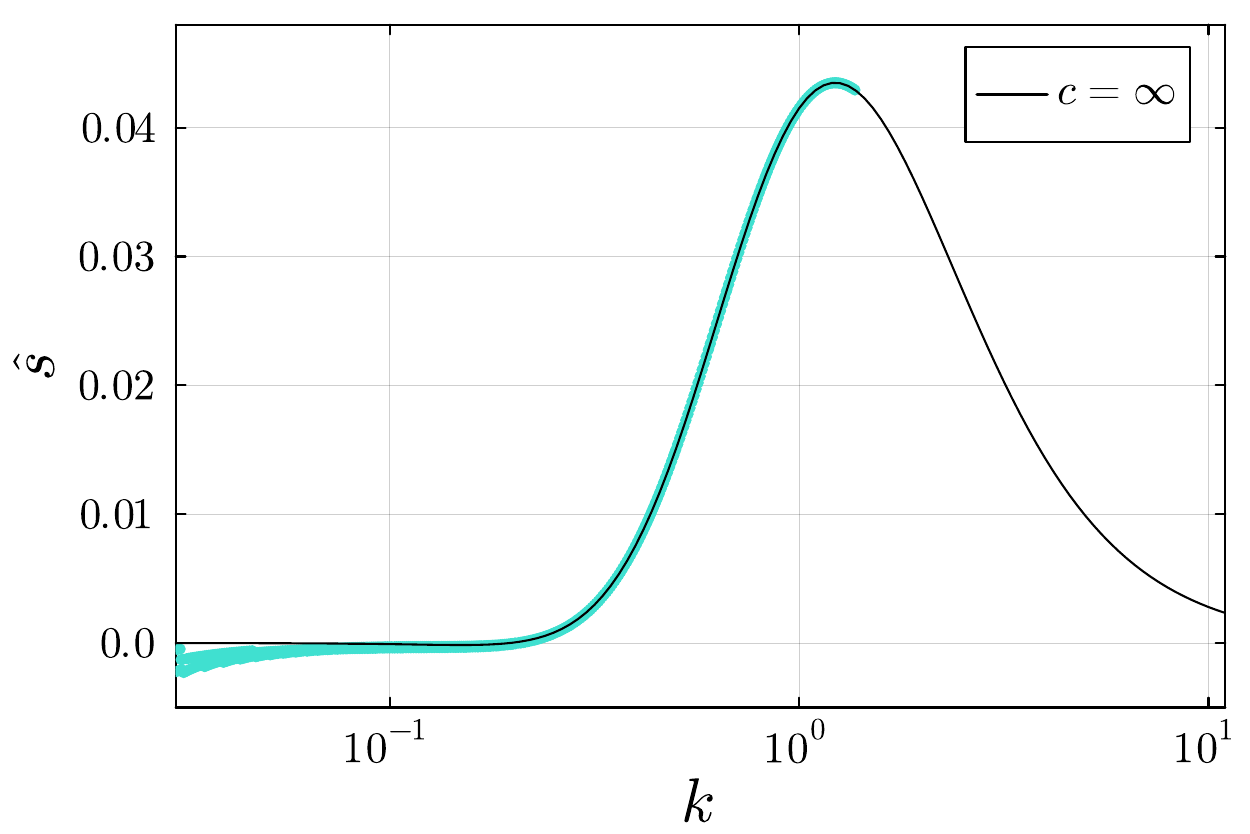}
  \caption{\label{pseudoPVlogpoly}Predictions \eqref{shatpseudoPV} (lines) and
  direct data \eqref{shatpseudodirect} (dots) for the normalized entropy in
  the log-polynomial pseudo Pauli-Villars scheme. In the top panel, the
  normalized entropy for various cutoff ratios $k$ is plotted vs. the radial
  index $n$, while in the bottom panel the normalized entropy is plotted
  vs. $k$ for the radial indices $n \in \{ 65, \ldots, 255 \}$. Note that
  since we have only tracked $512$ angular momentum modes in our simulation,
  we only have data for radial indices $n \leqslant 512 / (\kappa_{\max} k)
  \simeq 89 / k$.}
\end{figure}

Since the log-polynomial Pauli-Villars scheme fulfills all the cancellation
conditions of a polynomial one, we may use all the generic results from the
previous section. This includes the $\tilde{c} \rightarrow \infty$ limit
\eqref{polyvasy} of the single mode entropy, with the prefactor
\[ \sigma_L = - \sum_{i > 1} \frac{p_i \kappa_i^6}{24} \simeq 40.221 \]
extracted from Tab.~\ref{poltablog}, so
\begin{equation}
  \tilde{S}_k^L (\tilde{c}) \xrightarrow{\tilde{c} \rightarrow \infty} \sim
  \sigma_L k^6 \frac{24 \ln (2 \tilde{c}) - 7}{\tilde{c}^{10}}\;.
  \label{logpolyvasy}
\end{equation}
More importantly, the expression for the continuum limit of the single mode
entropy \eqref{genform} is still valid, as is the resulting normalized total
entropy \eqref{shatpvpoly}. Crucially, if we compute the latter for any
log-polynomial Pauli-Villars scheme, we find that it vanishes because of
\eqref{pm2log}. In fact, in deriving \eqref{shatpvpoly} we never used the
cancellation of the fourth powers of the mass, thus we find that for any
continuum Pauli-Villars scheme that fulfills
\[ \sum_{i = 1}^n p_i = \sum_{i = 1}^n p_i M_i^2 = \sum_{i = 1}^n p_i M_i^2
   \ln M_i^2 = 0 \]
the leading term in the entanglement entropy, which is proportional to the surface area, vanishes.

Let us explicitly introduce the exact continuum limit single mode entropy
\begin{equation}
  \hat{S}^L (\hat{c}) = - \frac{1}{12} \sum_{i > 1} p_i \ln \left(
  \frac{\kappa_i^2}{\hat{c}^2} + 1 \right) \label{logpolyform}
\end{equation}
as well as the approximate entropy function at finite cutoff ratio $k$ based
on \eqref{skint} for the simple Pauli-Villars subtraction
\begin{equation}
  \hat{S}_k^{L, i} (\hat{c}) = - \sum_{i > 1} p_i \hat{S}^i_{k_i} \left(
  \frac{\hat{c}}{\kappa_i} \right) \;,\label{lkint}
\end{equation}
where the $\kappa_i$ are given in Tab.~\ref{poltablog}. We compare these
functions to the direct numerical data in Fig.~\ref{logpolyvlim} and again
find excellent agreement. We can clearly see that there is a region in
$\tilde{c}$ where the single mode entropy contribution is negative, which is
necessary for the cancellation that results in a total vanishing entropy.

\begin{figure}[htb]\centering
  \includegraphics[width=\columnwidth]{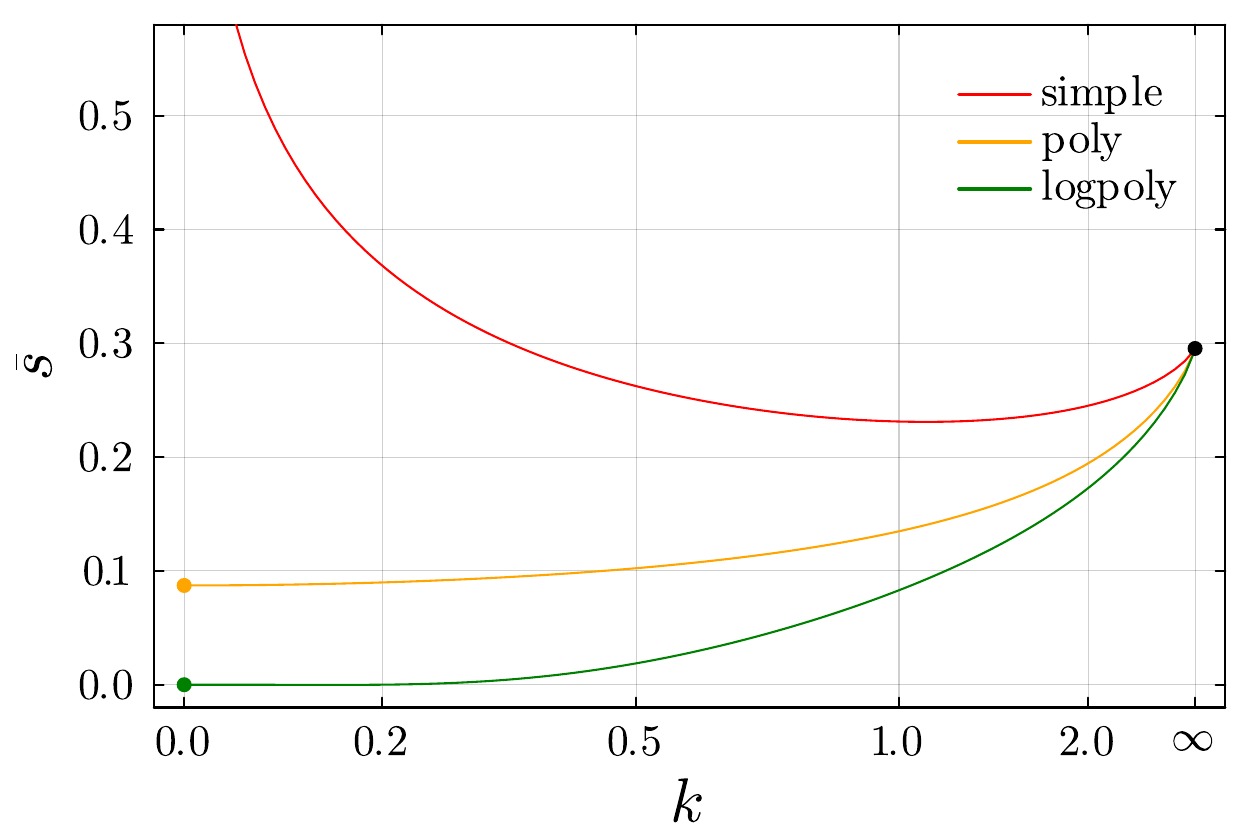}
  \caption{\label{finscal}Plot of the normalized entropy \eqref{sbar} vs. the
  cutoff ratio $k$ for three Pauli-Villars cutoff schemes at anisotropy factor
  $c = \infty$. The $x$-axis is scaled according to $\overline{k} = (1 + k^{-
  2})^{- 1 / 2}$, so the $k = 0$ line corresponds to the continuum limit while
  $k \rightarrow \infty$ is the brick wall limit. The black dot on the right
  hand side represents the Srednicki limit for brick wall regularization,
  while the orange and green dots on the left hand side correspond to the
  analytical continuum results for the polynomial and log-polynomial case,
  which are given by \eqref{polycontlimshat} and $\overline{s} = 0$
  respectively.}
\end{figure}

We can now proceed to obtain an approximation for the normalized entropy
$\hat{s}$ as a function of the cutoff ratio $k$ by integrating \eqref{lkint}
according to \eqref{shat}. In Fig.~\ref{logpolyshat} we display the result and
compare it to direct numerical data. There is again a good overall agreement
except for the expected boundary effects in the direct data at very small $k$.
We also plot the normalized entropy vs.~$k$ to show how it numerically
approaches zero for all anisotropy factors $c$. Although we are unable to
prove it analytically, there is numerical evidence that for an infinite
anisotropy factor $c$, i.e.~when taking all $l$-modes into account, the
normalized entropy is nonnegative and approaches zero very smoothly in the
continuum limit. For finite anisotropy factors, on the other hand, there seems
to be a region where the normalized entropy $\hat{s}$ turns negative and
reaches the vanishing continuum limit from below. For smaller anisotropy
factors $c$, this region seems to be shifted towards smaller $k$. Clearly a
negative entropy is unphysical. However, one should note that in the continuum
limit the entropy vanishes for all $c > 0$ and that the appearance of
intermittent regions of negative entropy depends on the order in which we take
field and regulator modes into account along our trajectory towards the
continuum. Ultimately, in light of the vanishing entropy in the continuum
limit, negative as well as positive entropies can be considered discretization
artifacts.

In full analogy to the polynomial Pauli-Villars case, we may also decompose
the normalized entropy into separate integrals over the physical and regulator
fields according to \eqref{shatpseudoPV} and obtain the pseudo Pauli-Villars
normalized entropy \eqref{shatpseudodirect} from brick wall data alone. The
resulting normalized entropy is presented in Fig.~\ref{pseudoPVlogpoly}. We
again see good agreement between various determination methods, except in the regions where boundary or discretization effects
are large. As \eqref{shatpseudolim} predicts, the normalized entropy does
indeed vanish in the $k \rightarrow 0$ limit.

Finally, we would like to compare the different regulator schemes. For that
purpose we define a normalized entropy
\begin{equation}
  \overline{s} = \hat{s} (k^2 + 1) = s (1 + k^{- 2}) \rightarrow
  \left\{\begin{array}{lll}
    s &  & k \rightarrow \infty\\
    \hat{s} &  & k \rightarrow 0
  \end{array}\right. \label{sbar}
\end{equation}
that goes over into the brick wall resp.~Pauli-Villars normalized entropies
$s$ and $\hat{s}$ in the respective limits. We evaluate (\ref{sapprox}) and (\ref{shatpseudoPV}) using the approximation to the scaling function (\ref{stili}) to obtain the leading (proportional to area) term of the respective entropies. The result is displayed in Fig.~\ref{finscal}, where we plot
$\overline{s}$ vs.~the cutoff ratio at infinite anisotropy factor $c$. One can
clearly see the qualitative difference between the various Pauli-Villars
schemes, with the continuum limit values of the normalized entropy diverging,
being finite or vanishing for the simple, polynomial and log-polynomial variants
respectively.

\section{Numerical results for dynamical spacetime}

In this section, we attempt a first investigation of the entropy evolution
during dynamical collapse. We will mainly focus on conceptual issues
regarding the different regularization schemes and defer a more thorough
investigation to future studies.

\subsection{Dynamical collapse with brick wall regularization}
\begin{figure}[htb]\centering
  {\includegraphics[width=\columnwidth]{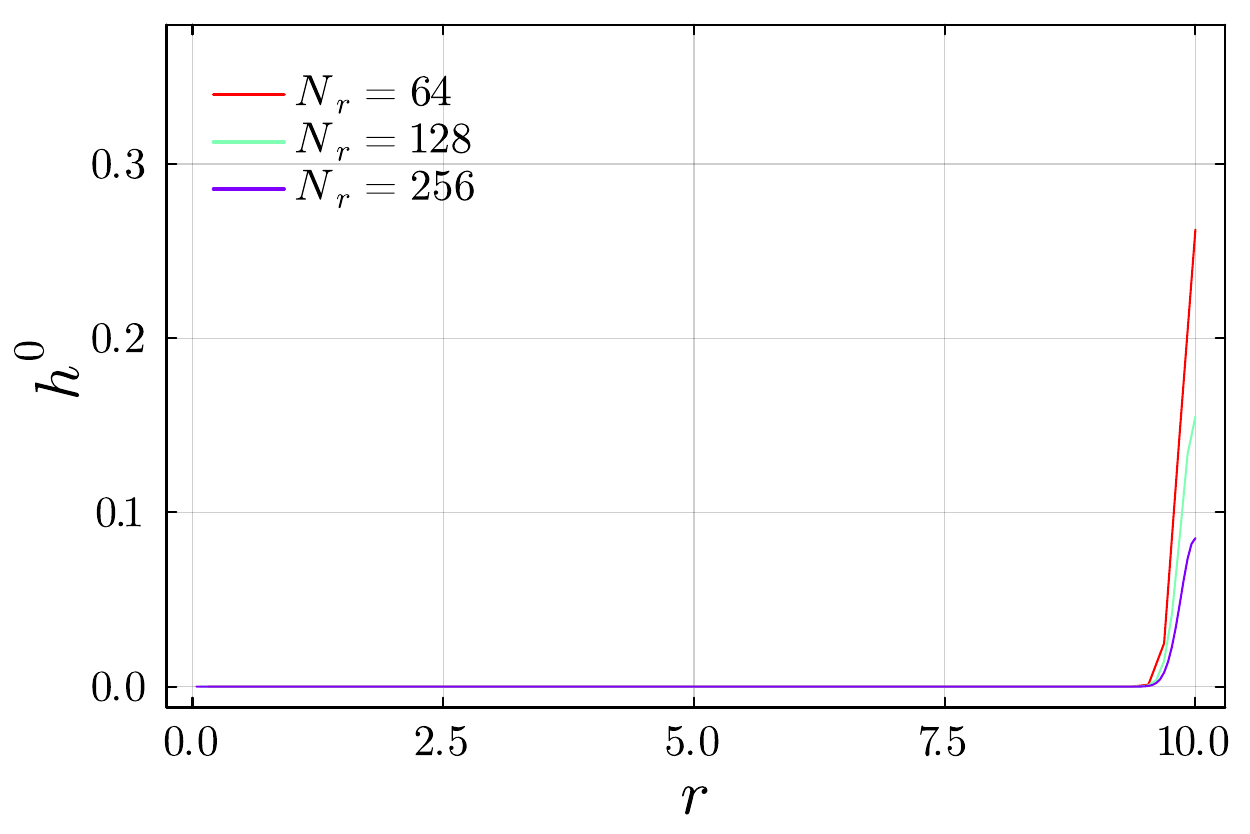}}
  \includegraphics[width=\columnwidth]{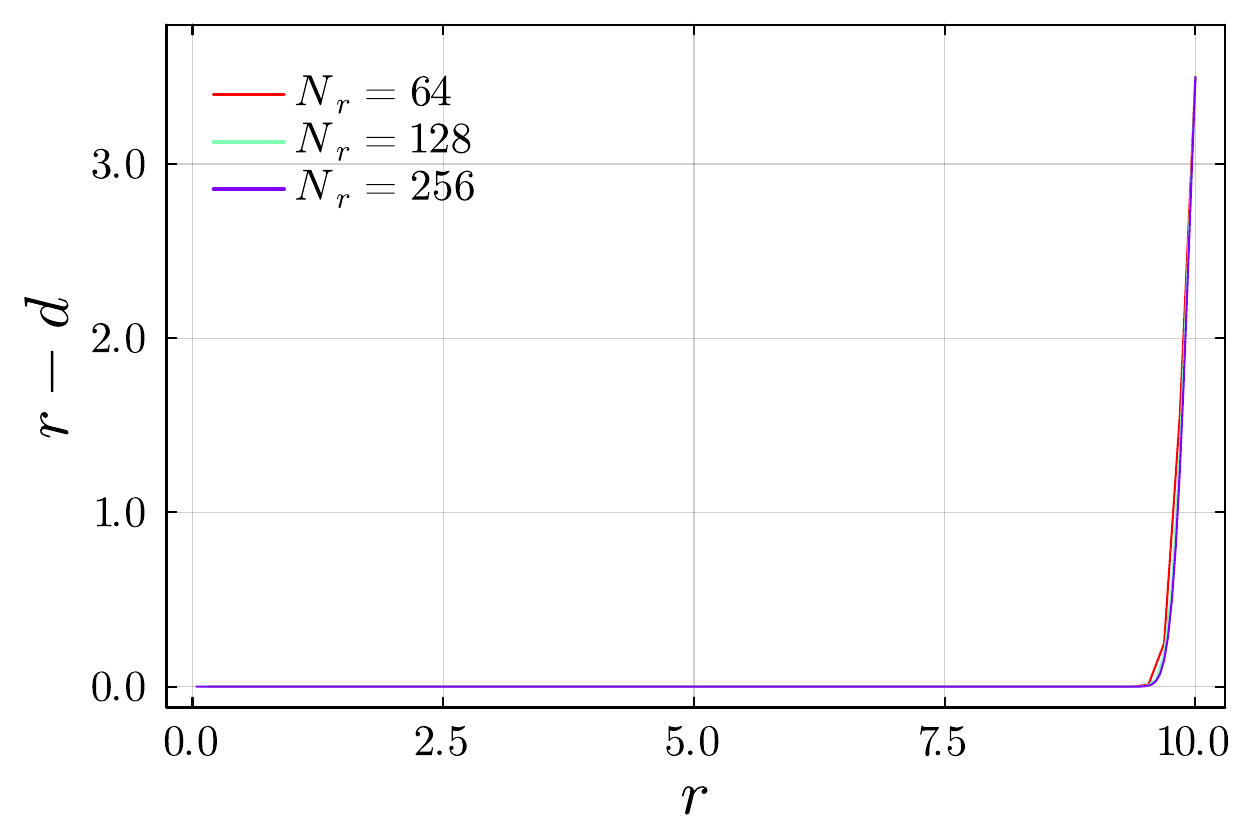}
  \includegraphics[width=\columnwidth]{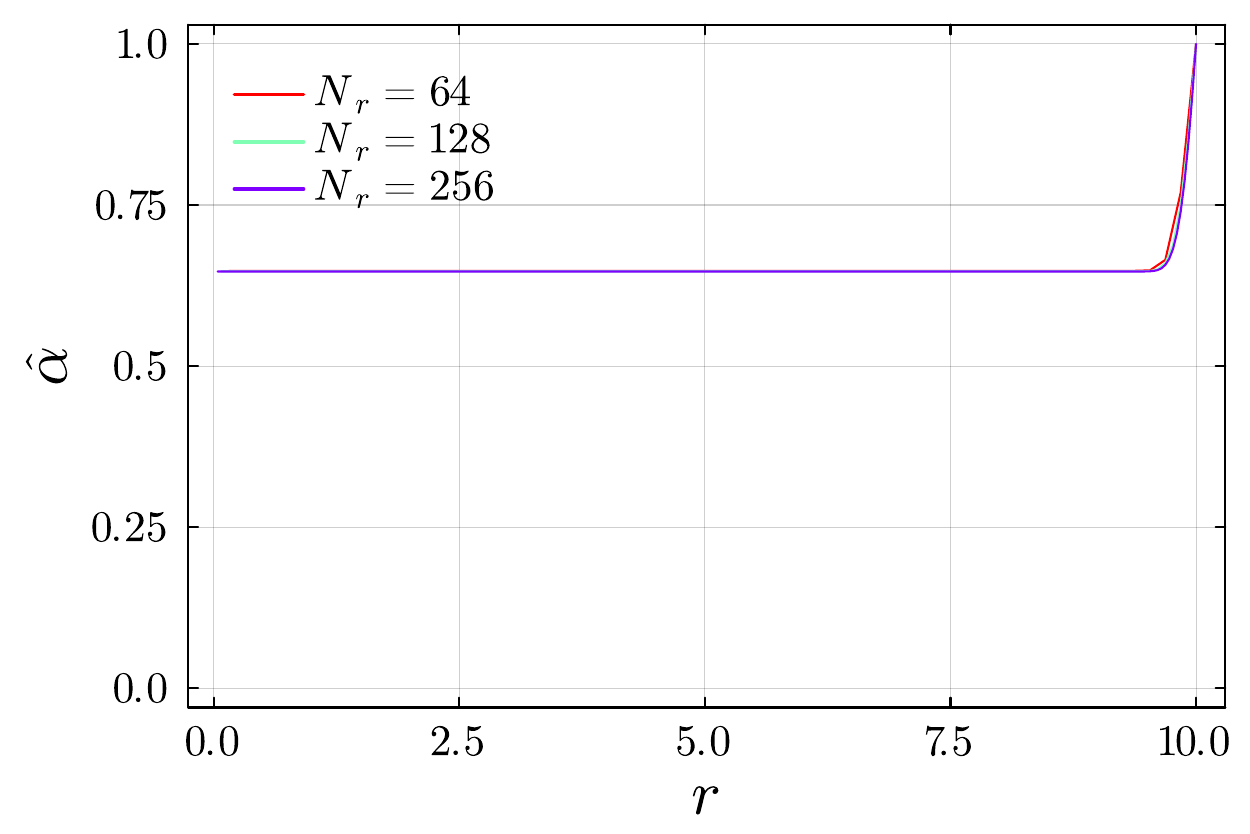}
  \caption{\label{metricinit}Shell integrated classical energy density
  \eqref{h0i} (top) and the metric parameters $r - d$ (middle) and
  $\hat{\alpha} $ (bottom) vs.~radial coordinate $r$ at the initial time
  $t = 0$ for three different discretizations.}
\end{figure}

\begin{figure*}[htb]\centering
  \includegraphics[width=\columnwidth]{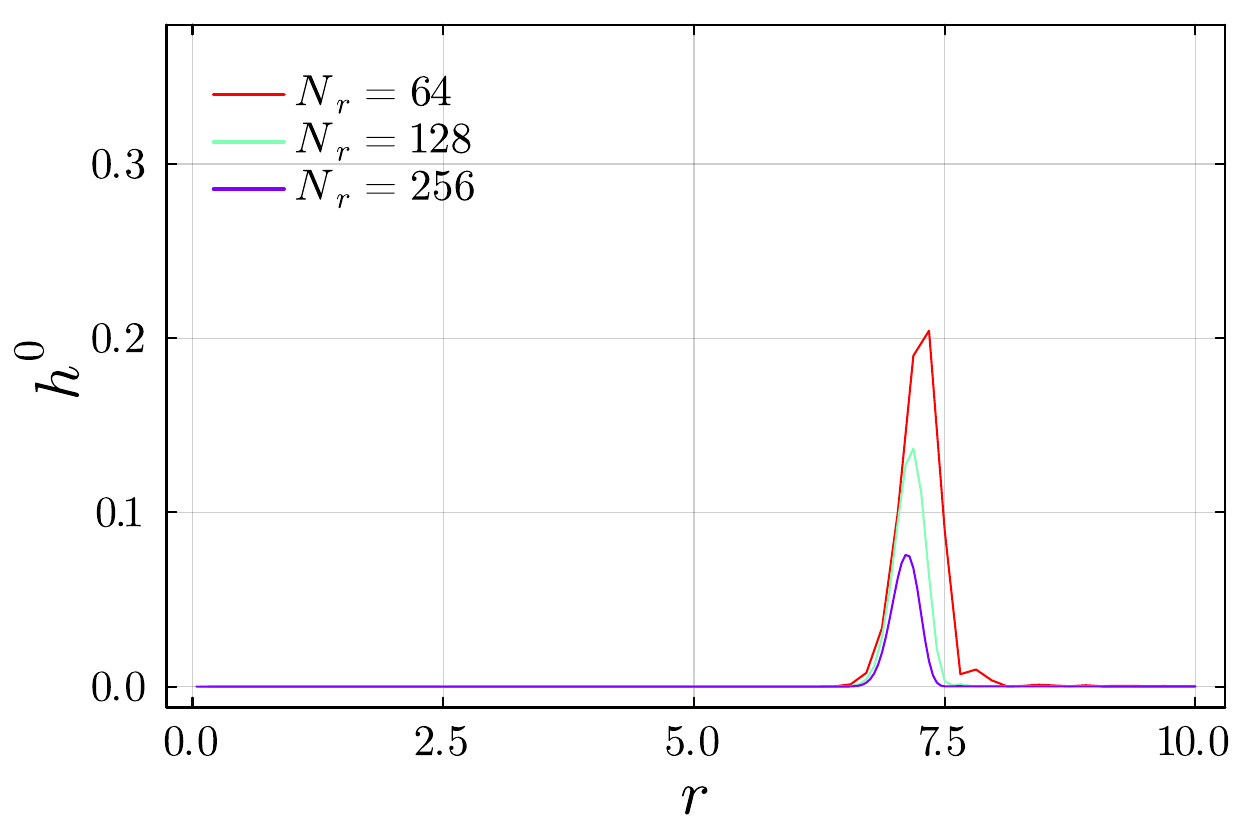}
  \includegraphics[width=\columnwidth]{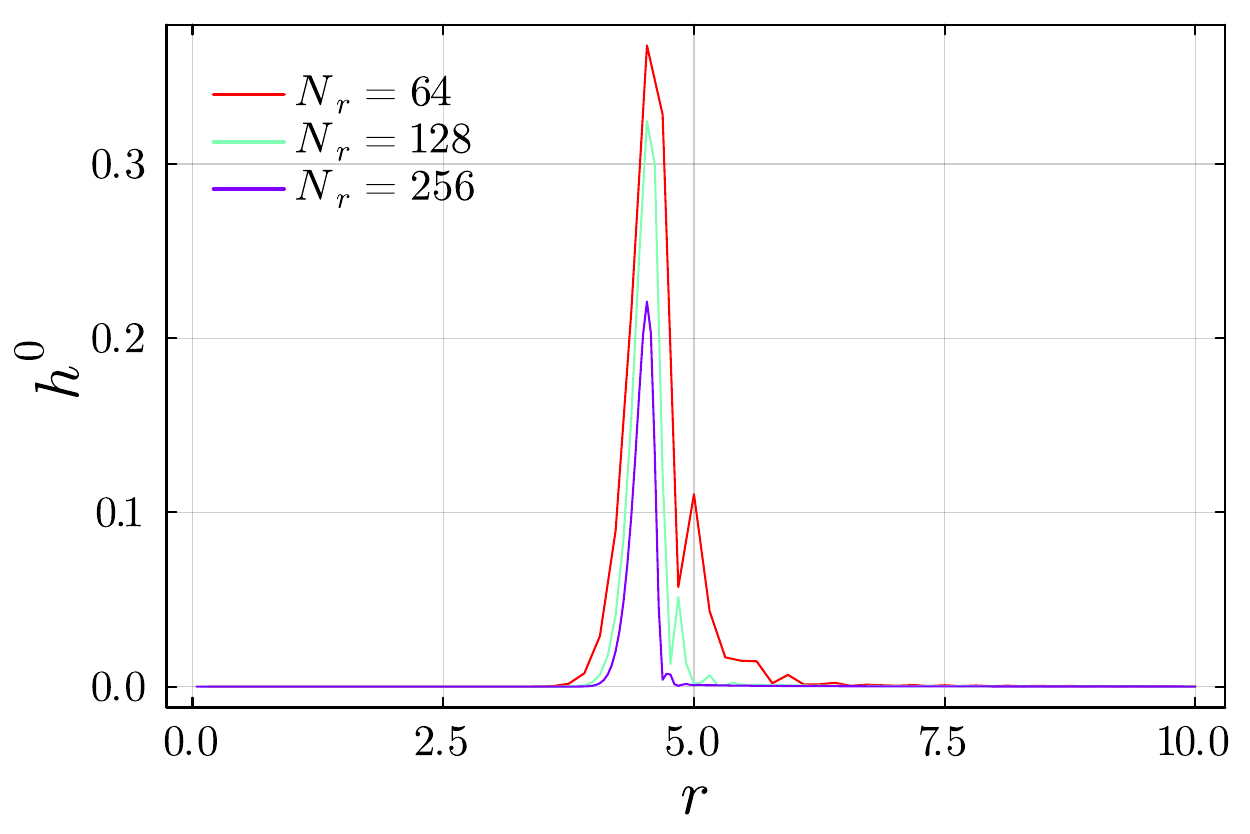}
  
  {\includegraphics[width=\columnwidth]{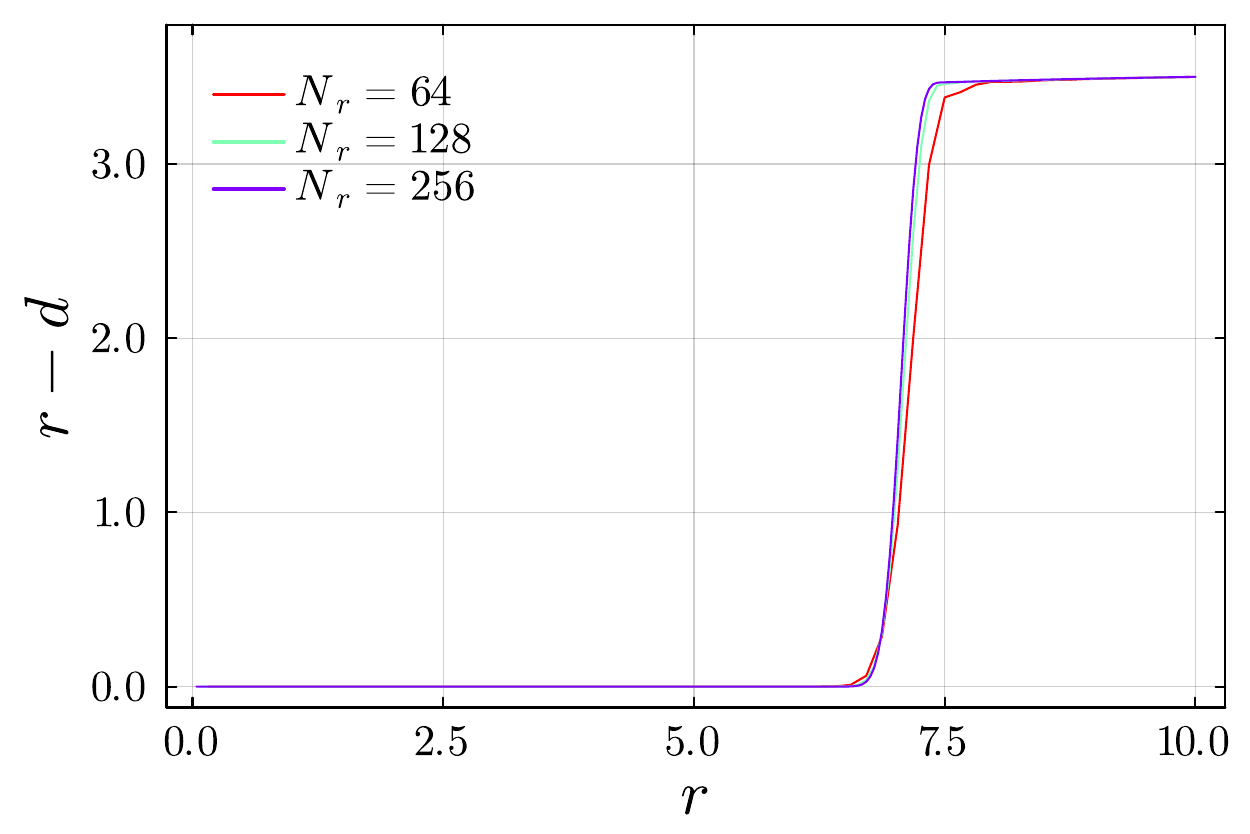}}
  {\includegraphics[width=\columnwidth]{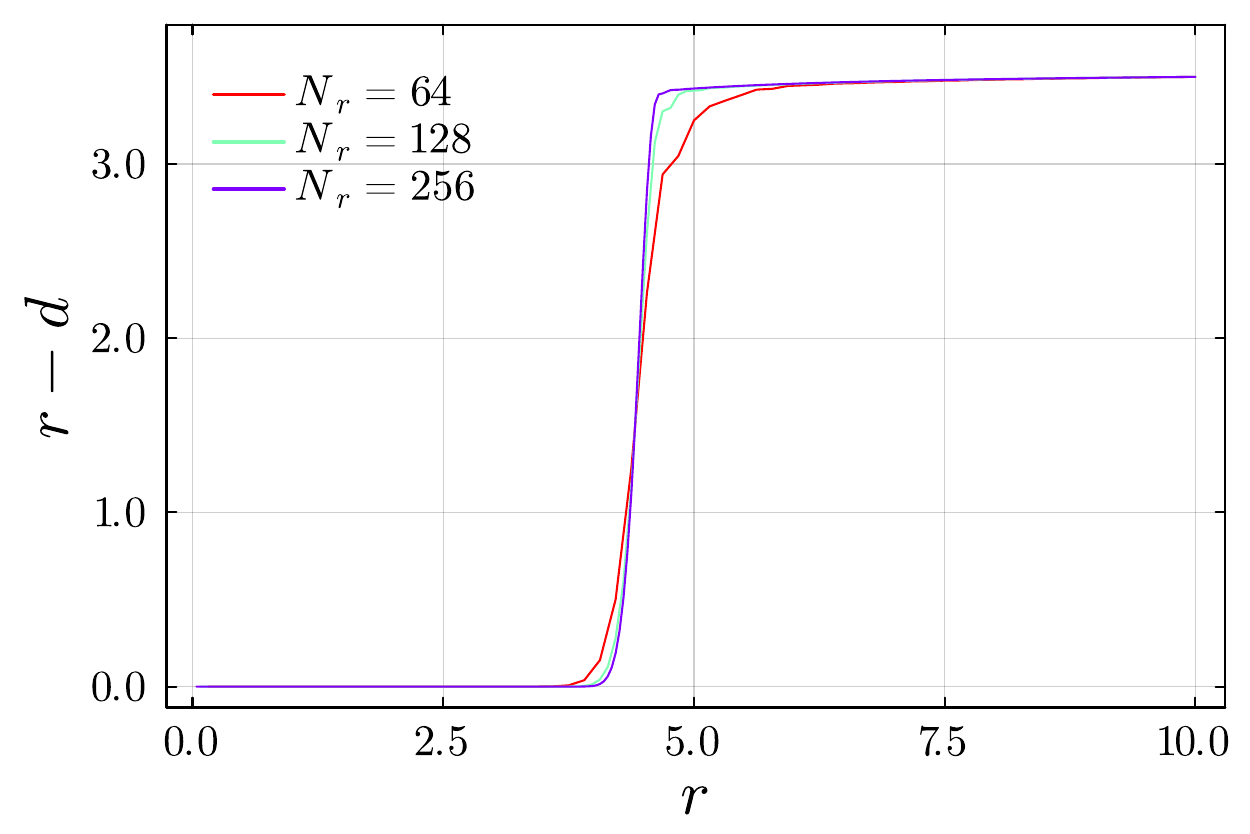}}

  {\includegraphics[width=\columnwidth]{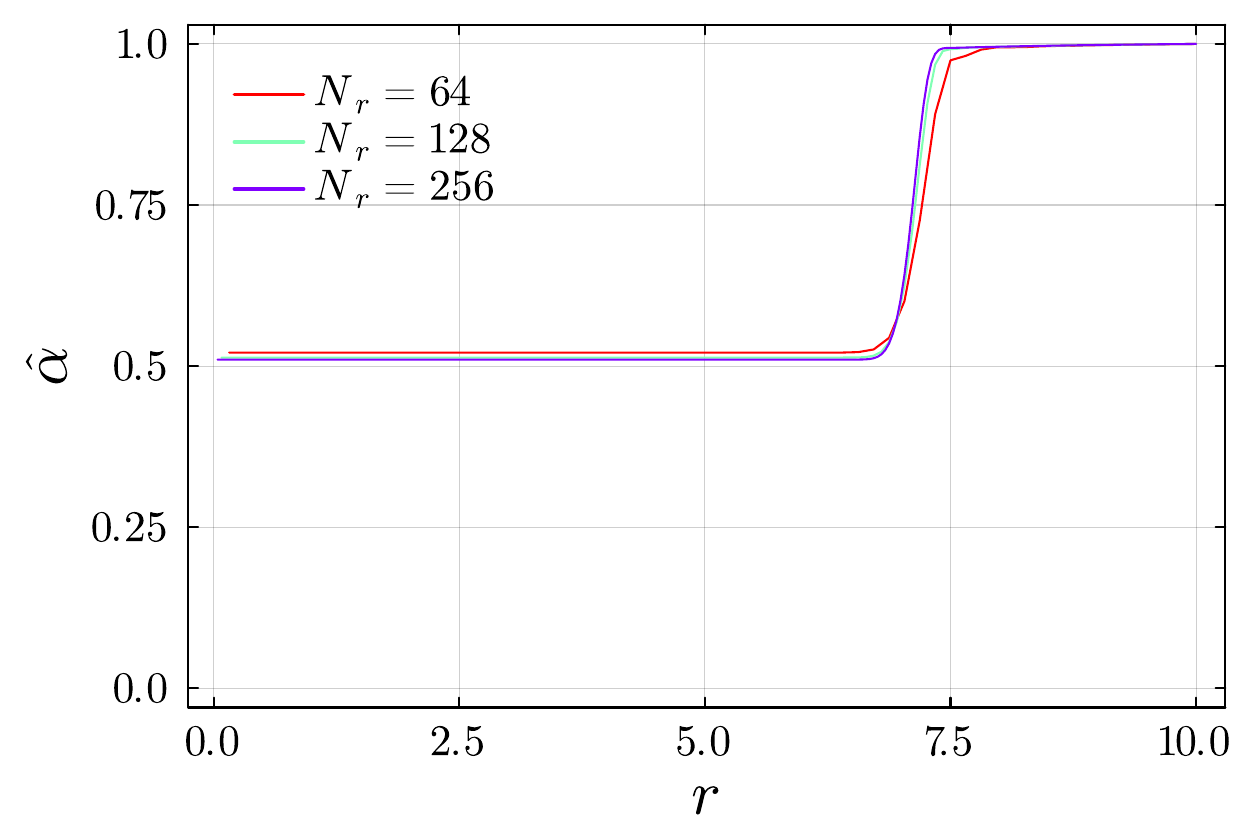}}
  {\includegraphics[width=\columnwidth]{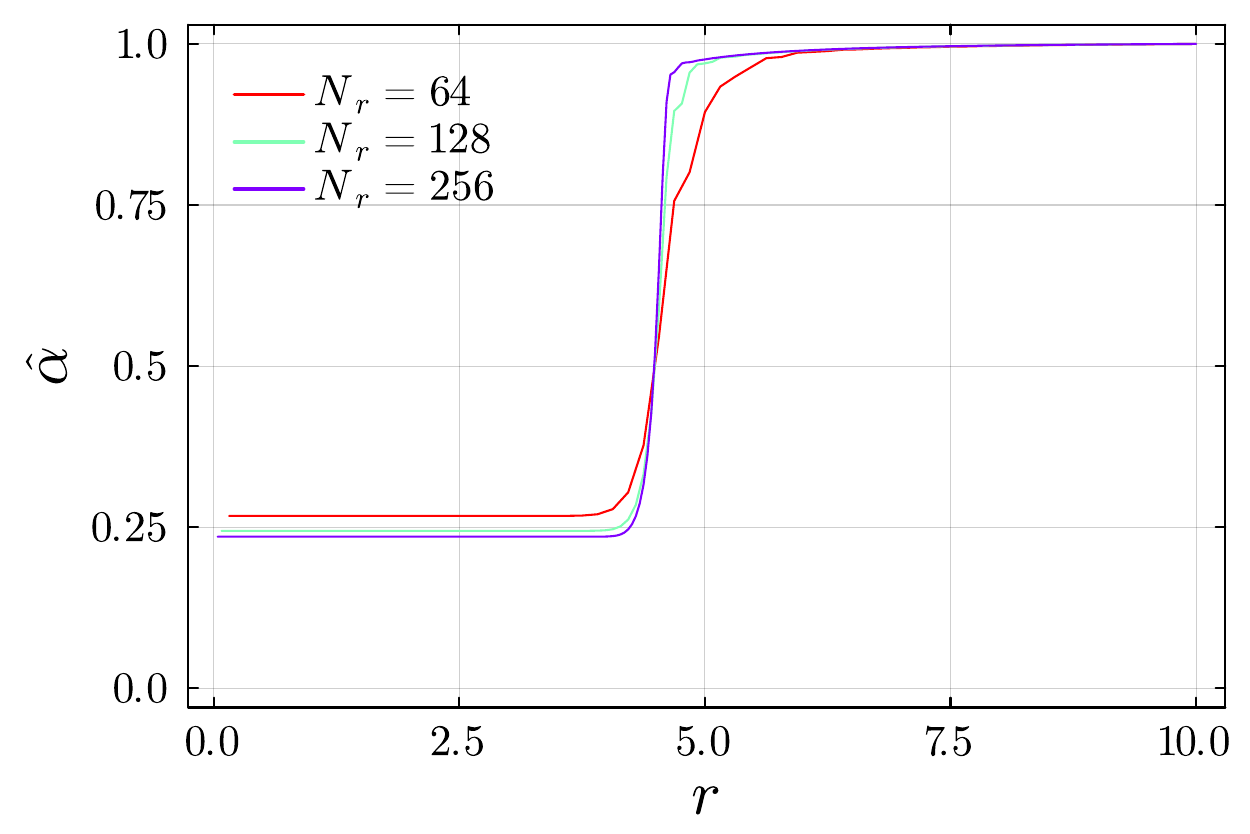}}
  \caption{\label{metricevo}Shell integrated classical energy density
  \eqref{h0i} (top row) and the metric parameters $r - d$ (middle row) and
  $\hat{\alpha}$ (bottom row) vs.~radial coordinate $r$ at $t =
  5$ (left panels) and $t = 12$ (right panels) for three different
  discretizations. The numerical condition $h^0 \ll 1$, which is necessary for small radial discretization errors, is only marginally fulfilled for course discretizations and later times.}
\end{figure*}

\begin{figure}[htb]\centering
  \includegraphics[width=\columnwidth]{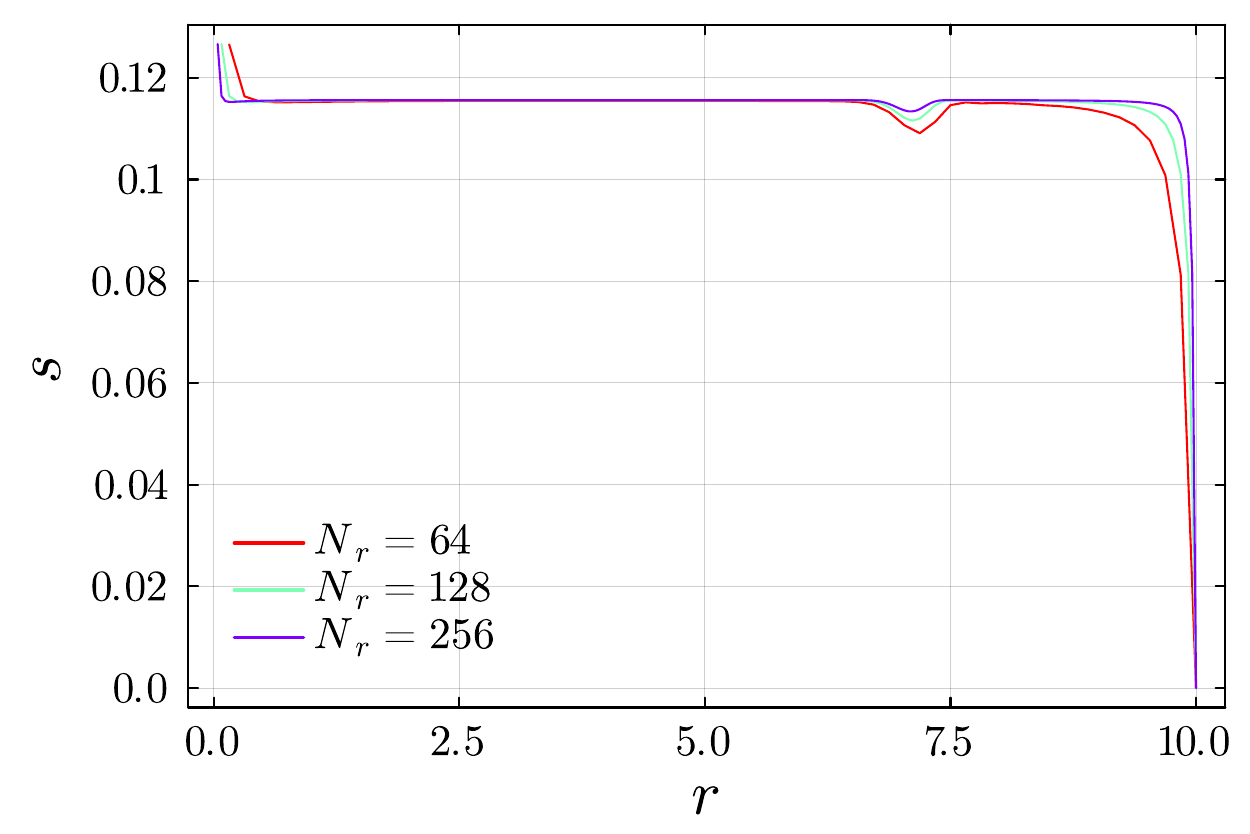}\par\medskip
  \includegraphics[width=\columnwidth]{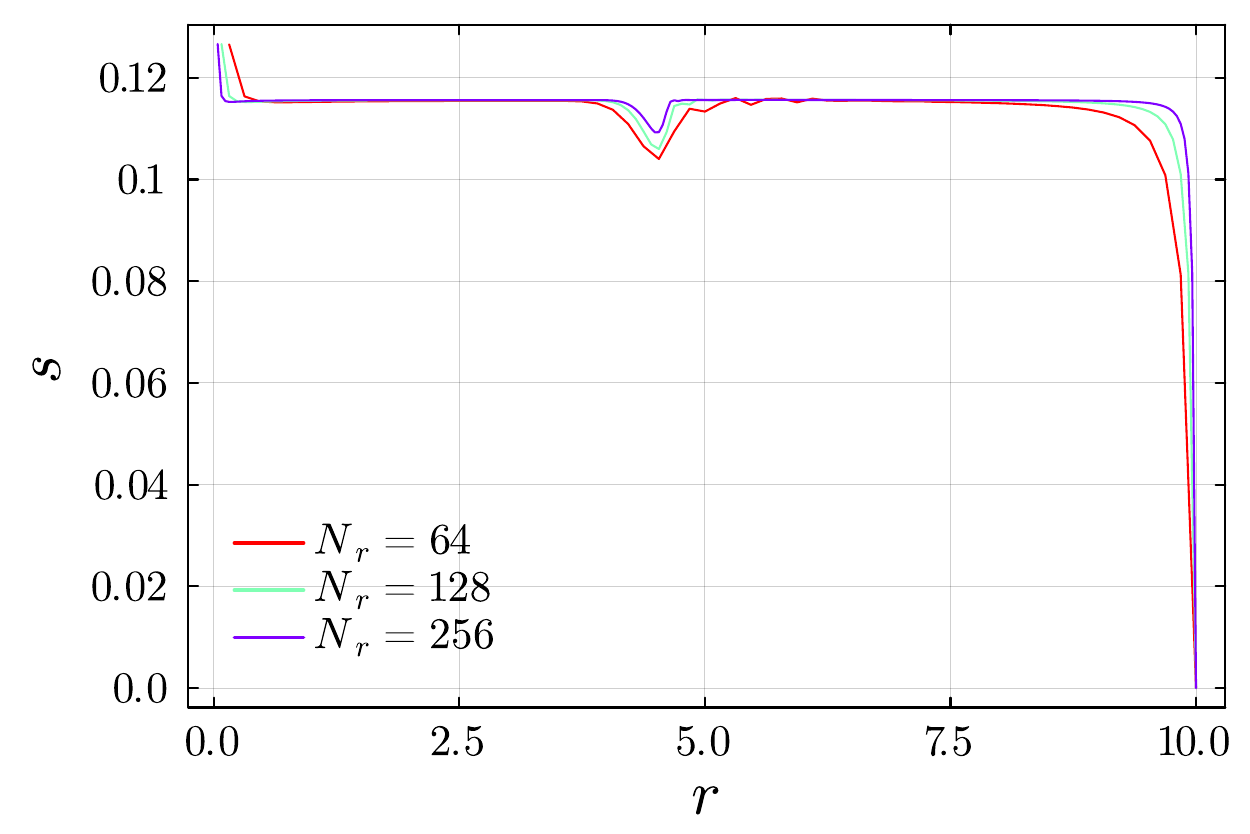}
  \caption{\label{bwsimp}Brick wall normalized entropy vs.~radial coordinate
  $r$ at $t = 5$ (top panel) and $t = 12$ (bottom panel) for three different
  discretizations at anisotropy factor $c = 1$.}
\end{figure}

\begin{figure}[htb]\centering
  \includegraphics[width=\columnwidth]{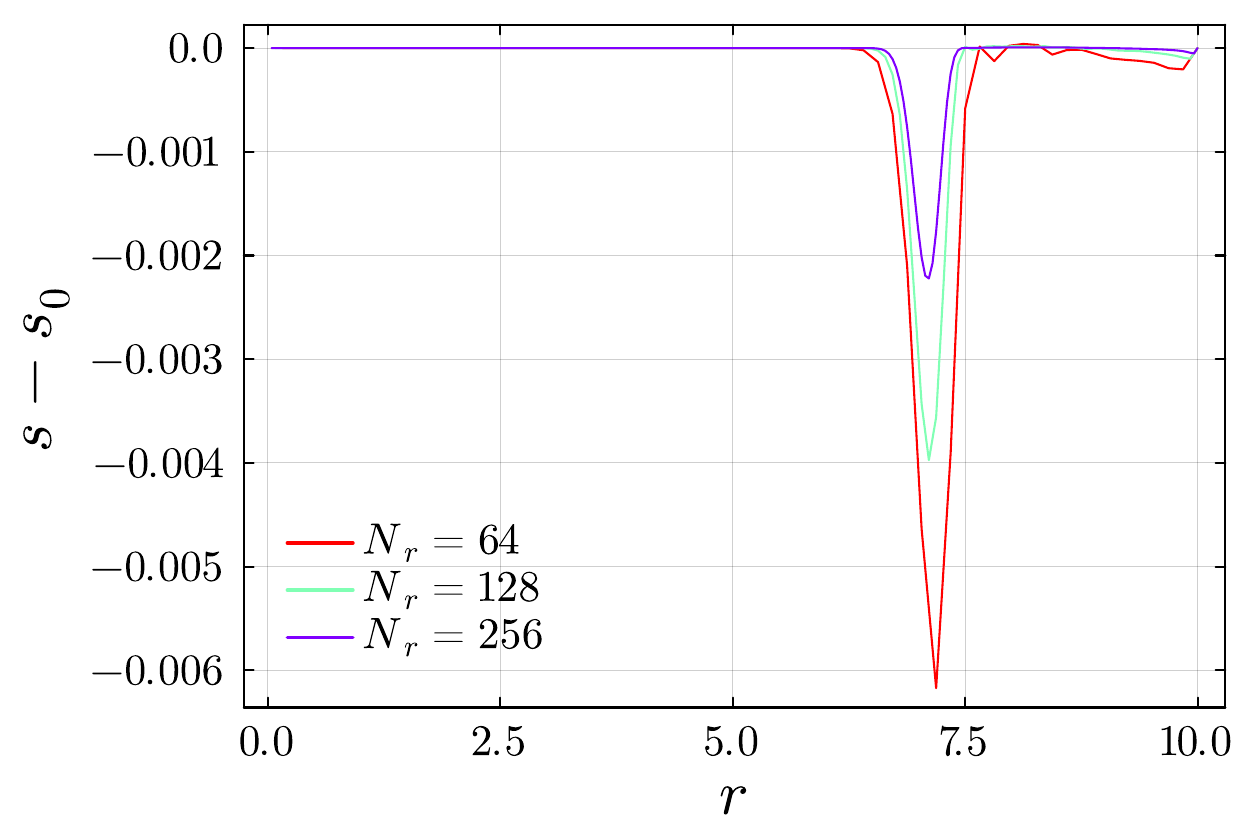}\par\medskip
  \includegraphics[width=\columnwidth]{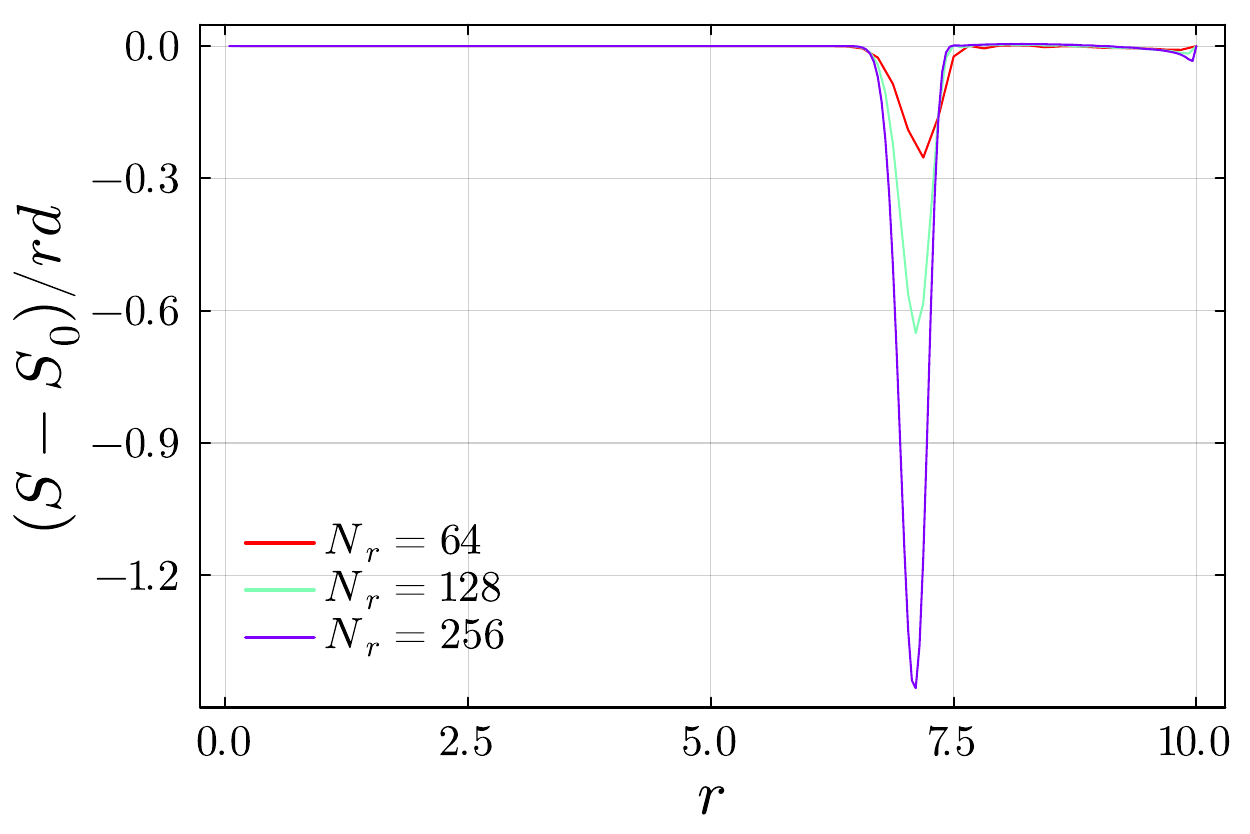}
  \caption{\label{bwzoom}Difference between brick wall normalized entropy
  (top panel) and entropy per surface area (bottom panel) of the dynamical
  system at simulation time $t = 5$ and the corresponding flat system. The
  dynamical data are the same as displayed in the top panel of Fig.~\ref{bwsimp}.}
\end{figure}
\begin{figure*}[htb]\centering
  {\includegraphics[width=\columnwidth]{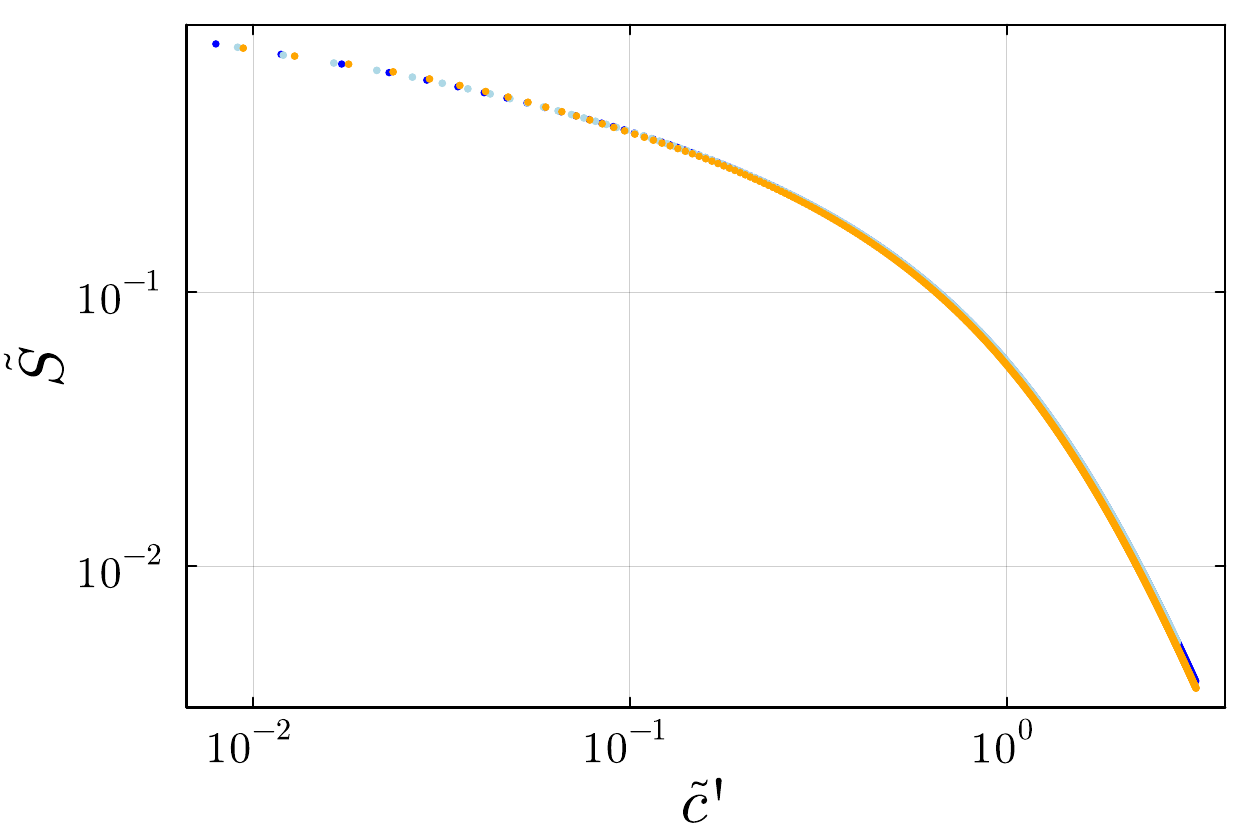}}{\includegraphics[width=\columnwidth]{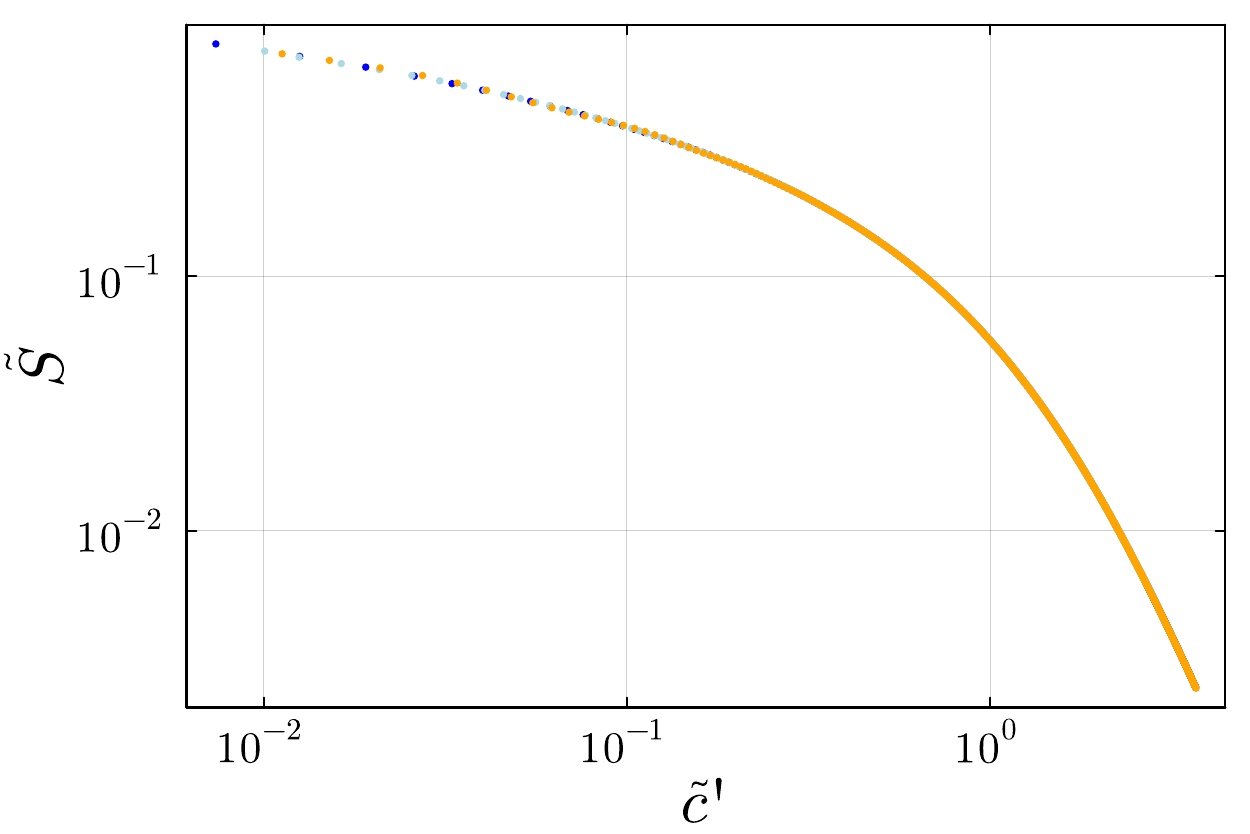}}
  
  {\includegraphics[width=\columnwidth]{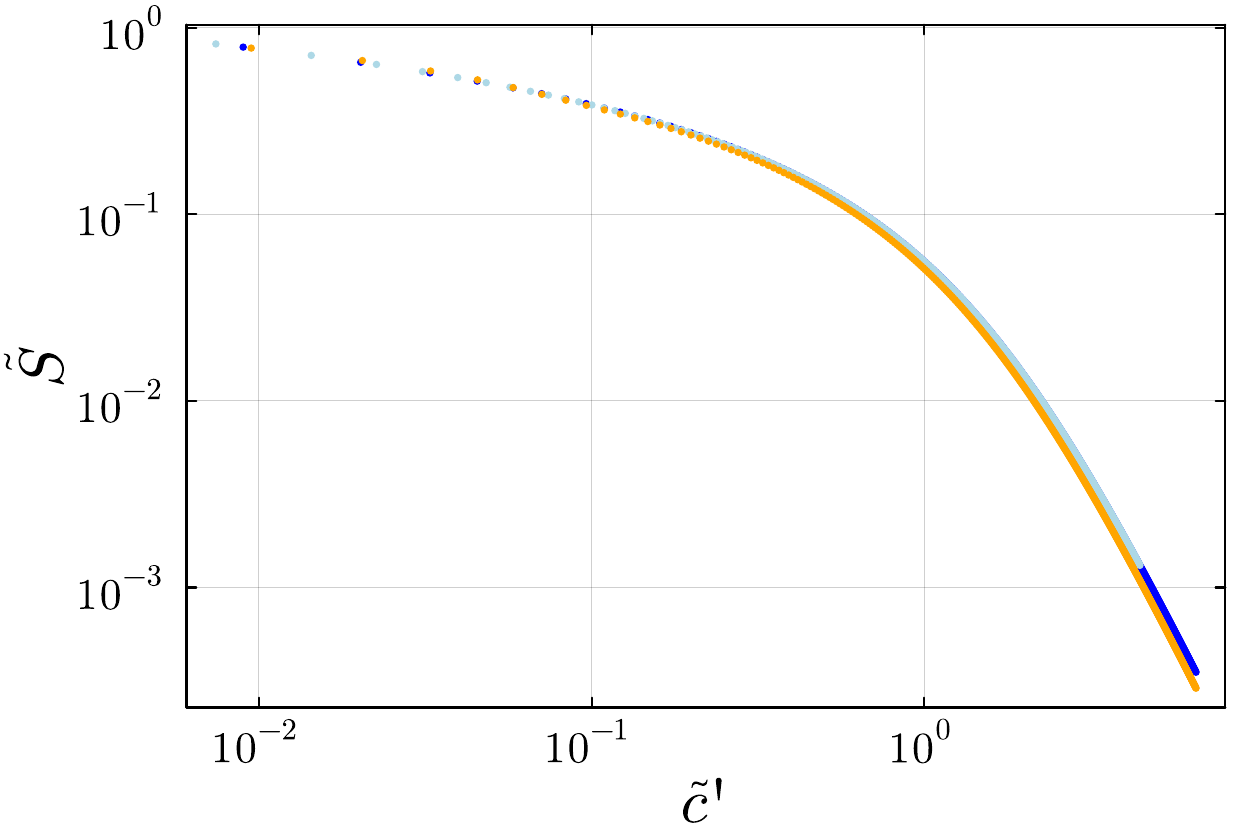}}{\includegraphics[width=\columnwidth]{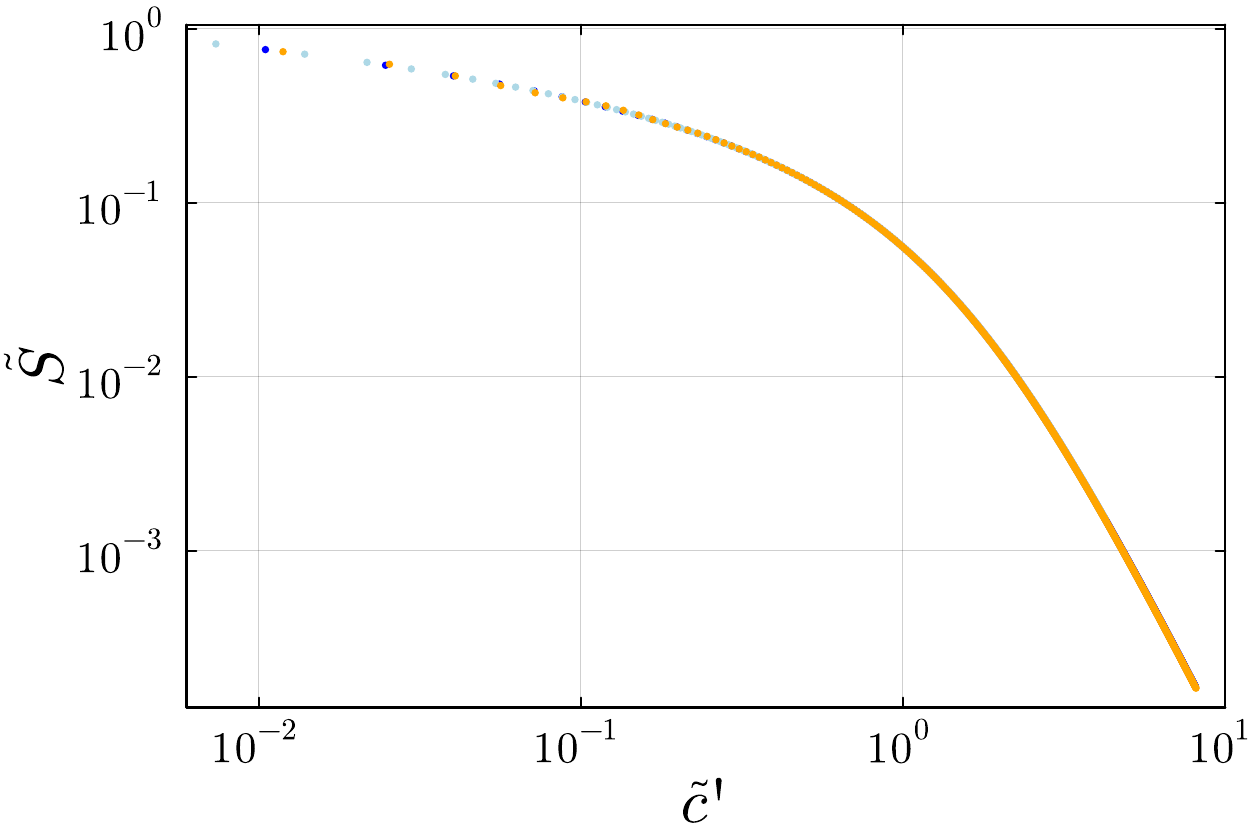}}
  \caption{\label{twomodes}Free (light blue, blue) and dynamical (orange)
  single mode entropies vs. the boundary corrected scaling variable at fixed
  radial coordinates for an $N_r = 256$ system at simulation times $t = 5$
  (top panels) and $t = 12$ (bottom panels). For the free case the entropy at
  two radial coordinates is plotted. The light blue points are taken at the
  same radial coordinate as in the dynamical case, while the dark blue points
  are taken at the radial coordinate most closely matching the dynamical
  $n_{\tmop{eff}}$. In the left panels, the radial coordinates are chosen to
  be close to the inmoving peak ($n = 180$, $n_{\tmop{eff}} = 161$ for $t = 5$
  and $n = 115$, $n_{\tmop{eff}} = 78$ for $t = 12$) while for the right
  panels they are just outside the peak region ($n = 190$, $n_{\tmop{eff}} =
  139$ for $t = 5$ and $n = 120$, $n_{\tmop{eff}} = 63$ for $t = 12$).}
\end{figure*}

Starting with the brick wall regularization,
we first note that the physical distance between adjacent shells is given by
\begin{equation}
  a \Delta = \sqrt{\frac{r}{d}} \Delta \label{physd}
\end{equation}
while the surface area of the sphere still is $4 \pi r^2$. Consequently, the
proper generalization of the normalized entanglement entropy \eqref{norment}
for the non-flat case is
\begin{equation}
  s = \frac{\Delta^2}{r d} S \;. \label{gennorment}
\end{equation}
Similarly, we need to use the general definition \eqref{constanisogen} of the
anisotropy factor $c$ instead of its flat space variant \eqref{lnscal}. From a
purely computational point of view, these quantities are a bit more cumbersome
than their flat space counterparts, since they involve the dynamical scale $d$
in addition to the purely parametric $n$ and $l_{\max}$. It is convenient to
introduce an ``effective radial coordinate''
\begin{equation}
  n_{\tmop{eff}} = \frac{\sqrt{d r}}{\Delta} \;,\label{neffdef}
\end{equation}
in terms of which the normalized entropy may be written as
\[ s = \frac{S}{n_{\tmop{eff}}^2} \]
and the anisotropy relation is
\begin{equation}
  (l_{\max} + 1) = n_{\tmop{eff}} c \;. \label{anisodyn}
\end{equation}

For our first numerical study, we look at a single component $N_c=1$ scalar field. The radial extent of our system is $L =r_{N_r}= 10$ and we use three 
different radial discretizations $N_r = 64, 128$ and $256$ as well as an
integration time step  of $\Delta t = 1 / 128$. We initialize the classical
hamiltonian density $h^0$ at the initial time $t = 0$ with a bump hat has
compact support and peak position $P$ directly at the boundary $P = L$.
Specifically, we choose a Nuttall bump \cite{1163506} that has the form
\[ h_i^0 = \lambda \Theta (\sigma - | P - r_i |)  \sum_{k = 0}^3 a_k \cos
   \left( k \frac{\pi (P - r_i + \sigma)}{\sigma} \right) \]
with the coefficients
\[ \begin{aligned}
   a_1 ={}& - 0.487396 & a_2 ={}& 0.144232\\
   a_3 ={}& - 0.012604 & a_0 ={}& - \sum_{i = 1}^3 a_i\;.
   \end{aligned} \]
We choose the half width of the bump $\sigma = 1$ and iterate the bump height
$\lambda$ so that at the outermost shell we have
\[ L - d_{N_r} = 3.5 \;.\]
The resulting $h^0$ and its corresponding metric parameters are displayed in Fig.~\ref{metricinit}.
\begin{figure*}[htb]\centering
  {\includegraphics[width=\columnwidth]{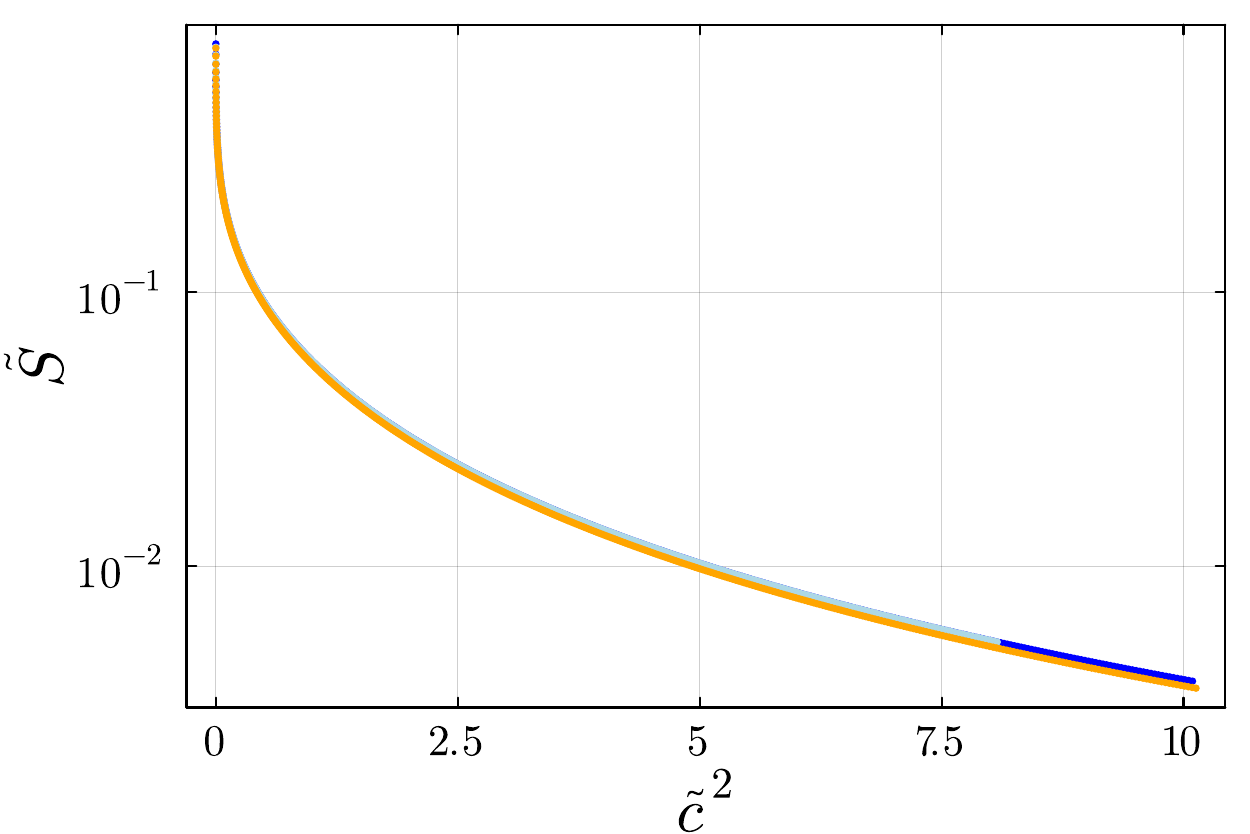}}{\includegraphics[width=\columnwidth]{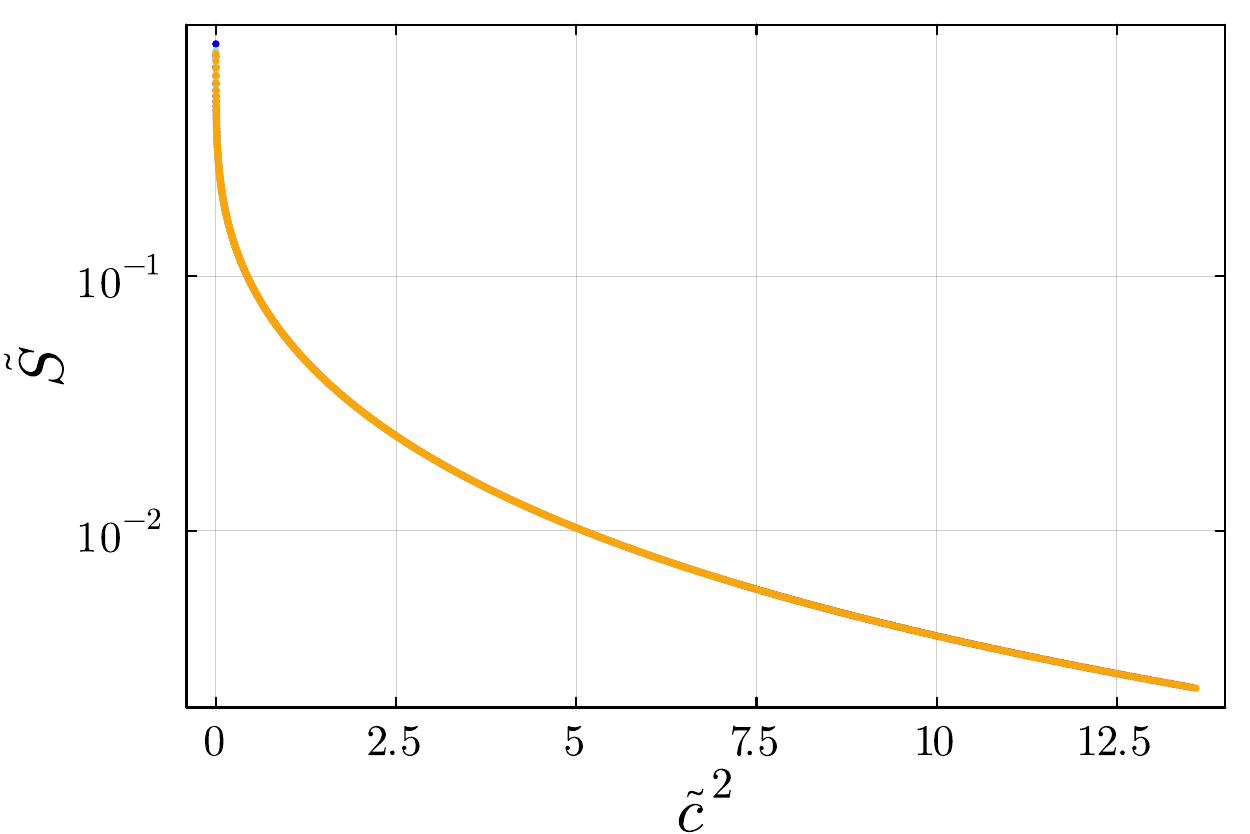}}
  
  {\includegraphics[width=\columnwidth]{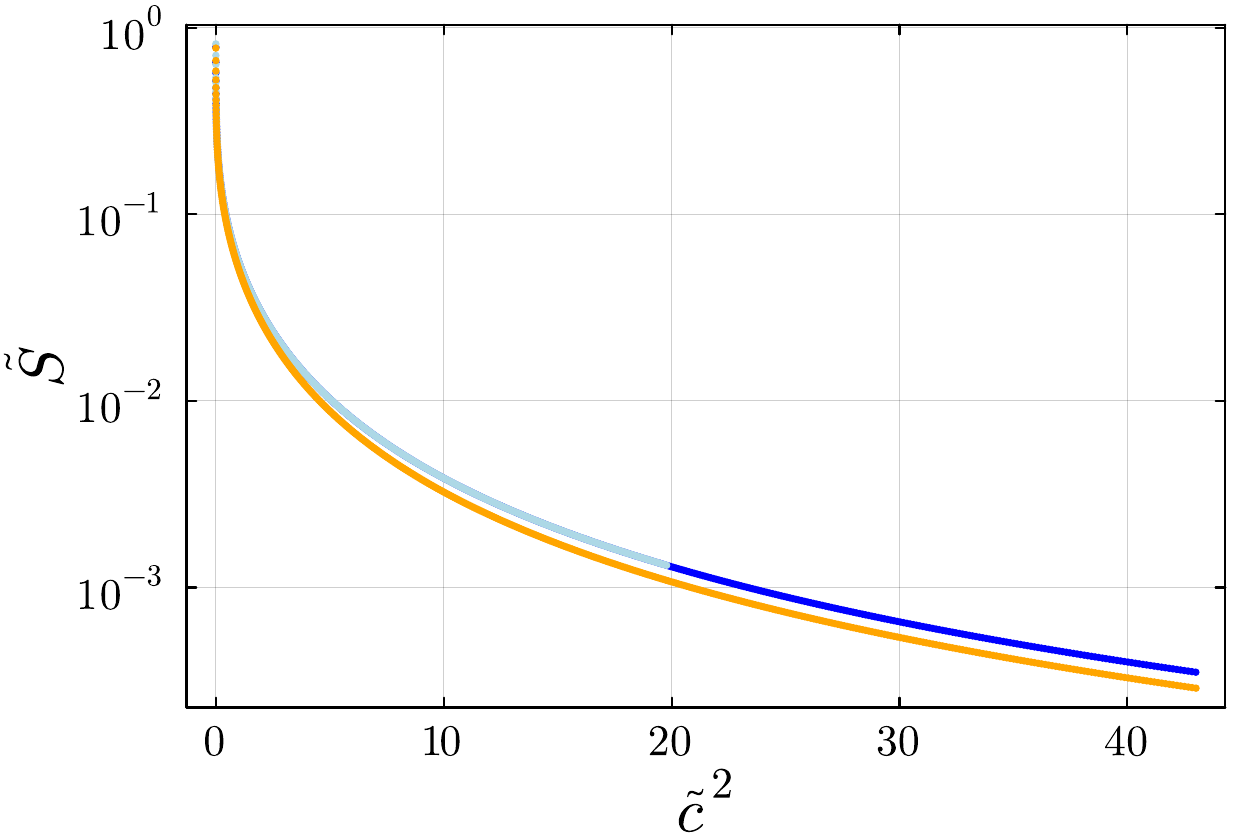}}{\includegraphics[width=\columnwidth]{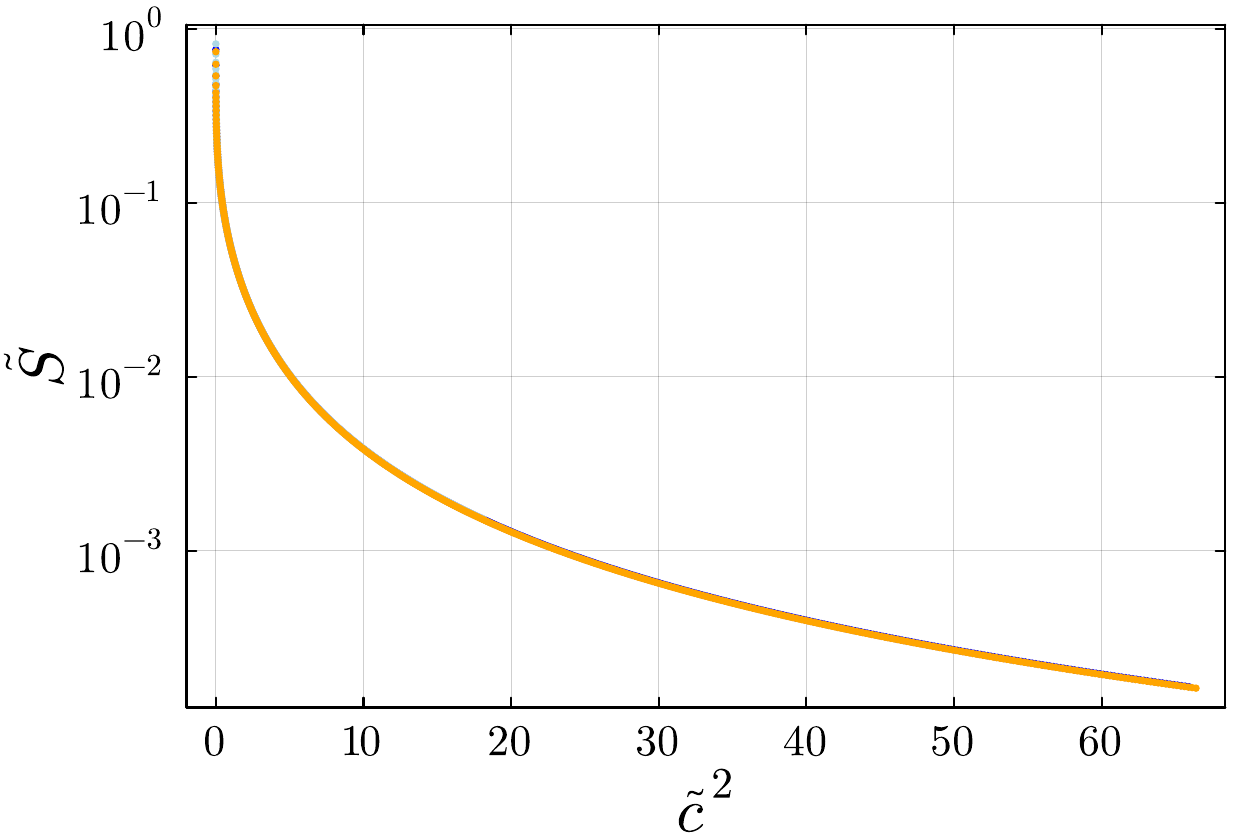}}
  \caption{\label{twomodeslin}The same data as in Fig.~\ref{twomodes}, but
  plotted vs. the uncorrected $\tilde{c}^2$.}
\end{figure*}

Since in this exploratory study we restrict ourselves to a classical time
evolution and neglect all backreaction effects, the choice of regulator does
not have influence on the time evolution of the metric. In Fig.~\ref{metricevo}
we show the evolution of the inmoving density $h^0$ and  the corresponding evolution of the metric. As one can observe, the condition $h^0 \ll 1$, which is necessary in order for radial
discretization effects to be negligible, is only marginally fulfilled for our
coarsest discretization and later times. We will accept this numerical shortcoming for this first exploratory study.

In Fig.~\ref{bwsimp}, we plot the brick wall normalized entropy
\eqref{gennorment} vs.~the radial coordinate at two different simulation times
and anisotropy factor $c = 1$. Note that according to \eqref{anisodyn}
$l_{\max}$ is generally not an integer, so the normalized entropies plotted
are linear interpolations between the closest integers. Remarkably, the
normalized entropy is still almost constant except for a small dip at the
position of the inmoving bump.

We may zoom in on this dip by subtracting the normalized entropy of the
corresponding free system, which is done in Fig.~\ref{bwzoom}.  The
dynamical effects seem to reduce the entropy in the bump region without affecting the rest of the system in a substantial way. It is also notable that the bump
height of the entropy difference seems to diverge as one goes towards the
continuum, as shown in the bottom panel of Fig.~\ref{bwzoom}. This result
suggests that in regions where the metric is not dynamical, i.e.~inside and
outside the bump region, the subtraction of the normalized free entropy is
valid. In the dynamical region, the Hadamard form predicts additional
divergences that cannot be subtracted with a procedure that is essentially a
local normal ordering. A brick wall cutoff therefore seems to be a particularly
unsuitable regularization scheme to study the evolution of entanglement
entropy during a gravitational collapse.

As in the free case, we may decompose these results into their single angular
momentum mode entropy contributions. By performing the mode expansion into
spherical harmonics, we have decomposed our system into 1+1 dimensional
harmonic oscillator chains with angular momentum dependent effective mass
terms
\[ \mu^2 = M^2 + \frac{l (l + 1)}{r^2} \;.\]
Taking into account the physical separation between radial shells $\Delta r$
\eqref{radss}, the mass term in dimensionless units becomes

\begin{figure}[htb]\centering
  \includegraphics[width=\columnwidth]{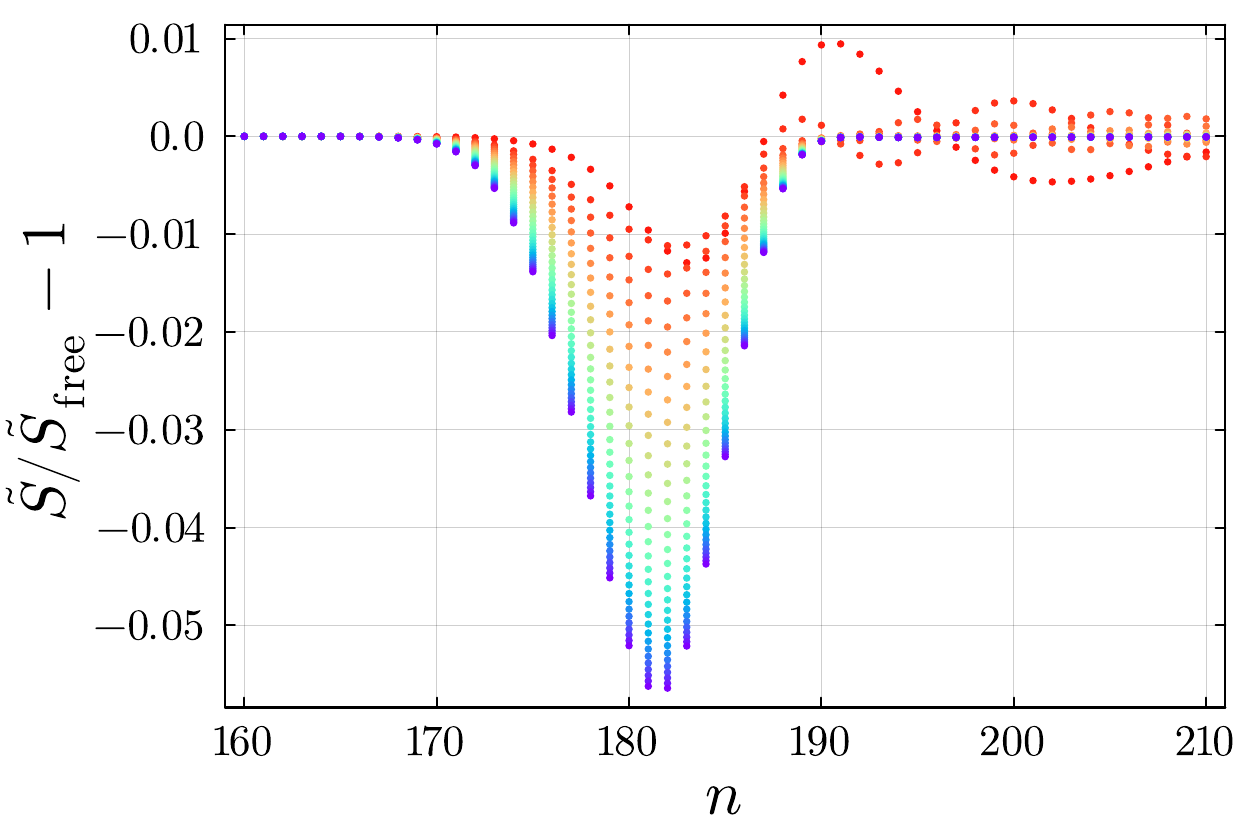}\par\medskip
  \includegraphics[width=\columnwidth]{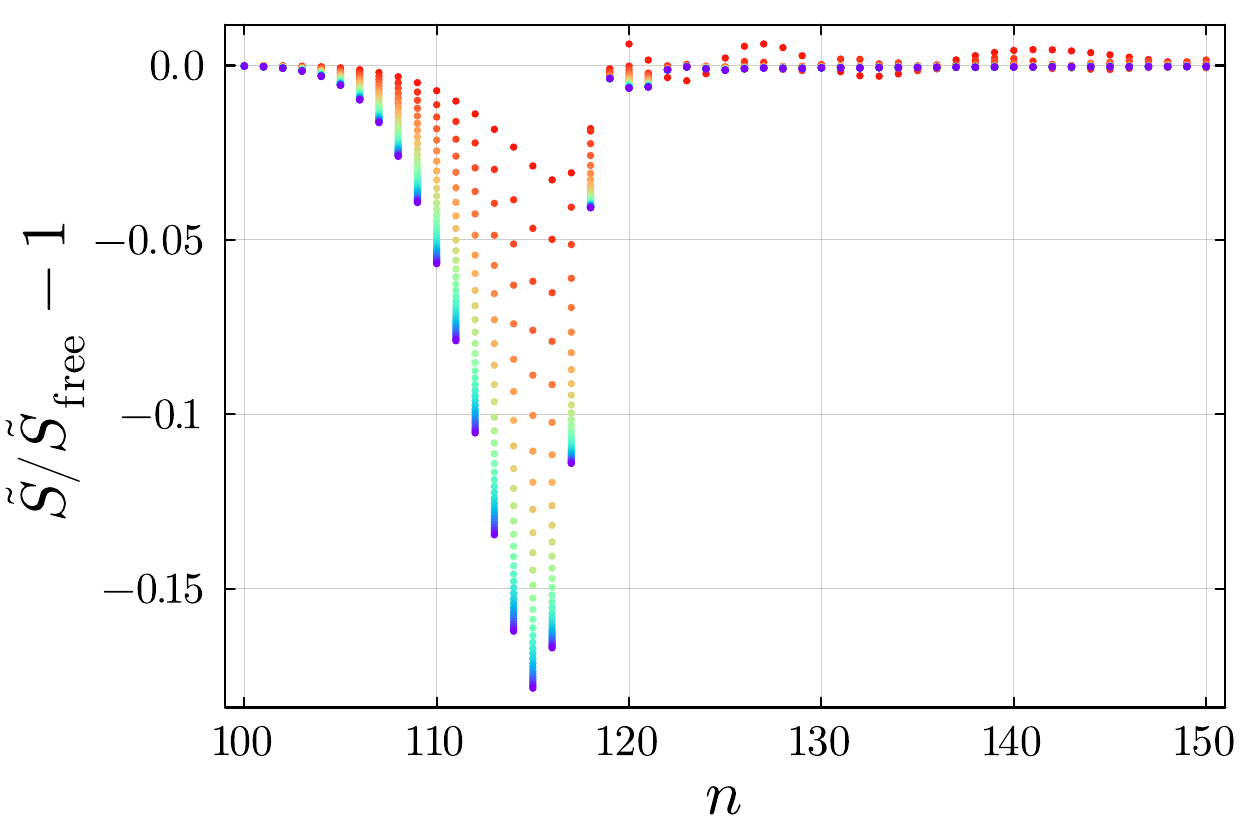}
  \caption{\label{stildif}Relative difference between the dynamical and the
  free entropy per angular momentum mode vs. radial index at times $t = 5$
  (top panel) and $t = 12$ (bottom panel) for the $N_r = 256$ system. Every
  $16^{\tmop{th}}$ angular momentum mode is plotted from $l = 15$ (red) to $l
  = 511$ (violet). Note that the comparison is not performed at a fixed radial
  coordinate $n$, but rather between the dynamical data at a given $n$ and the free data interpolated to the corresponding $\tilde{c}'$
  as explained in the text.}
\end{figure}
\begin{figure}[htb]\centering
  \includegraphics[width=\columnwidth]{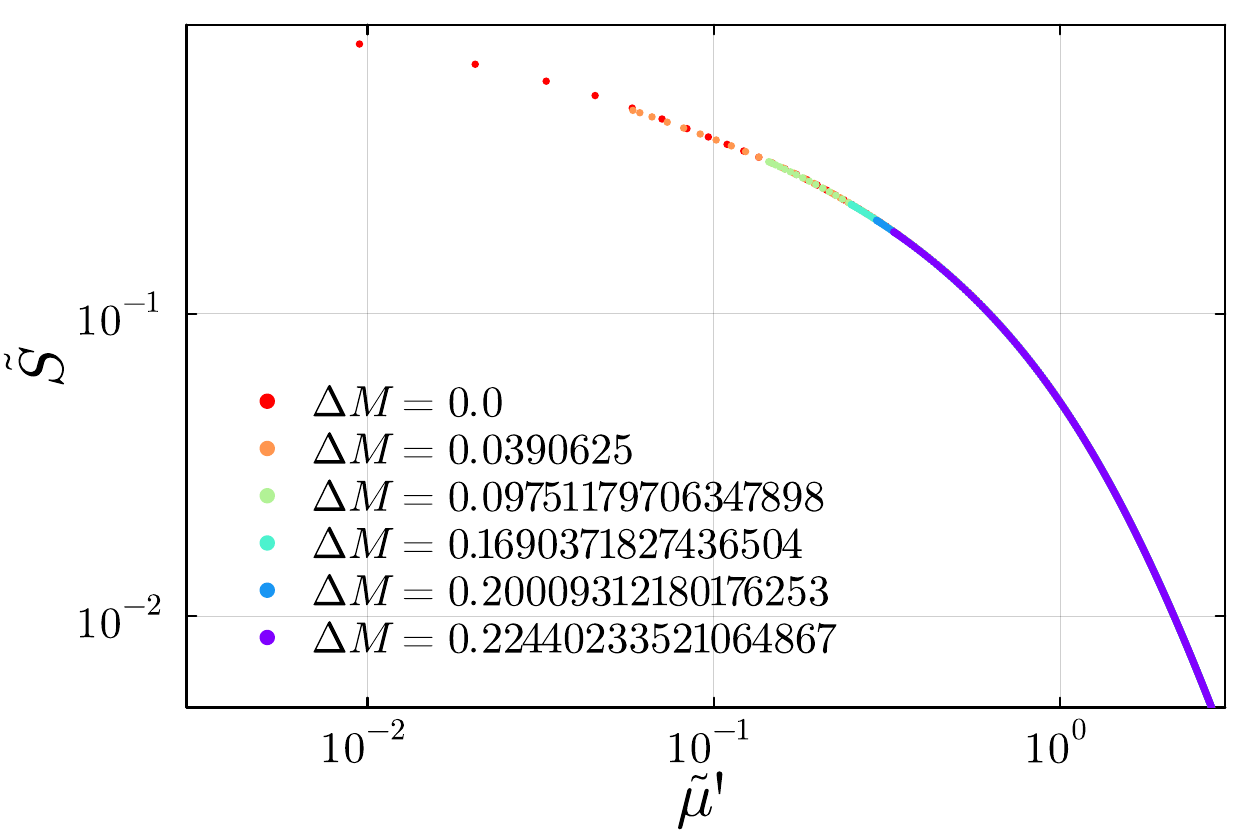}\par\medskip
  \includegraphics[width=\columnwidth]{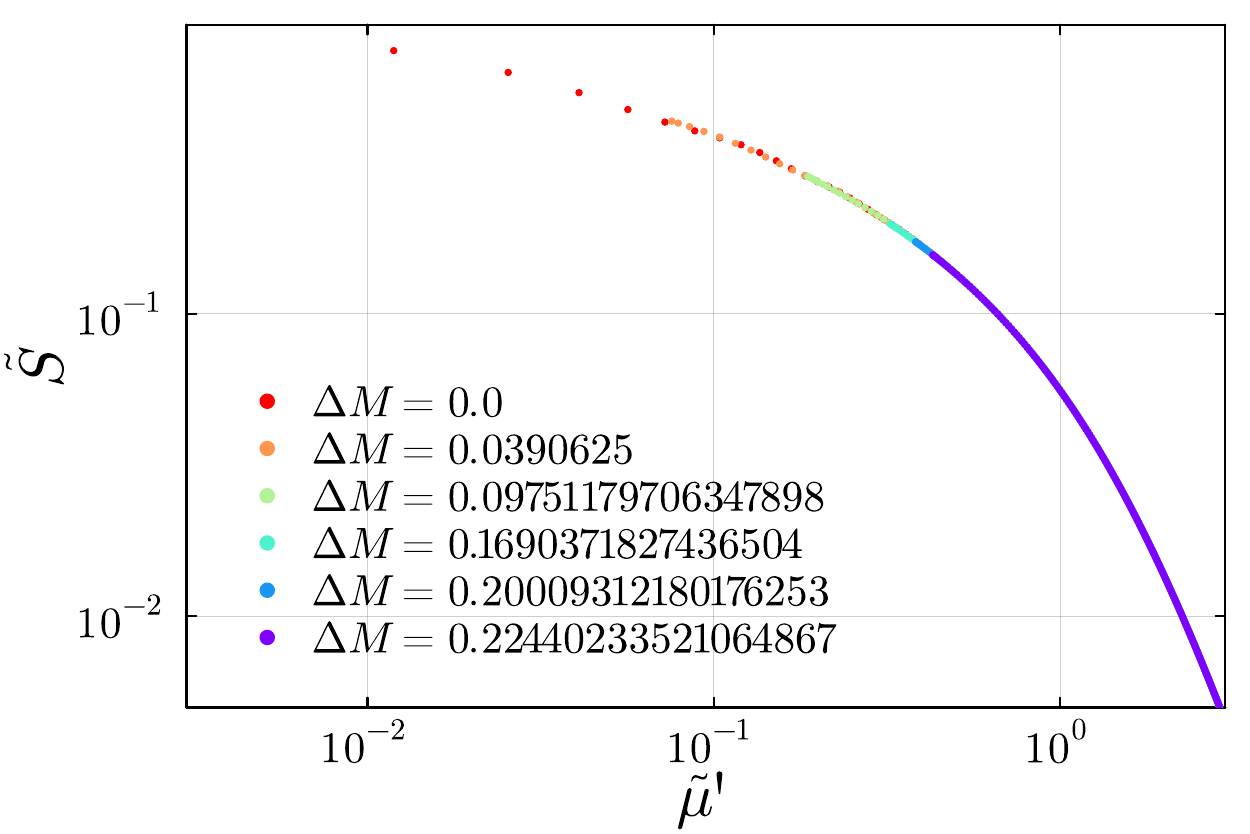}
  \caption{\label{stilmass}Single mode entropy vs. boundary corrected
  effective mass term \eqref{bcmeff} for the $N_r = 256$ system at radial
  coordinates $n = 115$ (top panel) and $n = 120$ (bottom panel) that are
  inside and just outside the peak region at $t = 12$ and for various masses.}
\end{figure}
\begin{figure*}
  {\includegraphics[width=\columnwidth]{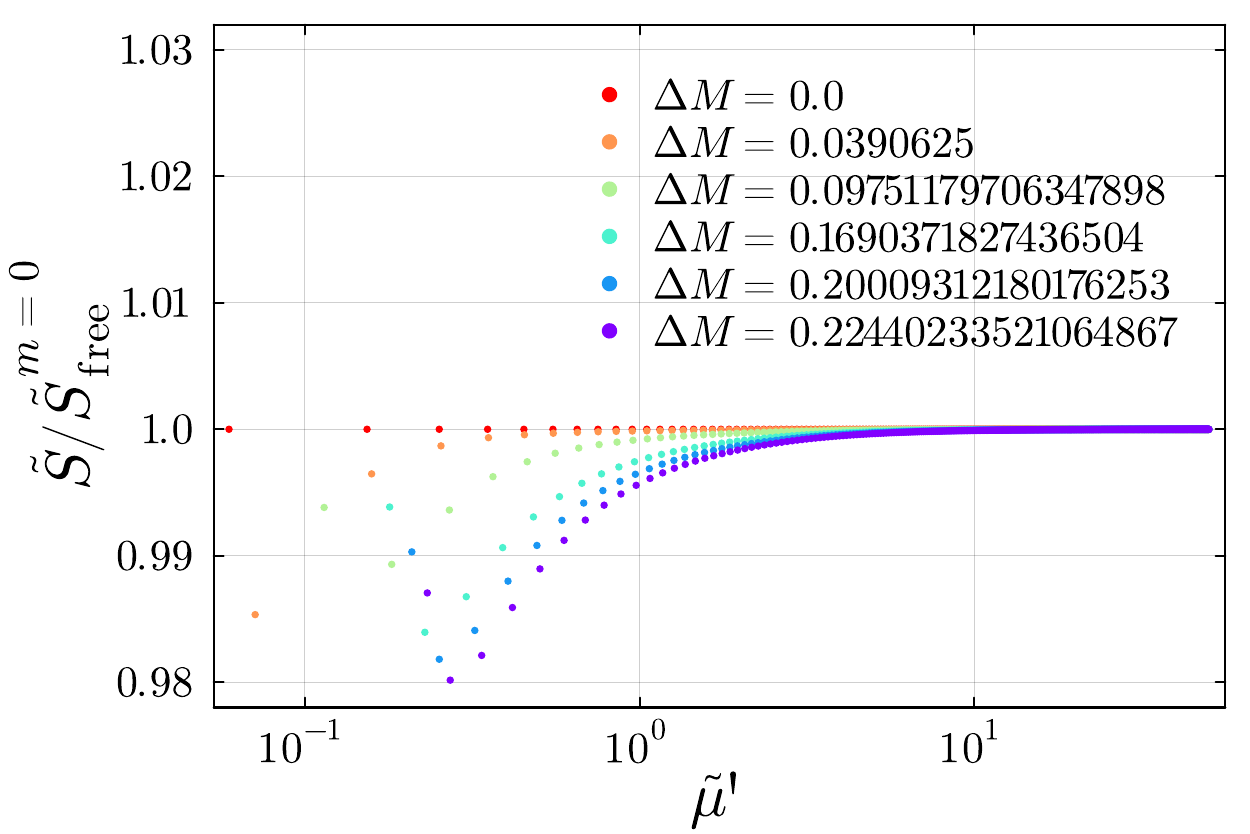}}{\includegraphics[width=\columnwidth]{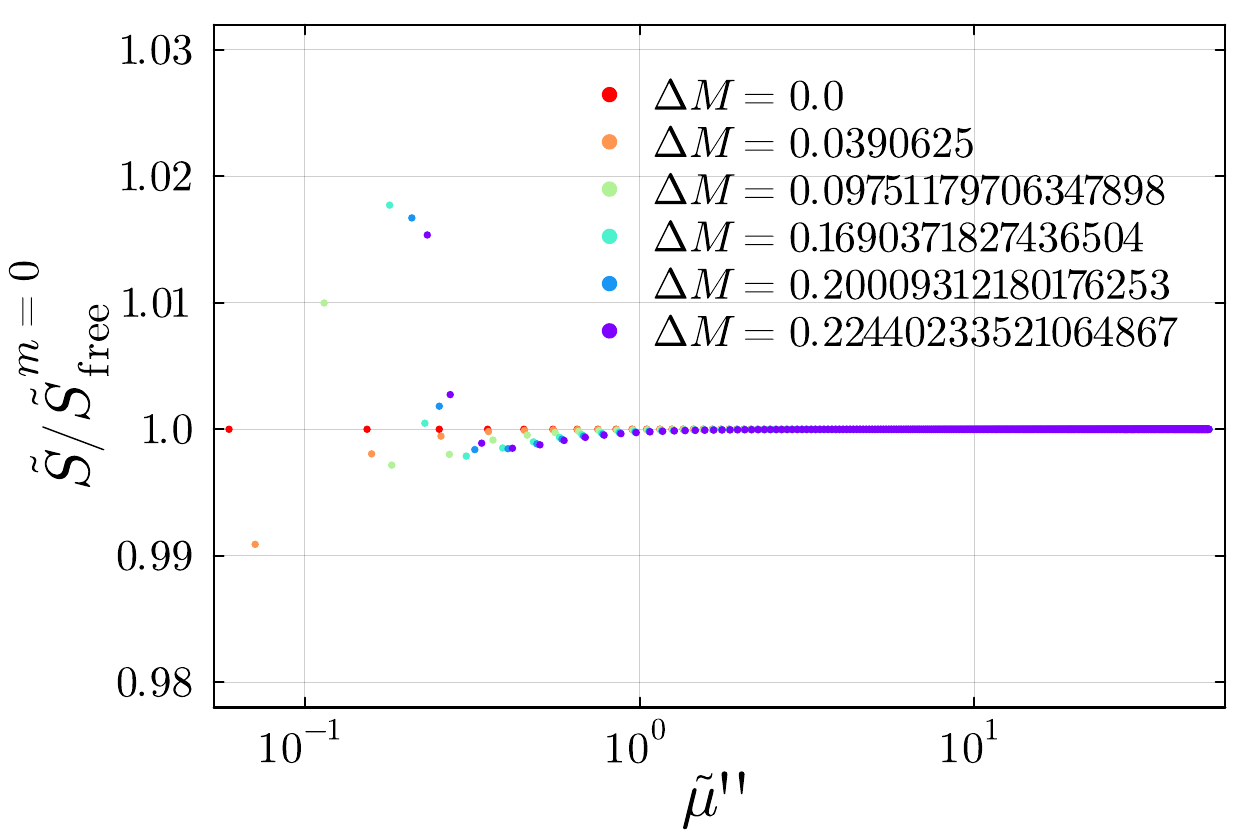}}
  {\includegraphics[width=\columnwidth]{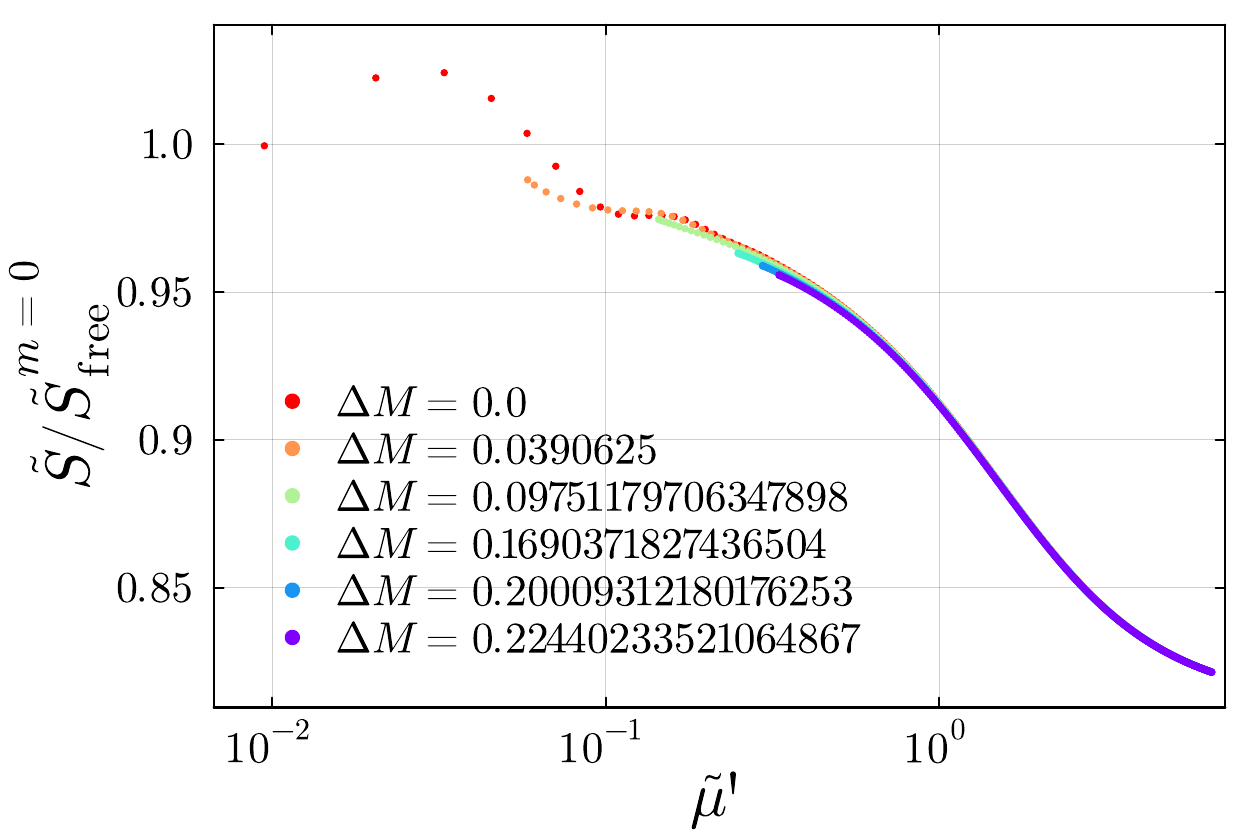}}{\includegraphics[width=\columnwidth]{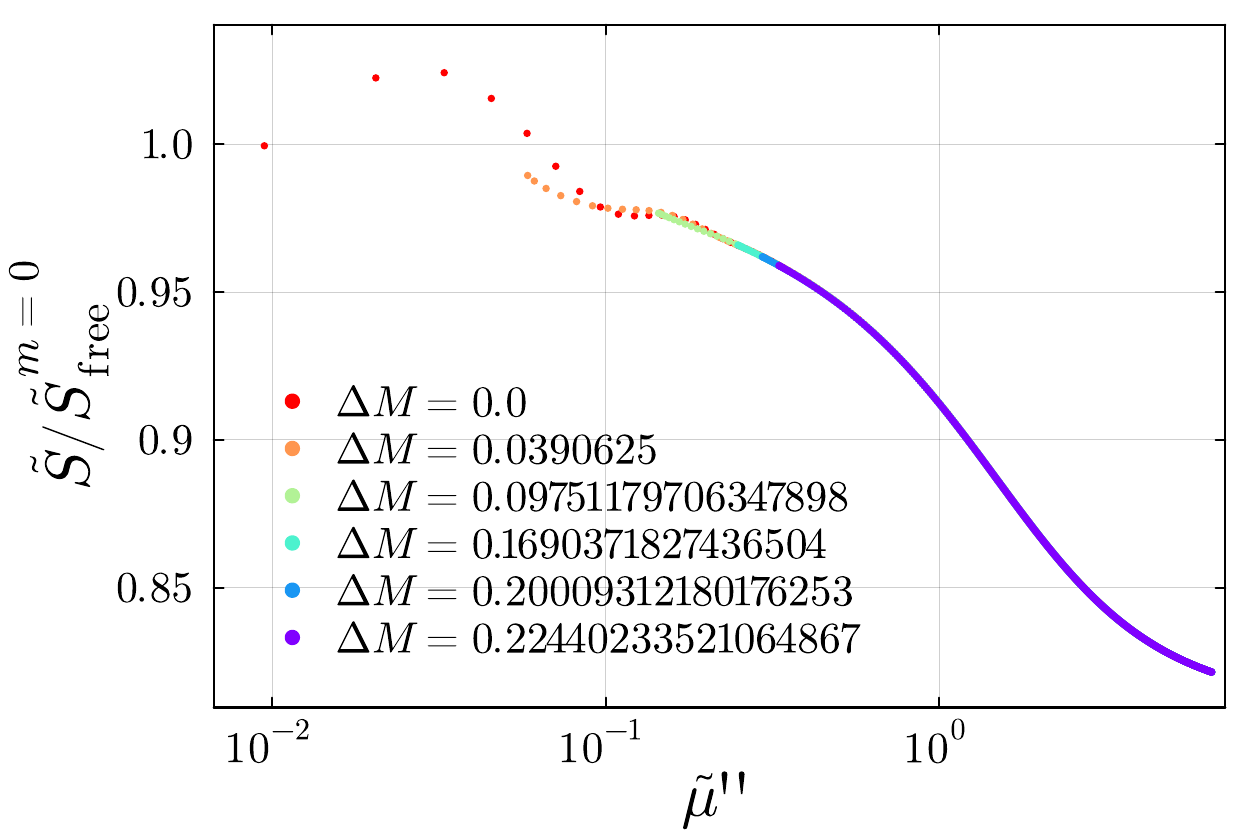}}
  {\includegraphics[width=\columnwidth]{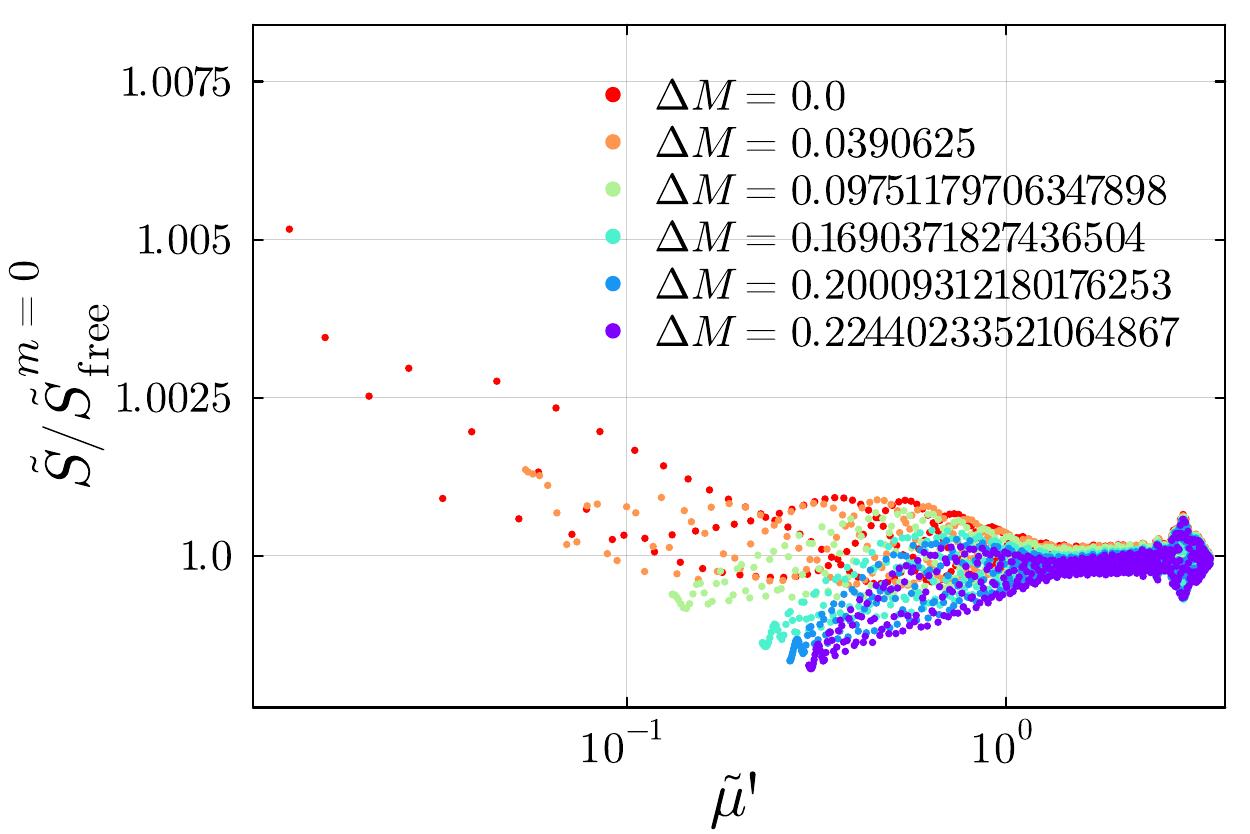}}{\includegraphics[width=\columnwidth]{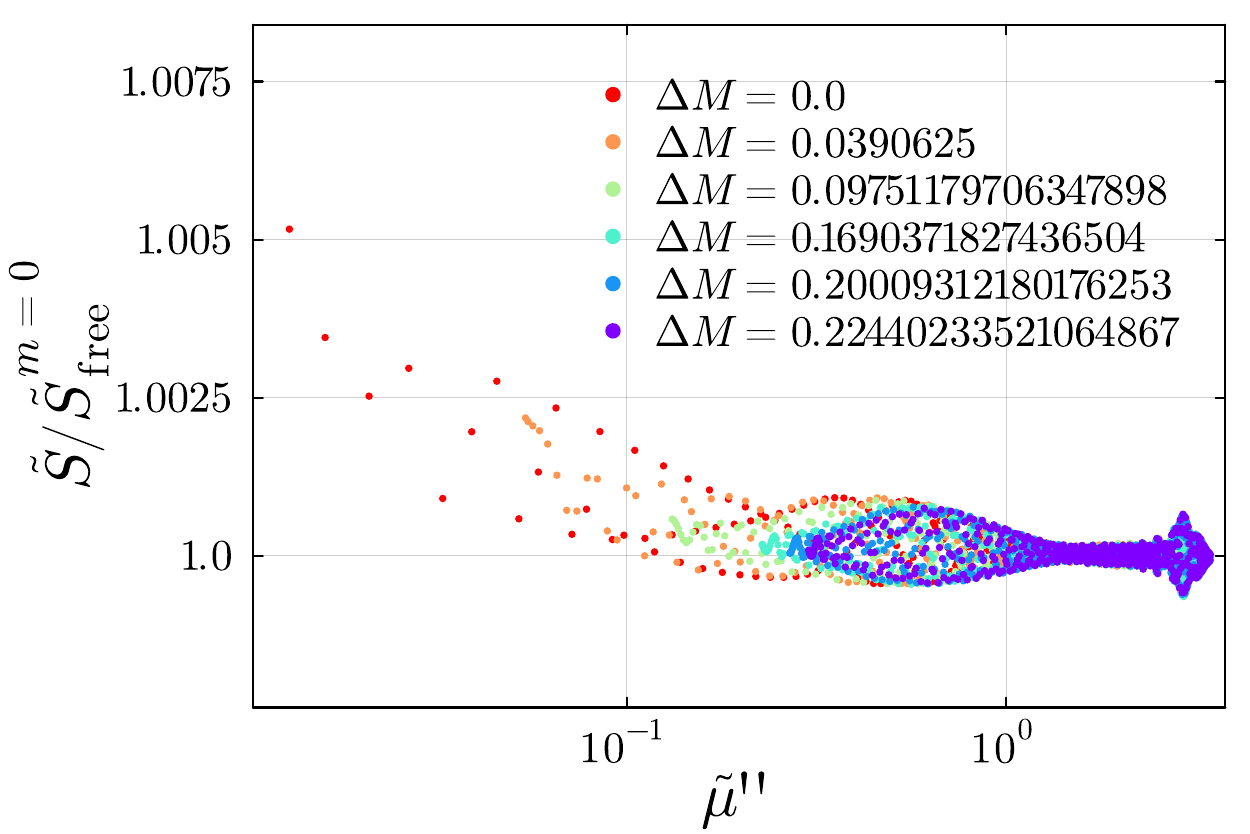}}
  \caption{\label{massdetail}Normalized single mode entropy vs. boundary
  corrected effective mass terms $\tilde{\mu}'$ (left column) and $\mu''$ \eqref{bcmeff} (right column) for the $N_r = 256$ system at simulation time $t=12$ and radial coordinates $n = 10$, $n=115$ and $n=200$. Normalization is provided by the spline interpolated zero mass free entropy at the corresponding $\tilde{\mu}'$ resp.~$\mu"$. At the coordinates that the bump has not yet reached ($n=10$) or has already passed over ($n=200$), deviations from the free universal behavior are small, while at the peak of the bump ($n=115$), there is a large drop of the entropy compared to the free case for large $l$. Note that at $n=10$ the dynamical and free entropies agree exactly.}
\end{figure*}

\begin{equation}
  \begin{aligned}
    \tilde{\mu}^2 ={}& (\Delta r)^2 \mu^2\\
    ={}& \frac{r}{d} \Delta^2
    \left( \frac{l (l + 1)}{r^2} + M^2 \right)\\
    ={}& \frac{l (l + 1)}{n_{\tmop{eff}}}
    + \frac{r}{d} \Delta^2 M^2
  \end{aligned}
  \label{dynmeff}
\end{equation}
which is the generalization of the dimensionless free effective mass
\eqref{meff}. To correct for boundary effects, we look at the $l = 0$ mode
separately. As detailed in App.~\ref{l0approx}, a good heuristic approximation
may be found even in the interacting case by replacing the effective boundary
mass term in \eqref{sourunits} with a modified version \eqref{btildyn}. This
in turn motivates us to approximate the boundary corrections with an effective
mass term $\tilde{b}'$ that we extract from the $l = 0$ data via
\eqref{boundcr}, as in the free case to define a boundary corrected effective
mass term
\begin{equation}
  \tilde{\mu}'{}^2 = \tilde{\mu}^2 + \tilde{b}'{}^2 \;.\label{bcmeff}
\end{equation}
For the massless case, we define
\[ \tilde{c}^2 = \Delta^2 \frac{l (l + 1)}{d r} = \frac{l (l +
   1)}{n_{\tmop{eff}}^2} \;,\]
which is the proper generalization of the free scaling variable $\tilde{c}$,
and its boundary corrected version
\[ \tilde{c}'{}^2 = \tilde{c}^2 + \tilde{b}'{}^2\;. \]
Noting that $n_{\tmop{eff}}$ is a constant for a given radial coordinate $n$
and simulation time $t$, we find that the total normalized entropy in the
continuum limit is still connected to the single mode entropy via a relation
of the form \eqref{fusimo}, namely
\begin{equation}
  s = \bigintlim_0^{c^2} \mathd \tilde{c}'{}^2 \, \tilde{S}_n \left(
  \tilde{c}'{}^2 \right) \;.\label{fusimodyn}
\end{equation}
Of course the universal scaling function $\tilde{S} (\tilde{c}^2)$ is replaced
with $\tilde{S}_n \left( \tilde{c}'{}^2 \right)$ that depends on the radial
coordinate $n$. For small coordinates that are not yet affected by the
incoming bump, there is indeed no difference between the free and dynamical
cases. What happens in and beyond the bump region is displayed in Fig.~\ref{twomodes}. As one can see, there is a little systematic shift in the bump
region so that the dynamical points do not follow the free universal function
anymore. Outside the bump region, however, it seems that except for a trivial
shift along the universal curve there is very little deviation except for some
small jitter at low $\tilde{c}'$.

In Fig.~\ref{twomodeslin} we plot the same data vs. the uncorrected
$\tilde{c}^2$ on a linear scale. Plotted in this form, the boundary correction
term $\tilde{b}'{}^2$ corresponds to an additive shift in the $x$-axis.
Clearly, the difference between the free and dynamical results cannot be
attributed to such a shift and its numerical value in the central region of the
lattice $\tilde{b}'{}^2 = e^{- 12 S_0} \sim 10^{- 4}$ is negligible. We thus
conclude that in the dynamical case there is a deviation from the free
universal behavior, which is mostly confined to the bump region for large
$\tilde{c}$.

In order to obtain a clearer picture, we perform a cubic spline interpolation
of the free per mode entropy for each of the radial coordinates, which we call
$\tilde{S}_{\tmop{free}}$. We then compute the relative difference of the
dynamical free mode entropy to the interpolated $\tilde{S}_{\tmop{free}}$ at
the same $\tilde{c}'$, which we plot in Fig.~\ref{stildif} for all radial
coordinates close to the bump region at two different simulation times. A
clear systematic deviation can only be seen in the bump region, which is
consistent with the entropy dip of Fig.~\ref{bwzoom}. On the other hand, the
oscillating deviations for coordinates outside the bump region do seem to
cancel out in the mode sum.

Finally, we investigate the effect of adding a mass term. As in the free case,
the Hamiltonian density \eqref{classhamdens} only depends on the effective
mass \eqref{dynmeff}. Including boundary corrections, we thus expect the
entropy contribution $\tilde{S}$ at a certain radial coordinate to only depend
on the boundary corrected effective mass \eqref{bcmeff} with the boundary
correction term given by \eqref{boundcr}. As seen in Fig.~\ref{stilmass}, this
is indeed the case to a relatively good accuracy.

In order to see how accurately this universality holds, we divide the massive
entropies at a certain radial coordinate $n$ by the spline interpolated free
massless entropy at the same boundary corrected effective mass \eqref{bcmeff}.
The result for three radial coordinates of the $N_r = 256$ system at $t =
12$ are displayed in the left column of Fig.~\ref{massdetail}.  As in the free case, we see a systematic deviation of the
massive entropies from the massless ones that can be addressed with a multiplicative correction of the mass term. In strict analogy to the free case \eqref{cmefffree} we thus define a new scaling variable
\begin{equation}
  \tilde{\mu}''{}^2 = \frac{r}{d} \Delta^2 \left( \frac{l (l + 1)}{r^2} +
  \left( 1 + \frac{1}{n} \right) M^2 \right) + \tilde{b}'{}^2 \;.\label{cmeff}
\end{equation}
The entropy ratios with this new scaling variable are displayed in the right column of Fig.~\ref{massdetail}. We see that universality violations due to the mass term appear predominantly in the lowest $l$ modes, similar to what we observed in the free case, although there is some additional high frequency jitter in the region that the bump has already passed.
Because of this near perfect universality, we can generalize the integral
representation \eqref{fusimodyn} to the massive case. To a very good
approximation, the brick wall normalized entropy at radial coordinate index
$n$ and anisotropy factor $c$ is thus given by
\begin{equation}
  s_{n, \tilde{k}} (c^2) \sim \bigintlim_{\tilde{k}^2}^{c^2 + \tilde{k}^2}
  \mathd \tilde{c}^2 \, \tilde{S}_n (\tilde{c}^2) \;,\label{murd}
\end{equation}
where, according to \eqref{cmeff}
\[ \tilde{k}^2 = \frac{r}{d} \Delta^2 \left( 1 + \frac{1}{n} \right) M^2\;. \]

\subsection{Dynamical collapse with Pauli-Villars regularization}

\begin{figure}[htb]\centering
  \includegraphics[width=\columnwidth]{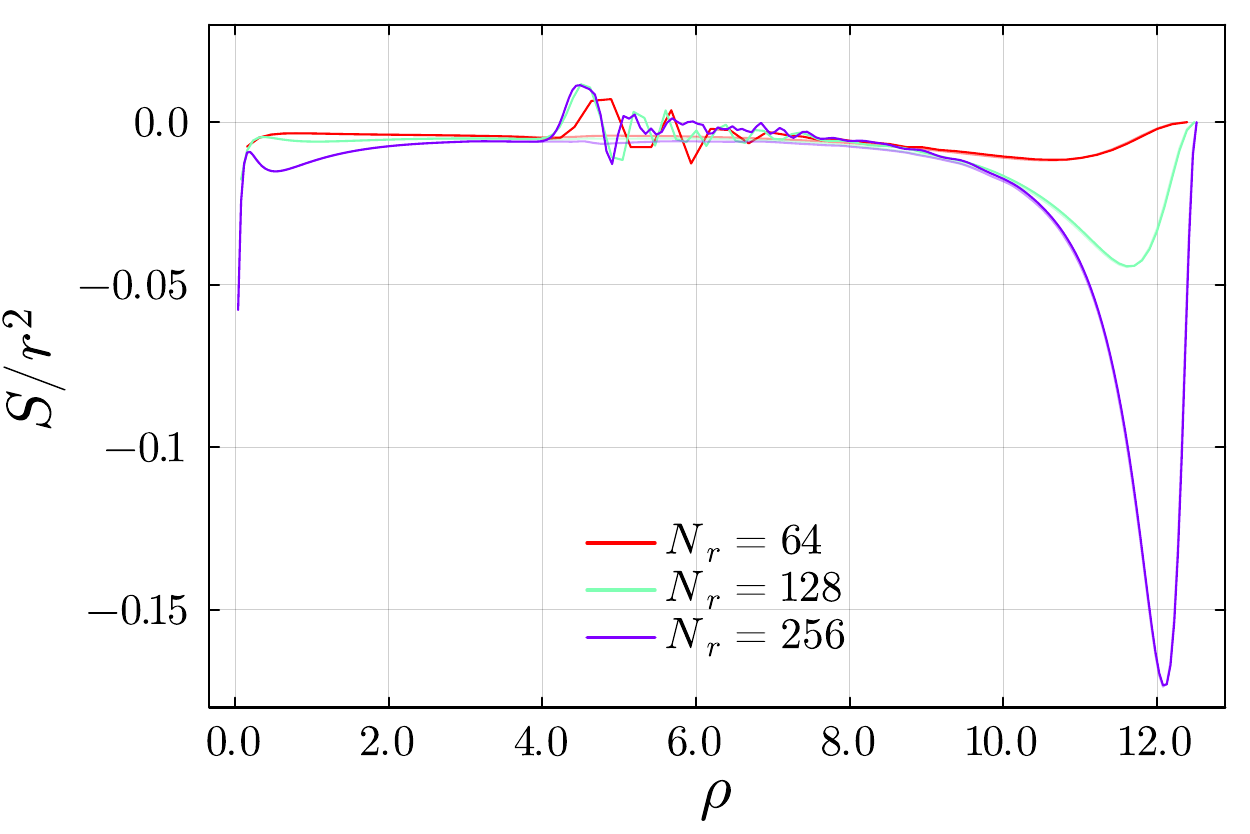}\par\medskip
  \includegraphics[width=\columnwidth]{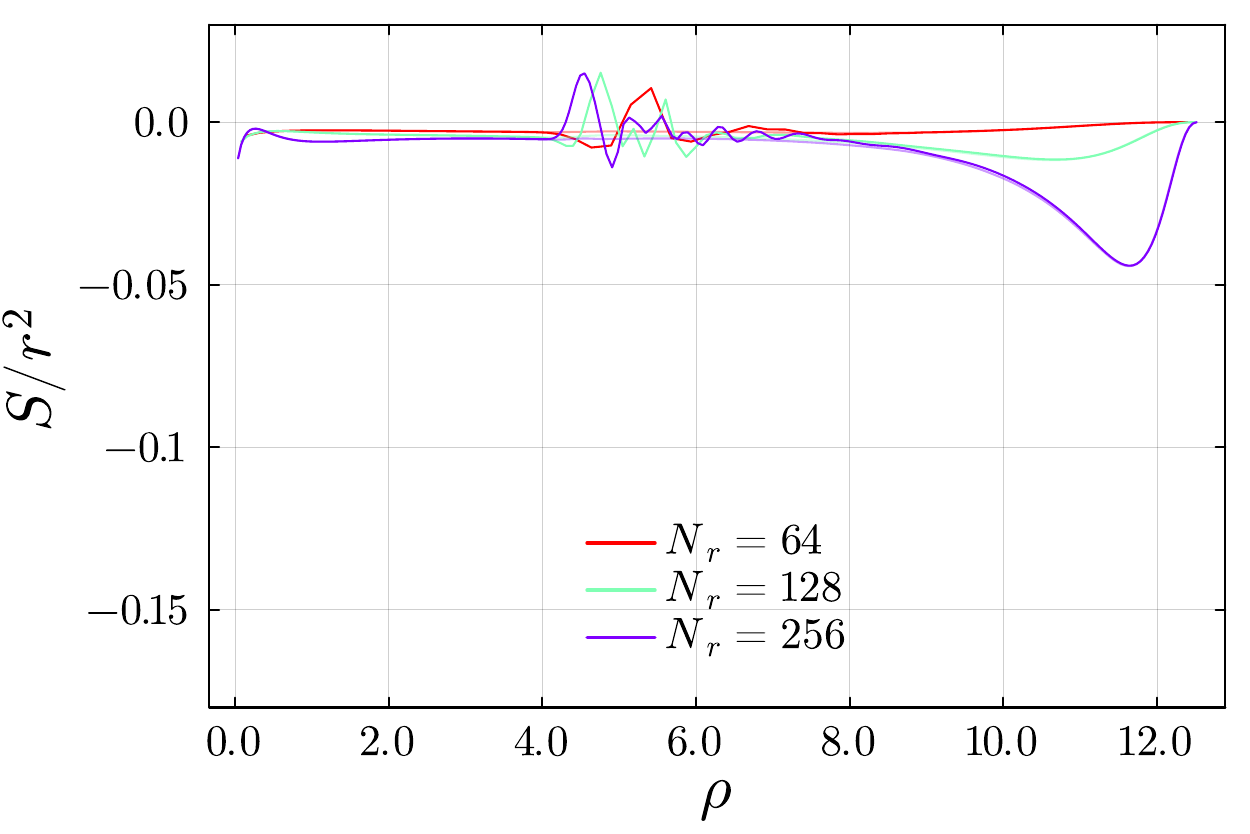}
  \caption{\label{dynpv}Area normalized entropies vs.~the physical center distance $\rho$ for the log-polynomial Pauli-Villars scheme at cutoff ratio
  $k = 0.078125$ (top) and $k = 0.0390625$ (bottom) and time $t = 12$ with
  $l_{\max} = 511$. The dark lines represent dynamical data, while the transparent lines of the same color
  correspond to the free subtraction term $S^0 (L_0 - (L_n - \rho_n)) / r^2$ from
  \eqref{freesub}.}
\end{figure}

\begin{figure}[htb]\centering
  \includegraphics[width=\columnwidth]{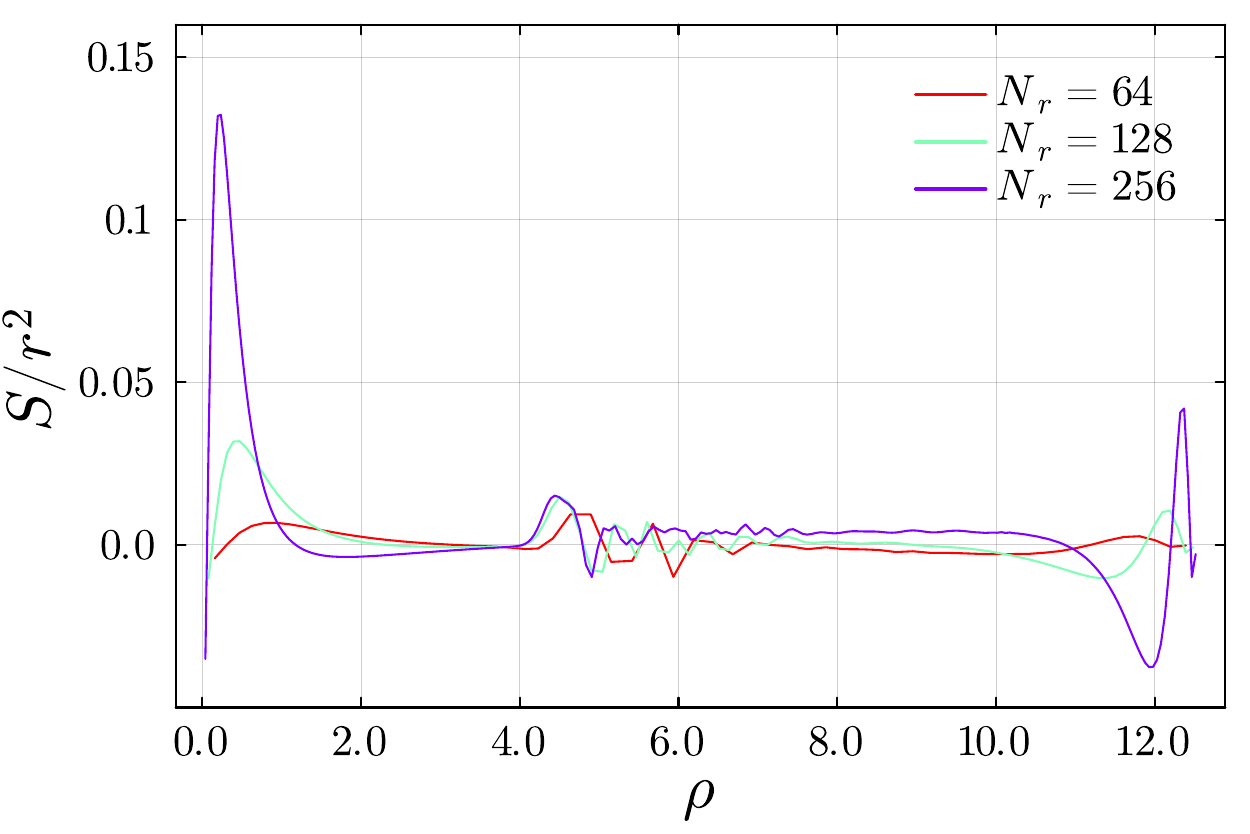}\par\medskip
  \includegraphics[width=\columnwidth]{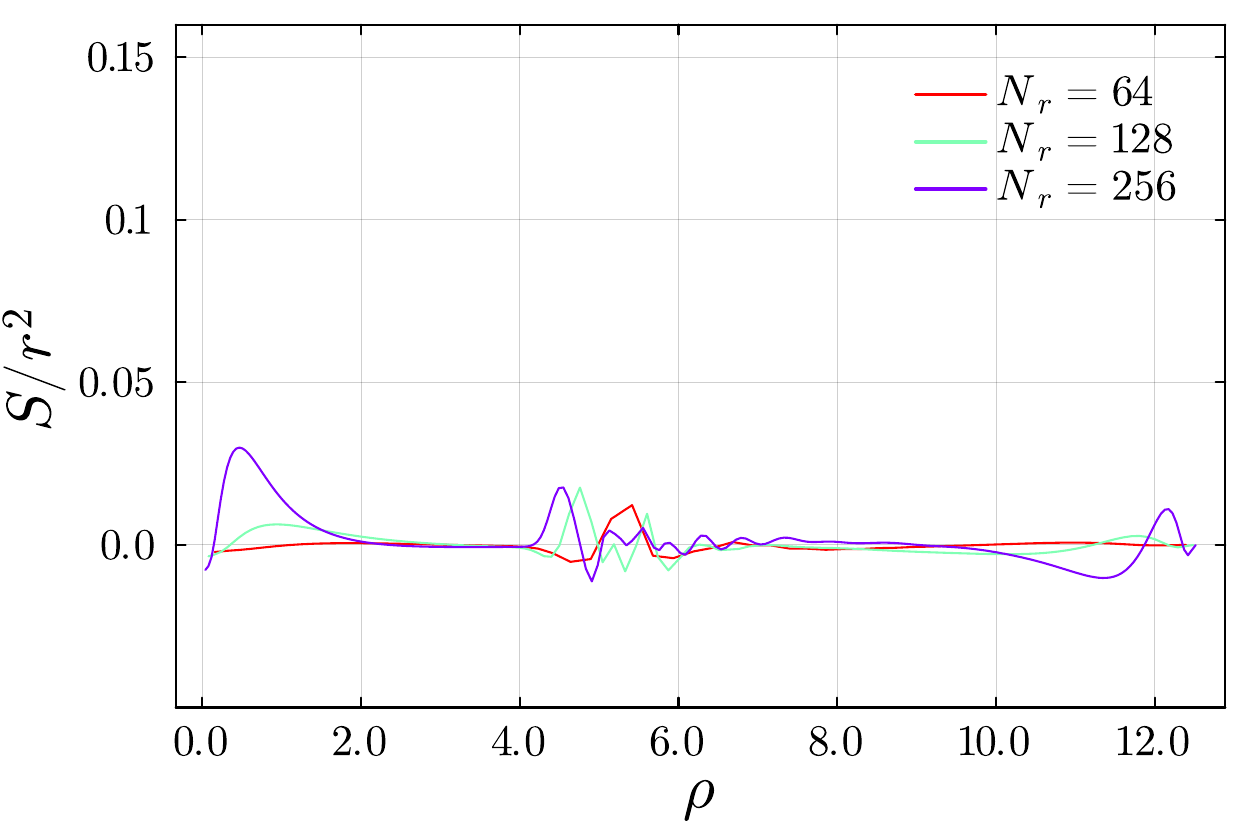}
  \caption{\label{dynpvc}The dynamical data of Fig.~\ref{dynpv} with the free
  boundary correction term \eqref{fbct} added.}
\end{figure}

\begin{figure*}[htb]\centering
  {\includegraphics[width=\columnwidth]{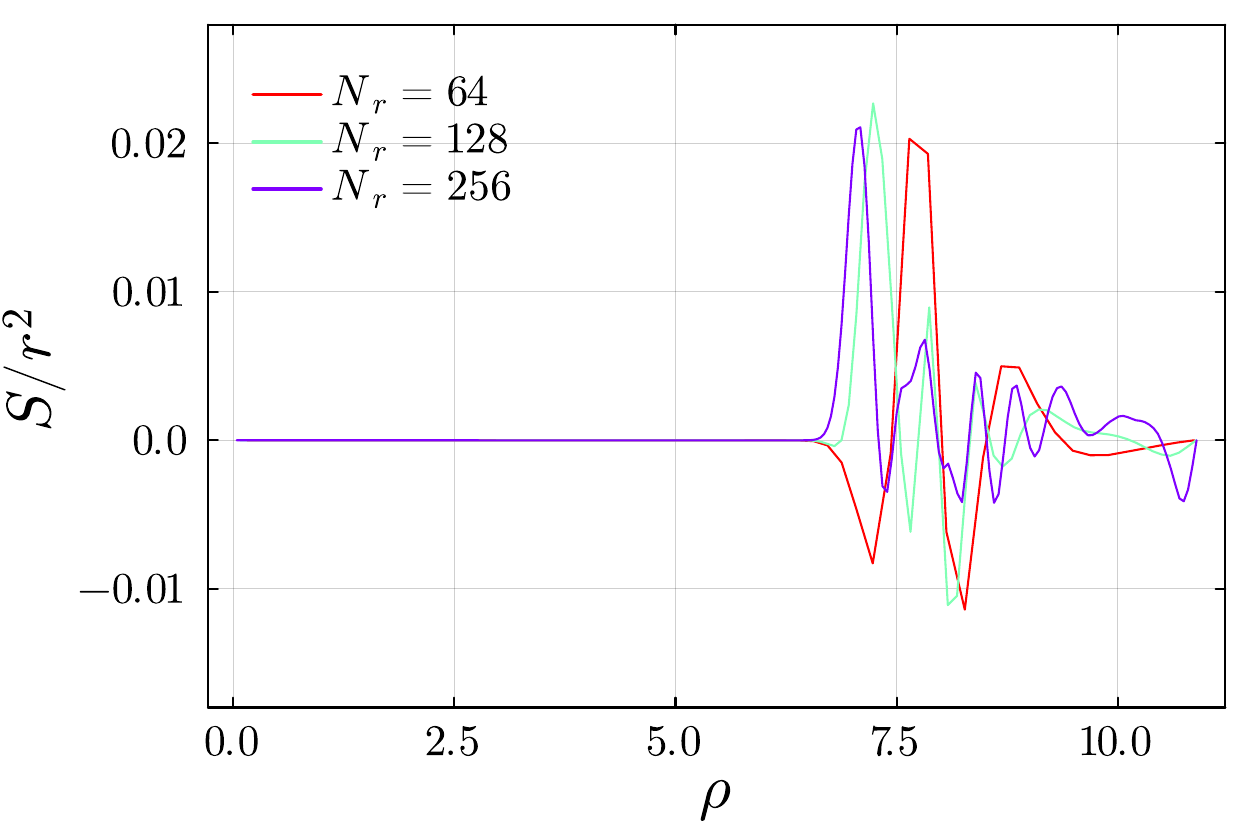}}{\includegraphics[width=\columnwidth]{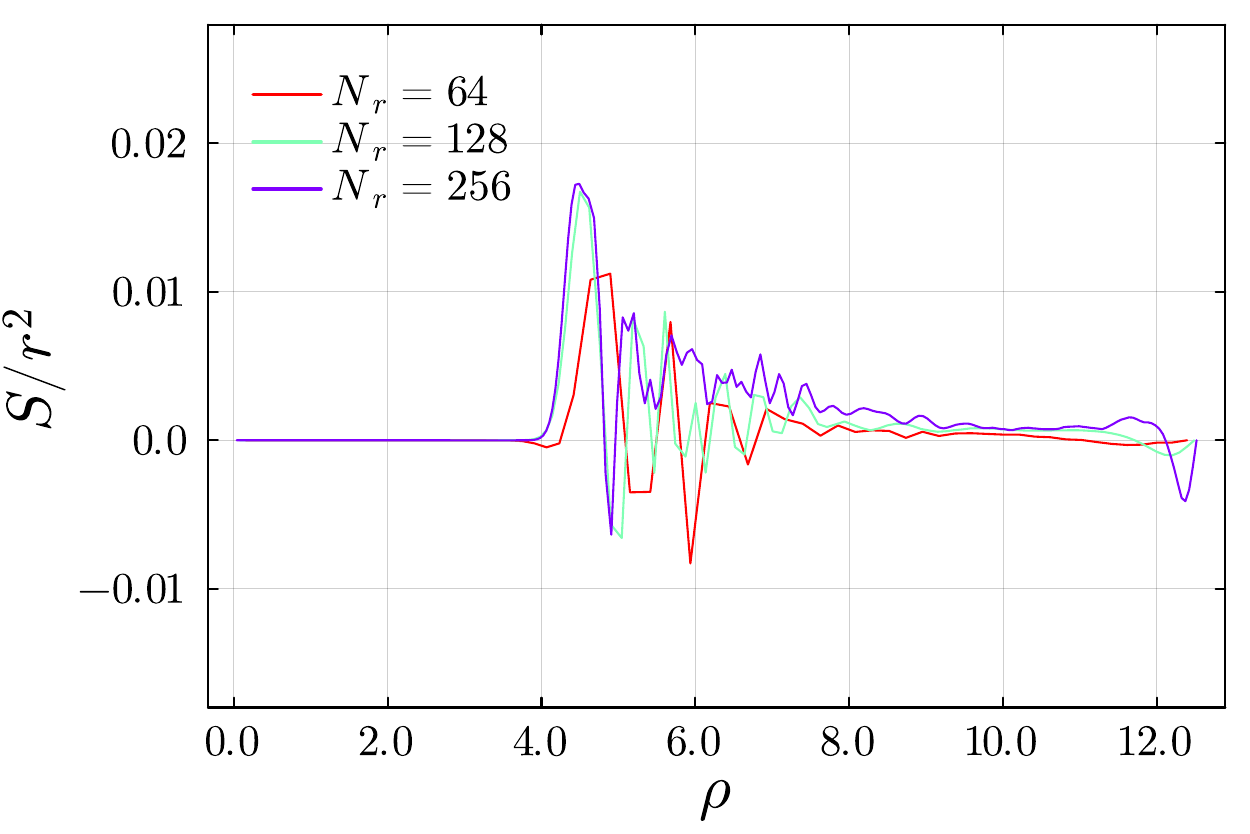}}
  
  {\includegraphics[width=\columnwidth]{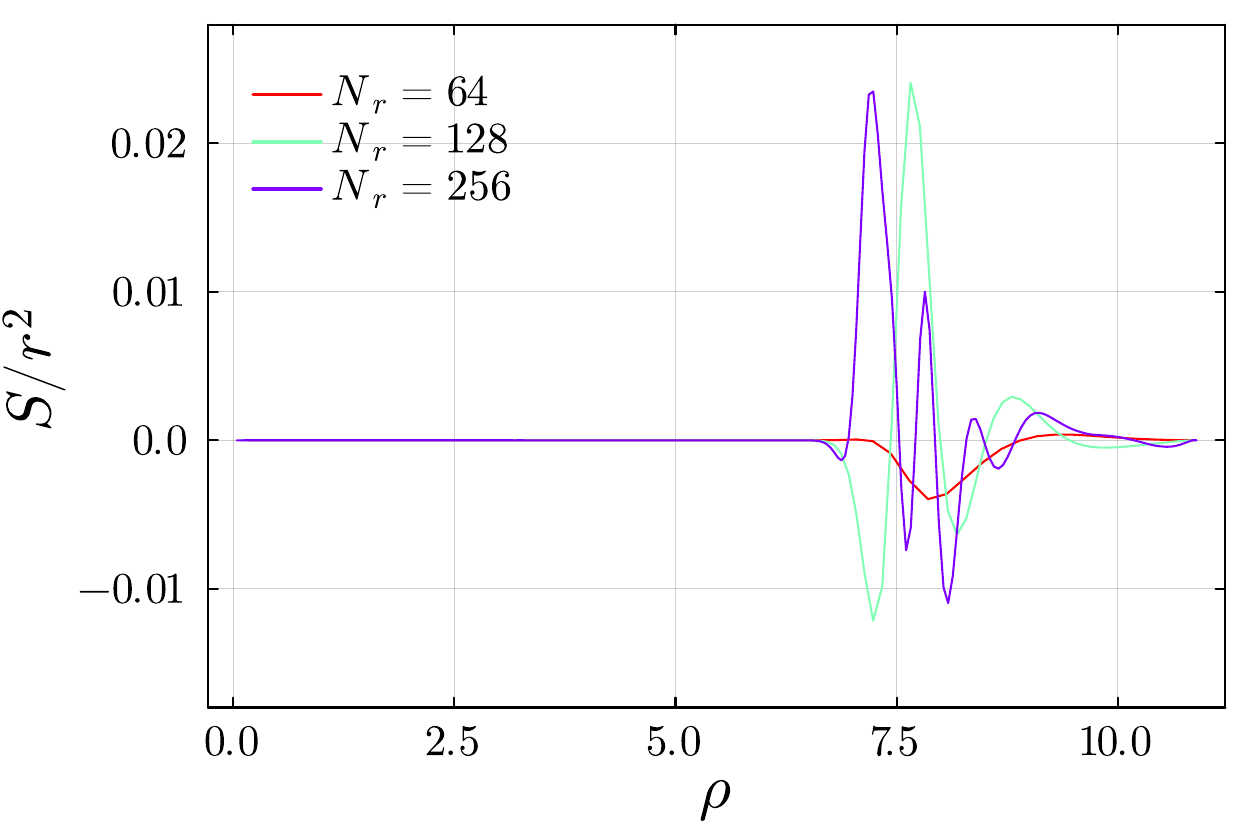}}{\includegraphics[width=\columnwidth]{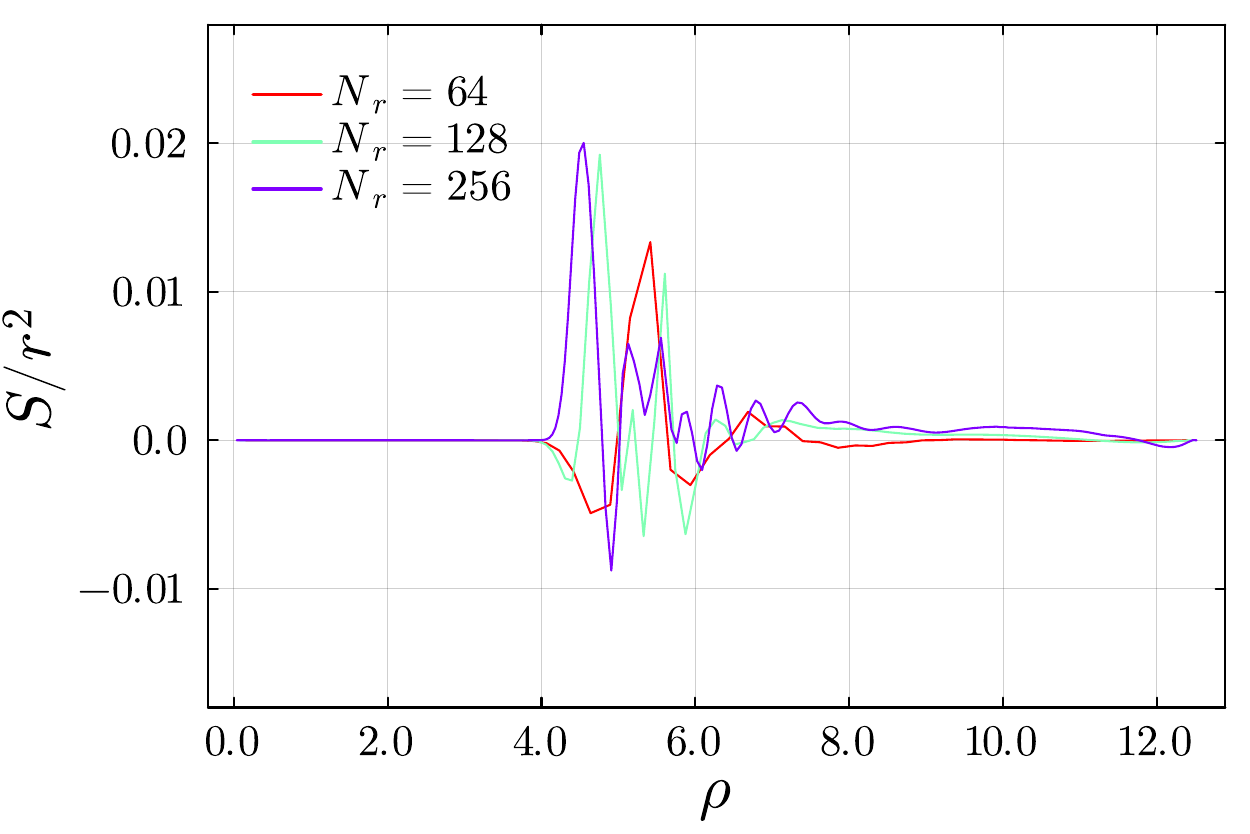}}

  \includegraphics[width=\columnwidth]{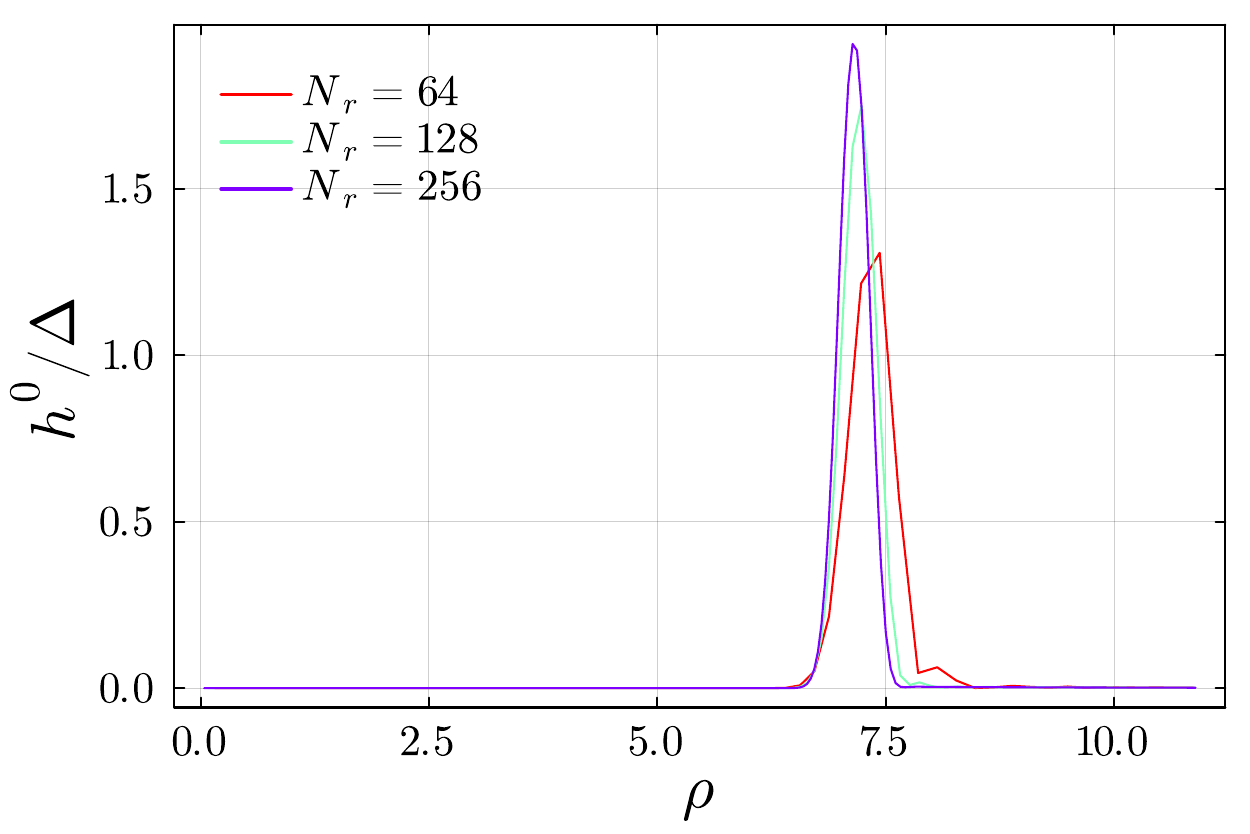}
  \includegraphics[width=\columnwidth]{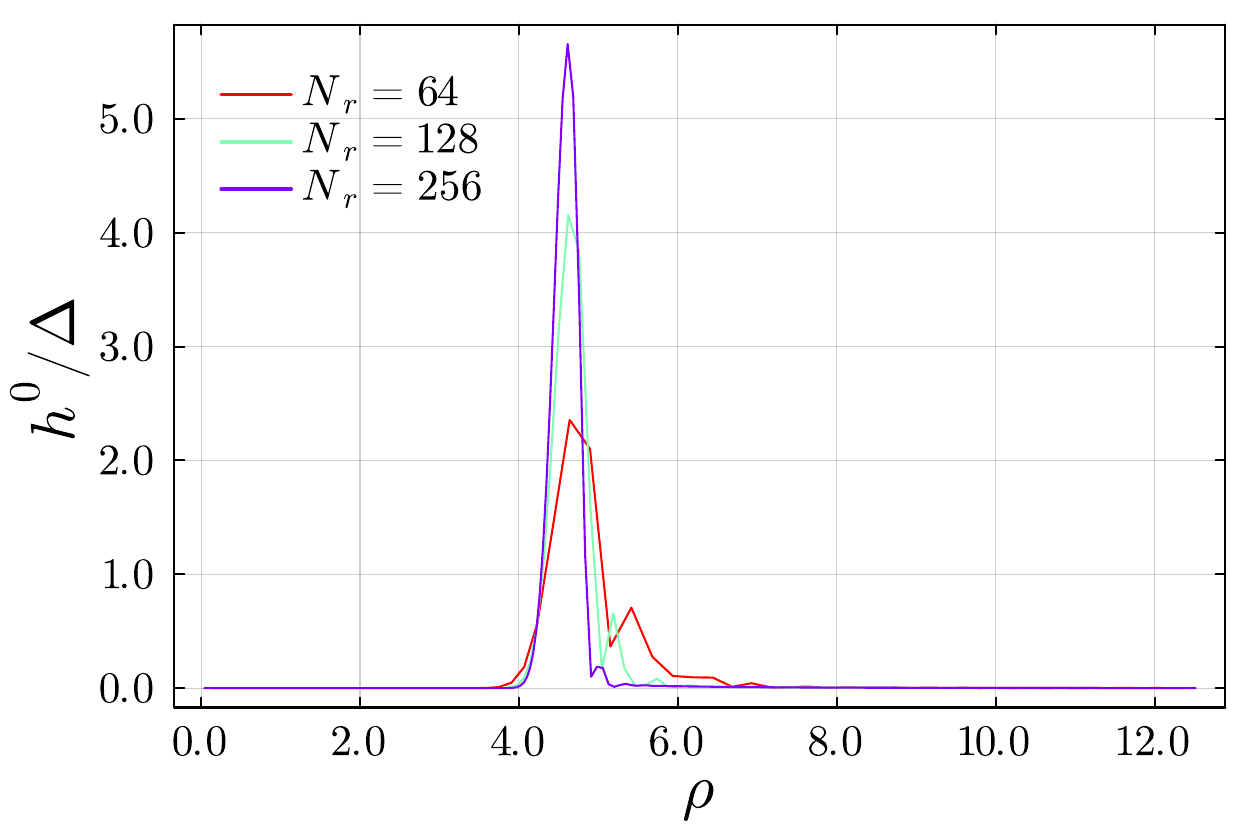}
  \caption{\label{k2evo}Boundary corrected, area normalized entropy for the
  log-polynomial Pauli-Villars scheme at cutoff ratio $k = 0.078125$ (top
  panels) and $k = 0.0390625$ (middle panels), times $t = 5$ (left panels)
  and $t = 12$ (right panels) and three different radial discretizations vs. physical distance
  from the center $\rho$. The corresponding classical energy densities at the respective times are given in the bottom panels as a reference. Although the data are noisy, it seems that
  there is a finite entropy in the continuum limit.}
\end{figure*}

\begin{figure*}[htb]\centering
  {\includegraphics[width=\columnwidth]{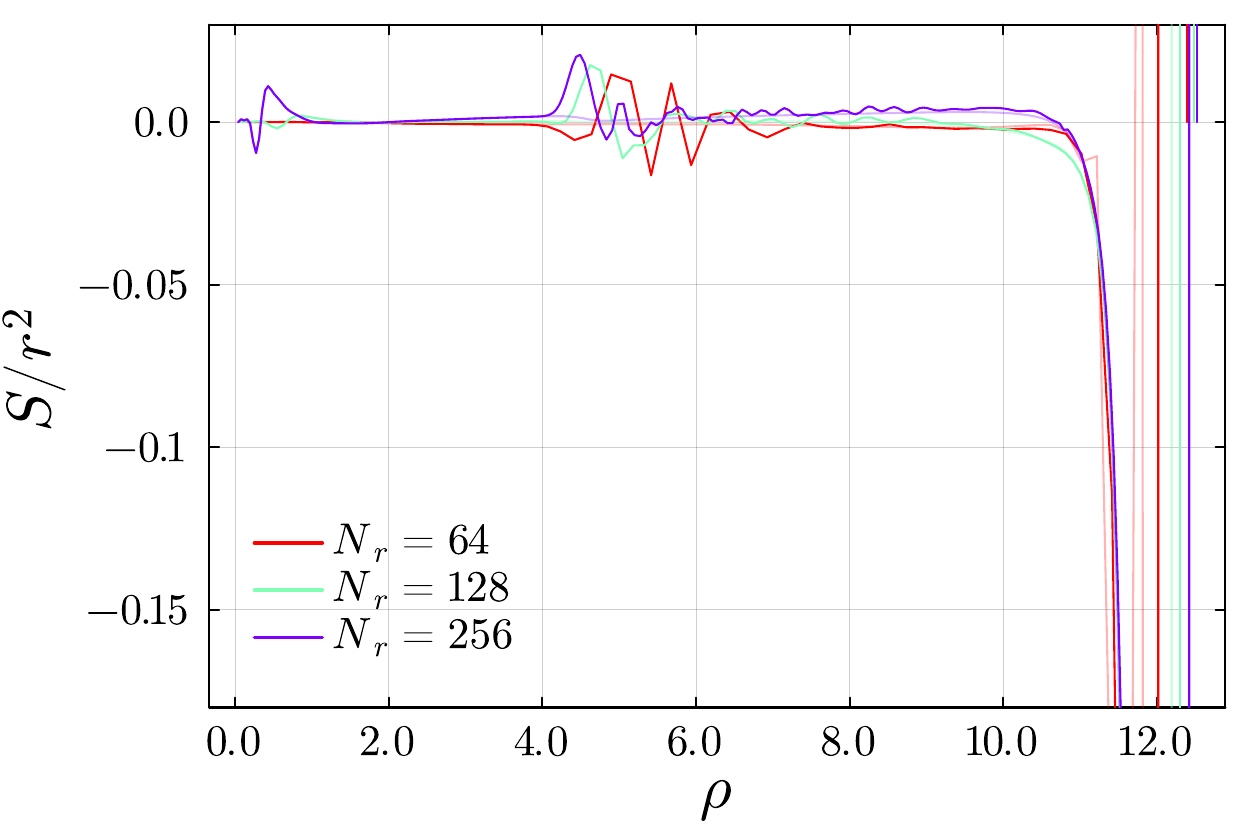}}{\includegraphics[width=\columnwidth]{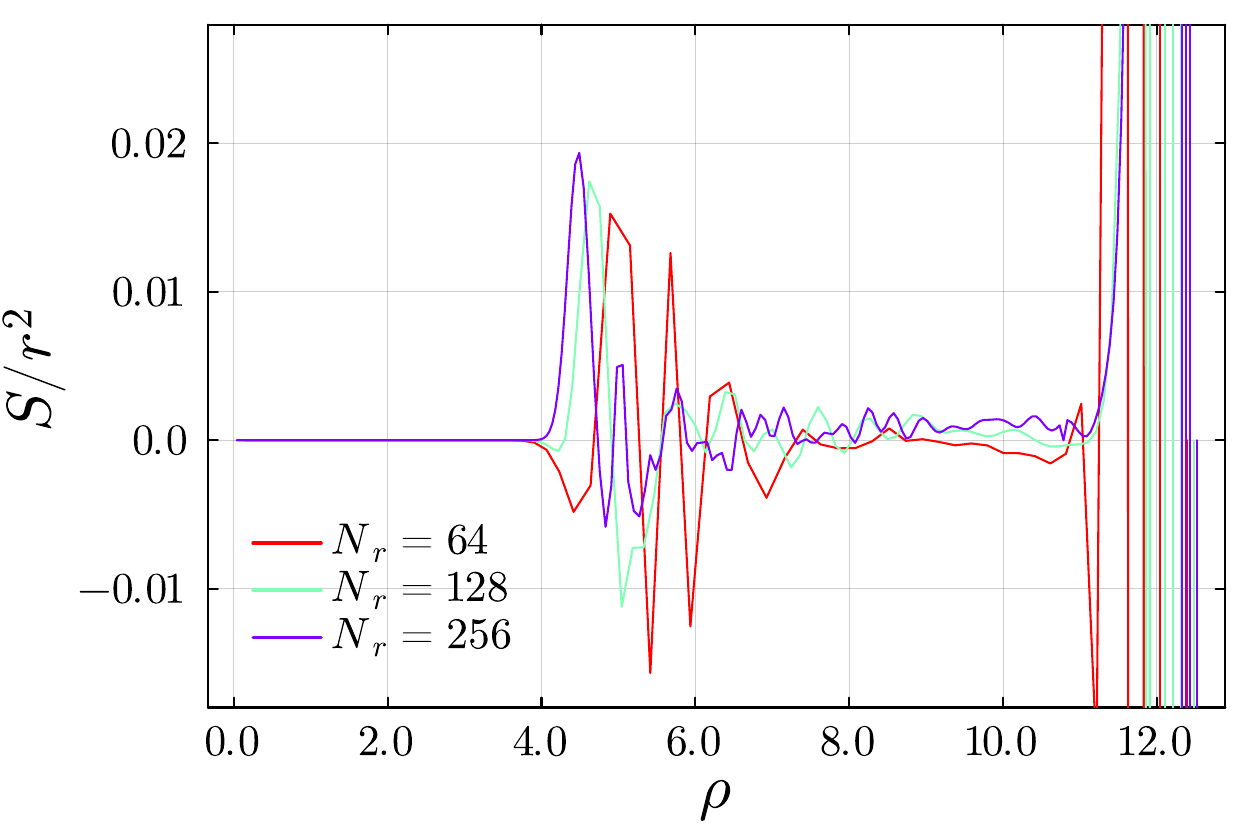}}
  
  {\includegraphics[width=\columnwidth]{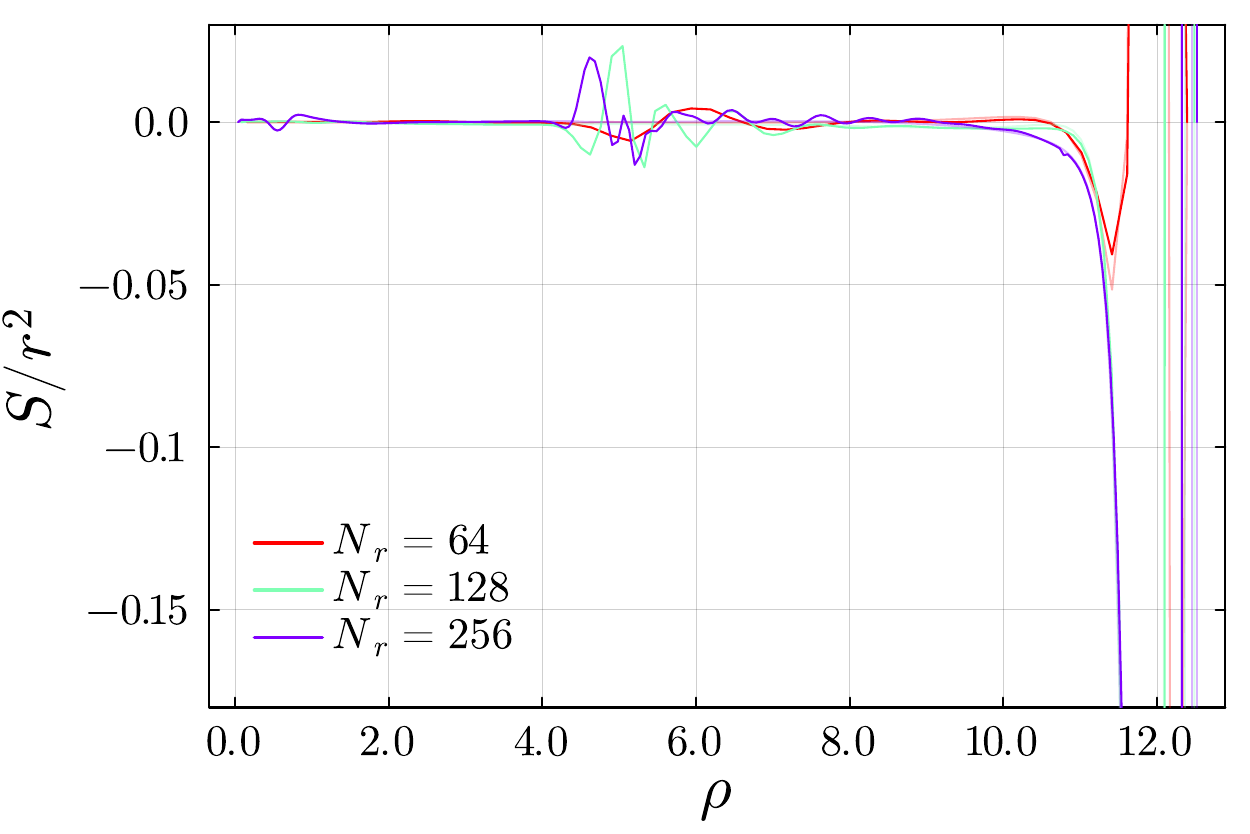}}{\includegraphics[width=\columnwidth]{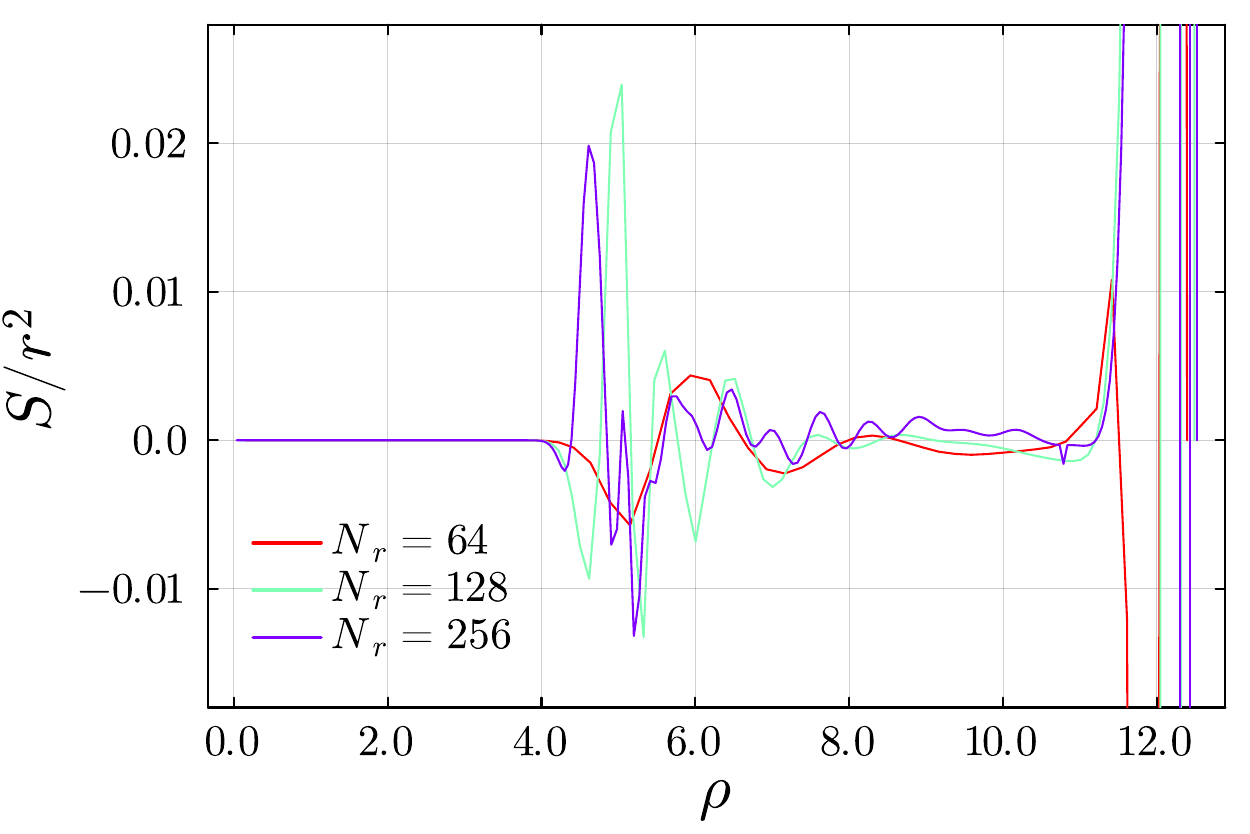}}
  \caption{\label{pPVsub}Area normalized entropy for the log-plolynomial
  pseudo Pauli-Villars scheme at cutoff ratio $k = 0.078125$ (top panels) and
  $k = 0.0390625$ (bottom panels) and time $t = 12$. The left panels show the
  unsubtracted entropies (dark lines) and the corresponding subtraction terms
  (transparent lines), whereas the right panels display the result of the
  subtraction.}
\end{figure*}

Finally, we turn our attention to the dynamical data with a Pauli-Villars
regulator. We will only consider the log-polynomial scheme because of its
vanishing area term in the entanglement entropy for $k\rightarrow0$ in the free case. In contrast
to the brick wall normalized entropy \eqref{gennorment}, the Pauli-Villars
normalized entropy \eqref{normentpv} does not explicitly depend on the metric
and can thus be used in the dynamical case without modification. Note however
that the relation between brick wall normalized entropy $s$ and Pauli-Villars
normalized entropy $\hat{s}$
\[ \hat{s} = \frac{s}{\Delta^2 M_{\tmop{PV}}^2} \frac{d}{r} \]
is no longer a constant but depends on the metric. Keeping our definition
\eqref{cur} of the cutoff ratio from the free case, we find
\[ \hat{s} = \frac{s}{k^2} \frac{d}{r} \;.\]
All results we present in this section are at an effectively infinite
anisotropy factor $c \rightarrow \infty$. This is easily realized due to the
mass universality relation \eqref{murd}, which is very accurate for large
$\tilde{c}^2$. If we chose the anisotropy factor $c$ such that $c^2 = c_0^2 -
\tilde{k}^2$ where $c_0$ is a constant, we may write the normalized entropy at infinite anisotropy factor as
\begin{equation}
  s_{n, \tilde{k}} (\infty) \sim s_{n, \tilde{k}} (c_0^2 - \tilde{k}^2) +
  \bigintlim_{c_0^2}^{\infty} \mathd \tilde{c}^2 \, \tilde{S}_n (\tilde{c}^2) \;.
  \label{saniso}
\end{equation}
For Pauli-Villars regularization, we need the difference between massless and
massive entropies. Using \eqref{saniso}, we may reduce the difference
$\tilde{s}_{n, \tilde{k}}$ at infinite anisotropy factor to a difference at finite anisotropy factors via
\[ \tilde{s}_{n, \tilde{k}} = s_{n, 0} (\infty) - s_{n, \tilde{k}} (\infty)
   \sim s_{n, 0} (c_0^2) - s_{n, \tilde{k}} (c_0^2 - \tilde{k}^2)\;. \]
Effectively, this leads to a mass dependent modification of the anisotropy
relation \eqref{anisodyn}
\[ (l_{\max} + 1)_{\tilde{k}} = \sqrt{c_0^2 - \tilde{k}^2} n_{\tmop{eff}} \]
which we utilize, with an obvious interpolation in the case of non-integer
$l_{\max}$. To minimise the contribution of the correction integral in
\eqref{saniso}, we use the full dataset for the massless case, so
\[ c_0 = \frac{(l_{\max} + 1)_0}{n_{\tmop{eff}}} = \frac{512}{n_{\tmop{eff}}}
\]
We thus compute the dynamical Pauli-Villars normalized entropy via
\[ \hat{s} = \frac{1}{k^2} \frac{d}{r} \sum_i p_i s_{n, \tilde{k}_i} (c_0^2 -
   \tilde{k}^2_i) \;,\]
where
\begin{equation}
  \tilde{k}^2_i = \frac{r}{d} \left( 1 + \frac{1}{n} \right) k^2 \kappa_i^2
  \label{kapidyn}
\end{equation}
with the $\kappa_i$ from Tab.~\ref{poltablog}. In Fig.~\ref{dynpv} we plot the
area normalized entropy
\[ \frac{S}{r^2} = \hat{s} M_{\tmop{PV}}^2 \]
for two different cutoff ratios and three different discretizations at one
simulation time $t = 12$ vs.~the physical distance $\rho$ from the origin. The
physical distance from the origin at radial coordinate index $n$ is obtained
by summing the radial separation \eqref{physd}
\[ \rho_n = \Delta \sum_{i = 1}^n \sqrt{\frac{r_i}{d_i}}\;. \]
Apart from boundary and discretization effects, which closely resemble those
of the free case (see e.g.~Fig.~\ref{logpolyshat}) the inmoving bump region is
clearly visible.

To account for boundary effects, we first note that they can be
modelled by adding a boundary term $\tilde{b}'{}^2$ to the effective mass
\eqref{cmeff}. Consequently, the integral representation that most closely
resembles the measured entropy $s^{(m)}_{n, \tilde{k}} (c^2)$ is in fact
\[ s^{(m)}_{n, \tilde{k}} (c^2) \sim \bigintlim_{\tilde{k}^2 +
   \tilde{b}'{}^2}^{c^2 + \tilde{k}^2 + \tilde{b}'{}^2} \mathd \tilde{c}^2 \,
   \tilde{S}_n (\tilde{c}^2) \]
where the integral bounds are shifted with respect to the idealized case
\eqref{murd}. The shift in the upper bound can be ignored since it is
universal for all $\tilde{k}^2$ and thus leads to a simple modification of
$c_0^2$. The shift in the lower bound is more problematic since there are no
data points in that region. Taking the most simpleminded approach of adding the
integral
\begin{equation}
  \bigintlim_{\tilde{k}^2}^{\tilde{k}^2 + \tilde{b}'{}^2} \mathd
  \tilde{c}'{}^2 \, \tilde{S} \left( \tilde{c}'{}^2 \right) \label{fbct}
\end{equation}
to $s^{(m)}_{n, \tilde{k}} (c^2)$ where $\tilde{S} \left( \tilde{c}'{}^2
\right)$ is the free universal scaling function, one obtains the results
displayed in Fig.~\ref{dynpvc}. While some qualitative improvement, especially
at large radial coordinates, can be seen, it is evident that divergent terms
remain, which seem to scale proportional to $M_{\tmop{PV}}^2$.

\begin{figure*}[htb]\centering
  {\includegraphics[width=\columnwidth]{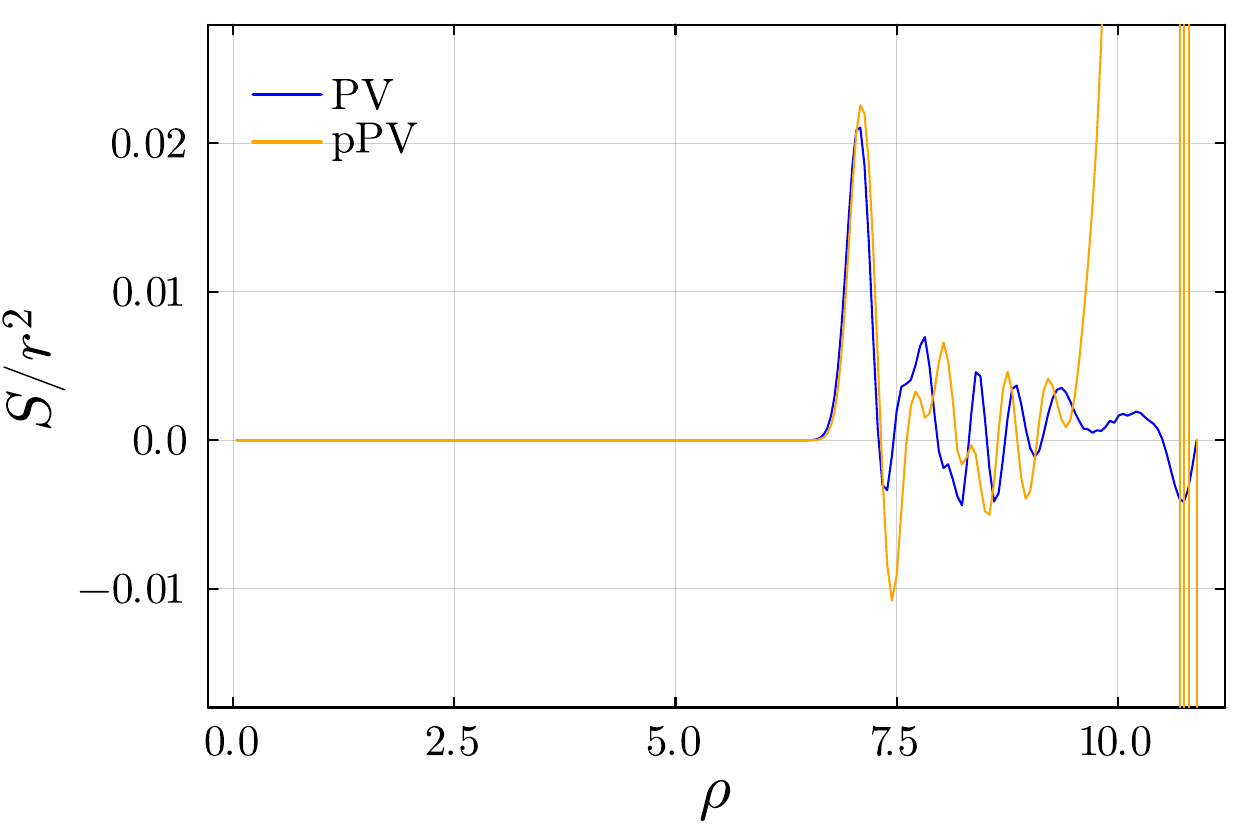}}
  {\includegraphics[width=\columnwidth]{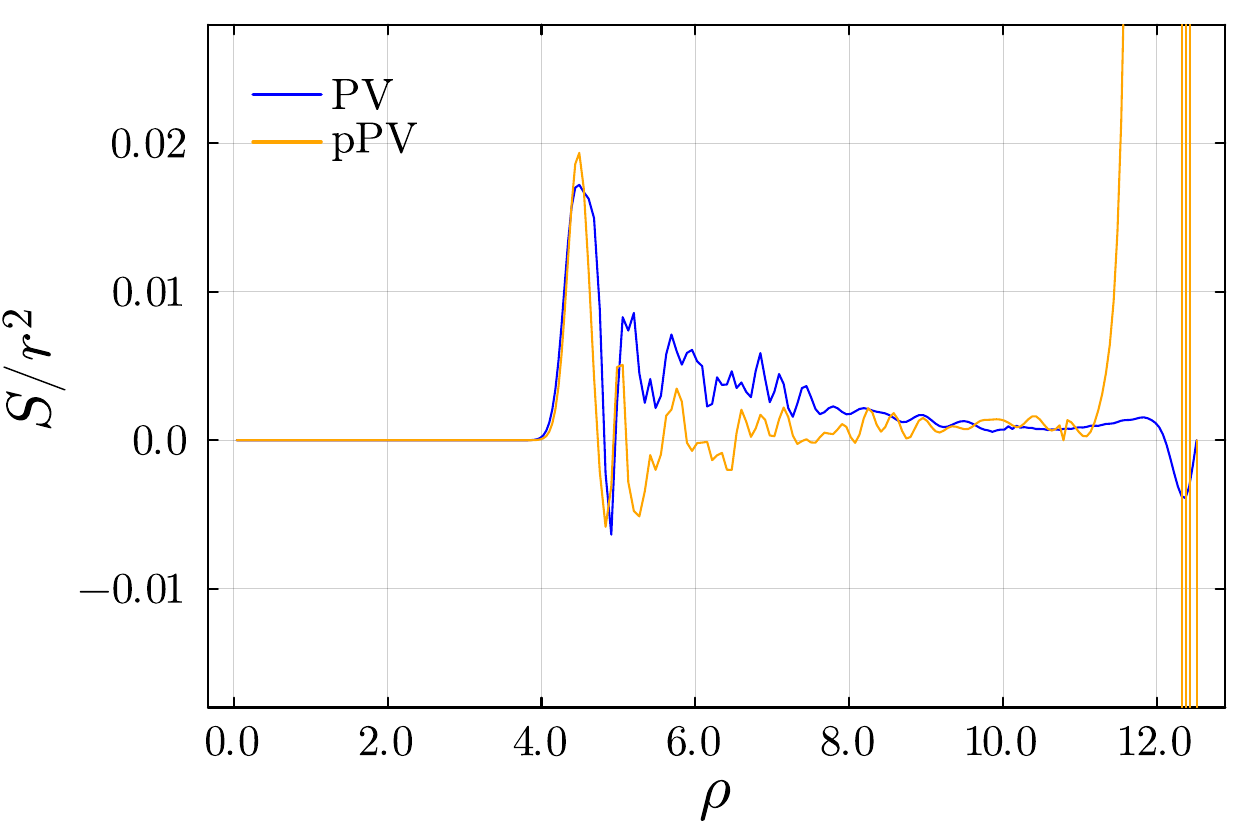}}
  
  {\includegraphics[width=\columnwidth]{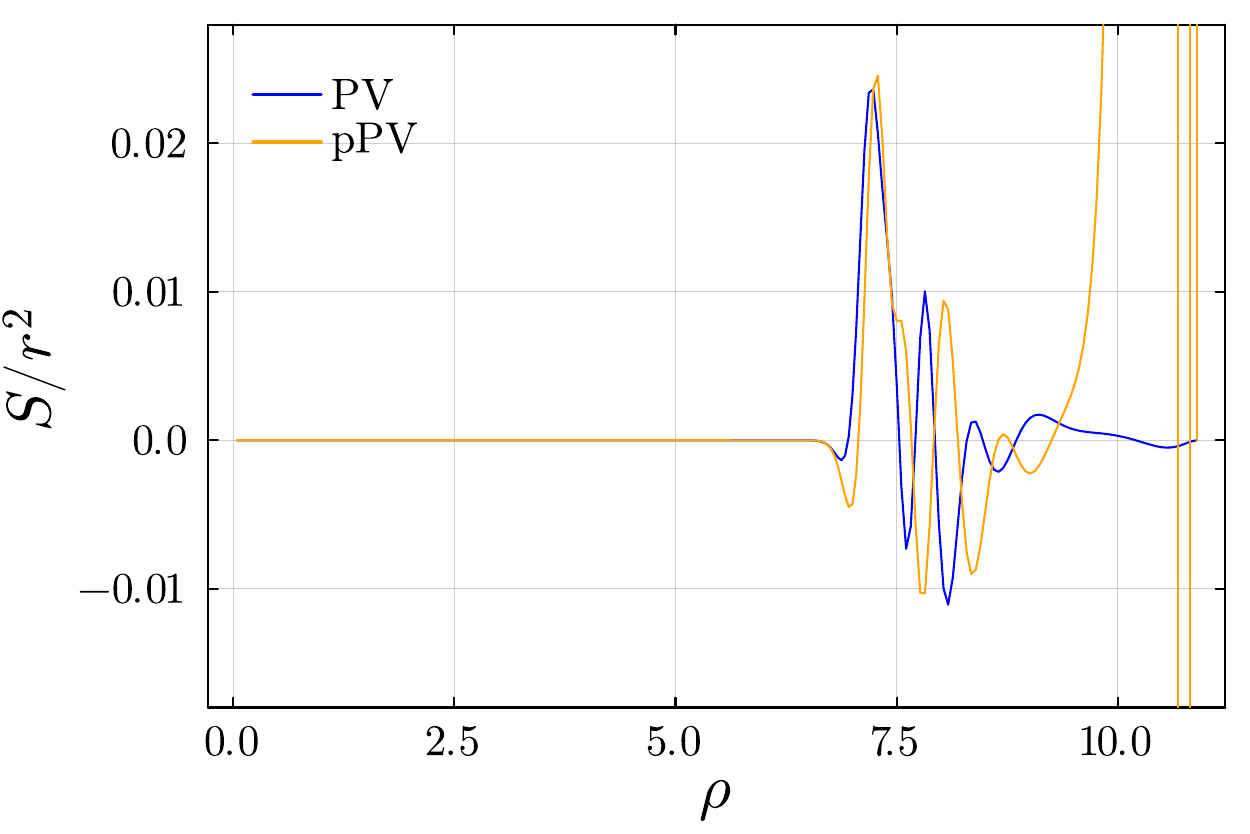}}
  {\includegraphics[width=\columnwidth]{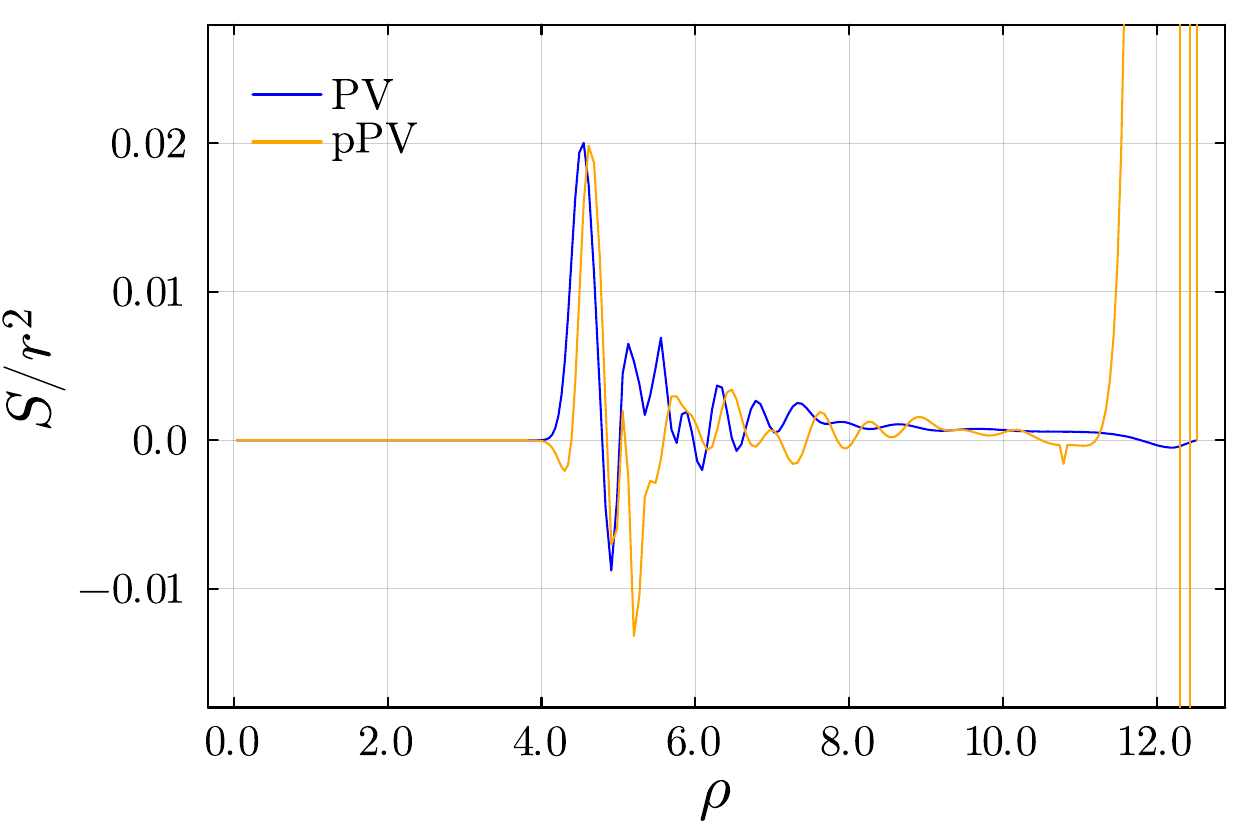}}
  \caption{\label{pPVcomp}Comparison of area normalized, subtracted
  Pauli-Villars (PV) and pseudo Pauli-Villars (pPV) entropies for our finest
  $N_r = 256$ system at simulation times $t = 5$ (left panels), $t = 12$
  (right panels) and cutoff ratios $k = 0.078125$ (top panels) resp.~$k =
  0.0390625$ (bottom panels).}
\end{figure*}

Fortunately, there is a much more straightforward method of subtracting these
unphysical divergences. Since any nonzero value of the free entropy $S^0$ in
the log-polynomial scheme is due to discretization or boundary effects or a possible logarithmic correction that was too small to numerically resolve in the free case, we may
simply subtract from the bare dynamical value $S$ its free counterpart at the
appropriately matched radial coordinate. In the inside region that has not yet
been reached by the inmoving bump, this matching is trivial. Since the entropy
can be computed via its inside definition $S_-$, which depends only on the points
inside the cut, there is no
difference to the free case, and entropies at matching radial coordinates match identically. Similarly, in the region outside the bump and
close to the outer boundary, the natural comparison seems to be the one where
the physical distance to the outer boundary matches between the interacting
and the free theory. To interpolate between these two cases, we note, in the
spirit of the argument of App.~\ref{l0approx}, that the inside definition of
the entropy $S_-$ on the $n^{\tmop{th}}$ radial shell is unaware of any
changes in the metric due to nonvnishing energy densities outside that shell.
Consequently, we can define a percieved system length at radial coordinate
index $n$
\[ L_n = \rho_n + \Delta \sum^{N_r}_{i = n + 1} \sqrt{\frac{r_i}{d_{n i}}} \;,\]
where $\rho_n$ is the physical distance from the center at radial coordinate $r_n$, and
\[ d_{n i} = d_n + (i - n) \Delta \]
is the fictitous metric parameter $d$ at radial coordinate $i > n$ that only
takes into account the energy densities in the inside region. 
By matching the perceived distance from the outer boundary in the interacting case $L_n - \rho_n$ to the actual outer boundary distance $L - r$ of the free case, we obtain the corresponding free radial coordinate
\begin{equation}
  r = L - (L_n - \rho_n) \label{bmc}
\end{equation}
that most closely matches the boundary effects of the interacting theory at
radial index $n$. Correspondingly, we define the boundary corrected entropy at
radial distance $\rho_n$ from the origin as
\begin{equation}
  S_{\tmop{bc}} (\rho_n) = S (\rho_n) - S^0 (L - (L_n - \rho_n))\;.
  \label{freesub}
\end{equation}
The area normalized subtraction terms $S^0 (L - (L_n - \rho_n)) / r^2$ are
displayed in Fig.~\ref{dynpv}, while the resulting area normalized subtracted
entropies at two different times during the evolution are displayed in Fig.~\ref{k2evo}. Fig.~\ref{k2evo}  also shows the corresponding classical
energy density $\mathcal{h}_0 = h^0 / \Delta$ of the inmoving bump as a reference. As one can see, an inmoving peak
region is clearly visible in the entropy data. Ignoring for the moment the
considerable noise and focusing on the main peak, the data seem at least
compatible with a positive bump in the area normalized entanglement entropy that is approximately constant during time evolution. Furthermore, there are no hints in our data of a divergence either in the continuum limit $N_r \rightarrow \infty$ at constant cutoff ratio $k$ (which implies removing the Pauli-Villars regulator together with the brick wall cutoff scale) or when removing the brick wall cutoff $k \rightarrow 0$ (while keeping $M_{\tmop{PV}}$ constant). Of course, one would need substantially more data to draw any firm conclusion, especially regarding the appearance of possible logarithmic divergencies. 

Due to the near perfect universality with respect to the mass of the scaling
function $\tilde{S}_n (\tilde{c}^2)$ at radial index $n$, we may define a
pseudo Pauli-Villars scheme also in the dynamical theory. We construct
$\tilde{S}_n (\tilde{c}^2)$ via a cubic spline interpolation of the ratio $S_l
/ \tilde{S}^i \left( \tilde{\mu}''{}^2 \right)$ of the massless single mode
entropy over the free universal scaling function at the corresponding boundary
corrected effective mass \eqref{cmeff}, with a simple continuation to
$\tilde{c}^2 = 0$ by using the first spline out of range. In analogy to the
free case, we then compute the pseudo Pauli-Villars normalized entropy via
\[ \hat{s} = - \frac{1}{k^2} \frac{d}{r} \sum_{i > 1} p_i \tilde{s}_{n,
   \tilde{k}_i} \;,\]
where the $\tilde{k}^2_i$ are given by \eqref{kapidyn} and
\[ \tilde{s}_{n, \tilde{k}} = \bigintlim_0^{\tilde{k}^2} \mathd \tilde{c}^2 \,
   \tilde{S}_n (\tilde{c}^2) \]
is the numerical integral over the scaling function. The result for three
lattice discretizations, two cutoff ratios and one time $t = 12$ is presented
in the left hand panels of Fig.~\ref{pPVsub}. Although the most prominent
feature are some huge artifacts close to the outer boundary, the bump region
is also clearly visible. As in the Pauli-Villars case, we compute the
corresponding entropy for the free system to perform a free subtraction
\eqref{freesub}. The subtraction terms are plotted as transparent lines in the
left hand panels of Fig.~\ref{pPVsub} and the resulting subtracted, area
normalized entropies are displayed in the right hand panels. Although the
signal region gives the impression of being dominated by the actual physical
signal, the region near the outer system boundary remains dominated by what
appears to be noise.

Finally, we present in Fig.~\ref{pPVcomp} a comparison of the area
normalized, subtracted entropies obtained via the Pauli-Villars and the pseudo
Pauli-Villars regularization for two cutoff ratios, two simulation times and
our finest discretization. It is reassuring that the qualitative features of
the main peak are very similar, although the pseudo Pauli-Villars method seems
to produce huge artifacts in the region close to the outer boundary. Since
this is a first exploratory study, we do not seek to improve further on the
pseudo Pauli-Villars method here. It should be noted, however, that for
potential future numerical investigations this normalization scheme is
attractive due to its far lower numerical cost. In contrast to the full
Pauli-Villars regulator it requires only the massless field. Additionally, the
number of angular momentum modes needed is drastically reduced, depending on
the chosen system size and desired cutoff ratio. One further advantage, at
least in our case where the backreaction is ignored, is the potential
flexibility in choosing the cutoff ratio a posteriori, which eliminates the
need of redoing the simulation for multiple cutoff ratios.

\section{Discussion and outlook\label{conc}}

In this paper, we presented a method to numerically study the gravitational
collapse of a scalar quantum field in a coherent state for the radially
symmetric case. We propose to use a Pauli-Villars regulator that cancels all
divergences in the energy-momentum tensor, fixing the residual Wald ambiguity by a choice of renormalization scale. We then studied the
entanglement entropy of the scalar field upon tracing out a spherical region à la
Srednicki \cite{Srednicki:1993im}. We reproduce his seminal result  for the free case and with a brick wall
regulator and generalize it slightly for finite anisotropy factors. 
More generally, we show that the area term of the entanglement entropy is an integral over a universal function of an effective mass, allowing us to compute it for any Pauli-Villars scheme. 
With our
new log-polynomial Pauli-Villars regulator we find that the area term of the entanglement
entropy vanishes identically in the free case. 

We interpret the vanishing of the area law term of the entanglement entropy in the log-polynomial renormalization scheme in a similar way to Susskind and Uglum \cite{Susskind:1994sm}.  In their picture, the divergent one-loop contribution of matter fields to the black
hole entropy, originating from modes arbitrarily close to the horizon, is not an independent physical entropy but part of the renormalization of Newton's constant $G$. The area coefficient we obtain in our analysis agrees with the one-loop shift of $1/G$ found by Demers, Lafrance and Myers \cite{Demers:1995dq}.  
The conditions of the log-polynomial Pauli-Villars scheme set this shift to zero, i.e.~in our scheme $G_{\mathrm{ren}}=G_{\mathrm{bare}}$ and the contribution to the area law coming from the matter entanglement entropy vanishes. In this framework, only the total generalized entropy $S_{\mathrm{gen}}= A/4 G_{_\mathrm{ren}} + S_{\mathrm{matter}}$ is physical (hence scheme independent), and our regulator moves the entire area law into the first term.

In a first exploratory study, we then looked at the dynamical case without
backreaction. Some numerical evidence in our new regularization scheme suggests that during gravitational collapse of a thin shell the entanglement
entropy per surface area stays approximately constant on the shell, while it
seems to vanish both inside and outside, although the latter statement comes with significant numerical uncertainty. Our exploratory study further
shows that the value of the constant has a very mild dependence on the system size and the cutoff ratio $k$ and may even be finite in both the continuum limit at fixed $k$ and for $k \rightarrow 0$.

There are several ways in which we would like to improve on our current, exploratory study.
On the more phenomenological side, we would like to perform a systematic study of the volume, cutoff and bump height dependencies and evolve our system closer to horizon formation. We also intend to vary the initial conditions so that we start with a Minkowski vacuum and feed in the collapsing shell from outside. On the more conceptual side,  we would like to clarify the conditions for universality in the free theory and the related issue of subleading contributions to the entanglement entropy. In the dynamical theory, we want to further investigate the treatment of boundary and discretization effects as well as the feasibility of the pseudo Pauli-Villars scheme.
The most crucial point, however, is the obvious addition of the energy-momentum tensor, both as an observable and as part of a fully dynamical update that includes backreaction effects. The ultimate aim of this effort is, of course, to test if our regularization scheme is finite in the continuum limit and free of unexpected ambiguities. 

\begin{acknowledgments}
We thank Antoine Rignon-Bret, Stephan Greiner, Felix Orda, Michel Krause, Vivien Müller, Lars-Hendrik Torspechen, Tobias Velten and Stephan Dürr for helpful discussions.
SM has received funding from the European Research Council (ERC) from the QFT.zip project (grant agreement No. 101040260).
\end{acknowledgments}
\clearpage
\onecolumngrid
\appendix

\section{Radial symmetry\label{apprs}}

Let us assume that we have the expectation value of a bilinear in a radially
symmetric, Gaussian state
\[ \langle \tmmathbf{\psi}^{\dag} (\theta, \varphi) \tmmathbf{\psi} (\theta,
   \varphi) \rangle = \sum_{l = 0}^{\infty} \sum_{m = - l}^l \sum_{l' =
   0}^{\infty} \sum_{m' = - l'}^{l'} \langle \tmmathbf{\psi}_{l m}^{\dag}
   \tmmathbf{\psi}_{l' m'} \rangle Y^{\ast}_{l m} (\theta, \varphi) Y_{l' m'}
   (\theta, \varphi) \]
that is characterized by the averages
\[ \psi_{l m} = \langle \tmmathbf{\psi}_{l m} \rangle \]
and the covariance matrix\tmcolor{red}{}\footnote{We may neglect the
commutator term $i \Omega / 2$ here, as $\tmmathbf{\psi}$ refers to either
fields or momenta only, and never a combination of both.}
\[ C_{l l' m m'} = 2 (\langle \tmmathbf{\psi}_{l m}^{\dag} \tmmathbf{\psi}_{l'
   m'} \rangle - \psi^{\dag}_{l m} \psi_{l' m'}) \]
in the basis of spherical harmonics. Since we do not point split, we will
suppress the angular coordinates $\theta$ and $\varphi$. We also suppress
all additional indices, such as radial coordinates. Writing the bilinear in
terms of averages and the covariance matrix, we find
\[ \langle \tmmathbf{\psi}^{\dag} \tmmathbf{\psi} \rangle = \sum_{l =
   0}^{\infty} \sum_{m = - l}^l \psi^{\dag}_{l m} Y^{\ast}_{l m} \sum_{l' =
   0}^{\infty} \sum_{m' = - l'}^{l'} \psi_{l' m'} Y_{l' m'} + \frac{1}{2}
   \sum_{l = 0}^{\infty} \sum_{m = - l}^l \sum_{l' = 0}^{\infty} \sum_{m' = -
   l'}^{l'} C_{l l' m m'} Y^{\ast}_{l m} Y_{l' m'} \;.\]
Since the averages and correlation functions are independent, we can
immediately see, that radial symmetry requires all averages except the
$\psi_{00}$ to vanish. Using the shorthand $\psi = \psi_{00}$, we obtain for
real fields
\[ \langle \tmmathbf{\psi}^{\dag} \tmmathbf{\psi} \rangle = \frac{1}{4 \pi}
   \psi^T \psi + \frac{1}{2} \sum_{l = 0}^{\infty} \sum_{m = - l}^l \sum_{l' =
   0}^{\infty} \sum_{m' = - l'}^{l'} C_{l l' m m'} Y^{\ast}_{l m} Y_{l' m'} \;.\]
Because of the identity
\begin{equation}
  Y^{\ast}_{l m} Y_{l' m'} = (- 1)^{m + m'} Y_{l (- m)} Y^{\ast}_{l'  (- m')}
  \label{shid}
\end{equation}
not all terms in this expansion are independent. In fact, the covariance
matrix coefficients $C_{l l' m m'}$ and $C_{l' l (- m')  (- m)}$ multiply the
same combination of spherical harmonics. We may thus, without loss of
generality, demand
\begin{equation}
  C_{l l' m m'} = (- 1)^{m + m'} C_{l' l (- m')  (- m)} \label{csym}
\end{equation}
to remove the spurious degrees of freedom. Using the identities
\[ \partial_{\varphi} Y_{l m} (\theta, \varphi) = i m Y_{l m} (\theta,
   \varphi) \qquad \partial_{\varphi} Y^{\ast}_{l m} (\theta, \varphi) =
   - i m Y^{\ast}_{l m} (\theta, \varphi) \]
of the spherical harmonics, we may cast the azimuthal symmetry condition as
\[ \partial_{\varphi} \langle \tmmathbf{\psi}^{\dag} \tmmathbf{\psi} \rangle =
   \frac{i}{2} \sum_{l = 0}^{\infty} \sum_{m = - l}^l \sum_{l' = 0}^{\infty}
   \sum_{m' = - l'}^{l'} C_{l l' m m'} (m' - m) Y^{\ast}_{l m} Y_{l' m'} = 0 \;.
\]
Apart from the identity \eqref{shid}, all terms in the sum have different
angular dependencies. Therefore, the vanishing of the sum implies
\[ (C_{l l' m m'} + (- 1)^{m + m'} C_{l' l (- m')  (- m)}) (m' - m) = 0 \]
which, together with the symmetry condition \eqref{csym} requires the
vanishing of the covariance matrix coefficients $C_{l l' m m'} = 0$ unless $m'
= m$. We may now define $C_{l l' m} = C_{l l' m m}$, for which the symmetry
condition \eqref{csym} implies
\begin{equation}
  C_{l l' m} = C_{l' l (- m)} \label{css}
\end{equation}
and write
\[ \langle \tmmathbf{\psi}^{\dag} \tmmathbf{\psi} \rangle = \frac{1}{4 \pi}
   \psi^T \psi + \frac{1}{2} \sum_{l = 0}^{\infty} \sum_{l' = 0}^{\infty}
   \sum_{m = - \min (l, l')}^{\min (l, l')} C_{l l' m} Y^{\ast}_{l m} Y_{l' m}\;.
\]
For the $\theta$-dependence, we look at the ladder operator
\[ D_+ = \partial_{\theta} + i \cot (\theta) \partial_{\varphi} \]
which has the property
\[ D_+ Y_{l m} (\theta, \varphi) = e^{- i \varphi} \sqrt{(l + m + 1) (l - m)}
   Y_{l (m + 1)} \]
from which it follows that
\[ D_+ Y^{\ast}_{l m} (\theta, \varphi) = - e^{- i \varphi} \sqrt{(l - m + 1)
   (l + m)} Y_{l (m - 1)}^{\ast} \;.\]
Rotational invariance implies
\begin{align*}
  0 ={}& D_+ \langle \tmmathbf{\psi}^{\dag} \tmmathbf{\psi} \rangle\\
    ={}& e^{- i \varphi} \sum_{l' = 0}^{\infty} \sum_{l = 0}^{l' - 1} C_{l
    l' l} \sqrt{(l' + l + 1) (l' - l)} Y^{\ast}_{l l} Y_{l'  (l + 1)}\\
    & + \frac{e^{- i \varphi}}{2} \sum_{l = 0}^{\infty} \sum_{l' =
    0}^{\infty} \sum_{m = - \min (l, l')}^{\min (l, l') - 1} Y^{\ast}_{l m}
    Y_{l'  (m + 1)} \\
    & \quad {}\left( C_{l l' m} \sqrt{(l' + m + 1) (l' - m)}
    - C_{l l' (m + 1)} \sqrt{(l - m) (l + m + 1)} \right)\;.
\end{align*}
All terms in the sum have different angular dependencies, so their
coefficients have to vanish individually. This implies, for all $- \min (l,
l') \leqslant m < \min (l, l')$,
\[ C_{l l' m} \sqrt{(l' + m + 1) (l' - m)} = C_{l l'  (m + 1)} \sqrt{(l - m)
   (l + m + 1)} \]
and, for all $l < l'$
\[ C_{l l' l} \sqrt{(l' + l + 1) (l' - l)} = 0 \;.\]
Thus, for $l < l'$, rotational invariance implies $C_{l l' m} = 0$, and, due
to \eqref{css}, the same is true for $l > l'$. For $l = l'$, we
may set $C_{l l m} = C_l$ and write
\[ \langle \tmmathbf{\psi}^{\dag} \tmmathbf{\psi} \rangle = \frac{1}{4 \pi}
   \psi^T \psi + \frac{1}{2} \sum_{l = 0}^{\infty} C_l \sum_{m = - l}^l
   Y^{\ast}_{l m} Y_{l m} \;.\]
Using Uns{\"o}ld's theorem
\[ \sum_{m = - l}^l Y^{\ast}_{l m} Y_{l m} = \frac{2 l + 1}{4 \pi} \]
and reinstating the angular coordinates, we obtain
\[ \langle \tmmathbf{\psi}^{\dag} (\theta, \varphi) \tmmathbf{\psi} (\theta,
   \varphi) \rangle = \frac{1}{4 \pi} \left( \psi^T \psi + \sum_{l =
   0}^{\infty} \frac{2 l + 1}{2} C_l \right) \]
for an arbitrary, rotationally invariant bilinear. Integrating over the
angular variables we finally arrive at
\[ \langle \tmmathbf{\psi}^{\dag} \tmmathbf{\psi} \rangle = \psi^T \psi +
   \sum_{l = 0}^{\infty} \frac{2 l + 1}{2} C_l \;.\]
Clearly, the $C_l$ need to be real, symmetric matrices for a real, scalar
field.

\section{Review of some thermodynamic properties of Gaussian
states}\label{thermorev}

Let us assume that we have a generic Gaussian state given by $2 N$ averages
$\psi$ and a $2 N \times 2 N$ covariance matrix $C$. Since $C$ is even
dimensional and symmetric, we may bring it into Williamson form by a
symplectic transformation $D$ such that
\[ C = D \left(\begin{array}{cc}
     \sigma & 0\\
     0 & \sigma
   \end{array}\right) D^T \;,\]
where $\sigma$ is a diagonal matrix. With the correspondingly transformed
averages
\[ \psi = D \chi \]
we may write the density matrix of the Gaussian state as
\[ \tmmathbf{\rho}= \frac{e^{- \beta \tmmathbf{\mathcal{H}} (\chi)}}{\tmop{Tr}
   (e^{- \beta \tmmathbf{\mathcal{H}} (\chi)})} \qquad
   \underset{}{\tmmathbf{\mathcal{H}} (\chi)} = \frac{1}{2} (\tmmathbf{\chi}-
   \chi)^T \left(\begin{array}{cc}
     \mathrm{\omega} & 0\\
     0 & \mathrm{\omega}
   \end{array}\right) (\tmmathbf{\chi}- \chi) \]
with the relation
\begin{equation}
  \beta \omega_j = \ln \left( \frac{\sigma_j + 1}{\sigma_j - 1} \right)
  \label{besi}
\end{equation}
between the diagonal elements of the diagonal matrices $\sigma$ and $\omega$.
It is thus a thermal state in the canonical ensemble corresponding to the
Hamiltonian $\tmmathbf{\mathcal{H}} (\chi)$, where $\tmmathbf{\mathcal{H}}
(\chi)$ is unique up to an additive constant and a multiplicative factor,
which may be absorbed into $\beta$. The trace in the denominator of the
density matrix may be evaluated to give
\[ \tmop{Tr} (e^{- \beta \tmmathbf{\mathcal{H}} (\chi)}) = \sqrt{\det \left(
   \frac{\hat{\sigma}}{2} \right)} \qquad \hat{\sigma} =
   \left(\begin{array}{cc}
     \sigma & 0\\
     0 & \sigma
   \end{array}\right) + i \Omega \;,\]
so that the density matrix may be written as
\begin{equation}
  \tmmathbf{\rho}= \frac{e^{- \beta \tmmathbf{\mathcal{H}} (\chi)}}{\sqrt{\det
  \left( \frac{\hat{\sigma}}{2} \right)}} \;. \label{demafi}
\end{equation}
For a pure state, $\tmop{Tr} (\tmmathbf{\rho}^2) = \tmop{Tr} (\tmmathbf{\rho})
= 1$. Generically, we may define the purity of a state as $\gamma = \tmop{Tr}
(\tmmathbf{\rho}^2)$, which evaluates to
\begin{equation}
  \gamma = \tmop{Tr} (\tmmathbf{\rho}^2) = \prod_{j = 1}^N \frac{1}{\sigma_j}
  = \frac{1}{\sqrt{\det C}} \;,\label{pos}
\end{equation}
where the last identity is due to the unit determinant of a symplectic
transformation. The average occupation number of the $j^{\tmop{th}}$
oscillator is given by
\begin{equation}
  \overline{n}_j = \frac{1}{e^{\beta \omega_j} - 1} = \frac{\sigma_j - 1}{2}
  \label{aon}
\end{equation}
and thus the average energy
\begin{equation}
  E_{\tmmathbf{\rho}} = \tmop{Tr} (\tmmathbf{\mathcal{H}} (\chi)
  \tmmathbf{\rho}) = \sum_{j = 1}^N \frac{\omega_j \sigma_j}{2} \;.\label{avge}
\end{equation}
The von Neuman entropy is defined as
\begin{equation}
  S_{\tmmathbf{\rho}} = - \tmop{Tr} (\tmmathbf{\rho} \ln \tmmathbf{\rho})\;.
  \label{vnedef}
\end{equation}
Starting from \eqref{demafi} we have
\[ - \ln \tmmathbf{\rho}= \beta \tmmathbf{H}+ \frac{1}{2} \tmop{Tr} \left( \ln
   \left( \frac{\hat{\sigma}}{2} \right) \right) \]
and since $\tmop{Tr} (\tmmathbf{\rho}) = 1$, we can write
\[ S_{\tmmathbf{\rho}} = \frac{\beta \tmop{Tr} (e^{- \beta \tmmathbf{H}}
   \tmmathbf{H})}{\sqrt{\det \left( \frac{\hat{\sigma}}{2} \right)}} +
   \frac{1}{2} \ln \left( \det \left( \frac{\hat{\sigma}}{2} \right) \right)\;.
\]
In the Williamson basis, we have
\[ \tmop{Tr} (e^{- \beta \tmmathbf{H}} \tmmathbf{H}) = - \frac{1}{2}
   \sqrt{\det \left( \frac{\hat{\sigma}}{2} \right)} \sum_{j = 1}^N
   \partial_{\beta} \ln \left( \frac{\sigma_j^2 - 1}{4} \right) \]
which, using \eqref{besi}, evaluates to
\[ \partial_{\beta} \ln \left( \frac{\sigma_j^2 - 1}{4} \right) = - \omega_j
   \sigma_j \]
so that all together
\begin{equation}
  S_{\tmmathbf{\rho}} = \sum_{j = 1}^N \left( \left( \frac{\sigma_j}{2} +
  \frac{1}{2} \right) \ln \left( \frac{\sigma_j}{2} + \frac{1}{2} \right) -
  \left( \frac{\sigma_j}{2} - \frac{1}{2} \right) \ln \left(
  \frac{\sigma_j}{2} - \frac{1}{2} \right) \right) \;.\label{vne}
\end{equation}
Note that the variation of the entropy is given by
\[ \mathd S_{\tmmathbf{\rho}} = \frac{1}{2} \sum_{j = 1}^N \ln \left(
   \frac{\sigma_j + 1}{\sigma_j - 1} \right) \mathd \sigma_j = \frac{\beta}{2}
   \sum_{j = 1}^N \omega_j \mathd \sigma_j \]
so that, in conjunction with \eqref{avge}, the Clausius relation $\mathd
S_{\tmmathbf{\rho}} = \beta \mathd E_{\tmmathbf{\rho}}$ is fulfilled. By
expanding the logarithm in \eqref{vnedef} to first order as $\ln
\tmmathbf{\rho} \simeq \tmmathbf{\rho}-\mathbbm{1}$, we obtain the linear
entropy
\begin{equation}
  \hat{S}_{L, \tmmathbf{\rho}} = 1 - \tmop{Tr} (\tmmathbf{\rho}^2) = 1 -
  \gamma = 1 - \frac{1}{\sqrt{\det C}} \label{slin}
\end{equation}
as a simple function of the state purity \eqref{pos}. Since in the Williamson
basis the state decomposes into a direct product over the individual modes, we
may in fact compute the state purity $\gamma_j$ and the corresponding linear
entropy $S_{L j}$ for the $j^{\tmop{th}}$ mode as
\[ \gamma_j = \frac{1}{\sigma_j} \qquad S_{L j} = 1 - \frac{1}{\sigma_j}\;.
\]
Summing the linear entropies of all modes, we may define a cumulative linear
entropy
\[ S_{L, \tmmathbf{\rho}} = \sum_j \left( 1 - \frac{1}{\sigma_j} \right) \;.\]

\section{Connection to the Heisenberg picture formalism}

According to (\ref{cortm}, \ref{avtm}), the real space average and covariance
matrix at time $t$ may be expressed in terms of a time evolution matrix $B^t
T_t$. According to (\ref{tmatp}, \ref{tevp}), the time evolution of $T^t$ is
given by
\[ T_{t + 2 \Delta t} = T_{t + 2 \Delta t, t} T_t = O^{t + 2 \Delta t} (B^{t +
   2 \Delta t})^{- 1} B^t O^t T_t \]
so
\[ B^{t + 2 \Delta t} T_{t + 2 \Delta t} = B^{t + 2 \Delta t} O^{t + 2 \Delta
   t} (B^{t + 2 \Delta t})^{- 1} B^t O^t (B^t)^{- 1} B^t T_t \;.\]
Let us now define a time evolution matrix with respect to a certain reference
metric $\tau$
\begin{equation}
  \hat{U}^{\tau}_t = Q^{\tau} B^t T_t \label{urbt}
\end{equation}
which evolves as
\[ \hat{U}^{\tau}_{t + 2 \Delta t} = \hat{O}^{\tau}_{t + 2 \Delta t}
   \hat{O}^{\tau}_t \hat{U}^{\tau}_t \]
where
\[ \hat{O}^{\tau}_t = Q^{\tau} B^t O^t (B^t)^{- 1} (Q^{\tau})^{- 1}\;. \]
We also define the intermediate
\begin{equation}
  \tilde{U}^{\tau}_{t + \Delta t} = \hat{O}^{\tau}_t \hat{U}^{\tau}_t
  \label{tevs1}
\end{equation}
so
\begin{equation}
  \hat{U}^{\tau}_t = \hat{O}^{\tau}_t \tilde{U}^{\tau}_{t - \Delta t}
  \label{tevs2}
\end{equation}
and write the transfer operator $\hat{O}^{\tau}_t$ in block form as
\[ \hat{O}^{\tau}_t = Q^{\tau} B^t O^t (B^t)^{- 1} (Q^{\tau})^{- 1} =
   \left(\begin{array}{cc}
     c^{t, \tau} & s^{t, \tau}_1\\
     - s^{t, \tau}_2 & (c^{t, \tau})^T
   \end{array}\right) \;,\]
where we have defined
\[ \begin{split}
     c^{t, \tau} = & \sqrt{\frac{\hat{\alpha}^t d^t}{\hat{\alpha}^{\tau}
     d^{\tau}}} V^t \cos (\omega^t \Delta t) (V^t)^T
     \sqrt{\frac{\hat{\alpha}^{\tau} d^{\tau}}{\hat{\alpha}^t {d^t} }}\\
     s_1^{t, \tau} = & \sqrt{\frac{\hat{\alpha}^t d^t}{\hat{\alpha}^{\tau}
     d^{\tau}}} V^t \frac{1}{\omega^t} \sin (\omega^t \Delta t) (V^t)^T
     \sqrt{\frac{\hat{\alpha}^t d^t}{\hat{\alpha}^{\tau} d^{\tau}}}\\
     s_2^{t, \tau} = & \sqrt{\frac{\hat{\alpha}^{\tau}
     d^{\tau}}{\hat{\alpha}^t {d^t} }} V^t \omega^t \sin (\omega^t \Delta t)
     (V^t)^T \sqrt{\frac{\hat{\alpha}^{\tau} d^{\tau}}{\hat{\alpha}^t {d^t} }}\;.
   \end{split} \]
Writing the time evulution operator $\hat{U}^{\tau}_t$ and the intermediate
$\tilde{U}^{\tau}_{t + \Delta t}$ in block form
\begin{equation}
  \hat{U}^{\tau}_t = \left(\begin{array}{cc}
    u^T_{t, Q} & u^T_{t, I}\\
    - v^T_{t, I} & v^T_{t, Q}
  \end{array}\right) \qquad \tilde{U}^{\tau}_{t + \Delta t} =
  \left(\begin{array}{cc}
    \tilde{u}^T_{t + \Delta t, Q} & \tilde{u}^T_{t + \Delta t, I}\\
    - \tilde{v}^T_{t + \Delta t, I} & \tilde{v}^T_{t + \Delta t, Q}
  \end{array}\right) \label{ucomp}
\end{equation}
we find, that at the initial time $t_0$
\[ \hat{U}_{t_0}^{\tau} = Q^{\tau} B^{t_0} = Q^{\tau} \left( {R^{t_0}} 
   \right)^{- 1} W^{t_0}  \left( {S^{t_0}}  \right)^{- 1} =
   \left(\begin{array}{cc}
     \sqrt{\frac{\hat{\alpha}^0 d^0}{\hat{\alpha}^{\tau} d^{\tau}}} V^{t_0}
     \frac{1}{\sqrt{\omega^0}} & 0\\
     0 & \sqrt{\frac{\hat{\alpha}^{\tau} d^{\tau}}{\hat{\alpha}^0 d^0}}
     V^{t_0} \sqrt{\omega^0}
   \end{array}\right) \]
so that we have the initial conditions
\[ \begin{split}
     u_{t, R} = & \frac{1}{\sqrt{\omega}} (V^{t_0})^T
     \sqrt{\frac{\hat{\alpha}^0 d^0}{\hat{\alpha}^{\tau} d^{\tau}}}\\
     v_{t, R} = & \sqrt{\omega} (V^{t_0})^T \sqrt{\frac{\hat{\alpha}^{\tau}
     d^{\tau}}{\hat{\alpha}^0 d^0}}\\
     u_{t, I} = & 0\\
     v_{t, I} = & 0\;.
   \end{split} \]
From \eqref{tevs1}, we find the time evolution
\[ \left(\begin{array}{cc}
     \tilde{u}_{t + \Delta t, R} & - \tilde{v}_{t + \Delta t, I}\\
     \tilde{u}_{t + \Delta t, I} & \tilde{v}_{t + \Delta t, R}
   \end{array}\right) = \left(\begin{array}{cc}
     u_{t, R} & - v_{t, I}\\
     u_{t, I} & v_{t, R}
   \end{array}\right) \left(\begin{array}{cc}
     (c^{t, \tau})^T & - s^{t, \tau}_2\\
     s^{t, \tau}_1 & c^{t, \tau}
   \end{array}\right) \]
so
\[ \begin{split}
     \tilde{u}_{t + \Delta t, R} = & u_{t, R} (c^{t, \tau})^T - v_{t, I}
     s^{t, \tau}_1\\
     \tilde{u}_{t + \Delta t, I} = & u_{t, I} (c^{t, \tau})^T + v_{t, R}
     s^{t, \tau}_1\\
     \tilde{v}_{t + \Delta t, R} = & - u_{t, I} s^{t, \tau}_2 + v_{t, R}
     c^{t, \tau}\\
     \tilde{v}_{t + \Delta t, I} = & u_{t, R} s^{t, \tau}_2 + v_{t, I} c^{t,
     \tau}\;,
   \end{split} \]
which we can write in the compact form
\[ \begin{array}{lll}
     \tilde{u}_{t + \Delta t} & = & u_t (c^{t, \tau})^T - i v_t s^{t,
     \tau}_1\\
     \tilde{v}_{t + \Delta t} & = & v_t c^{t, \tau} - i u_t s^{t, \tau}_2
   \end{array} \]
when we identify the $R$ and $I$ components as real and imaginary parts
respectively. Similarly, from \eqref{tevs2} we obtain
\[ \begin{split}
     u_t = & \tilde{u}_{t - \Delta t} (c^{t, \tau})^T - i \tilde{v}_{t -
     \Delta t} s^{t, \tau}_1\\
     v_t = & \tilde{v}_{t - \Delta t} c^{t, \tau} - i \tilde{u}_{t - \Delta
     t} s^{t, \tau}_2\;.
   \end{split} \]
If we set the reference time equal to the initial time $\tau = t_0$, we
recover the time evolution equations (8) and initial conditions (3) of the
Heisenberg picture formalism presented in \cite{Hoelbling:2021axl}.

Next, we would like to express the densities \eqref{h0m} in therms of the $u$
and $v$. We first rewrite them using discretized notation, substituting the
covariance matrix \eqref{cortm} and also introducing vacuum subtraction terms
\begin{equation}
  \begin{split}
    h^0_i = & \frac{1}{r_i^3} \left( \overline{\Pi}^2_i + N_c \sum_{l =
    0}^{\infty} \frac{2 l + 1}{2} (\overline{C}_l^{(\tmop{sub}) \Pi \Pi})_{i
    i} \right) + r_i \left( (\overline{\phi}'_i)^2 + N_c \sum_{l = 0}^{\infty}
    \frac{2 l + 1}{2} (\overline{D}_l^{(\tmop{sub}) \phi \phi})_{i i}
    \right)\\
    m_i = & r_i^2 M^2 \overline{\phi}_i^2 + N_c \sum_{l = 0}^{\infty}
    \frac{2 l + 1}{2} (l (l + 1) + r_i^2 M^2) (\overline{C}_l^{(\tmop{sub})
    \phi \phi})_{i i}\\
    & \\
    \overline{C}_l^{(\tmop{sub}) \Pi \Pi} = & c_l^{\Pi} (c_l^{\Pi})^T -
    c_l^{(\tmop{vac}) \Pi} (c_l^{(\tmop{vac}) \Pi})^T\\
    \overline{D}_l^{(\tmop{sub}) \phi \phi} = & (\partial_r c_l^{\phi})
    (\partial_r c_l^{\phi})^T - (\partial_r c_l^{(\tmop{vac}) \phi})
    (\partial_r c_l^{(\tmop{vac}) \phi})^T\\
    \overline{C}_l^{(\tmop{sub}) \phi \phi} = & c_l^{\phi} (c_l^{\phi})^T -
    c_l^{(\tmop{vac}) \phi} (c_l^{(\tmop{vac}) \phi})^T\;,
  \end{split} \label{fude}
\end{equation}
where we have defined the real space time evolution matrices
\[ c^{\phi} = P^{\phi} B^t T_t \qquad c^{\Pi} = P^{\Pi} B^t T_t
   \qquad c^{(\tmop{vac}) \phi} = P^{\phi} B^t T^{(\tmop{vac})}_t
   \qquad c^{(\tmop{vac}) \Pi} = P^{\Pi} B^t T^{(\tmop{vac})}_t \]
with the projectors
\[ P^{\phi} = \left(\begin{array}{cc}
     \mathbbm{1} & 0
   \end{array}\right) \qquad P^{\Pi} = \left(\begin{array}{cc}
     0 & \mathbbm{1}
   \end{array}\right) \;.\]
From \eqref{urbt} and \eqref{ucomp} we find
\[ B^t T_t = (Q^{\tau})^{- 1} \left(\begin{array}{cc}
     u^T_{t, R} & u^T_{t, I}\\
     - v^T_{t, I} & v^T_{t, R}
   \end{array}\right) \;, \]
so that
\[ c^{\phi} (c^{\phi})^T = \sqrt{\frac{\hat{\alpha}^{\tau} d^{\tau}}{r^3}}
   u_t^{\dag} u_t \sqrt{\frac{\hat{\alpha}^{\tau} d^{\tau}}{r^3}} \qquad
   c^{\Pi} (c^{\Pi})^T = \sqrt{\frac{r^3}{\hat{\alpha}^{\tau} d^{\tau}}}
   v_t^{\dag} v_t \sqrt{\frac{r^3}{\hat{\alpha}^{\tau} d^{\tau}}} \]
and thus
\[ (\partial_r c^{\phi}) (\partial_r c^{\phi})^T = \sqrt{\frac{1}{r
   \hat{\alpha}^{\tau} d^{\tau}}} q^{\tau} u_t^{\dag} u_t (q^{\tau})^T
   \sqrt{\frac{1}{r \hat{\alpha}^{\tau} d^{\tau}}}\;, \]
where we have defined
\[ q = \sqrt{\hat{\alpha} d r} \partial_r \sqrt{\frac{\hat{\alpha} d}{r^3}} \;.\]
Let us now turn to the vacuum subtraction terms. In the current paper, we only
use the quenched approximation, where $c^{(\tmop{vac})} = c$. In
\cite{Hoelbling:2021axl}, we have used normal ordering, which corresponds to
initial state vacuum subtraction. In this scheme, the transfer operator
\eqref{tmatp} is substituted by a simple basis transformation without time
evolution of the scalar field, thus $T^{(\tmop{IVS})}_{t + 2 \Delta t, t} =
(B^{t + 2 \Delta t})^{- 1} B^t$. This, in turn, leads to a time evolution
matrix $T_t^{(\tmop{IVS})} = (B^t)^{- 1} B^{t_0}$, so
\[ c^{(\tmop{IVS}) \phi} (c^{(\tmop{IVS}) \phi})^T =
   \sqrt{\frac{\hat{\alpha}^{\tau} d^{\tau}}{r^3}} u_{t_0}^{\dag} u_{t_0}
   \sqrt{\frac{\hat{\alpha}^{\tau} d^{\tau}}{r^3}} \qquad
   c^{(\tmop{IVS}) \Pi} (c^{(\tmop{IVS}) \Pi})^T =
   \sqrt{\frac{r^3}{\hat{\alpha}^{\tau} d^{\tau}}} v_{t_0}^{\dag} v_{t_0}
   \sqrt{\frac{r^3}{\hat{\alpha}^{\tau} d^{\tau}}}\;. \]
We can thus express the vacuum terms in \eqref{fude} as
\[ \begin{split}
     \overline{C}^{(\tmop{sub}) \Pi \Pi} = &
     \sqrt{\frac{r^3}{\hat{\alpha}^{\tau} d^{\tau}}} (v_t^{\dag} v_t -
     v_{t_0}^{\dag} v_{t_0}) \sqrt{\frac{r^3}{\hat{\alpha}^{\tau} d^{\tau}}}\\
     \overline{D}^{(\tmop{sub}) \phi \phi} = & \sqrt{\frac{1}{r
     \hat{\alpha}^{\tau} d^{\tau}}} q^{\tau} (u_t^{\dag} u_t - u_{t_0}^{\dag}
     u_{t_0}) (q^{\tau})^T \sqrt{\frac{1}{r \hat{\alpha}^{\tau} d^{\tau}}}\\
     \overline{C}^{(\tmop{sub}) \phi \phi} = &
     \sqrt{\frac{\hat{\alpha}^{\tau} d^{\tau}}{r^3}} (u_t^{\dag} u_t -
     u_{t_0}^{\dag} u_{t_0}) \sqrt{\frac{\hat{\alpha}^{\tau} d^{\tau}}{r^3}}\;.
   \end{split} \]
The averages, on the other hand, can be expressed as
\[ \begin{split}
     \overline{\phi} = & \sqrt{\frac{\hat{\alpha}^{\tau} d^{\tau}}{r^3}}
     \left( u_t^{\dag} \frac{\sqrt{\omega^0} \hat{\phi}^0 + i
     \frac{1}{\sqrt{\omega^0}} \hat{\Pi}^0}{2} - u^T_t \frac{- \sqrt{\omega^0}
     \hat{\phi}^0 + i \frac{1}{\sqrt{\omega^0}} \hat{\Pi}^0}{2} \right)\\
      & \\
     \overline{\Pi} = & - i \sqrt{\frac{r^3}{\hat{\alpha}^{\tau} d^{\tau}}}
     \left( v_t^{\dag} \frac{\sqrt{\omega^0} \hat{\phi}^0 + i
     \frac{1}{\sqrt{\omega^0}} \hat{\Pi}^0}{2} + v^T_t \frac{- \sqrt{\omega^0}
     \hat{\phi}^0 + i \frac{1}{\sqrt{\omega^0}} \hat{\Pi}^0}{2} \right)\\
      & \\
     \overline{\phi}' = & \frac{1}{\sqrt{\hat{\alpha}^{\tau} d^{\tau} r}}
     q^{\tau} \left( u_t^{\dag} \frac{\sqrt{\omega^0} \hat{\phi}^0 + i
     \frac{1}{\sqrt{\omega^0}} \hat{\Pi}^0}{2} - u^T_t \frac{- \sqrt{\omega^0}
     \hat{\phi}^0 + i \frac{1}{\sqrt{\omega^0}} \hat{\Pi}^0}{2} \right)\;.
   \end{split} \]
Identifying
\[ l_{\pm} = \frac{\pm \sqrt{\omega^0} \hat{\phi}^0 + i
   \frac{1}{\sqrt{\omega^0}} \hat{\Pi}^0}{\sqrt{2}} \]
and defining
\[ l_u = u_t^{\dag} l_+ - u^T_t l_- \qquad l_v = - i (v_t^{\dag} l_+ +
   v^T_t l_-) \]
as in \cite{Hoelbling:2021axl}, we may ultimately rewrite \eqref{fude} as
\[ \begin{split}
     h^0_i = & \frac{1}{2 \hat{\alpha}^{\tau}_i d^{\tau}_i} \left( | l_{v,
     i} |^2 + | (q^{\tau} l_u)_i |^2 + N_c \sum_{l = 0}^{\infty} (2 l + 1)
     \left( v_{l, t}^{\dag} v_{l, t} - v_{l, t_0}^{\dag} v_{l, t_0} + q^{\tau}
     (u_t^{\dag} u_t - u_{t_0}^{\dag} u_{t_0}) (q^{\tau})^T \right)_{i i}
     \right)\\
     m_i = & \frac{\hat{\alpha}^{\tau}_i d^{\tau}_i}{2 r_i} \left( M^2 |
     l_{u, i} |^2 + N_c \sum_{l = 0}^{\infty} (2 l + 1) \left( \frac{l (l +
     1)}{r_i^2} + M^2 \right) (u_t^{\dag} u_t - u_{t_0}^{\dag} u_{t_0})
     \right)\;.
   \end{split} \]
These expressions agree with (6) in \cite{Hoelbling:2021axl}, except for a
factor $1 / 2$, which is due to the difference between real and complex
fields.

\section{Relation between radial and angular resolutions}
\label{aitken}

One of the main limiting factors in our numerical calculations is the number
of angular momentum modes $l_{\max}$ taken into account. Generically, at a
given $l_{\max}$ there are $(l_{\max} + 1)^2$ spherical harmonics available,
so the resolution in solid angle $\Delta \sigma$ is limited to
\[ \Delta \sigma = \frac{4 \pi}{(l_{\max} + 1)^2}\;. \]
The linear resolution in angular direction $\Delta_{\perp}$ is therefore
\[ \Delta_{\perp} = r \sqrt{\Delta \sigma} = \frac{2 \sqrt{\pi} r}{l_{\max} +
   1}\;. \]
We can keep the radial resolution $\Delta r$ proportional to the angular
resolution by demanding
\[ \Delta r = \frac{c}{2 \sqrt{\pi}} \Delta_{\perp} \]
with a constant anisotropy factor $c$. Taking into account that the radial
distance between two shells that are separated by a coordinate distance
$\Delta = r_n - r_{n - 1}$ is given by
\begin{equation}
  \Delta r = a \Delta = \sqrt{\frac{r}{d}} \Delta \label{radss}
\end{equation}
we find that the
\begin{equation}
  (l_{\max} + 1) = c \sqrt{d r} \frac{1}{\Delta}\;. \label{constanisogen}
\end{equation}
For the special case of flat spacetime $d = r$, so the maximum angular
momentum has to be increased linearly with the radius
\begin{equation}
  (l_{\max} + 1) = c \frac{r}{\Delta} \label{constaniso}
\end{equation}
to keep the ratio between angular and radial resolution constant.

\section{Asymptotic scaling from the harmonic oscillator chain
approximation}\label{hocapp}

For the free case $a = \alpha = 1$ and large radii $r$, a single $l$-mode may
effectively be described as a harmonic oscillator chain or $1 + 1$ dimensional
scalar field theory with effective mass
\begin{equation}
  \mu^2 = M^2 + \frac{l (l + 1)}{r^2} \;,\label{meffdim}
\end{equation}
that is generically dependent on the radial coordinate. For the case of vanishing effective mass $\mu = 0$, there are detailed results in the literature
\cite{Larsen:1994yt,Katsinis:2024gef,Holzhey:1994we}.  Specifically, the
entanglement entropy of cutting the vacuum of a continuous $1 + 1$
dimensional, real scalar field theory in an interval of length $L$ at position
$r$ is given by
\begin{equation}
  S_0^H = \frac{1}{6} \ln \left( \frac{2 L}{\pi \varepsilon'} \sin \left(
  \frac{\pi r}{L} \right) \right) \;,\label{l0e}
\end{equation}
where $\varepsilon'$ is a necessary UV cutoff. For the massive case, Callan
and Wilczek \cite{Callan:1994py} have demonstrated the logarithmic
divergence of the entropy in the cutoff $\varepsilon$ using a heat kernel
approach. In this approach, the entanglement entropy is computed by first
defining the Euclidean theory on a conic surface with deficit angle $\delta$.
The entropy is then obtained by evaluating the integral
\[ S = \left. \frac{1}{2} \bigintlim_{\varepsilon^2}^{\infty} \mathd t \,
   \frac{1}{t} \left( 2 \pi \frac{\mathd}{\mathd \delta} + 1 \right) \zeta (t)
   e^{- \mu^2 t} \right|_{\delta = 0} \]
at vanishing deficit angle with the heat kernel
\[ \zeta (t) = \bigintlim_{\mathbb{R}} \mathd E \, \rho (E) e^{- E t} \;.\]
We use the result by Balian and Bloch \cite{BALIAN1971271} that, up to
finite volume corrections, the eigenmode density for any curved two
dimensional manifold is approximated by
\[ \rho (E) \approx \Theta (E) \frac{1}{4 \pi} \left( A \pm \frac{1}{2
   \sqrt{E}} \partial A \right) + \frac{1}{6} \delta (E) \chi\;, \]
where $A$, $\partial A$ and $\chi$ are its area, circumference and Euler
characteristic respectively. Noting that both area and circumference of a cone
scale as $1 - \delta / (2 \pi)$ with the deficit angle, we find
\[ \rho (E) \approx \Theta (E) \frac{1}{4} \left( 1 - \frac{\delta}{2 \pi}
   \right) \left( L^2 \pm 2 \frac{L}{\sqrt{E}} \right) + \frac{1}{6} \delta
   (E) \]
and thus
\[ \zeta (t) = \left( 1 - \frac{\delta}{2 \pi} \right) \frac{1}{4} \left(
   \frac{L^2}{t} \pm \sqrt{\pi} \frac{L}{\sqrt{t }} \right) + \frac{1}{6} + O
   \left( \frac{t}{L^2} \right) \;.\]
Since
\[ \left. \left( 2 \pi \frac{\mathd}{\mathd \delta} + 1 \right) \left( 1 -
   \frac{\delta}{2 \pi} \right) \right|_{\delta = 0} = 0 \]
the area and circumference terms vanish in the entropy and we are left with
\begin{equation}
  S \approx \frac{1}{12} \bigintlim_{\varepsilon^2}^{\infty} \mathd t \,
  \frac{1}{t}  \left( 1 + O \left( \frac{t}{L^2} \right) \right) e^{- \mu^2 t}
  = \frac{1}{12} \Gamma (0, \varepsilon^2 \mu^2) + O \left( \frac{1}{L^2
  \mu^2} \right) \;,\label{hocbare}
\end{equation}
where we have removed the cutoff in the finite volume corrections since they
are finite. Inserting the effective mass term \eqref{meff}, we thus find an
asymptotic solution for the entropy of an angular momentum $l$ mode in the
continuum and infinite volume limits and for points far away from the
boundary. As the physical scale of our system is given by the inverse of the
the effective mass $\mu$ the continuum limit entails
\[ \Delta \mu \rightarrow 0 \]
while the condition that a point is far from the boundary requires
\[ r \mu \rightarrow \infty \hspace{4em} (L - r) \mu \rightarrow \infty\;. \]
Summing the last two conditions implies the infinite volume limit
\[ L \mu \rightarrow \infty \]
so it does not need to be specified separately. In these limits, we find
\begin{equation}
  S \rightarrow \frac{1}{12} \Gamma \left( 0, \varepsilon^2 \left( M^2 +
  \frac{l (l + 1)}{r^2} \right) \right) \;.\label{sassyinf}
\end{equation}
The similarity between the expressions for the massive and the massless finite
volume case suggest that one may find a reasonable approximation for the
finite volume, massive case by interpreting the argument
\[ \frac{2 L}{\pi \varepsilon'} \sin \left( \frac{\pi r}{L} \right) \]
of the finite volume result \eqref{l0e} as an inverse effective mass term.
Combining it with the infinite volume effective mass term from
\eqref{sassyinf}, one may conjecture a continuum entanglement entropy for
finite effective mass and volume
\begin{equation}
  S = \frac{1}{12} \Gamma \left( 0, \varepsilon^2 \left( M^2 + \frac{l (l +
  1)}{r^2} \right) {+ \varepsilon'}^2 e^{- \gamma} \left( \frac{2 L}{\pi} \sin
  \left( \frac{\pi r}{L} \right) \right)^{- 2} \right) \label{sugint}
\end{equation}
which has the correct infinite volume \eqref{sassyinf} and effective mass zero
\eqref{l0e} limits.

\section{Hopping expansion}\label{hop}

For large effective masses, the initial time covariance matrix
$\overline{C}_{t_0}$ may, according to \eqref{cortm} be expressed as
\[ \overline{C}_{t_0} = B  B^T = Q^{- 1}  \left(\begin{array}{cc}
     K^{- \frac{1}{2}} & 0\\
     0 & K^{\frac{1}{2}}
   \end{array}\right) Q^{- 1} \;,\]
where the angular momentum $l$ components of $K$ are given by \eqref{kdef} and $Q$ is the diagonal rescaling operator \eqref{rescop}. We will in the following concentrate on a single $l$ mode, drop all explicit $l$ indices and take the obvious projectors where appropriate.

In order to compute the entanglement entropy upon cutting the system at a radial coordinate $r_c$, we
need the symplectic eigenvalues of the projected covariance matrix
\[ \overline{C} = (P^c \otimes \mathbbm{1}) \overline{C}_{t_0} (P^c \otimes
   \mathbbm{1}) \;,\]
where $P^c$ is the projector onto radial coordinates $r_i \leqslant
r_c$\footnote{The entire argument works equivalently if we use the
complementary projector $\mathbbm{1}- P^c$.}. Since the rescaling operator $Q$
is diagonal, it commutes with the projector and we may write the truncated
covariance matrix as
\[ \overline{C} = Q^{- 1}  \left(\begin{array}{cc}
     K_c^{- \frac{1}{2}} & 0\\
     0 & K_c^{\frac{1}{2}}
   \end{array}\right) Q^{- 1}\;, \]
where $K^{\pm \frac{1}{2}}_c = P^c K^{\pm \frac{1}{2}} P^c$ are the
projections of $K^{\pm \frac{1}{2}}$. 
We find the symplectic eigenvalues of $\overline{C}$ by bringing it into Williamson form. This can be achieved by the transformation
\[ \overline{C} = Q^{- 1} \hat{Q} \hat{W} \hat{S} \left(\begin{array}{cc}
     \sigma_{\pm} & 0\\
     0 & \sigma_{\pm}
   \end{array}\right) \hat{S} \hat{W}^T \hat{Q} Q^{- 1} \]
with
\[ \hat{R} = \left(\begin{array}{cc}
     \left. \left( K_c^{\frac{1}{2}} \right. \right)^{- \frac{1}{2}} & 0\\
     0 & \left( K^{\frac{1}{2}}_c \right)^{\frac{1}{2}}
   \end{array}\right) \qquad \hat{W} = \left(\begin{array}{cc}
     \hat{V} & 0\\
     0 & \hat{V}
   \end{array}\right) \qquad \hat{S} = \hat{S}^T = \left(\begin{array}{cc}
     \sqrt{\sigma} & 0\\
     0 & \frac{1}{\sqrt{\sigma}}
   \end{array}\right) \;,\]
where $\sigma^2$ is the diagonal operator of eigenvalues and $\hat{V}$ the
corresponding eigenvectors of the symmetric matrix
\[ \Sigma = \left( K^{\frac{1}{2}}_c \right)^{\frac{1}{2}} K^{- \frac{1}{2}}_c
   \left( K^{\frac{1}{2}}_c \right)^{\frac{1}{2}} \]
obtained via the diagonalization
\[ \Sigma = \hat{V} \sigma^2 \hat{V}^T \qquad \hat{V}^T = \hat{V}^{- 1}\;.
\]
Defining
\[ \hat{U} = \left( K^{\frac{1}{2}}_c \right)^{\frac{1}{2}} \hat{V} \]
we find the eigenvalue equation
\begin{equation}
  K^{\frac{1}{2}}_c K^{- \frac{1}{2}}_c = \hat{U} \sigma^2 \hat{U}^{- 1}
  \label{kkev}
\end{equation}
and thus the symplectic eigenvalues $\sigma$ of $\overline{C}$ may be obtained
by diagonalizing $K^{\frac{1}{2}}_c K^{- \frac{1}{2}}_c$. The corresponding
entropy may then be computed according to \eqref{vne}.

We now need the explicit form of the kernel operator $K$. As in our numerical
studies, we discretize the radial coordinate in regular intervals $r_i =
\Delta i$ and use the forward difference operator
\[ \partial_{i j} = \frac{\delta_{i + 1, j} - \delta_{i, j}}{\Delta} \;.\]
This leads to an explicit expression
\[ K = L + \frac{1}{\Delta^2} \Xi \]
for the kernel operator with a diagonal term
\[ L_{i j} = L_i \delta_{i j} \qquad L_i = \alpha_i^2 \mu_i^2 +
   \frac{1}{\Delta^2} \Lambda_i \qquad \Lambda_i = \frac{\alpha_i
   \alpha_{i + 1}}{a_i a_{i + 1}} \left( \frac{i }{i + 1} \right)^2 +
   \frac{\alpha_i^2}{a_i^2} \qquad \mu_i^2 = \frac{l (l + 1)}{r_i^2} +
   M^2 \]
and an offdiagonal term
\[ \Xi_{i j} = \Xi_i \delta_{i + 1, j} + \Xi_{i - 1} \delta_{i, j + 1}
   \qquad \Xi_i = - \sqrt{\frac{\alpha_i \alpha_{i + 1}^3}{a_i a^3_{i +
   1}}} \frac{i}{i + 1}\;. \]
This tridiagonal structure of $K$ forms the basis for a hopping expansion. In
order to construct it, we first define the expansion parameter
\[ \varepsilon = \frac{1}{\tilde{\mu}_c^2} \qquad \tilde{\mu}_i^2 =
   \Delta^2 \mu_i^2 \]
which allows us to write the kernel operator as
\[ K = \mu^2 \left( \alpha^2 + \frac{1}{\tilde{\mu}^2} \Lambda + \varepsilon
   \frac{\tilde{\mu}_c^2}{\tilde{\mu}^2} \Xi \right) \]
with the obvious definition of the diagonal operators $\mu$, $\alpha$ and
$\tilde{\mu}$. Although by strict power counting arguments we could categorize
the $\Lambda$ term as first order in $\varepsilon$, we treat it as $O (1)$
since it is diagonal.

We now make a Taylor series ansatz for $K^{\pm \frac{1}{2}}$
\[ J = \sum_{k = 0}^{\infty} \varepsilon^k J^{(k)} \qquad J^{- 1} =
   \sum_{k = 0}^{\infty} \varepsilon^k \overline{J}^{(k)} \qquad K =
   \mu^2 J^2 \;.\]
According to our power counting, which treats all diagonal terms as leading
order, we have
\begin{equation}
  J^{(0)} = {\overline{J}^{(0)}}^{- 1} = \sqrt{\alpha^2 +
  \frac{1}{\tilde{\mu}^2} \Lambda} \;. \label{j0}
\end{equation}
Furthermore, we have
\[ {J^{(0)}}^2 + \varepsilon \frac{\tilde{\mu}_c^2}{\tilde{\mu}^2} \Xi = J^2 =
   \sum_{n = 0}^{\infty} \varepsilon^n \sum_{k = 0}^n J^{(n - k)} J^{(k)} \]
and by equating coefficients we find
\[ \{ J^{(1)}, J^{(0)} \} = \frac{\tilde{\mu}_c^2}{\tilde{\mu}^2} \Xi \]
as well as
\[ \{ J^{(n)}, J^{(0)} \} + \sum_{k = 1}^{n - 1} J^{(n - k)} J^{(k)} = 0
   \qquad n \geqslant 2 \;. \]
Since $J^{(0)}$ is diagonal, it is easy to find the solutions
\begin{equation}
  J^{(1)}_{i j} = \frac{\tilde{\mu}_c^2}{\tilde{\mu}_i^2} \frac{\Xi_{i
  j}}{J^{(0)}_i + J^{(0)}_j} \label{j1}
\end{equation}
and
\begin{equation}
  J^{(n)}_{i j} = - \frac{\left( \sum_{k = 1}^{n - 1} J^{(k)} J^{(n - k)}
  \right)_{i j}}{J^{(0)}_i + J^{(0)}_j} \qquad n \geqslant 2 \;.\label{jn}
\end{equation}
The $\overline{J}^{(n)}$ may be found via
\[ \mathbbm{1}= J J^{- 1} = \sum_{n = 0}^{\infty} \varepsilon^n \sum_{k = 0}^n
   J^{(n - k)} \overline{J}^{(k)} \;,\]
which implies
\begin{equation}
  \overline{J}^{(n)} = - \overline{J}^{(0)} \sum_{k = 0}^{n - 1} J^{(n - k)}
  \overline{J}^{(k)} \;.\label{jbarn}
\end{equation}
Note that $J^{(0)}$ is diagonal and according to \eqref{j1} $J^{(1)}$ is
tridiagonal with vanishing diagonal, or, in terms of radial indices,
represents a matrix with strictly one hop in either direction. Consequently,
\eqref{jn} and \eqref{jbarn} imply that $J^{(n)}$ and $\overline{J}^{(n)}$ are
both banded matrices with hops from $- n$ to $n$ in spacings of 2.

Let us now construct the matrix
\[ M_{i j} = \left( K^{\frac{1}{2}}_c K^{- \frac{1}{2}}_c - P^c \right)_{(c -
   i) (c - j)} \]
which, according to \eqref{kkev} has eigenvalues $\sigma_k^2 - 1$. With the
restrictions $i, j \geqslant 0$, we find
\[ M_{i j} = (J P^c \overline{J})_{(c - i) (c - j)} - \delta_{i j} \;.\]
Since $J \overline{J} =\mathbbm{1}$, we may also write this as
\begin{equation}
  \begin{split}
    M_{i j} = & \sum_{k \leqslant 0} J_{(c - i) (c + k)} \overline{J}_{(c +
    k) (c - j)} - \sum_k J_{(c - i) (c + k)} \overline{J}_{(c + k) (c - j)}\\
     = & - \sum_{k \geqslant 1} J_{(c - i) (c + k)} \overline{J}_{(c + k) (c
    - j)}\;.
  \end{split} \label{mijx}
\end{equation}
For each term in the sum we have a total of at least $2 k + i + j$ hops,
which, together with $k \geqslant 1$, implies that the leading term towards
$M_{i j}$ appears at order $2 + i + j$ of the hopping expansion.

In principle the relations (\ref{j0}-\ref{mijx}) can be used to recursively
build $M_{i j}$ in the general case to any desired order. We will from now on
restrict ourselves to flat space, large cutoff coordinate $c \gg 1$ and a
small stencil $\max (i, j) \ll c$, which in physical terms corresponds to the
continuum limit $\Delta \rightarrow 0$ at a fixed, finite distance from each
boundary. In this limit, we have
\[ J_i^{(0)} \rightarrow \sqrt{1 + 2 \varepsilon} \qquad
   \frac{\tilde{\mu}_c^2}{\tilde{\mu}_i^2} \rightarrow 1 \qquad \Xi_i
   \rightarrow - 1 \]
so that
\[ J^{(1)}_{i j} = - \frac{\delta_{i + 1, j} + \delta_{i, j + 1}}{2 \sqrt{1
   + 2 \varepsilon}} \]
and \eqref{jn} can be written in matrix form as
\begin{equation}
  J^{(n)} = - \frac{\sum_{k = 1}^{n - 1} J^{(k)} J^{(n - k)}}{2 \sqrt{1 + 2
  \varepsilon}} \qquad n \geqslant 2\;. \label{jne}
\end{equation}
Interestingly, each term in the sum is just a split of an $n$ hop matrix into
a $k$ and a $n - k$ hop submatrix. Since each of them is a sum of the product
of one hop matrices, it follows that the individual terms are equal up to a
multiplicative prefactor. Let us thus define $\varepsilon^n J^{(n)} = p_n
\hat{J}^{(n)}$ as
\[ \hat{J}^{(n)} = - \frac{\hat{J}^{(m)} \hat{J}^{(n - m)}}{2 \sqrt{1 + 2
   \varepsilon}} = - \frac{{{}\hat{J}^{(1)}}^n}{(4 (1 + 2
  \varepsilon))^\frac{n-1}{2}} 
   \]
for $n\ge2$ and $1\le m<n$ and $p_1 = 1$.  Plugging this into \eqref{jne} and defining
\[ x = \frac{\varepsilon}{4 (1 + 2 \varepsilon)} = \frac{1}{4 (2 +
   \tilde{\mu}^2)} \]
we find
\begin{equation}
  J^{(n)} = - \frac{ {J^{(1)}}^n}{(4 (1 + 2
   \varepsilon))^\frac{n-1}{2}}\sum_{k = 1}^{n - 1}p_{k}p_{n-k}  \qquad n \geqslant 2 
\end{equation}
implying
\[ p_n = \sum_{k = 1}^{n - 1} p_k p_{n - k} \]
which is one of the recurrence relations for Catalan numbers $C_{n - 1} =
p_n$. So
\[ \varepsilon^n J^{(n)} = C_{n - 1} \hat{J}^{(n)} \]
and any single $n$-hop path in $J$, which may include alternating forward and backward hops, incurs a factor
\begin{equation}
  H_n = - \sqrt{\frac{\varepsilon}{x}} C_{n - 1} x^n = -
  \sqrt{\frac{\varepsilon}{x}} \frac{1}{n} \left(\begin{array}{c}
    2 n - 2\\
    n - 1
  \end{array}\right) x^n\;. \label{nhop}
\end{equation}
For $\overline{J}$ we note that for the leading two orders are proportional to
their $J$ counterparts
\[ 
\varepsilon \overline{J}^{(n)} = 4 x J^{(n)} \qquad n \in \{ 1, 2 \}\;.
\]
Furthermore, the recurrence relation \eqref{jbarn} exactly mirrors the one for
$J$ except for prefactors, which leads us to define
\[ \widehat{\overline{J}}^{(n)} = - \overline{J}^{(0)} \hat{J}^{(n)}
   \overline{J}^{(0)} = - \frac{4 x}{\varepsilon}  \hat{J}^{(n)} \;.\]
Paralleling the construction for $J$, we have $\varepsilon^n
\overline{J}^{(n)} = \overline{p}_n \widehat{\overline{J}}^{(n)}$ where
$\overline{p}_0 = \overline{p}_1 = 1$ and the rest of the prefactors
$\overline{p}_n$, $n \geqslant 2$ have to be determined. Plugging this into
\eqref{jbarn} one can show that
\[ \varepsilon^n \overline{J}^{(n)} =  \widehat{\overline{J}}^{(n)} \left( p_n
   + 2 \sum_{k = 1}^{n - 1} p_{n - k}  \overline{p}_k \right) \;,\]
which implies the recursion relation
\[ \overline{p}_n = p_n + 2 \sum_{k = 1}^{n - 1} p_{n - k}  \overline{p}_k \]
that resolves into
\[ \overline{p}_n =  \left(\begin{array}{c}
     2 n - 1\\
     n - 1
   \end{array}\right)=(2n-1)p_n
   \;. \]
In total, we thus find a factor
\begin{equation}
  \overline{H}_n = 4 \overline{p}_n \sqrt{\frac{x}{\varepsilon}} x^n
  \label{nbarhop}
\end{equation}
for every $n$-hop path in $\overline{J}$.

\begin{figure}[htb]\centering
  {\center{\includegraphics[width=0.5\textwidth]{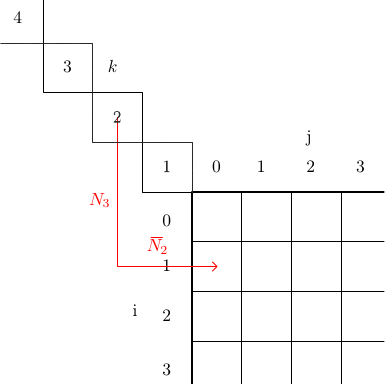}}}
    \caption{\label{hopfig}Illustration of one contribution to the matrix
  element $M_{i j}$.}
\end{figure}

Let us now compute the matrix element $M_{i j}$ in the hopping expansion.
According to \eqref{mijx}, we may write
\begin{equation}
  M_{i j} = \sum_{k \geqslant 1} N_{k + i} \overline{N}_{k + j}\;, \label{mij}
\end{equation}
where the $N_l$ and $\overline{N}_l$ are the sums over all paths in $- J$ and
$\overline{J}$ respectively with a distance of $l$ hops between the start- and
endpoint, weighted with the weighting factor of each path. Due to possible
backward hops, the length of the path might actually be larger than $l$. In
fact, if there are $m$ backward hops, the path has a length of $l + 2 m$.
Accordingly, there is more than one path which leads to an additional
combinatorial factor except in the minimal case. In fact, the number of paths
of length $l + 2 m$ covering a distance of $l$ hops is equal to the number of
paths from the top of Pascals triangle to row $l + 2 m$ and column $m$.
Combining this with \eqref{nhop} and \eqref{nbarhop} and considering that the
factors $\sqrt{x / \varepsilon}$ and $\sqrt{\varepsilon / x}$ always cancel,
we may thus write
\begin{equation}
  N_l = 2 \sum_{m \geqslant 0} c_m^l x^{l + 2 m} \qquad \overline{N}_l =
  2 \sum_{\overline{m} \geqslant 0} \overline{c}_{\overline{m}}^l x^{l + 2
  \overline{m}} \label{nlnbl}
\end{equation}
with
\[ c_m^l = \frac{1}{l + 2 m} \left(\begin{array}{c}
     l + 2 m\\
     m
   \end{array}\right) \left(\begin{array}{c}
     2 l + 4 m - 2\\
     l + 2 m - 1
   \end{array}\right) \qquad \overline{c}_{\overline{m}}^l =
   \left(\begin{array}{c}
     l + 2 \overline{m}\\
     \overline{m}
   \end{array}\right) \left(\begin{array}{c}
     2 l + 4 \overline{m} - 1\\
     l + 2 \overline{m} - 1
   \end{array}\right) \;.\]
This setup is illustrated in Fig.~\ref{hopfig}. It should be noted that we
have assumed an unbounded range of coordinates in both directions. A bound on
either side would alter the higher order combinatorial factors, while a lower
bound would additionally put an upper limit on the sum over $k$. It is
straightforward in principle to include these finite size effects, but we will
ignore them here.

In order to find the convergence radius for $N_l$ and $\overline{N}_l$ we note
that the asymptotic coefficient ratios are
\[ \frac{c_m^l}{c_{m - 1}^l} \xrightarrow{m \rightarrow \infty} 64
   \qquad
   \frac{\overline{c}_{\overline{m}}^l}{\overline{c}_{\overline{m} - 1}^l}
   \xrightarrow{\overline{m} \rightarrow \infty} 64\;. \]
Since the associated expansion parameter is $x^2$, the convergence region is
given by $x < 1 / 8$, or, equivalently, $\tilde{\mu} > 0$. Similarly, for the
convergence of the matrix element \eqref{mij} we note that the asymptotic
coefficient ratios
\[ \frac{c_m^l}{c_m^{l - 1}} \xrightarrow{l \rightarrow \infty} 4 \qquad
   \frac{\overline{c}_{\overline{m}}^l}{\overline{c}_{\overline{m}}^{l - 1}}
   \xrightarrow{l \rightarrow \infty} 4 \]
and
\[ \frac{c_m^l}{c_m^{l - 1}} \xrightarrow{m \rightarrow \infty} 8 \qquad
   \frac{\overline{c}_{\overline{m}}^l}{\overline{c}_{\overline{m}}^{l - 1}}
   \xrightarrow{\overline{m} \rightarrow \infty} 8 \]
also imply convergence for $x < 1 / 8$ or $\tilde{\mu} > 0$.

Since we were unable to find an analytic expression for \eqref{mij} and
\eqref{nlnbl}, we evaluated the truncated sums numerically. There are several
ways of performing this truncated sum, the most straightforward of which is a
fixed order truncation. If we denote the upper bound of the sums in
\eqref{mij} and \eqref{nlnbl} by $k_{\max}$, $m_{\max}$ and
$\overline{m}_{\max}$ for $M_{i j}$, $N_l$ and $\overline{N}_l$ respectively,
then truncation to a fixed order $n$ in $x$ corresponds to
\[ k_{\max} = \frac{n - i - j}{2} \qquad m_{\max} = k_{\max} - k
   \qquad \overline{m}_{\max} = m_{\max} - m \]
For example, the fixed second order truncation is given by
\[ M = 4 x^2 \left(\begin{array}{ccc}
     1 + 39 x^2 & 3 x & 10 x^2\\
     x & 3 x^2 & 0\\
     2 x^2 & 0 & 0
   \end{array}\right)\;. \]
From a computational point of view, this truncation is expensive because the
dependence of $\overline{m}_{\max}$ on $m_{\max}$ prohibits a factorization of
the individual terms in $M_{i j}$. Furthermore, since populating and then
diagonalizing a large matrix is numerically expensive and the largest
eigenmodes dominate the entropy, a resummation of higher order terms might
accelerate convergence substantially. We therefore explore two alternative
truncations: The first, which we call full resummation, is characterized by
\[ k_{\max} = O - i - j \qquad \overline{m}_{\max} = m_{\max} = O \qquad
   i, j \leqslant n - 1 \;,\]
where $O$ is a large constant. The second scheme, which we call partial
resummation, is instead given by
\[ k_{\max} = \frac{n - i - j}{2} \qquad m_{\max} = \overline{m}_{\max}
   = O \qquad i, j \leqslant \frac{n}{2} \;.\]
Numerically diagonalizing the respective matrices we obtain their eigenvalues
$\sigma^2 - 1$, which we can then use to obtain the entropy $S$ according to
\eqref{vne}.

\begin{figure}[htb]\centering
  \includegraphics[width=0.47\columnwidth]{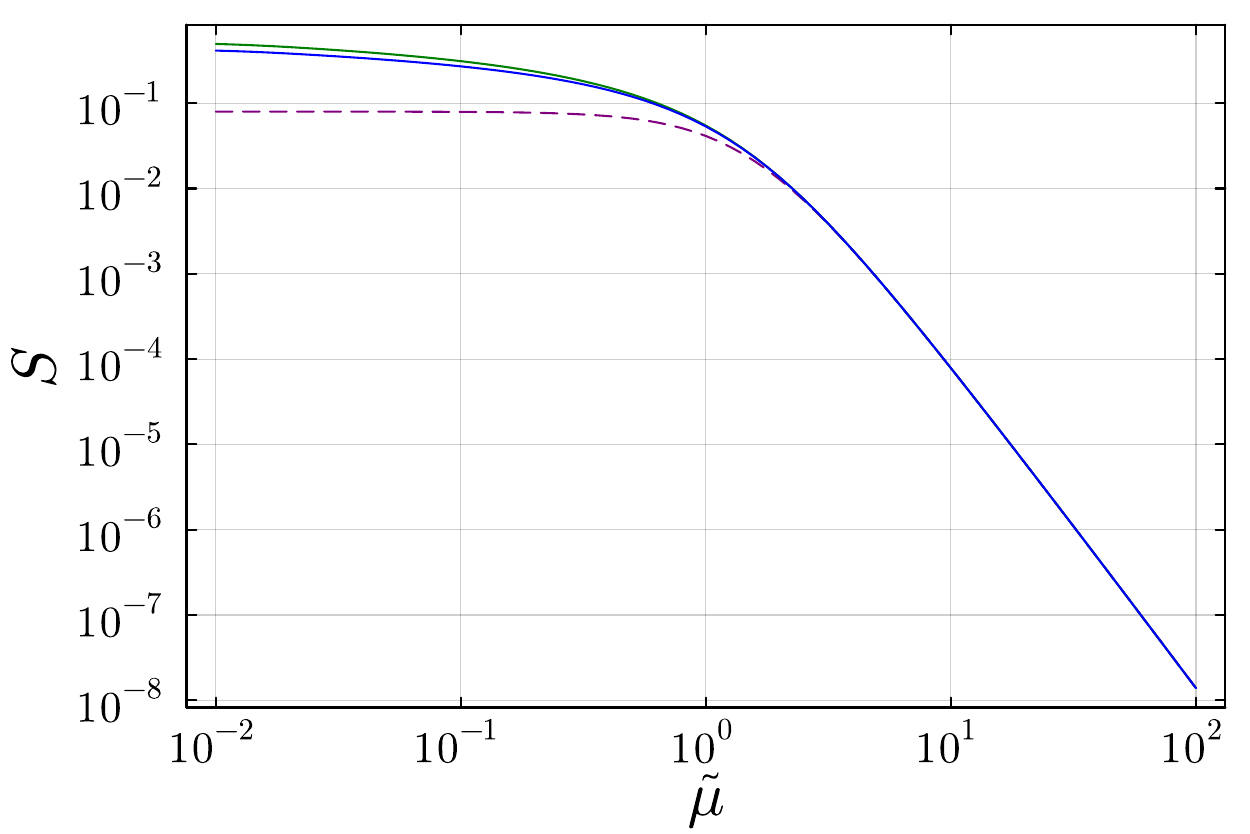}
  \includegraphics[width=0.47\columnwidth]{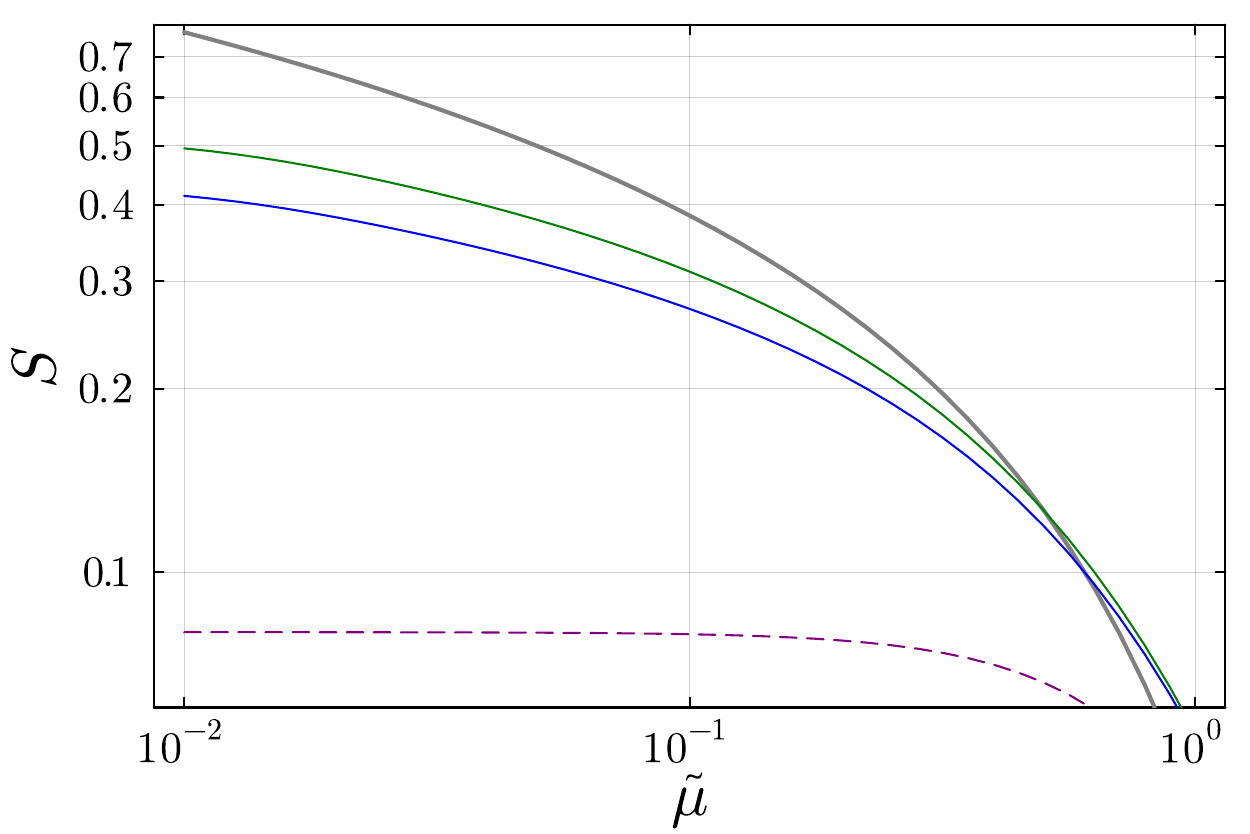}
  \caption{\label{fighop2}Entanglement entropy in the hopping expansion at
  order $n = 2$. Dashed lines represent the fixed order truncation, while the
  blue and green lines show the partial and full resummation, respectively.
  The thick gray line denotes the small $\tilde{\mu}$ expansion
  \eqref{hocforhop}.}
\end{figure}

In Fig.~\ref{fighop2} we compare the entropy in various truncations at order
$n = 2$ and for $O = 2^{14}$. A comparison with the harmonic oscillator chain
approximation
\begin{equation}
  \tilde{S}^{\tmop{HOC}} = \frac{1}{12} \Gamma (0, e^{- \gamma} \tilde{\mu}^2)
  \label{hocforhop}
\end{equation}
which is obtained by combining \eqref{hocbare} with \eqref{epsdelrat} and is
valid for small $\tilde{\mu}$, is also presented.

\begin{figure}[htb]\centering
  \includegraphics[width=0.47\columnwidth]{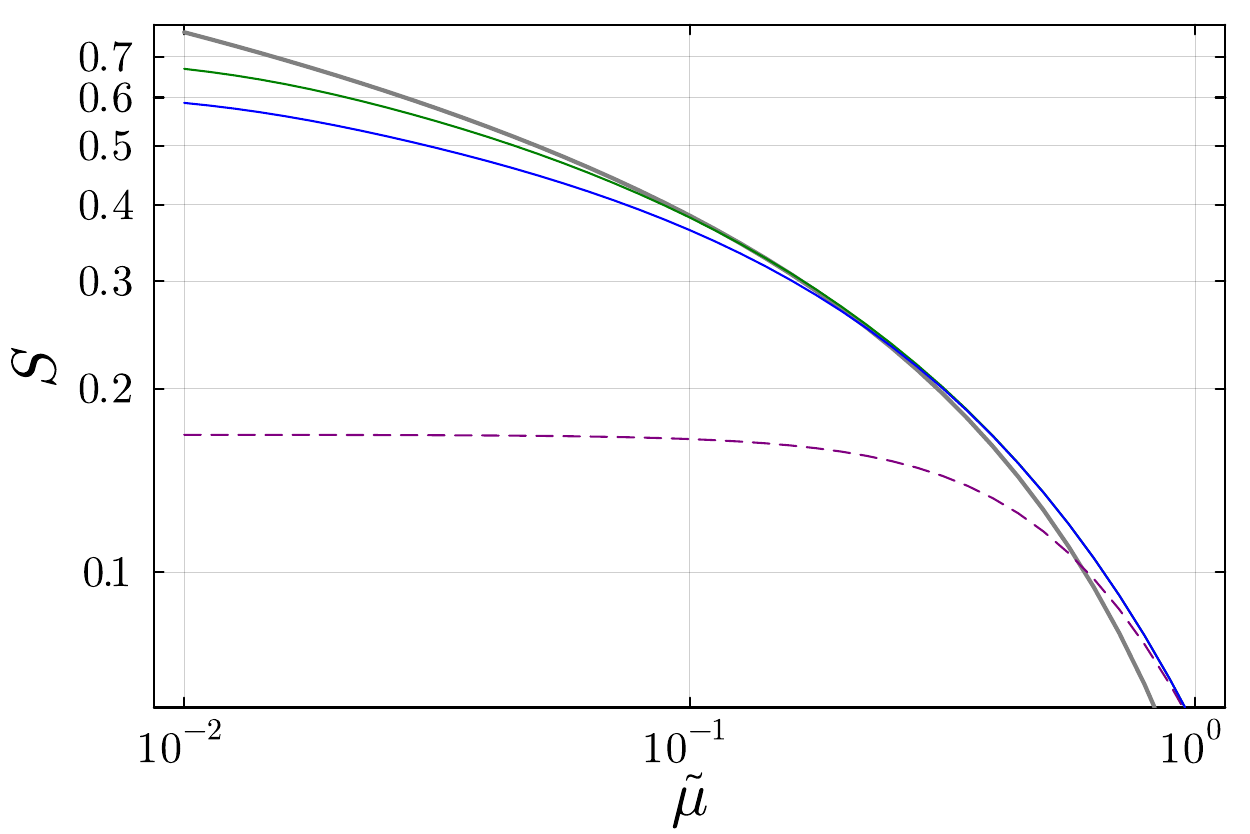}
  \includegraphics[width=0.47\columnwidth]{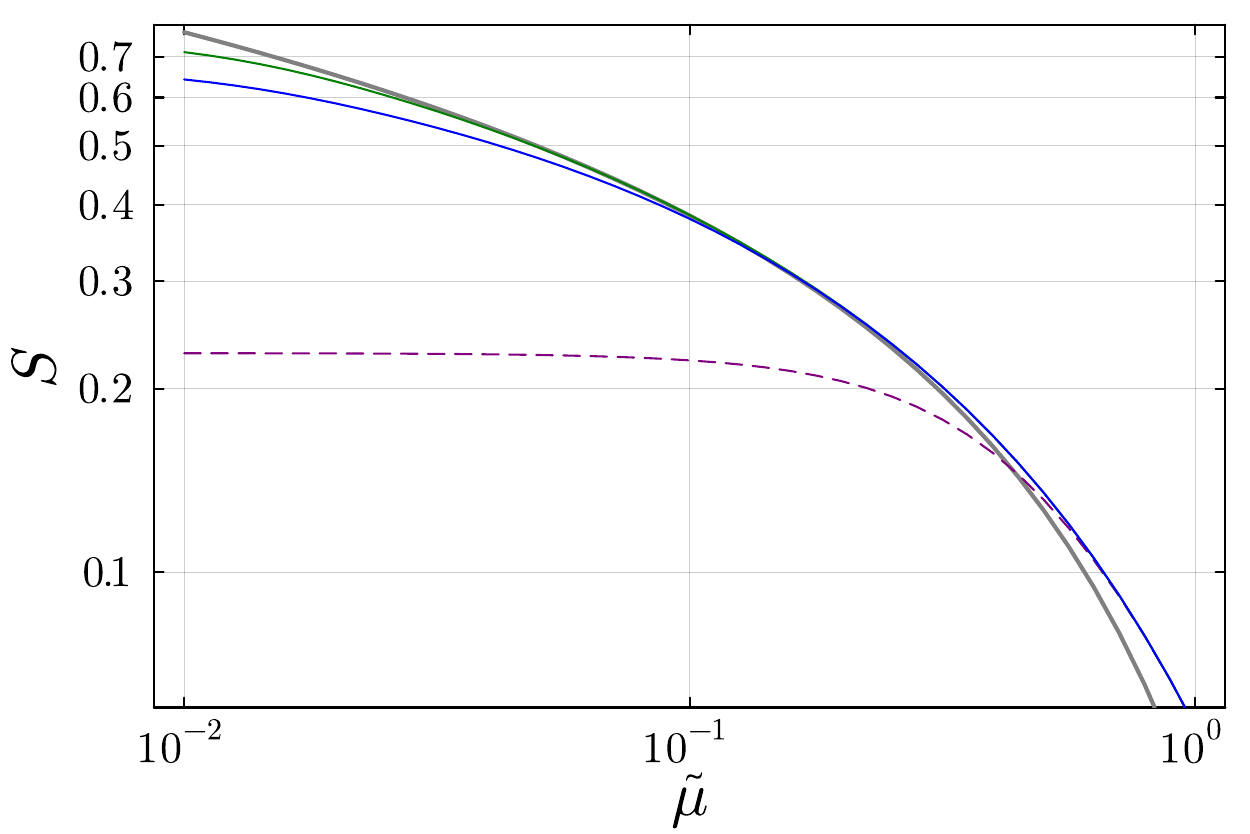}
  \caption{\label{fighop618}Entanglement entropy in the hopping expansion at
  order $n = 8$ (left panel) and $n = 18$ (right panel). Dashed lines
  represent the fixed order truncation, while the blue and green lines show
  the partial and full resummation, respectively. The thick gray line denotes
  the small $\tilde{\mu}$ expansion \eqref{hocforhop}.}
\end{figure}

Fig.~\ref{fighop618} presents higher order truncations, again using $O=2^{14}$. Clearly, the convergence at small $\tilde{\mu}$ is substantially improved by resummation. Although it is clear that at a given order $n$ the full resummation gives the best result, it is numerically expensive. For a given computational cost, the partial resummation thus gives the best results for the parameters we chose.

\begin{figure}[htb]\centering
  \includegraphics[width=0.47\columnwidth]{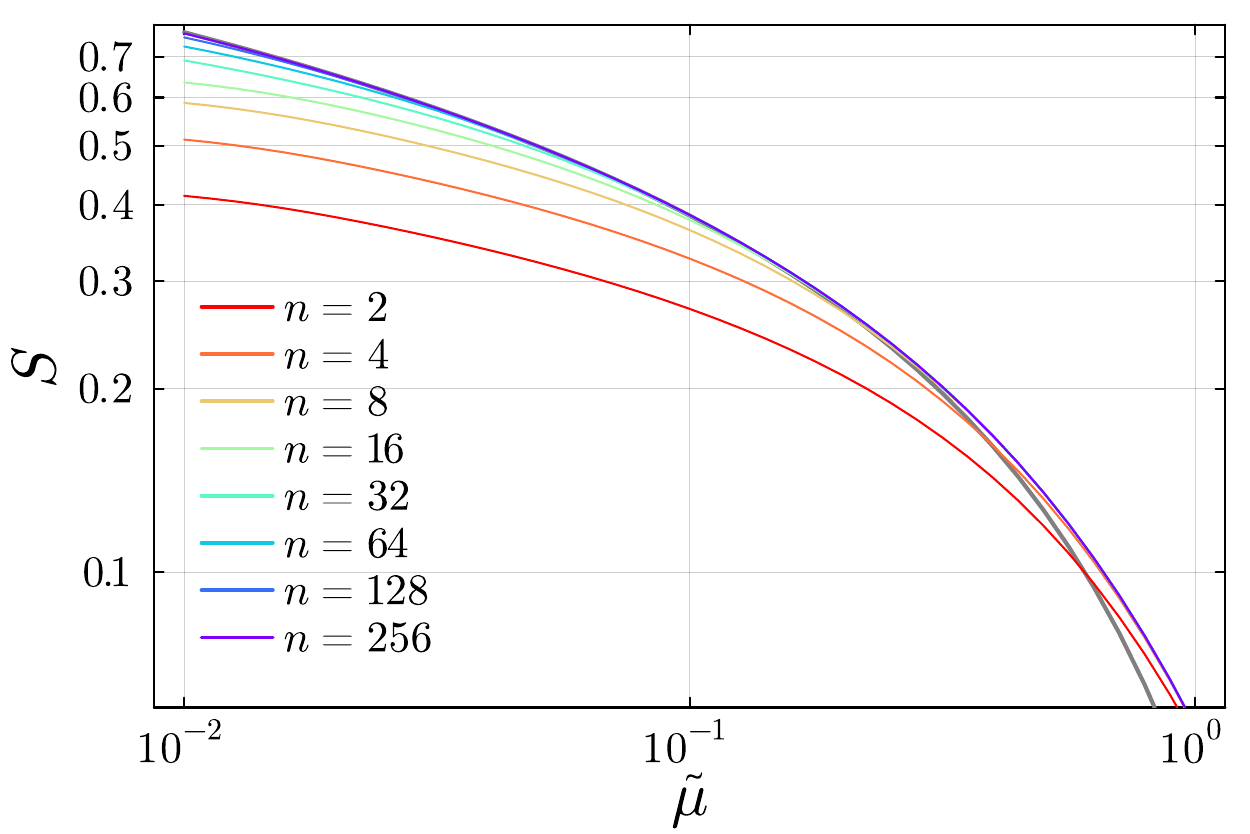}
  \includegraphics[width=0.47\columnwidth]{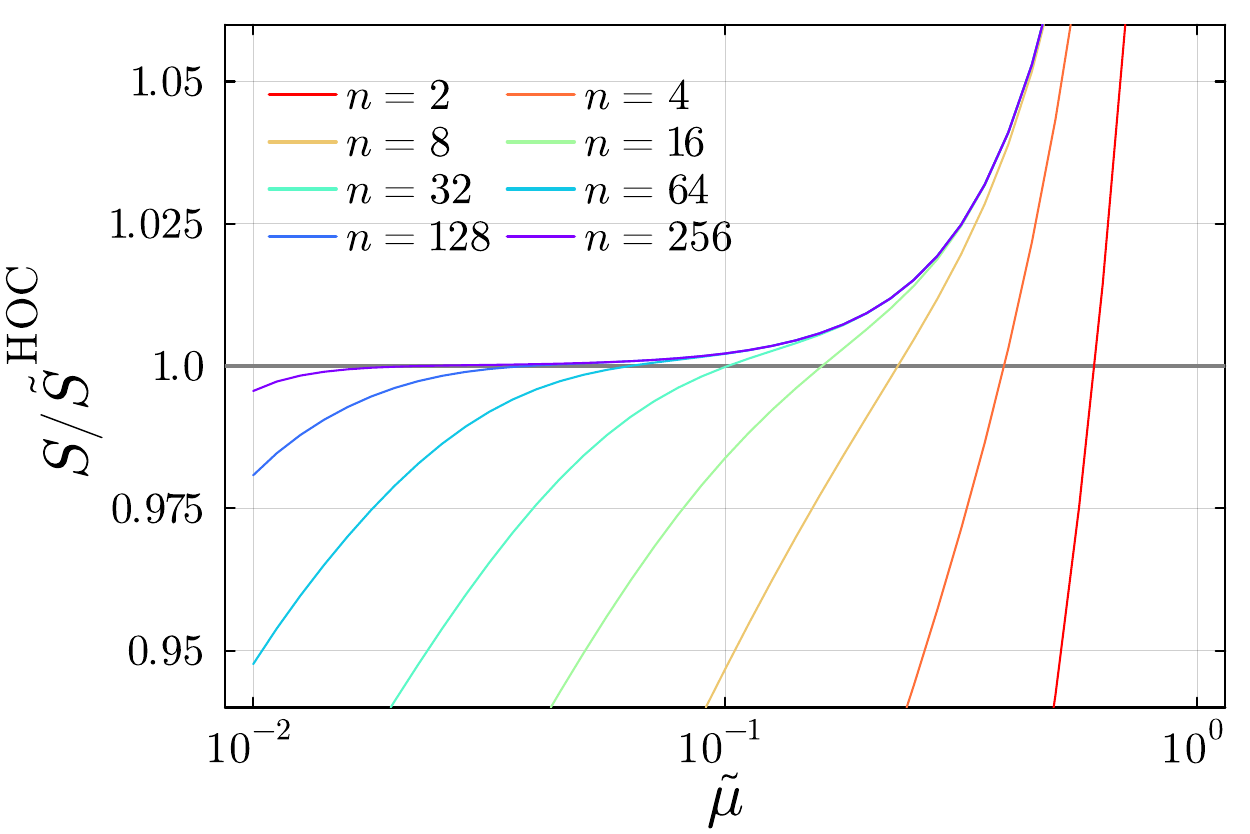}
  \caption{\label{fighopres}Entanglement entropy in the partially resummed hopping expansion (left panel) and its ratio to the small $\tilde{\mu}$
  expansion \eqref{hocforhop} (right panel) at various orders $n$. The thick
  gray line denotes the small $\tilde{\mu}$ expansion.}
\end{figure}

\begin{figure}[htb]\centering
  \includegraphics[width=0.47\columnwidth]{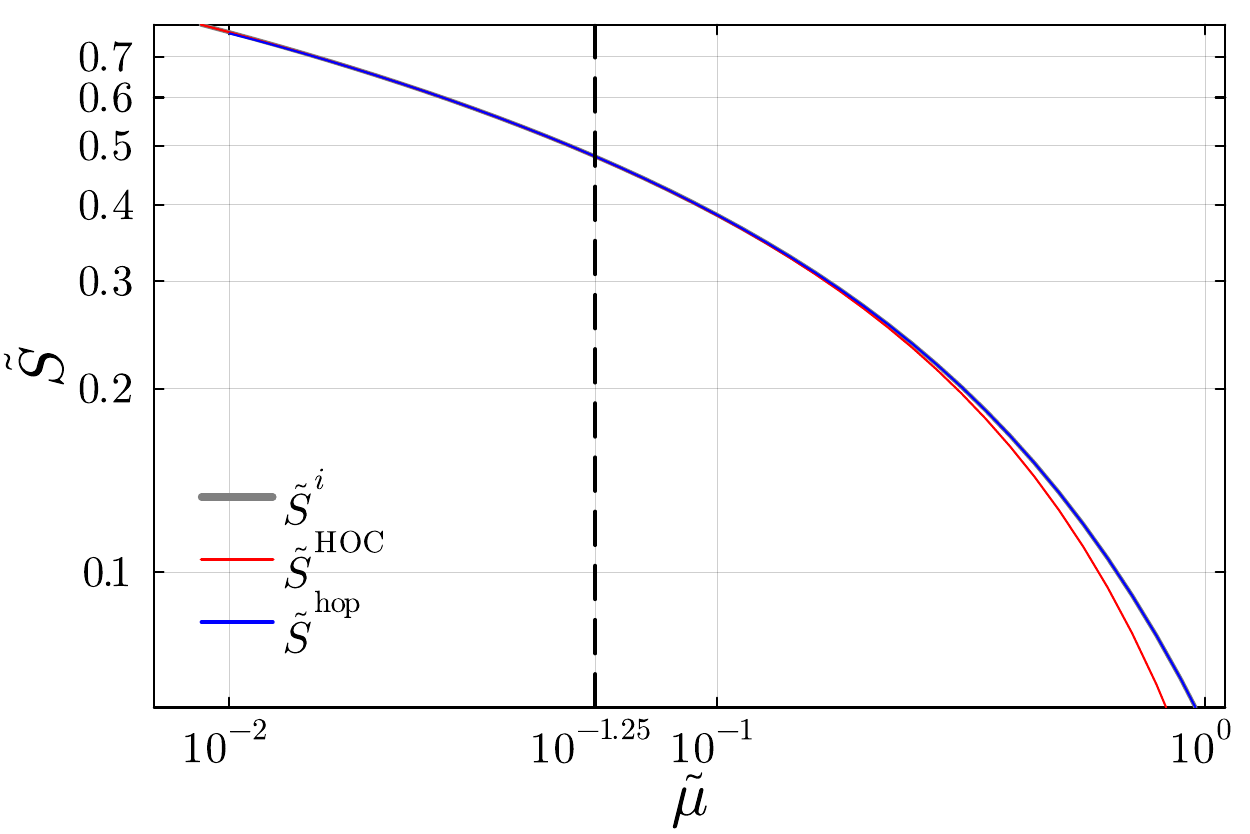}
  \includegraphics[width=0.47\columnwidth]{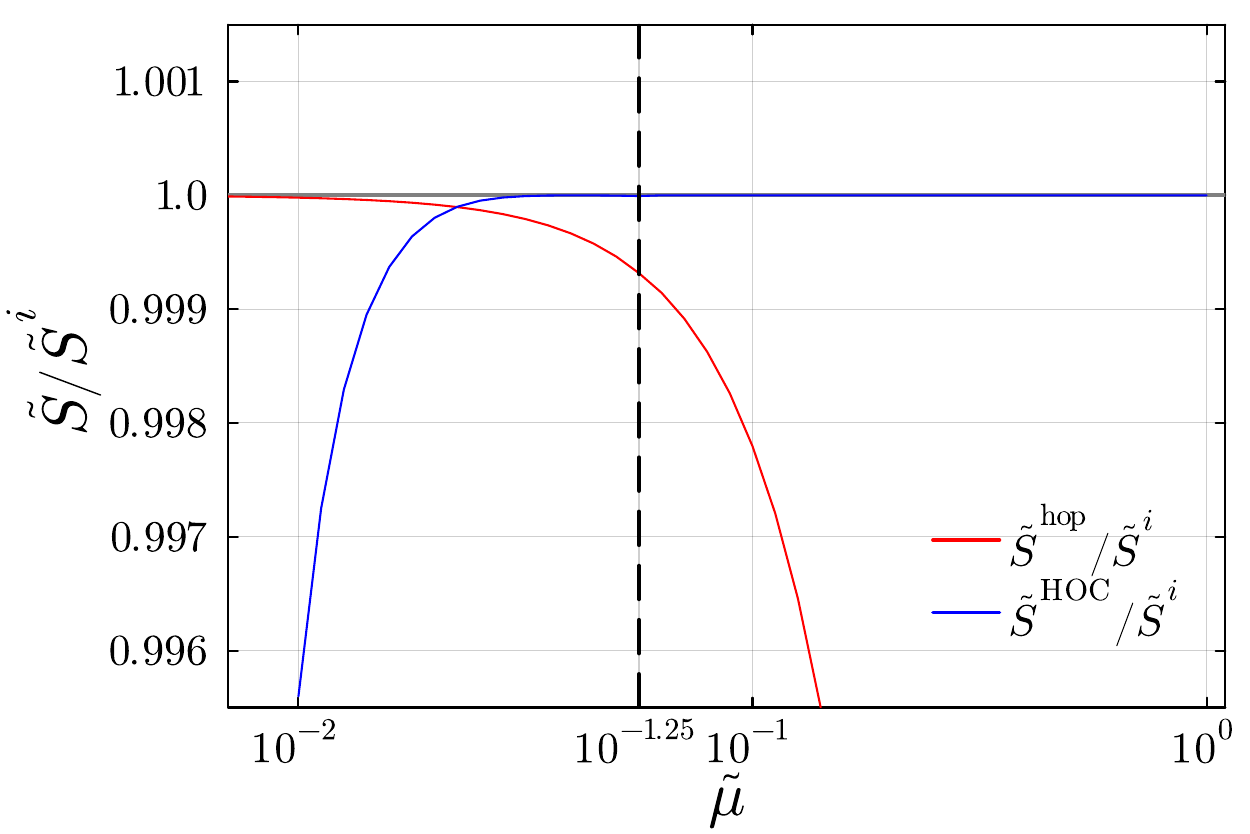}
  \caption{\label{figstildi}The approximation \eqref{stili} compared to both asymptotic forms $\tilde{S}^{\mathrm{HOC}}$ \eqref{hocforhop} and the $256^\mathrm{th}$ order resummed hopping expansion $\tilde{S}^{\mathrm{hop}}$. The left hand plot shows a direct comparison while the right hand plot shows the ratios of the asymptotic forms to the approximation. The dashed vertical line denotes the transition point in the interpolator \eqref{stili}.}
\end{figure}

In Fig.~\ref{fighopres} we compare the results of the partial resummation at different orders. We use a resummation limit of $O=2^{14}$ for approximations up to and including $16^{\mathrm{th}}$ order and $O=2^{16}$ from $32^{\mathrm{nd}}$ order on. One can clearly see convergence over a large range of $\tilde{\mu}$. Additionally, the highest order approximations seem to
make contact with the small $\tilde{\mu}$ expansion \eqref{hocforhop} at the
per mil level. We use the partially resummed $256^{\tmop{th}}$ order hopping
approximation $\tilde{S}^{\mathrm{hop}}(\tilde{\mu}^2)$ of the entropy as an asymptotic form in the main part of the paper. In addition, we define an interpolating function
\begin{equation}
    \tilde{S}^i(\tilde{\mu}^2)=\left\{
\begin{array}{ll}
     \frac{1}{6a}\ln\frac{1}{\tilde{\mu}^a+1}&\quad \tilde{\mu}\leqslant 10^{-\frac{5}{4}} \\
    \tilde{S}^{\mathrm{hop}}(\tilde{\mu}^2) & \quad \tilde{\mu} > 10^{-\frac{5}{4}}
\end{array}
    \right.
    \qquad
    a=1.8252
    \label{stili}
\end{equation}
that agrees with the asymptotic form \eqref{hocforhop} in the small $\tilde{\mu}$ limit at the sub per mil level. This precise numerical agreement also supports \eqref{epsdelrat} and thus the conjecture that \eqref{hocforhop} is the correct asymptotic form for small $\tilde{\mu}$.

\section{Heuristic description of the $l = 0$ mode entropy in the interacting
theory\label{l0approx}}

\begin{figure}[htb]\centering
  \includegraphics[width=0.47\columnwidth]{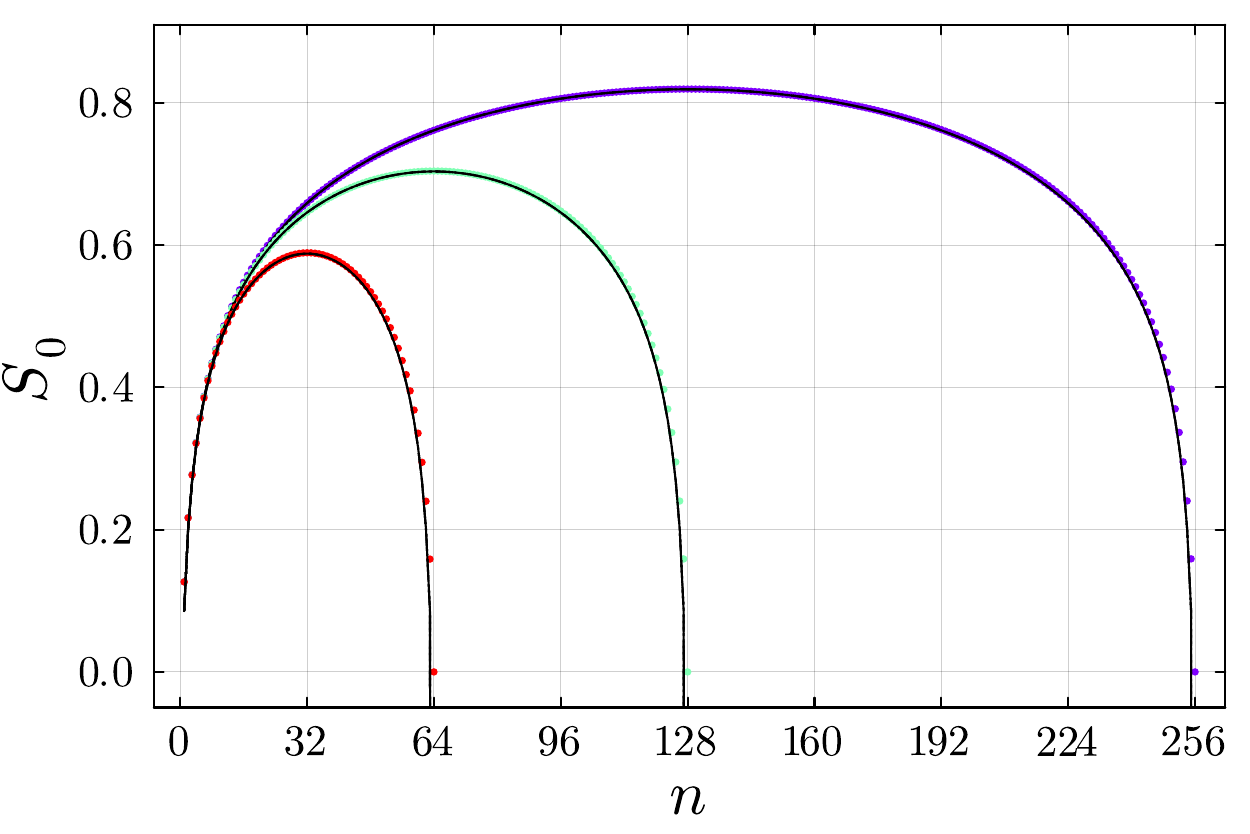}
  \includegraphics[width=0.47\columnwidth]{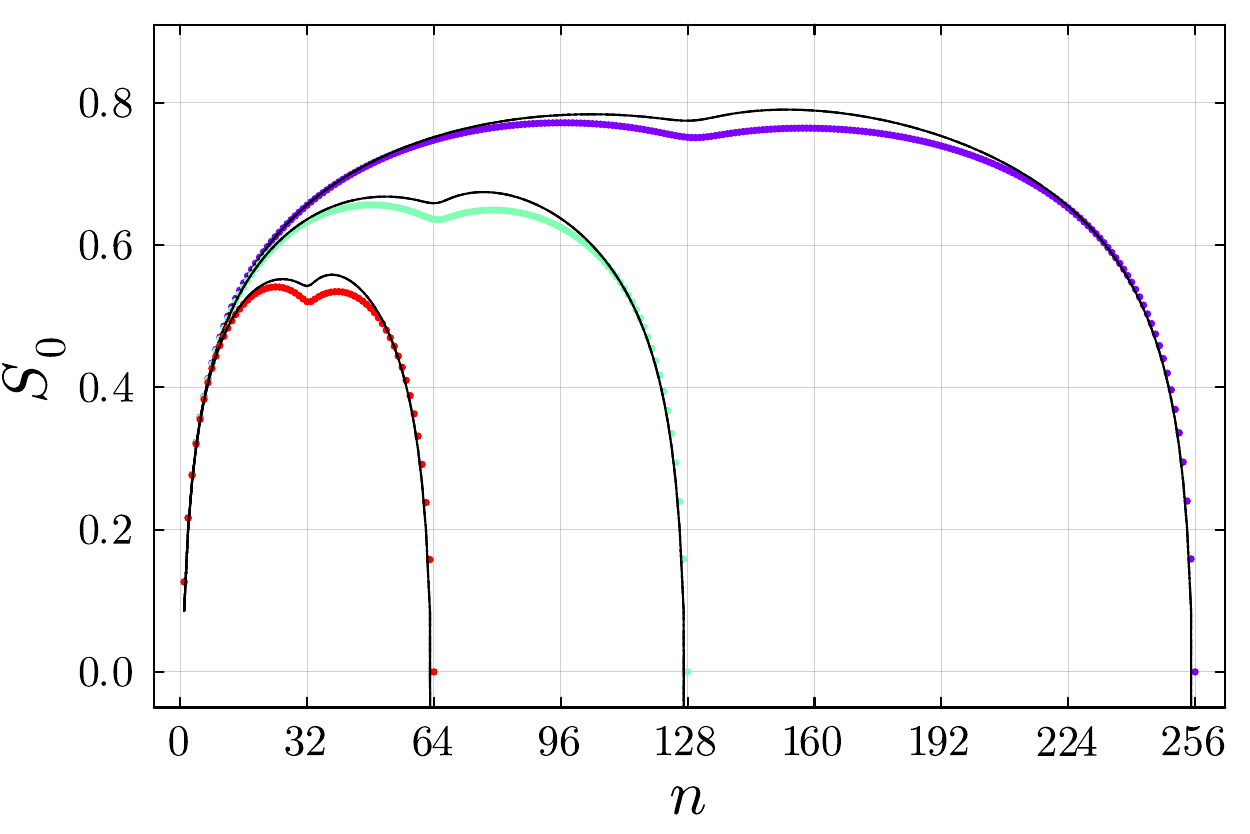}
  \caption{\label{inisl0}The $l = 0$ mode contribution to the entanglement
  entropy vs.~radial coordinate at initial time $t = 0$ for our usual initial
  condition that is peaked at the boundary (left panel) and an alternative
  initial state that is peaked at radial coordinate $L / 2$ (right panel). Red, green and
  violet dots are numerical data for $N_r = 64, 128$ and $256$ respectively,
  while solid black lines represent the approximation \eqref{hocdynapprox}.}
\end{figure}

We would like to find a heuristic approximation to the $l = 0$ component of
the dynamical entropy, which is based on the harmonic oscillator
approximation. In the interacting case, we can write the $l = 0$ part of the
interacting Hamiltonian \eqref{hamori} as
\[ \mathcal{H}_{l = 0} = \frac{1}{2} ((q \phi)^2 + \Pi^2) \]
with the operator $q$ given by \eqref{qop} and the fields and momenta given by
the canonical transformation
\[ \phi = \sqrt{\frac{a}{\alpha}} r \overline{\phi} =
   \sqrt{\frac{r}{\hat{\alpha} d}} \overline{\phi} \qquad \psi =
   \sqrt{\frac{\alpha}{a}} \frac{1}{r} \overline{\psi} =
   \sqrt{\frac{\hat{\alpha} d}{r^3}} \overline{\psi} \;.\]
Defining an effective coordinate
\[ x_r = \bigintlim_0^r \mathd \rho \, \frac{\rho}{d (\rho) \hat{\alpha} (\rho)}\;,
\]
we find that the operator $q$ may be written as
\[ q = \partial_x + m_{\tmop{eff}} \qquad m_{\tmop{eff}} = -
   \frac{\hat{\alpha} d}{r^2} + \partial_r \left( \frac{\hat{\alpha} d}{r}
   \right) \]
with a position dependent effective mass term $m_{\tmop{eff}}$. Neglecting
$m_{\tmop{eff}}$, which is not a good approximation in the generic case, we
find
\[ \mathcal{H}_{l = 0} \sim \frac{1}{2} ((\partial_x \phi)^2 + \Pi^2) \;.\]
In this approximation, the vacuum state entanglement entropy
\begin{equation}
  S_{l = 0} \sim \frac{1}{6} \ln \left( \frac{2 X_r}{\pi \varepsilon_x'} \sin
  \left( \frac{\pi x_r}{X_r} \right) \right) \label{hocdynapprox}
\end{equation}
follows from the harmonic oscillator chain approximation \eqref{l0e} where $X$
is the length of the system in effective coordinates and
\[ \varepsilon_x' = \varepsilon' \frac{\mathd x}{\mathd r} =
   \frac{\varepsilon' r}{d \hat{\alpha}} \]
is the UV cutoff in the effective coordinate. As displayed in Fig.~\ref{inisl0}, this approximation is reasonably good for our particular initial
state that is peaked at the boundary, while for an initial state that is
peaked in the middle of the region the correction terms are sizable.

\begin{figure*}[htb]\centering
  \includegraphics[width=0.47\columnwidth]{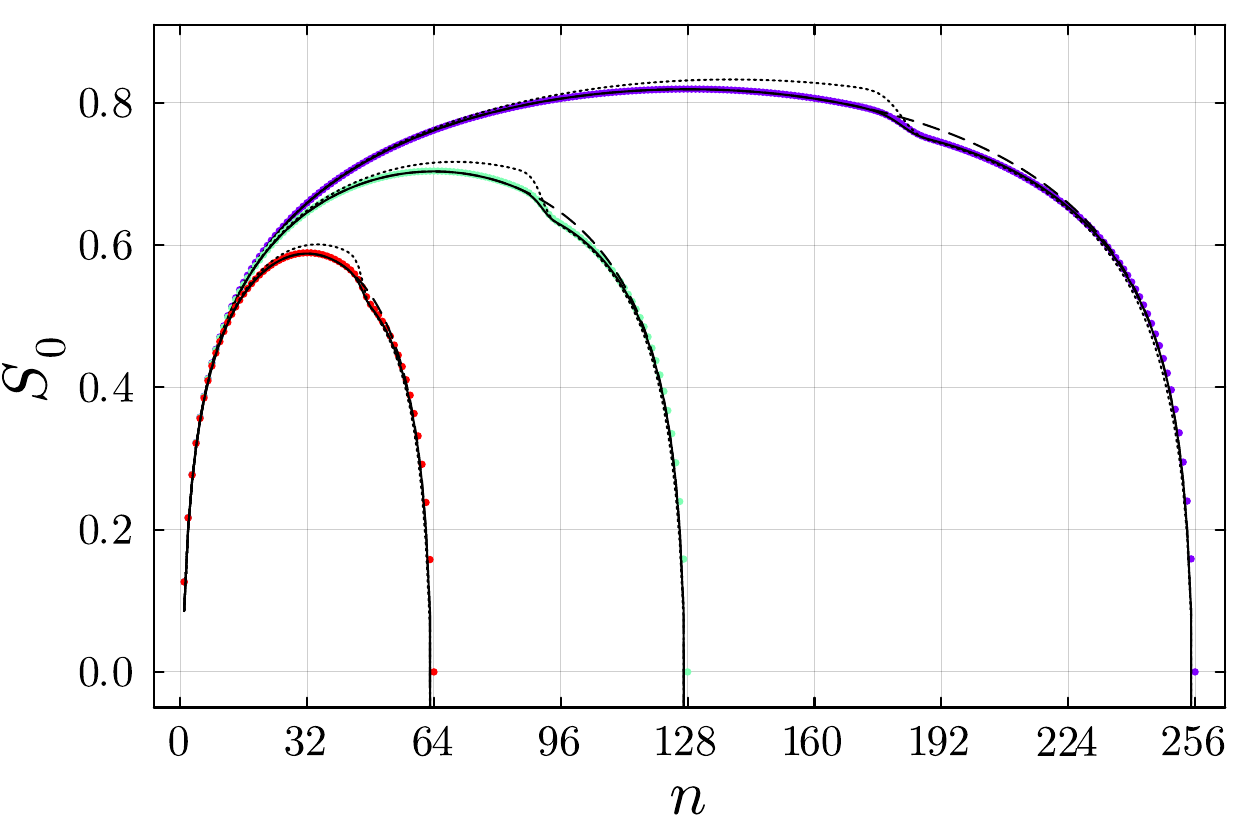}
  \includegraphics[width=0.47\columnwidth]{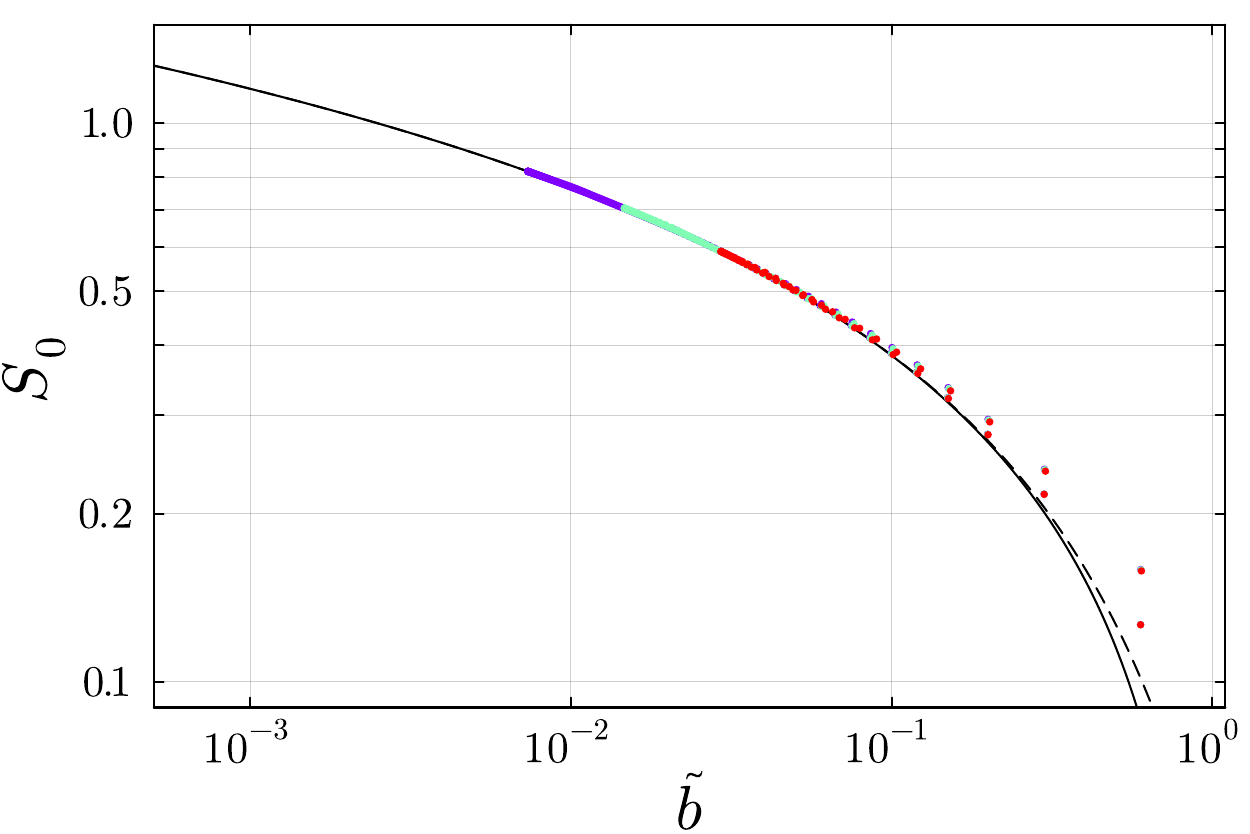}
  
  \includegraphics[width=0.47\columnwidth]{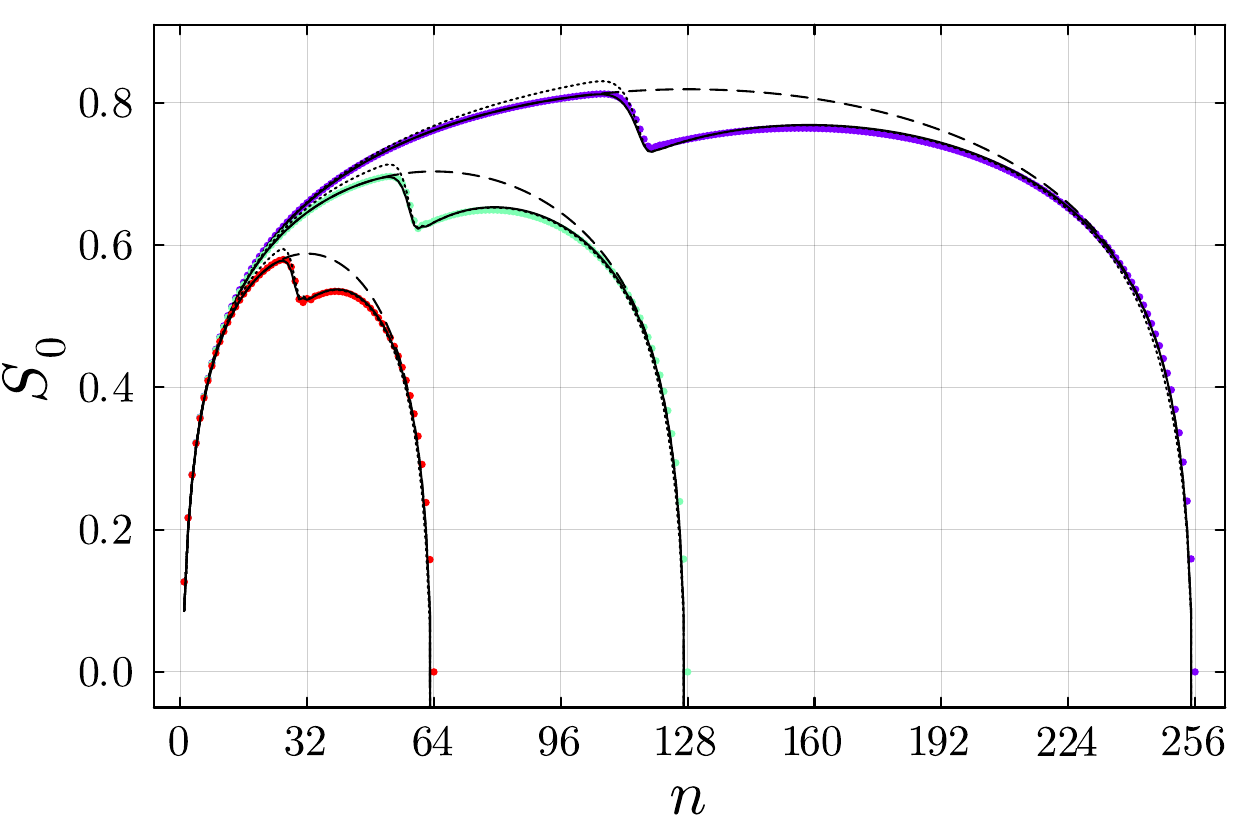}
  \includegraphics[width=0.47\columnwidth]{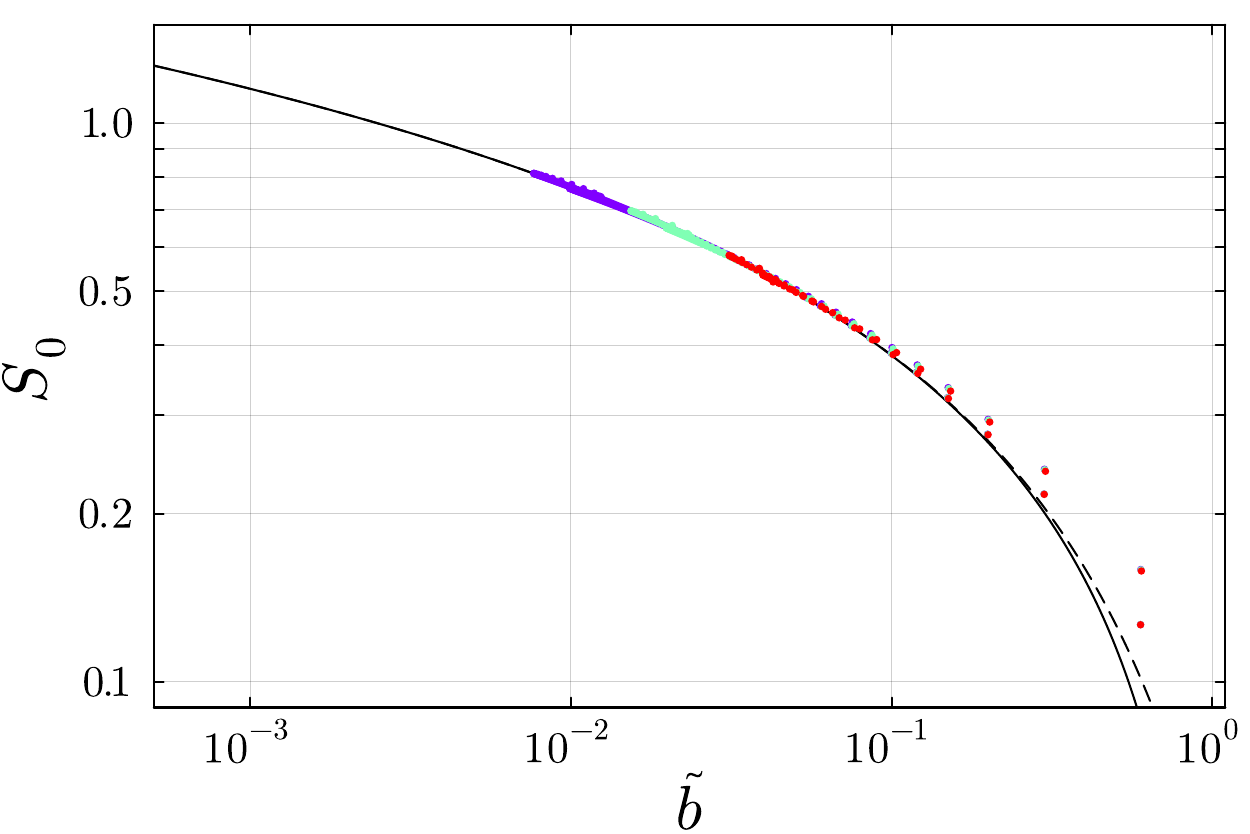}
  \caption{\label{l0dynbw}The $l = 0$ mode contribution to the entanglement
  entropy vs. radial coordinate (left panels) and the dynamical boundary
  correction term \eqref{btildyn} (right panels) at simulation times $t = 5$
  (upper panels) and $t = 12$ (lower panels). Red, green and violet dots are
  numerical data for $N_r = 64, 128$ and $256$ respectively. In the left
  panels, dashed and dotted lines represent the inside \eqref{hocdynapprox}
  and outside \eqref{hocout} approximations respectively, while the solid
  lines correspond to the interpolation \eqref{hocint}. In the right panels,
  the dashed and solid lines represent the interpolation \eqref{masterformraw}
  with $\mu = 0$ and the prediction \eqref{sourunits} where in both cases
  $\tilde{b} $ is given by \eqref{btildyn}.}
\end{figure*}

At later times the harmonic oscillator chain approximation does no longer
apply since the effective one dimensional system is no more in a vacuum state.
It is thus quite remarkable that one may find a phenomenological description
which still captures the qualitative behavior of the $l = 0$ mode entropy, at
least for the particular setup that we are studying. To find this
approximation, we first note that the entanglement entropy can be calculated
from the inner or outer regions alone. Using the inner definition $S_-$, we
expect the inside region that has not yet been reached by the inmoving bump to
be unaltered and thus still described by \eqref{hocdynapprox} with the initial
metric parameters. As can be seen from the left panels of Fig.~\ref{l0dynbw},
this is indeed the case. On the other hand, for coordinates outside of the
bump region our numerical data suggest that after a proper rescaling of the
radial distance by the appropriate metric parameter $a = \sqrt{r / d}$ the
entropy is again approximated well by the free case. If this is to hold for
arbitrary anisotropy factors $c$, it needs to hold on a mode by mode basis. We
thus define the rescaled effective coordinate $y$ via
\[ \mathd y = \frac{a}{a_0} \mathd x = \sqrt{\frac{d_0}{d}} \mathd x =
   \frac{r}{\sqrt{d_0 d} \hat{\alpha}_0} \mathd r\;, \]
where the subscript zero on the metric parameters indicates their values at
the initial time. Of course the effective coordinate was matched at the
initial time, so
\[ x_r = \bigintlim_0^r \mathd \hat{r} \, \frac{\hat{r}}{d_0 (\hat{r})
   \hat{\alpha}_0 (\hat{r})} \]
and the UV cutoff also needs to be appropriately rescaled as
\[ \varepsilon_y' = \varepsilon' \frac{\mathd y}{\mathd r} =
   \frac{\varepsilon' r}{\sqrt{d_0 d} \hat{\alpha}_0} \;.\]
The rescaled distance at radial coordinate $r$ from the outer boundary is then
given by
\[ y_r = \bigintlim_r^L \mathd \hat{r} \, \frac{\hat{r}}{\sqrt{d_0 (\hat{r}) d
   (\hat{r})} \hat{\alpha}_0 (\hat{r})} \]
and we define the perceived rescaled system size
\[ Y_r = x_r + y_r \;,\]
where only the distances outside the reference radial coordinate $r$ are
rescaled. Since the outside definition $S_+$ of the entropy at the radial
coordinate $r$ is only aware of the rescaling in the outside region, we expect
this to be a good estimate of the effective length of the harmonic oscillator
chain that describes the outside region where the bump has already passed. As
can be seen in the left panels of Fig.~\ref{l0dynbw}, the corresponding
expression
\begin{equation}
  S_{l = 0} \sim \frac{1}{6} \ln \left( \frac{2 Y_r}{\pi \varepsilon_y'} \sin
  \left( \frac{\pi y_r}{Y_r} \right) \right) \label{hocout}
\end{equation}
provides a good approximation in the outside region.

In order to find an
approximation that is also valid in the bump region, we note that
\eqref{hocout} may be written as
\[ S_{l = 0} \sim \frac{1}{6} \ln \left( \frac{2 Y_r}{\pi \varepsilon_y'} \sin
   \left( \frac{\pi x_r}{Y_r} \right) \right) \]
with the effective coordinate $x$ of the inside region. We cannot, however,
interpret this as an inside entropy definition $S_-$, since the effective
system length $Y$ still contains information from the outside region. This may
be remedied by defining an effective outer coordinate
\[ z_r = \bigintlim_r^L \mathd \hat{r} \, \frac{\hat{r}}{\sqrt{d_0 (\hat{r}) d_r
   (\hat{r})} \hat{\alpha}_0 (\hat{r})} \;,\]
where
\[ d_r (\hat{r}) = d (r) + \hat{r} - r \]
is the effective metric parameter at radial coordinate $\hat{r}$ if no source
of curvature occurs between radial coordinates $r$ and $\hat{r}$. The
resulting effective system size
\[ Z_r = x_r + z_r \]
and the corresponding entropy
\begin{equation}
  S_{l = 0} \sim \frac{1}{6} \ln \left( \frac{2 Z_r}{\pi \varepsilon_y'} \sin
  \left( \frac{\pi x_r}{Z_r} \right) \right) \label{hocint}
\end{equation}
includes information from the inside region only. Assuming that we have a
localized bump region between effectively free inside and outside regions,
this expression is an interpolation between the inside \eqref{hocdynapprox}
and outside \eqref{hocout} expressions, as can be seen from the left panels of
Fig.~\ref{l0dynbw}. Given the rather accurate description of the data by \eqref{hocint}, it seems
reasonable to propose an effective boundary mass term
\begin{equation}
  \tilde{b} = \left( \frac{2 Z_r}{\pi \varepsilon_y'} \sin \left( \frac{\pi
  x_r}{Z_r} \right) \right)^{- 1} \;.\label{btildyn}
\end{equation}

\section{The log-polynomial renormalization scheme}\label{App:renormalizationSE}

The problem of defining a consistent and covariant renormalization
prescription for the energy-momentum tensor in quantum field theory on curved
spacetime (QFTCS) is a long-standing one (for a nice summary, see for instance
\cite{Fulling_1989,Wald:1995yp,Birrell:1982ix,Parker_Toms_2009}).

The difficulty of defining $\langle T_{\mu \nu} \rangle$ arises from the fact
that it involves expectation values of local quadratic field operators, which
are formally divergent. In flat spacetime, these divergences can be removed by
subtracting the vacuum expectation value (for free theories or certain classes
of interacting QFTs, normal ordering is enough), but such a preferred vacuum
does not exist in a generic curved background. The notion of ``vacuum''
becomes observer-dependent and, in general, cannot be defined globally.

Moreover, while in flat spacetime only differences in energy are physically
observable, once backreaction effects are included, this approach fails: the
energy-momentum tensor acts as a source of curvature through Einstein's
equations, and therefore the absolute value of $\langle T_{\mu \nu} \rangle$
acquires physical meaning. As a consequence, the divergent terms cannot simply
be ignored, and a renormalization prescription that is both covariant and
consistent with the local dynamics of the gravitational field is required.

These divergences are ultraviolet in origin and therefore arise from the
behavior of correlation functions at arbitrarily short spacetime separations.
A crucial result of quantum field theory in curved spacetime is that these
short-distance singularities are universal: they depend only on the local
geometry and not on the quantum state. This universal short-distance structure
motivates the point-splitting (or ``local normal ordering'') procedure, first
introduced by DeWitt \cite{DeWitt:1975ys} and further developed by
Christensen and Wald
\cite{PhysRevD.14.2490,PhysRevD.17.946,Wald:1978ce,PhysRevD.17.1477}. The
basic idea is to define the two-point function $G^{(1)} (x, x')$ with the field arguments
slightly separated to isolate and then subtract the purely geometric singular
term before taking the coincidence limit:
\[ \langle T_{\mu \nu} (x)_{\textrm{\tmop{ren}}} \rangle = \lim_{x \rightarrow
   x'} \mathcal{T}_{\mu \nu} [G^{(1)} (x, x') - G_{\textrm{\tmop{sing}}}^{(1)}
   (x, x')] \;.\]
One takes this limit along the (for sufficiently close points) unique geodesic between $x$ and $x'$. Here $\mathcal{T}_{\mu \nu}$ is a suitable
differential operator constructed from the metric, and
$G_{\textrm{\tmop{sing}}}^{(1)}$ contains only the universal Hadamard
singularities. Christensen \cite{PhysRevD.14.2490} demonstrated how this
procedure can be carried out in a manifestly covariant way. In some sense, this can therefore be regarded as a generalization of normal ordering to curved space-time. Instead of subtracting a global Minkowski vacuum contribution, one subtracts the universal short-distance singularity of the two-point function (whose leading behavior is locally Minkowskian by the equivalence principle, with subleading curvature-dependent
corrections). 
These additional terms are needed, because 
the ultraviolet structure in curved spacetime is not exactly that of flat spacetime but includes
universal curvature corrections, i.e.~they only depend on the local geometry and not on the quantum state. Consequently, the divergent part of the
two-point function is entirely fixed by the metric and its derivatives. 
By subtracting these purely geometric singular terms before taking the
coincidence limit one therefore obtains a covariant and local renormalization
prescription.

For vanishing curvature, the leading short-distance behavior reduces to that of
Minkowski spacetime by construction. The subtraction therefore
guarantees that $\langle T_{\mu \nu} \rangle = 0$ in flat spacetime, ensuring
that Minkowski space remains a solution of the semiclassical Einstein
equations.

Fulling, Sweeny and Wald \cite{Fulling:1978ht,Fulling:1981cf} showed that
this universal short-distance structure is precisely captured by the Hadamard
condition, which characterizes physically admissible states in QFTCS by the
singular structure of their two-point functions. A quantum state is of
Hadamard form if its two-point function $G^{(1)} (x, x') = \langle \{\phi (x),
\phi (x')\} \rangle$ has, in four spacetime dimensions, the short-distance
expansion
\[ G^{(1)} (x, x') \sim U (x, x') \sigma (x, x')^{- 1} + V (x, x') \log
   (\sigma (x, x')) + W (x, x') \;,\]
where $\sigma (x, x')$ is Synge's world function, equivalent to half the
squared geodesic distance between $x$ and $x'$, and $U$, $V$, and $W$ are
smooth symmetric biscalars. The functions $U$ and $V$ encode the universal
(state-independent) geometric singularity structure, while $W$ contains the
state-dependent finite part. Importantly, all Hadamard states share the same
singularity structure, and the difference between two such Hadamard states is smooth in the
coincidence limit. In order to renormalize the energy-momentum tensor one therefore subtracts the universal $U$ and $V$ terms before taking the coincidence limit. This yields a finite and locally covariant result.

These general properties that any renormalized energy-momentum tensor should
satisfy were formalized by Wald in a set of axioms
\cite{PhysRevD.17.1477,Wald:1978ce}. These axioms require local covariance,
energy-momentum conservation, and vanishing in Minkowski-space, among other conditions. Wald's analysis shows that any two renormalization schemes
satisfying these requirements can differ only by a local, conserved curvature
term constructed from $g_{\mu \nu}$ and its derivatives up to fourth order. This finite ambiguity is the only freedom left in the definition of $\langle T_{\mu\nu}\rangle$. 

A variety of covariant renormalization prescriptions for the energy-momentum
tensor in curved spacetime have been developed, including point--splitting
\cite{PhysRevD.14.2490,PhysRevD.17.946,Davies:1977ze,PhysRevD.15.2088,Adler:1976jx,Adler:1977ac,PhysRevD.17.1477,Decanini:2005eg},
dimensional regularization
\cite{PhysRevD.15.1469,PhysRevD.15.2810,PhysRevD.18.1844}, $\zeta$-function
regularization \cite{Dowker:1975tf,Hawking:1976ja}, adiabatic regularization
\cite{PhysRevD.10.3905,PhysRevD.18.1844} and Pauli-Villars regularization
\cite{BERNARD1977201,Vilenkin:1978wc}. Although these schemes differ at the technical level, they were shown to remove the same universal Hadamard singularities and therefore yield renormalized energy-momentum tensors that differ at most by local, conserved curvature terms, as required by Wald's
axioms. 

To isolate the ultraviolet divergences of the energy-momentum tensor, we
employ the covariant point-splitting method developed by DeWitt and
Christensen \cite{DeWitt:1975ys,PhysRevD.14.2490,PhysRevD.17.946}. This
allows us to identify the divergent terms in $\langle T_{\mu \nu} \rangle$ and
to design a Pauli-Villars regularization scheme in which a set of massive
regulator fields cancels those divergences as $x \rightarrow x'$. If the limit
$M_i \rightarrow \infty$ is well defined, the cutoff can be removed and a
finite renormalized energy-momentum tensor obtained.

Our discussion remains general, since this renormalization prescription is not
tied to a specific background and can be applied in other contexts. As far as
we are aware, this constitutes a new implementation of energy-momentum tensor
renormalization in curved spacetime. We start from the action of the massless
scalar field, adding $n - 1$ massive Pauli-Villars fields $\chi_i$
\[ \begin{split}
     S [g, \phi, \chi_i] = & - \frac{1}{16\pi} \int \mathrm{d}^4 x \sqrt{- g} 
     \left[ R - g^{\mu \nu} \phi_{, \mu} \phi_{, \nu} + \xi_1 R \phi^2 +
     \sum_{i = 2}^n p_i  (-g^{\mu \nu} \chi_{i, \mu} \chi_{i, \nu} + M_i^2
     \chi_i^2 + \xi_i R \chi_i^2) \right]\\
     = & S^{M_1 = 0} [g, \phi] + \sum^n_{i = 2} p_i S^{M_i \neq 0} [g,
     \chi_i]\;,
   \end{split} \]
where $\phi$ is the scalar field and $\xi$ is the nonminimal coupling (later
set to $\xi = 0$). The multiplicities $p_i$ may be both positive or negative,
hence some regulator fields have opposite sign kinetic and mass terms.

The energy-momentum tensor follows from the metric variation, and because the
physical and regulator fields $\phi$ and $\chi_i$ do not interact with each
other, it decomposes into a sum of individual contributions
\[ T_{\mu \nu} = T_{\mu \nu}^{M_1 = 0} + \sum_{i = 2}^n p_i T_{\mu \nu}^{M_i
   \neq 0} \;.\]

Our strategy will be to identify the UV divergences of $T_{\mu \nu}$ by using
Christensen's covariant point splitting strategy based on DeWitt's proper time
formalism. The full derivation and explicit expressions can be found in
\cite{PhysRevD.14.2490,PhysRevD.17.946}, here we summarize the essential
steps needed to construct the PV cancellation scheme.

The goal is to determine the conditions on the coefficients $p_i$ and masses
$M_i$ that ensure the cancellation of all divergent terms as $x' \rightarrow
x$ arising for the physical massless field. We first take the coincidence
limit $x \rightarrow x'$ to obtain a finite result, and then remove the PV
cutoff $M_i \rightarrow \infty$, which turns out to be well-defined, too.

We follow Christensen's derivation and work directly with the one-loop
effective action. For a free scalar field, the path integral over the field is
Gaussian, and integrating it out yields
\[ W_{\textrm{\tmop{eff}}} [g] = - \frac{i}{2} \textrm{\tmop{Tr}} \log (- G_F
   [g, m]) \;,\]
where $G_F$ is the Feynman propagator in the background metric $g_{\mu \nu}$.
The proper-time representation of $W_{\textrm{\tmop{eff}}}$ involves exactly
the same local geometric coefficients $a_n $that appear in the Hadamard
expansion. In this form,
\[ W_{\textrm{\tmop{eff}}} (g) = - \lim_{x \rightarrow x'} \frac{\Delta^{1 /
   2}}{32 \pi^2} \int \mathrm{d}^4 x \, \sqrt{- g} \int_{0 }^{\infty}
   \frac{\mathrm{d} s}{s^3} \exp \left[ - i \left( m^2 s - \frac{\sigma}{2 s}
   \right) \right] \sum_{n = 0}^{\infty} a_n (x, x') (i s)^n \;,\]
where $\Delta (x, x')$ is the Van-Vleck-Morette determinant. Crucially, all
ultraviolet divergences of the theory arise from the small $s$ region of the
integral, and hence are determined solely by the local geometric coefficients
$a_0, a_1, a_2$. This is the same set of coefficients that appears
in the Hadamard parametrix for the two-point function. Therefore the divergent
part of the effective action is a local functional of the metric built from
these curvature invariants.

Since the expectation value of the energy-momentum tensor is obtained by the functional
derivative
\[ \langle T_{\mu \nu} \rangle = \frac{2}{\sqrt{- g}} \frac{\delta
   W_{\textrm{\tmop{eff}}}}{\delta g^{\mu \nu}} \]
the divergences in $\langle T_{\mu \nu} \rangle$ arise precisely from the
functional variation of these same local terms. Thus, subtracting the
divergent part of $W_{\textrm{\tmop{eff}}}$ removes all divergences in the
energy-momentum tensor as well. This is the content of Christensen's analysis and
underlies the equivalence between the effective-action and Hadamard
renormalization schemes.

For a general massive scalar field, Christensen's analysis shows that the
divergent pieces of $\langle T_{\mu \nu} \rangle$ organize into quartic,
quadratic, and logarithmic terms in the geodesic separation $\sigma$. In our
case, this results in a number of cancellation conditions on the
multiplicities of the Pauli-Villars fields, $p_i$, and their mass ratios $M_i^2
= r_i \Lambda^2$, where $\Lambda$ is a common UV cutoff with dimensions of mass.
Here we sketch the form of the divergences, their explicit expressions can be
found in \cite{PhysRevD.17.946}, Eq.~(5.6-5.8) and (6.2). The quartic
divergence,
\[ \langle T^{\mu \nu} \rangle_{\textrm{\tmop{quartic}}} = \frac{\sqrt{- g}}{2
   \pi^2} \frac{1}{(\sigma_{\rho} \sigma^{\rho})^2} \left[ g^{\mu \nu} - 4
   \frac{\sigma^{\mu} \sigma^{\nu}}{\sigma_{\rho} \sigma^{\rho}} \right] \]
is identical for massive and massless fields, leading to the condition
\begin{equation}
  \sum^n_{i = 2} p_i = - 1 \;.\label{eq:sumpi}
\end{equation}
There is a quadratic divergence of the schematic form
\[ \langle T^{\mu \nu} \rangle_{\textrm{\tmop{quadratic}}} = \frac{\sqrt{-
   g}}{4 \pi^2} \frac{1}{\sigma} \left( \mathcal{A}_2^{\mu \nu} (g, R) - \frac{1}{2}
   M_i^2 \mathcal{B}_2^{\mu \nu} (g) - \left( \frac{1}{6} - \xi_i \right) \mathcal{C}_2^{\mu \nu}
   (g, R) \right) \]
where $\mathcal{B}_2^{\mu \nu} (g)$ depends only on the metric, while $\mathcal{A}_2^{\mu \nu} (g, R)$ and $\mathcal{C}_2^{\mu \nu}(g,R)$ are curvature dependent. 
To cancel this divergent contribution we hence need the same condition
\eqref{eq:sumpi} (which removes $\mathcal{A}_2^{\mu \nu} (g, R)$) and additionally, for the minimally coupled case $\xi_i = 0$,
\begin{equation}
  \sum^n_{i = 2} p_i M_i^2 = 0 \;, \label{eq:sumMisq}
\end{equation}
which removes the mass term $\mathcal{B}_2^{\mu \nu}$ on its own, independent of $\xi_i$. 
There is also a logarithmic divergence, which in the massless case takes the
form
\[ \langle T^{\mu \nu} \rangle^{M_1 = 0}_{\mathrm{log}} = \left\{
   \mathcal{A}_{\mathrm{log}}^{\mu \nu} (g,R) - \left( \frac{1}{6} - \xi_1 \right)^2
   \mathcal{B}_{\mathrm{log}}^{\mu \nu} (g,R)  \right\} \left[ \gamma + \frac{1}{2} \ln \left|
   \frac{1}{4} \mu^2 \sigma  \right| \right]\;, \] 
where $\mu$ is the (arbitrary) renormalization scale introduced above and $\mathcal{A}_{\mathrm{log}}^{\mu \nu}$, $\mathcal{B}_{\mathrm{log}}^{\mu \nu}$ are curvature-squared tensors independent of $\sigma$ (distinct from $\mathcal{A}_2^{\mu \nu}$, $\mathcal{B}_2^{\mu \nu}$) \footnote{$\mathcal{A}_{\mathrm{log}}^{\mu \nu}$ and $\mathcal{B}_{\mathrm{log}}^{\mu \nu}$ have the same structure as the tensors $H^{(2)\mu\nu}$ and $H^{(1)\mu\nu}$ of Christensen \cite{PhysRevD.17.946}, obtained by varying $\int \sqrt{-g}R_{\mu\nu}R^{\mu\nu}$ and $\int \sqrt{-g}R^2$, respectively.}.
In the massive case, one gets
\[ \langle T^{\mu \nu} \rangle^{M_i \neq 0}_{\mathrm{log}} = \left[
   \mathcal{A}_{\mathrm{log}}^{\mu \nu} (g,R) -\left(\tfrac16 - \xi_i\right)^2\mathcal{B}_{\mathrm{log}}^{\mu \nu} (g,R) +M_{i }^2 \left( \frac{1}{6} - \xi_i \right) 
    \mathcal{C}_{\mathrm{log}}^{\mu \nu} (g,R) + M_i^4 \mathcal{D}^{\mu \nu}(g) \right]
   \left[ \gamma + \frac{1}{2} \ln \left| \frac{1}{4} M_i^2 \sigma \right|
   \right] \;,\]
   where $\mathcal{C}_{\mathrm{log}}^{\mu \nu} (g,R)$ is proportional to the Einstein tensor $G^{\mu\nu}=R^{\mu\nu}-1/2Rg^{\mu\nu}$. 
For the minimally coupled case, we therefore impose
\begin{equation}
  \sum_{i = 2}^n p_i \ln \left| \frac{1}{4} M_i^2 \sigma\right| + \ln
  \left| \frac{1}{4} \mu^2 \sigma \right| = 0 \label{eq:sumpilog} 
\end{equation}
together with condition \eqref{eq:sumpi} to cancel the divergences that are
present both in the massless and in the massive case (the $\mathcal{A}_{\mathrm{log}}^{\mu\nu}$ piece), and
\begin{equation}
  \sum_{i = 2}^n p_i M_i^2 \ln | M_i^2 \sigma| = 0 \qquad \sum_{i = 2}^n p_i M_i^4
  \ln | M_i^2 \sigma| = 0 \qquad \sum_{i = 2}^n p_i M_i^4 = 0 \label{eq:logcond}
\end{equation} 
in order to cancel the divergences only present in the massive case, together
with condition \eqref{eq:sumMisq}. These conditions remove both the universal
geometric logarithm and the new mass-dependent logarithms generated by the
massive PV fields, while terms suppressed by $\sigma$ and $M_i^{-
2}$ or higher vanish automatically as $\sigma \rightarrow 0$ and $M_i
\rightarrow \infty$. Thus we expect the full set of conditions \eqref{eq:sumpi}-\eqref{eq:logcond} to guarantee a finite coincidence limit $\sigma \rightarrow 0$ as well as a regulator-independent renormalized result in the limit $M_i \rightarrow \infty$.

In summary, the log-polynomial Pauli-Villars scheme removes all short-distance divergences
of the point-split energy-momentum tensor by imposing a set of algebraic relations
among the regulator multiplicities and masses. What is important to note is that
the full cancellation of divergences requires not only the usual power-law
constraints $\sum_i p_iM_i^{0, 2, 4} = 0$, but also the logarithmic and mixed
log-power constraints which we have summarized above. 
These conditions are rarely made explicit in the literature. 
Since the Pauli-Villars constraints mix polynomial and logarithmic dependencies on the regulator masses, the system is transcendental and cannot be solved in closed analytic form. In practice, one must determine a set of suitable regulator masses numerically, but once such a set is found the scheme consistently cancels all UV divergences identically in the coincidence limit, and once the limit $M_i\rightarrow \infty$ is taken, the renormalized tensor is independent of the details of the regulator fields. 

Finally, let us comment on a subtlety in the cancellation of the logarithmic
divergence. Introducing the parameterization $M_{i }^2 = r_i \Lambda^2$ with
fixed coefficients $r_i$, the unphysical regulator fields are removed by
sending the Pauli-Villars cutoff $\Lambda \rightarrow \infty$. Condition
\eqref{eq:sumpilog} then becomes
\[ \sum_{i = 2}^n p_i \ln | r_i | + \ln \left| \frac{\mu^2}{\Lambda^2} \right|
   = 0\;. \]
Only if we choose to identify the renormalization scale $\mu$ with the Pauli-Villars
cutoff scale $\Lambda$ does this condition reduce to
\[ \sum_{i = 2}^n p_i \ln | r_i | = 0\;. \]

We leave explicit verification of Wald's axioms \cite{PhysRevD.17.1477,Wald:1978ce} for future work, but we give some arguments below. First, the renormalized energy-momentum tensor is covariant even at finite $\Lambda$, since the Pauli-Villars fields are covariantly coupled scalar fields and the energy-momentum tensor is simply a sum of these terms. Since all $\langle T^{M_i}_{\mu\nu}\rangle$ are separately conserved on shell and by construction the sum is finite in the coincidence limit, we have $\langle T^{\mu\nu}\rangle_{;\mu}=0$. Furthermore,  since in Minkowski vacuum $\Lambda$ is the only relevant, independent scale, by dimensional analysis, 
the point-split energy-momentum tensor contains only terms that are canceled by the log-polynomial Pauli-Villars conditions.
Wald's theorem would then state that different sets $\{p_i,r_i\}$ differ at most by a local conserved curvature tensor, which is fixed by the condition $\sum_i p_i \ln r_i=0$ together with the identification $\mu=\Lambda$ that we discussed above. 
\bibliography{references}

\end{document}